\documentclass[a4paper,fleqn,10pt]{article}
\pdfoutput=1
\usepackage{amsmath}
\usepackage{mathtools}
\usepackage{cancel}
\usepackage{amssymb}
\usepackage{bbm}
\usepackage{array}
\usepackage{calc}
\usepackage{longtable}
\usepackage{multirow}
\usepackage{slashed}
\usepackage{pstricks}
\usepackage{graphicx}
\usepackage{xspace}
\usepackage{units}
\usepackage{tikz}
\usepackage{textcomp}

\numberwithin{equation}{section}
\usepackage{cite}
\usepackage[pdfborder={0 0 0}]{hyperref}
\usepackage[format=hang,labelfont=bf,hypcap=true]{caption}
\usepackage{subcaption}
\usepackage{sectsty}
\usepackage{enumitem}
\allsectionsfont{\sffamily}
\subsubsectionfont{\mdseries\itshape\large}
\makeatletter
\DeclareRobustCommand*{\bfseries}{%
  \not@math@alphabet\bfseries\mathbf
  \fontseries\bfdefault\selectfont
  \boldmath
}
\makeatother
\let\spreprint\empty
\newcommand{\preprint}[1]{\def\spreprint{\protect#1}}
\let\sinstitute\empty
\newcommand{\institute}[1]{\def\sinstitute{\protect#1}}
\makeatletter
\renewcommand{\maketitle}{\begingroup
  \null\thispagestyle{empty}%
    \ifx\spreprint\empty
      \vskip 5ex
    \else
      \flushright\large\spreprint\vskip 10ex
    \fi
    \vskip 5ex
    \flushleft
      {\sffamily\bfseries\huge\@title}\vskip 6ex
      \@author\vskip 2ex
      \ifx\sinstitute\empty
      \else
        {\small\sinstitute}
      \fi
    \vskip 5ex
  \endgroup
}
\makeatother
\renewenvironment{abstract}{\begin{center}
  {\large\sffamily\bfseries Abstract: }
  \begin{minipage}[t]{0.75\textwidth}
}{\end{minipage}\end{center}\vskip 10ex}

\numberwithin{equation}{section}
\allowdisplaybreaks[2]

\newcommand{\LHAPDF}{L\protect\scalebox{0.8}{HAPDF}\xspace}

\newcommand{\Rivet}{R\protect\scalebox{0.8}{IVET}\xspace}

\newcommand{\Sherpa}{S\protect\scalebox{0.8}{HERPA}\xspace}

\long\def\symbolfootnote[#1]#2{\begingroup%
\def\thefootnote{\fnsymbol{footnote}}\footnote[#1]{#2}\endgroup}

\newcommand{\done}{{\rm d}}
\newcommand{\order}{\mathcal{O}}

\newcommand{\nnb}{\nonumber}
\newcommand{\bea}{\begin{eqnarray}}
\newcommand{\eea}{\end{eqnarray}}
\newcommand{\bi}{\begin{itemize}}
\newcommand{\ei}{\end{itemize}}

\newcommand{\nbar}{{\ensuremath{\bar{n}}}}

\newcommand{\qT}{\ensuremath{q_{\mathrm{T}}}}
\newcommand{\qTvec}{\ensuremath{\vec{q}_{\mathrm{T}}}}

\newcommand{\mT}{\ensuremath{m_\mathrm{T}}}

\newcommand{\kTvec}{\ensuremath{\vec{k}_{\mathrm{T}}}}

\newcommand{\NLO}{\ensuremath{\text{NLO}}\xspace}
\newcommand{\NNLO}{\ensuremath{\text{N$^2$LO}}\xspace}
\newcommand{\NNNLO}{\ensuremath{\text{N$^3$LO}}\xspace}
\newcommand{\NNNNLO}{\ensuremath{\text{N$^4$LO}}\xspace}

\newcommand{\LL}{\ensuremath{\text{LL}}\xspace}

\newcommand{\NNNLLp}{\ensuremath{\text{N$^3$LL$'$}}\xspace}
\newcommand{\NNNNLL}{\ensuremath{\text{N$^4$LL}}\xspace}

\newcommand{\LP}{\ensuremath{\text{LP}}\xspace}
\newcommand{\NLP}{\ensuremath{\text{NLP}}\xspace}
\newcommand{\NNLP}{\ensuremath{\text{N$^2$LP}}\xspace}
\newcommand{\NNNLP}{\ensuremath{\text{N$^3$LP}}\xspace}
\newcommand{\FSmeth}{\ensuremath{\langle\mathrm{FS}\rangle}\xspace}
\newcommand{\NSmeth}{\ensuremath{\langle\mathrm{NS}\rangle}\xspace}

\newlist{myitemize}{itemize}{3}
\setlist[myitemize]{leftmargin=14em}

\newcolumntype{C}{>{\centering\arraybackslash}p{0.14\textwidth}}

\newlength{\unitcharwidth}
 \newcommand{\changed}[1]{#1}

\hypersetup{
  pdfauthor={Wan-Li Ju, Marek Schoenherr},
  pdftitle={Zero-bin subtraction and the qT spectrum beyond leading power}
}
\preprint{IPPP/23/83\\MCnet/23/21}
\title{Zero-bin subtraction and the $q_{\mathrm{T}}$ spectrum beyond leading power}
\author{Giancarlo Ferrera${}^{(a,b)}$, Wan-Li Ju${}^{(b)}$, Marek Sch{\"o}nherr${}^{(c)}$} 
\institute{
${}^{(a)}$~Dipartimento di Fisica, Universit\`a di Milano, Via Celoria 16, 20133 Milan, Italy\\
${}^{(b)}$~INFN, Sezione di Milano, Via Celoria 16, 20133 Milan, Italy\\
${}^{(c)}$~Institute for Particle Physics Phenomenology, Durham University, Durham DH1 3LE, United Kingdom\\
$~$\\
Emails: \href{ }{\color{black} giancarlo.ferrera@mi.infn.it}\,, \href{ }{\color{black} wanli.ju@mi.infn.it}\,, \href{ }{\color{black} marek.schoenherr@durham.ac.uk} }
\begin{document}
\vspace*{10mm}
\maketitle
\vspace*{20mm}
\begin{abstract}  
  In this paper, we present an algorithm to construct the \qT\
  distribution at \NLO accuracy to arbitrary power precision,
  including the assembly of suitable zero-bin subtrahends,
  in a mathematically well-defined way for a generic choice
  of rapidity-divergence regularisation prescription.
  In its derivation, we divide the phase space into two sectors,
  the interior of the integration domain as well as the integration
  boundary, which we include here for the first time.
  To demonstrate the applicability and usefulness of our algorithm,
  we calculate the \NNLP corrections for Higgs
  hadroproduction for the first time.
  We observe that our approximate \NNLP-accurate \qT~spectra replicate
  the asymptotic behaviour of the full QCD calculation to a much better degree
  than the previously available results, both within the $\qT\to 0$ limit
  and in the large-\qT\ domain for all the involved partonic processes.
  While playing a minor role at larger transverse momenta, we show that
  the newly incorporated boundary contribution  plays a vital role
  in the $\qT\to 0$ limit, where any subleading power accuracy would be
  lost without them.
  In particular, our \NNLP-accurate \qT\ expansion can approximate the
  exact \qT\ distribution up to $\qT\approx 30\,\text{GeV}$ at the
  percent level for rapidities  $|Y_H|\lesssim 3$.
\end{abstract}
\newpage
\tableofcontents
\section{Introduction}
\label{sec:intro}
    
Differential observables of the Drell-Yan processes and
Higgs hadroproduction play paramount roles in precisely
determining the Standard Model (SM) input parameters and
probing New Physics scenarios.
They have therefore drawn extensive experimental and
theoretical attention in the recent decades.
The latest measurements of the Drell-Yan processes have
been carried out by the ATLAS~\cite{ATLAS:2015iiu,ATLAS:2017rzl,
  ATLAS:2019zci,ATLAS:2023lsr}, CMS~\cite{CMS:2015hyl,CMS:2018ktx,
  CMS:2018mdl,CMS:2019raw,CMS:2020cph}~and~LHCb~\cite{LHCb:2021bjt}
collaborations at the LHC at colliding energies of $\sqrt{s}=7$, $8$,
and $13\,\text{TeV}$, as well as the CDF and D\O\ collaborations at
$\sqrt{s}=1.96~$TeV at the Tevatron~\cite{D0:2012kms,CDF:2012gpf,
  CDF:2013dpa,CDF:2022hxs}.
Inclusive and differential fiducial cross section measurements
for Higgs boson production, on the other hand, have been presented
in \cite{ATLAS:2014xzb,CMS:2015zpx,ATLAS:2017qey,CMS:2018gwt,
  ATLAS:2020wny,ATLAS:2022fnp,ATLAS:2022qef,ATLAS:2023hyd,ATLAS:2023pwa,
  CMS:2023gjz}.
Along with the progress in experimental precision, strides were also
made in the fixed-order calculations.
The QCD corrections are known up to third order for both the Drell-Yan
processes \cite{Melnikov:2006kv,Catani:2009sm,Camarda:2019zyx,Duhr:2020seh,
  Duhr:2021vwj,Chen:2021vtu,Chen:2022cgv,Neumann:2022lft,Baglio:2022wzu,
  Gehrmann-DeRidder:2023urf} and Higgs production \cite{Dawson:1990zj,
   Djouadi:1991tka,Spira:1995rr,Harlander:2002wh,Anastasiou:2002yz,
   Ravindran:2003um,Anastasiou:2015vya,Anastasiou:2016cez,
   Mistlberger:2018etf,Chen:2014gva,Boughezal:2015aha,Boughezal:2015dra,
   Chen:2016zka,Caola:2015wna,Cieri:2018oms,Billis:2021ecs,Chen:2021azt,
   Chen:2021isd,Cal:2023mib}.
Electroweak corrections have mostly been studied for Drell-Yan production
\cite{Li:2012wna,Dittmaier:2015rxo,Bonciani:2019nuy,Dittmaier:2020vra,
  Bonciani:2021zzf,Armadillo:2022bgm,Buccioni:2022kgy}, but are known for
Higgs production as well \cite{Actis:2008ug,Anastasiou:2008tj}.

Of the differential observables, the transverse momentum distribution
of the colourless final state is of the particular concern in this work.
Even though fixed order calculations are able to produce reliable
predictions in the majority of the phase space, substantial corrections
can emerge from higher perturbative order within the low~\qT~regime,
as a result of soft and collinear radiation,
where the bulk of the cross section resides.
Hence, in order to arrive at a sufficiently convergent result in
perturbation theory, a variety of the resummation techniques that are
capable of exponentiating \textit{the most singular}, so-called
leading power (\LP), behaviour of
the~\qT~distribution at every order have been developed in the past
decades.
Examples are the CSS formalism~\cite{Collins:1981uk,Collins:1981va,
  Collins:1984kg,Catani:2000vq,Bozzi:2005wk,Bozzi:2007pn}, the momentum
space resummation~\cite{Ebert:2016gcn,Monni:2016ktx,Bizon:2017rah,
  Bizon:2019zgf,Bizon:2018foh}, and the SCET-based analyses
\cite{Becher:2010tm,GarciaEchevarria:2011rb,Becher:2011dz,Chiu:2011qc,
  Chiu:2012ir,Li:2016axz,Li:2016ctv}.
The subsequent resummation improved \qT~distributions have been evaluated
up to approximate \NNNNLL accuracy \cite{Becher:2010tm,Bozzi:2010xn,
  Becher:2011xn,Banfi:2011dx,Banfi:2011dm,Banfi:2012du,Catani:2015vma,
  Scimemi:2017etj,Bizon:2018foh,Bacchetta:2019sam,Bizon:2019zgf,
  Becher:2020ugp,Ebert:2020dfc,Re:2021con,Camarda:2021ict,Ju:2021lah,
  Camarda:2023dqn,Neumann:2022lft,Moos:2023yfa} for the Drell-Yan process
and \NNNLLp in the case of Higgs production \cite{Belyaev:2005bs,
  Bozzi:2005wk,Becher:2020ugp,Becher:2012yn,Neill:2015roa,Bizon:2017rah,
  Chen:2018pzu,Bizon:2018foh,Gutierrez-Reyes:2019rug,Harlander:2014hya,
  Billis:2021ecs,Cal:2023mib} as to the Higgs production.
Recently, to further improve the precision as well as the phase space
coverage of the \qT~resummation, the factorisation pattern of
\textit{the subleading power} contribution has been investigated
within both the non-perturbative region $\Lambda_{H}\lesssim\qT\ll Q_H$ \cite{Balitsky:2017gis,Balitsky:2020jzt,Balitsky:2021fer,
  Balitsky:2017flc,Vladimirov:2021hdn,Ebert:2021jhy,Gamberg:2022lju,
  Rodini:2022wic,Rodini:2023plb,Vladimirov:2023aot,Rodini:2023mnh} and
the perturbative zone $\Lambda_{H}\ll\qT\ll Q_H$ \cite{Ebert:2018gsn,
  Inglis-Whalen:2021bea,Inglis-Whalen:2022vyn}, where $\Lambda_{H}$ and
$Q_H$ characterise the hadronic and hard scales in the process under
consideration, respectively.
At present, a resummation of those subleading power corrections has not
been accomplished yet beyond \LL.

Similarly, subleading power corrections also play a role in the limit
$\qT\to 0$.
Here, the \LP approximation can only recover the most singular
behaviour of the exact distribution at any given order in a fixed-order
expansion, leaving behind integrable (but numerically problematic)
singularities and constant terms.
These non-vanishing remainders are of particular concern in \qT-based
subtraction and slicing methods for higher-order calculations \cite{Bizon:2018foh,Chen:2018pzu,Camarda:2021ict}. 
Related results on power suppressed contribution can be found from
\cite{Oleari:2020wvt,Cieri:2019tfv,Camarda:2021jsw,Ebert:2020dfc,
  Ebert:2019zkb} concerning the slicing subtraction method, the endpoint
singularity in the Higgs production and decay~\cite{Liu:2022ajh,Liu:2020tzd,
  Liu:2020eqe,Liu:2020ydl,Liu:2019oav,Liu:2021chn,Liu:2018czl,Liu:2017vkm},
the event shapes in the leptonic and hadron colliders
\cite{Buonocore:2023mne,Agarwal:2023fdk,vanBeekveld:2019prq,Chen:2023wah,
  Ebert:2018lzn,Moult:2018jjd,Moult:2017jsg,Moult:2016fqy,Pal:2023vec}, and the
threshold resummation~\cite{vanBijleveld:2023vck,vanBeekveld:2021hhv,
  Laenen:2020nrt,vanBeekveld:2019cks,Bonocore:2016awd,Bonocore:2015esa,
  Bonocore:2014wua,Laenen:2010uz,Laenen:2008gt,Laenen:2008ux,
  Broggio:2023pbu,Beneke:2018gvs,Beneke:2022obx,Beneke:2019oqx,
  Beneke:2019mua,Bonocore:2021cbv,Bhattacharya:2021hae,AH:2020qoa,
  Bahjat-Abbas:2019fqa,Balsach:2023ema,Engel:2023rxp,Engel:2023ifn}.

To derive the factorisation formulae beyond the leading power
approximation, one of the prerequisites is to consistently combine
the contributions from the constituent dynamic regions, generally
comprising the hard, collinear, and soft modes \cite{Balitsky:2017gis,
  Balitsky:2020jzt,Balitsky:2021fer,Balitsky:2017flc,Vladimirov:2021hdn,
  Ebert:2021jhy,Gamberg:2022lju,Rodini:2022wic,Rodini:2023plb,
  Vladimirov:2023aot,Rodini:2023mnh,Ebert:2018gsn,Inglis-Whalen:2021bea,
  Inglis-Whalen:2022vyn}.%
\footnote{
   The irrelevance of the Glauber vertices is presumed in the subleading
   power factorisation of \cite{Balitsky:2017gis,Balitsky:2020jzt,
   Balitsky:2021fer,Balitsky:2017flc,Vladimirov:2021hdn,Ebert:2021jhy,
   Gamberg:2022lju,Rodini:2022wic,Rodini:2023plb,Vladimirov:2023aot,
   Rodini:2023mnh,Ebert:2018gsn,Inglis-Whalen:2021bea,Inglis-Whalen:2022vyn}.
   Up to now, a rigorous proof of the cancellation of the Glauber
   contributions beyond the leading power approximation is still absent,
   to our best knowledge.
   The leading power discussion on inclusive observables can be found
   in~\cite{Collins:1997sr,Collins:2004nx,Gaunt:2014ska,Schwartz:2018obd}.
}
Although these dynamic regions are well defined in given
segments of the phase space or the loop integrals, in practical
calculations one often extrapolates their contributions from their
respective intrinsic domains to the entire integration range,
necessitating a robust systematic subtraction process to remove
their overlap.
This procedure is frequently referred to as the \emph{zero-bin subtraction}
\cite{Manohar:2006nz,Idilbi:2007yi,Idilbi:2007ff,Chiu:2009yx}, the soft
subtraction \cite{Ji:2004xq,Ji:2004wu,Idilbi:2007yi,Idilbi:2007ff,
  Echevarria:2011epo}, or the overlap reduction~\cite{Goerke:2017ioi,
  Inglis-Whalen:2020rpi,Inglis-Whalen:2021bea}.
At leading power, zero-bin subtraction proceeds with deducting
the soft contribution from the collinear sectors, and has been
demonstrated to be valid up to \NNNLO~\cite{Echevarria:2015byo,
  Echevarria:2015usa,Li:2016axz,Li:2016ctv,Luo:2019hmp,Luo:2019bmw,
  Luo:2019szz,Luo:2020epw}.
Similar conclusions can also be found in the analysis of the subleading
angular coefficients (suppressed by a factor of $\qT/Q_H$ w.r.t.\
leading singular terms) in the Drell-Yan process and the
semi-inclusive deep-inelastic scatterings~\cite{Vladimirov:2021hdn,
  Ebert:2021jhy,Gamberg:2022lju,Rodini:2023plb}.
However, for the subleading inclusive observables of the Drell-Yan process
the zero-bin subtraction becomes more
involved, as it comprises mixed contributions in the
overlapping area from the lowest power accuracy up to \NLP
\cite{Inglis-Whalen:2020rpi,Inglis-Whalen:2021bea}.

Even though such recipes to construct zero-bin subtractions can
establish a factorisation at the \NLP level, a robust
algorithm that is capable of generating zero-bin subtrahends at an
arbitrary power and choice of a rapidity-divergence regularisation
prescription is still missing.
Investigations towards this goal were pioneered by \cite{Jantzen:2011nz},
in which a mathematically well-defined derivation was delivered for
the power series of the one-loop integrals in a variety of kinematical
limits.
Therein, the analysis of the massive Sudakov form factor is intimately
related to the asymptotic behaviour of the \qT~distribution.
Their power expansion begins with the introduction of a group of auxiliary
cutoff scales along the integration path, from which the relevant scales
of the process are well separated and the defining dynamic regions can
be prescribed accordingly.
Within these regions, the expansions are straightforward and always
well-defined.
This is then followed by a recombination procedure to lift all dependences 
on the auxiliary scales, thereby extrapolating the expanded integrands
from their intrinsic domains to the entire integration range.
During this procedure, however, a set of constituents emerge, consisting of
doubly and triply expanded integrands in line with their respective distinct
scaling laws which account for the overlapping contributions amongst different
modes and turn out to be in part the zero-bin subtrahends.
While integrals with a single definite scaling rule for every integration
variable are subjected only to a single expansion, multiple expansions can
induce ambiguous interpretations of the resulting integral as the integration
variables in general observe distinct scaling rules for each dynamic mode.
Subsequently, this ambiguity limits the applicability of the formalism in
\cite{Jantzen:2011nz} to only a special range of rapidity regularisation
schemes, such as the analytic regulator \cite{Becher:2010tm,Becher:2011dz}
and the pure rapidity regulator \cite{Ebert:2018gsn,Moult:2019vou}.
When using more generic regularisation schemes, e.g.\ the exponential
regulator \cite{Li:2016axz,Li:2016ctv}, doubly and triply expanded integrals
instead call for an unambiguous counting rule at each power accuracy to yield
results that are independent of the scheme.
As a matter of fact, any reliance on the choice of rapidity regulators
in carrying out the power expansion will hinder the establishment of
the rapidity renormalisation group \cite{Chiu:2011qc}, which, akin to
the renormalisation group governing the virtuality divergence, requires
the equivalence of the power series resulting from various
rapidity-divergence regularisation approaches.

In the following, we will make use of the \NLO \qT~distribution of the
process $pp\to H+X$ as a demonstrative example to accomplish this goal.
Since our derivation mainly concerns the pattern of the denominators of
the squared transition amplitudes, the conclusion here can also be
generalised to the analysis of the Drell-Yan processes, $pp\to V+X$ with
$V=W,Z$.
We commence by following the strategy outlined in \cite{Jantzen:2011nz}
to categorise the phase space so as to detach the emerging scales and
thereby carry out the expansion in momentum space.
Afterwards, instead of extrapolating the expanded integrands on the
cumulative level, in this work we shift the auxiliary boundaries at
a particular power accuracy only while maintaining the equality to
the power series derived from the momentum space.
As a result, a set of multiple expanded integrals emerge to
balance the overextended collinear sectors power by power.
We use these to construct the zero-bin subtrahends and thereby
retrieve the scalings for the occurring integration variables.
We will demonstrate that the zero-bin subtrahends proposed in this paper
are straightforwardly applicable to all rapidity regulators that preserve
their expressions before and after asymptotic expansions, including the
analytic regulator \cite{Becher:2010tm,Becher:2011dz}, the exponential
regulator \cite{Li:2016axz,Li:2016ctv}, and the pure rapidity regulator
\cite{Ebert:2018gsn,Moult:2019vou}.
In addition, with a few adaptations, our algorithm presented here can also
be generalised to use the $\Delta$-regulator \cite{Chiu:2009yx,
  Echevarria:2011epo,Echevarria:2015usa,Echevarria:2015byo,
  Echevarria:2016scs} and the $\eta$-regulator \cite{Chiu:2011qc,Chiu:2012ir}.
This accommodates the majority of the rapidity-divergence regularisation
schemes in use at this time.

Our paper is organised as follows.
In Sec.\ \ref{sec:asyexp:mom} we develop our framework to compute,
in a general way, all power corrections to \NLO accuracy.
To this end, we will use Sec.~\ref{sec:qTNLO} and \ref{sec:setups},
taking the example of Higgs hadroproduction at LHC, to review the fixed order calculation on the \qT\
spectrum and thereby categorise the phase space into two
sectors according to the origins of the power corrections.
The first sector, associated with the asymptotic expansion of the
boundary condition, will be investigated in Sec.~\ref{sec:red:qT:boundary}.
The other one takes in the bulk of phase space and thus forms the main
concern of this paper.
We will devote Sec.~\ref{sec:red:qT:central} to elaborate on its
expansion procedure, in particular highlighting the structural pattern
of the emerging zero-bin subtrahends.
At last, in Sec.~\ref{sec:asyexp:recap}, we combine the relevant
ingredients and present the power series of the \qT\ spectrum of
our example process $pp\to H+X$.
Sec.~\ref{sec:asyexp:outputs} then is dedicated to the application
of the framework derived in Sec.~\ref{sec:asyexp:mom}, where the
power corrections will be appraised up to \NNLP for the first time.
We calculate it in three different approaches, including the momentum
cutoff scales, the homogenous regularisation scheme, and the
inhomogeneous regulator, to regulate the rapidity divergences.
We will demonstrate that after appropriate combination, all three
methods result in the same analytic expressions at least up to \NNLP,
echoing the rapidity renormalisation group \cite{Chiu:2011qc}.
Finally, we will scrutinise our \NNLP results in
Sec.~\ref{sec:asyexp:numeric} by comparing to the full QCD calculation.
Eventually, we summarise our findings in Sec.~\ref{sec:conclusions}.

\section{Theoretical framework}\label{sec:asyexp:mom}

In this section, we introduce the formalism \changed{to be utilised in}
the next-to-leading and next-to-next-to-leading power
expansions, dubbed \NLP and \NNLP respectively.
We will use the case of Higgs production at hadron colliders,
$p+p \to H+X$, as an illustrative example.
The results we derive, however, are similarly applicable
to the production of any other colour-neutral final state.

\subsection{Analysis of the NLO fixed order results}\label{sec:qTNLO}

To introduce our notation,
we start our analysis by recalling the general expression 
for the differential cross section of the process $p+p \to H+X$
in proton-proton collisions.
According to the QCD factorization theorem~\cite{Collins:1989gx},
the differential \qT~spectra at NLO can be calculated as follows,
\begin{align}\label{eq:qcd:qT}
  \frac{\done \sigma_H}{\done Y_H\, \done \qT^2}
  \;=\;
  &
    \frac{1}{16\pi\, s^2}\sum_{i,j}
    \int^{k_+^{\mathrm{max}}}_{0}\!\!\!\!\done k_+\,
    \int^{k_-^{\mathrm{max}}}_{0}\!\!\!\!\done k_-\,
    \delta(k_+k_--\qT^2)\;
    \frac{f_{i/n}(\xi_n)}{\xi_n}\,
    \frac{f _{j/\nbar}(\xi_{\nbar})}{\xi_{\nbar}}\;
    \overline{\sum_{\mathrm{col},\mathrm{pol}} }
    \big|\mathcal{M}(i+j\to H+k)\big|^2\,\nonumber\\
  \;=\;
  &
    \frac{1}{16\pi\, s^2}\sum_{i,j}
    \int^{k_+^{\mathrm{max}}}_{k_+^{\mathrm{min}}}\,
    \frac{\done k_+}{k_+}
    \frac{f_{i/n}(\xi_n)}{\xi_n}
    \frac{f _{j/\nbar}(\xi_{\nbar})}{\xi_{\nbar}}\;
    \overline{\sum_{\mathrm{col},\mathrm{pol}} }
    \big|\mathcal{M}(i+j\to H+k)\big|^2\,,
\end{align}
where $\sqrt{s}$ denotes the collider centre-of-momentum energy
of the incoming protons.
$Y_H$ and $\qT$ stand for the rapidity and transverse momentum
of the Higgs boson measured in the laboratory reference frame,
respectively.
The boundaries of the phase space integral are given by the
variables $k_+^{\mathrm{max}}$ and $k_+^{\mathrm{min}}$, with
\begin{align}\label{eq:def:b.c.}
  k_+^{\mathrm{max}}=\sqrt{s}-\mT e^{-Y_H}\,,\qquad
  k_+^{\mathrm{min}}=\frac{\qT^2}{k_-^{\mathrm{max}}}=\frac{\qT^2}{\sqrt{s}-\mT e^{+Y_H}}\;.
\end{align}
Therein, $\mT^2\equiv m_H^2+\qT^2$ is the transverse
mass of the Higgs boson, $m_H$ being the mass of the Higgs boson.

In the integrand of Eq.~\eqref{eq:qcd:qT}, $f_{i/n}$ and
$f _{j/\nbar}$ represent the parton distribution functions (PDFs)
for the partons $i$ and $j$ in the protons $n$ and $\nbar$,
respectively, with the argument $\xi_{n(\nbar)}$ encoding
the momentum fractions carried by the incident particle
$i(j)$ with respect to its mother proton $n(\nbar)$.
In this paper, we will use the 5 flavour scheme, and
correspondingly, the partons
$\{i,j,k\}\in\{u,\bar{u},d,\bar{d},c,\bar{c},s,\bar{s},b,\bar{b},g\}$
will be considered massless and taken fully into account in
the following analysis.

$|\mathcal{M}|^2$ denotes the squared amplitudes for the contributing
partonic channel $i+j\to H+k$.
Since the main focus of this paper is on the region where $\qT$
approaches, but is not equal to, zero, only tree level contributions
will participate in Eq.~\eqref{eq:qcd:qT}.
\changed{In the large top mass limit, the amplitudes for the
processes $i+j\to H+k$ can be calculated in the low energy Higgs effective field
theory (HEFT) \cite{Kluberg-Stern:1974iel,Nielsen:1975ph,Wilczek:1977zn,
  Nielsen:1977sy,Inami:1982xt,Spiridonov:1984br,Chetyrkin:1997un}.
Therein, the creation and annihilation of the Higgs boson are governed
by the effective Lagrangian,
\begin{align}
  \mathcal{L}_{\text{eff}}
  &=
    \frac{\alpha_s(\mu)}{12 \pi v}\, C_t(m_t,\mu)\,
    G^{\mu\nu,a} G_{\mu\nu}^{a}\, H\, ,
\label{eq:Leff}
\end{align}
where $\alpha_s$ stands for the strong coupling constant and $v$ denotes
the Higgs vacuum expectation value, while
$H$ and $G_{\mu\nu}^a$ represent the Higgs field operator and the gluon
field strength tensor, respectively.
The closed top quark loop, coupling the gluons to the Higgs, is encoded
by the Wilson coefficient $C_t$,
which, up to now, has been calculated up to the four-loop order
\cite{Inami:1982xt, Djouadi:1991tk, Chetyrkin:1997iv,
  Chetyrkin:1997un, Chetyrkin:2005ia, Schroder:2005hy, Baikov:2016tgj}.
In this work, we only take into account its LO contribution in
accordance with the perturbative accuracy of Eq.~\eqref{eq:qcd:qT}.
In writing Eq.~\eqref{eq:Leff}, we have omitted the operators inducing
the Higgs-light-quark interaction, as the 5 active flavour scheme is
adhered to throughout our investigation.
}

\changed{
Combining Eq.~\eqref{eq:Leff} with the QCD Lagrangian enables the
generation of the squared amplitudes for the process $i+j\to H+k$.
Having summed and averaged, as appropriate, over all the colour and
polarisation configurations, the results
read \cite{Ravindran:2002dc,Glosser:2002gm},}
\begin{equation}
  \label{eq:msq}
  \begin{split}
    \overline{\sum_{\mathrm{col},\mathrm{pol}}}
    \big|\mathcal{M}(g+g\to H+g)\big|^2
    &=\,
      \frac{3\lambda^{\changed{2}}_t}{32}\,
      \frac{m_H^8+s_{ij}^4+s_{ik}^4+s_{jk}^4}{s_{ij}s_{ik}s_{jk}} \,,\\
    \overline{\sum_{\mathrm{col},\mathrm{pol}}}
    \big|\mathcal{M}(g+q(\bar{q})\to H+q(\bar{q}))\big|^2
    &=
      \,-\frac{\lambda^{\changed{2}}_t}{24}\,\frac{s^2_{ij}+s^2_{ik}}{s_{jk}}\,,\\
    \overline{\sum_{\mathrm{col},\mathrm{pol}}}
    \big|\mathcal{M}(q+\bar{q}\to H+g)\big|^2
    &=\,
      \frac{\lambda^{\changed{2}}_t}{9}\,\frac{s^2_{ik}+s^2_{jk}}{s_{ij}}\,,\\
  \end{split}
\end{equation}
\changed{where the parameter $\lambda_t$ is introduced to collect the
coupling constants, i.e.\ $\lambda^2_t\equiv {4\alpha_s^3C_t^2}/{(9\pi v^2)}$.
The missing expressions for $q(\bar{q})g$ and $\bar{q}q$
initial states can be derived from the above by swapping the
roles of $i$ and $j$. }

\changed{The expressions in Eq.~\eqref{eq:msq} involve} the scalar
products of the momenta of the initial and final particles,
\begin{equation}
  \begin{split}\label{eq:def:sji}
    s_{ij}
    =&\phantom{+}2p_i\cdot p_j
    =\left(k_++\mT e^{-Y_H}\right)\,
     \left(k_-+\mT e^{+Y_H}\right)\,,\\
    s_{ik}
    =&-2p_i\cdot p_k=  -\left(k_++\mT e^{-Y_H}\right)k_- \,, \\
    s_{jk}
    =&-2p_j\cdot p_k=  -\left(k_-+\mT e^{+Y_H}\right)k_+ \,.
  \end{split}
\end{equation}
Therein, $k_{\pm}$ denotes the light cone component of the momentum
of the emitted parton $k$.
To be precise, with the help of the light-like reference vectors
$n^{\mu}$ and $\nbar^{\mu}$, aligned with incoming beams and
satisfying $n\cdot\nbar=2$, it is defined as
\begin{align}
\label{eq:def:lcone}
  p_k^{\mu}
  =\frac{p_k\cdot n}{2}\nbar^{\mu}
   +\frac{p_k\cdot \nbar}{2}{n}^{\mu}
   +p_{k,\bot}^\mu
  \equiv\frac{k_-}{2}\nbar^{\mu}
        +\frac{k_+}{2}{n}^{\mu}
        +k_{\bot}^\mu
  \equiv\big[k_-,k_+,k_{\bot} \big]\,.
\end{align}
In this notation, the momentum fractions $\xi_n$  and $\xi_{\nbar}$
of Eq.~\eqref{eq:qcd:qT} can be written as
\begin{align}\label{eq:qcd:def:Xi}
  \xi_n
  =\frac{k_++\mT e^{-Y_H}}{\sqrt{s}}\qquad\text{and}\qquad
  \xi_{\nbar}
  =\frac{k_-+\mT e^{+Y_H}}{\sqrt{s}}\;.
\end{align}
In the following sections, we will investigate the asymptotic
properties of Eq.~\eqref{eq:qcd:qT} in the vicinity of $\qT=0$,
adhering to the following definition of the expansion accuracy,
counting powers of $\qT^2$, throughout,
\begin{equation}\label{eq:def:acc}
  \begin{split}
    \frac{\done \sigma_H}{\done Y_H\, \done \qT^2}=&
    \,
    \underbrace{\sum_m\,\frac{\Delta_{\LP}^{(m)}\,(L_H)^m}{\qT^2}}_{\LP}
    \,
    +
    \,
    \underbrace{\sum_m\,\Delta_{\NLP}^{(m)}\,(L_H)^m}_{\NLP}
    \,
    +
    \,
    \underbrace{\sum_m\, \qT^2\,\Delta_{\NNLP}^{(m)}\,(L_H)^m }_{\NNLP}
    \,
    +
    \,
    \dots\, .
  \\
  \end{split}
\end{equation}
Therein, the \qT~differential distribution in Eq.~\eqref{eq:qcd:qT}
is expanded in the low \qT~domain.
At a given power precision,~e.g.\ N$^{\omega+1}$LP, all non-logarithmic
\qT~dependences have been collected in the pre-factor
$\qT^{2\omega}$ with $(\omega\ge-1)$. 
They leave behind the logarithmic terms $L_H\equiv\ln[\qT^2/m_H^2]$
and the coefficient functions
$\Delta_{\mathrm{N}^{\omega+1}\mathrm{LP}}^{(m)}\equiv\Delta_{\mathrm{N}^{\omega+1}\mathrm{LP}}^{(m)}[m_H,\,s,\,Y_H;\; f_{i/n},\,f_{j/\nbar}]$ which accounts
for the convolution of the PDFs $f_{i/n}$ and $f_{j/\nbar}$
as well as the functions of the hard scales $s$ and $m_H$.
According to this definition, the LP contribution is expected to
accommodate the most singular behaviour in the low \qT~regime,
while the higher power corrections experience progressively stronger
suppression factor $\qT^{2(\omega-1)}$.
Correspondingly, $\Delta_{\LP}$ is known extremely precisely,
up to approximate fourth order \cite{Catani:2007vq,Catani:2012qa,
  Catani:2022sgr,Catani:2011kr,Luo:2020epw,Luo:2019szz,Ebert:2020yqt,
  Gehrmann:2012ze,Gehrmann:2014yya,Echevarria:2016scs,Luo:2019hmp,
  Luo:2019bmw,Gutierrez-Reyes:2019rug,Chetyrkin:1997iv,
  Chetyrkin:1997un,Gehrmann:2014vha,Gehrmann:2010ue,Baikov:2009bg,
  Gehrmann:2005pd,Harlander:2000mg,
  Duhr:2022yyp,Moult:2022xzt,Schroder:2005hy,Chetyrkin:2005ia,
  Lee:2022nhh,Chakraborty:2022yan}, while the $\Delta_{\NLP}$
has only been computed recently to \NLO accuracy \cite{Ebert:2018gsn}.
All further higher-power corrections are hitherto unknown,
and we will use the remainder of this section to introduce a
mathematically well-defined approach to develop a framework to
compute the $\Delta_{\mathrm{N}^{\omega+1}\mathrm{LP}}$ for all
$\omega\ge-1$ at \NLO accuracy before deriving the respective
\NLO-correct expression for $\Delta_\NNLP$ explicitly.

In the following subsections we will be evaluating the
coefficients $\Delta_{\mathrm{N}^{\omega+1}\mathrm{LP}}^{(m)}$.
In Sec.~\ref{sec:setups}, we will derive additional power
corrections originating from the kinematics of the process
and thereby categorise the phase space integral of
Eq.~\eqref{eq:qcd:qT} into two sectors.
The first sector encompasses the domains in the vicinity of
$k^{\mathrm{max}}_+$ and  $k^{\mathrm{max}}_-$,
accounting for the power suppressed contributions
induced by the boundary conditions of the phase space integral,
see Eq.~\eqref{eq:def:b.c.}.
We will make use of Sec.~\ref{sec:red:qT:boundary} to discuss the
asymptotic properties from this region.
The second sector consists of the remaining regions, entailing
a variety of scales along the integration path and is therefore
the main concern of this paper.
We will elaborate in Sec.~\ref{sec:red:qT:central} on the power
expansion of this contribution.
Eventually, with the results of the two, we are capable of deriving
the power series for the \qT~spectrum in Sec.~\ref{sec:asyexp:recap}.

\subsection{Prerequisites for the power expansion}
\label{sec:setups}

\changed{
We start the discussion with the transformation of the squared amplitudes
of Eq.~\eqref{eq:msq}.
As the power expansion in the low \qT\ domain primarily concerns the
relation of the kinematic variables $k_{\pm}$ and $\mT e^{\pm Y_H}$,
it is beneficial for us to regroup the results in Eq.~\eqref{eq:msq}
according to their $k_{\pm}$ and $\mT e^{\pm Y_H}$ dependences.
To this end, it merits reminding that the numerators of the $|\mathcal{M}|^2$
can be all recast in terms of the polynomials in the invariants
$s_{ij}$, $s_{ik}$, and $s_{jk}$, on account of the on-shell condition
$ m_H^2\,=\,s_{ij}\,+\,s_{ik}\,+\,s_{jk}$.
Hence, with Eq.~\eqref{eq:def:sji} and carrying out the polynomial
expansion, the numerators of the $|\mathcal{M}|^2$ are then cast into
a finite series of the products of $k_{\pm}$ and $\mT e^{\pm Y_H}$,
more specifically,
\begin{equation}\label{eq:qcd:def:H:num}
  \begin{split}
 \overline{\sum_{\mathrm{col},\mathrm{pol}}}
      \big|\mathcal{M}_{[\kappa]}\big|^2 \Bigg|_{\mathrm{Numerator}} \,\rightarrow\,
    \sum_{m,n}\,{(k_+)^{m}}\,{(k_-)^{n} }\,
   {c}^{m,n}_{[\kappa]} \left(\mT,Y_H\right)\,,
  \end{split}
\end{equation}
where the contributing partonic channels are denoted by the
subscript $\kappa=\{gg,gq(\bar{q}),q(\bar{q})g,q\bar{q},\bar{q}q\}$.
The coefficients ${c}^{m,n}_{[\kappa]}$ are a function of the hard
scales $\mT e^{\pm Y_H}$, with the superscripts $\{m,n\}\ge0$
corresponding to the powers of $k_{\pm}$.
Combining Eq.~\eqref{eq:qcd:def:H:num} with the denominators of the
$|\mathcal{M}|^2$ in Eq.~\eqref{eq:msq} and factoring out common prefactors
consisting of the coupling parameter $\lambda_t$ and the colliding
energy $\sqrt{s}$, it yields that,
\begin{equation}\label{eq:qcd:def:H}
  \begin{split}
        \overline{\sum_{\mathrm{col},\mathrm{pol}}}
      \big|\mathcal{M}_{[\kappa]}\big|^2
    \,\equiv&\;
  {16\pi\lambda_t^2 s}\,
    \sum_{\{\beta\}}
    \sum_{\rho,\sigma}\, 
      \frac{(k_+)^{\sigma}}{(k_++m_\mathrm{T}e^{-Y_H})^{\beta_n-1}}\,
      \frac{   (k_-)^{\rho} }{(k_-+m_\mathrm{T}e^{Y_H})^{\beta_\nbar-1}}\,
   \mathcal{H}^{\rho,\sigma}_{[\kappa],\{\beta\}} \left(\mT,Y_H,s\right)\,,
  \end{split}
\end{equation}
where a novel hard sector $\mathcal{H}^{\rho,\sigma}_{[\kappa],\{\beta\}}$
is introduced to assimilate the ${c}^{\rho,\sigma}_{[\kappa],\{\beta\}}$
of Eq.~\eqref{eq:qcd:def:H:num} as well as the colliding energy $\sqrt{s}$.
Based on Eq.~\eqref{eq:msq}, the indices $\rho$ and $\sigma$ as well as
$\beta_n$ and $\beta_\nbar$ in $\mathcal{H}^{\rho,\sigma}_{[\kappa],\{\beta\}}$
are always integers for the process $pp\to H+X$ at NLO.
At variance with $m$ and $n$ in Eq.~\eqref{eq:qcd:def:H:num}, the
superscripts $\rho$ and $\sigma$ in Eq.~\eqref{eq:qcd:def:H} can be
of negative value, as the squared amplitudes in the $g+g\to H+g$ and
$g+q(\bar{q})\to H+q(\bar{q})$ channels are both able to contribute
additional light-cone components $k_{\pm}$ from the denominators,
thereby lowering the powers of Eq.~\eqref{eq:qcd:def:H:num}.
$\beta_n$ and $\beta_\nbar$, however, also following Eq.~\eqref{eq:msq},
are always taken greater than or equal to one.
At last, it should be emphasised that the parameterisation in
Eq.~\eqref{eq:qcd:def:H} is not unique, since one can always
reweight the numerator and denominator of the results in Eq.~\eqref{eq:msq}
by a common factor, e.g.\ $(k_-+m_{\mathrm{T}} e^{+Y_H})^2$, and then
expand the numerator without modifying the fraction.
Although this arbitrariness may impact the expressions of individual
$\mathcal{H}^{\rho,\sigma}_{[\kappa],\{\beta\}}$ in Eq.~\eqref{eq:qcd:def:H},
but will not impact the squared amplitudes $|\mathcal{M}|^2$.
}

\changed{
Substituting Eq.~\eqref{eq:qcd:def:H} together with Eq.~\eqref{eq:qcd:def:Xi} into  Eq.~\eqref{eq:qcd:qT}, we recast the \qT~spectrum  as,
\begin{equation}\label{eq:qcd:qT:kpm}
  \begin{split}
    \frac{\done \sigma_H}{\done Y_H\, \done \qT^2}
    \,=&\;
      \lambda^2_t\,\sum_{\kappa}\,
      \sum_{\{\beta\}}\,
      \sum_{\rho,\sigma } \,
      \mathcal{H}^{\rho,\sigma}_{[\kappa],\{\beta\}}
      \left(\mT,Y_H,s\right) \,
    \\
    &\times\;
      \int^{k_+^{\mathrm{max}}}_{k_+^{\mathrm{min}}}
     \frac{\done k_+}{k_+}\;
      (k_-)^{\rho} \, (k_+)^{\sigma} \,
      {F}^{(0)}_{i/n,\beta_n}\left(k_++\mT e^{-Y_H}\right)\,
      {F}^{(0)}_{j/\nbar,\beta_\nbar}\left( k_-+\mT e^{+Y_H} \right)\,.
  \end{split}
\end{equation}
Herein, the integrand} contains the
derivatives of the scaled PDFs of the light-cone
components $(k_{\pm}+\mT e^{\mp Y_H})$, generically defined
as
\begin{align}\label{eq:def:FioN:FjoNbar}
  {F}^{(\alpha_n)}_{i/n,\beta_n}\left(Q_n\right)
  &\equiv
    \frac{\partial^{\alpha_n}}{\partial Q_n^{\alpha_n}}
    \left[
      \frac{f_{i/n}\left({Q_n}/{\sqrt{s}}\right)}{Q_n^{\beta_n}}
    \right]\,,
    \quad\quad
  {F}^{(\alpha_\nbar)}_{j/\nbar,\beta_\nbar}\left(Q_\nbar\right)
  \equiv
    \frac{\partial^{\alpha_\nbar}}{\partial Q_\nbar^{\alpha_\nbar}}
    \left[
      \frac{f_{j/\nbar}\left(Q_\nbar/{\sqrt{s}}\right)}
           {Q_\nbar^{\beta_\nbar}}
    \right]\,,
\end{align}
where ${Q}_{n}$ and $Q_\nbar$ represent the two kinematic variables,
$\alpha_n$ and $\alpha_\nbar$ determine the rank of the derivative,
and $\beta_n$ and $\beta_\nbar$ are the powers to which they scale their
respective PDFs.
Collectively, we denote the latter $\{\alpha\}$ and $\{\beta\}$.

For the results derived in full QCD, see Eq.~\eqref{eq:qcd:qT},
only the zero-rank scaled PDFs are involved.
Hence, only the $\alpha_n=\alpha_\nbar=0$ case is present in Eq.~\eqref{eq:qcd:qT:kpm}.
The $\alpha_n,\alpha_\nbar>0$ expressions, however, will enter in the higher-power \qT\ expansion.

In the following, we will concern ourselves with the analytic properties
of the functions ${F}^{(\alpha_n)}_{i/n,\beta_n}\left(Q_n\right)$ and
${F}^{(\alpha_\nbar)}_{j/\nbar,\beta_\nbar}\left(Q_\nbar\right)$.
However, due to the fact that the PDFs $f_{i/n}$ and $f_{j/\nbar}$
have their origin in the non-perturbative regime of QCD and, thus, their
analytic expressions are not fully established, we make the following
assumptions:
\begin{enumerate}
  \item[(a)]
    For a given $\alpha_n,\alpha_\nbar>0$ the
    ${F}^{(\alpha_n)}_{i/n,\beta_n}\left(Q_n \right)$ and
    ${F}^{(\alpha_\nbar)}_{j/\nbar,\beta_\nbar}\left(Q_\nbar\right)$,
    as well as their arbitrary order derivatives, exist and are bounded
    over the domain $\Lambda_{\mathrm{H}}\le{Q}_{n(\nbar)}\le\sqrt{s}$
    for all $\beta_n$ and $\beta_\nbar$,
    where $\Lambda_{\mathrm{H}}$ represent\changed{s} a hadronisation scale\;.
  \item[(b)]
    ${F}^{(\alpha_n)}_{i/n,\beta_n}\left(Q_n+\delta Q_n\right)$ and
    ${F}^{(\alpha_\nbar)}_{j/\nbar,\beta_\nbar}\left(Q_\nbar+\delta Q_\nbar\right)$
    have convergence radii $|\delta{Q}_{n(\nbar)}|<{Q}_{n(\nbar)}$
    as long as $\left({Q}_{n(\nbar)} + \delta{Q}_{n(\nbar)}\right)$
    and ${Q}_{n(\nbar)}$ are both well within the interval
    $\left[\Lambda_{\mathrm{H}},\sqrt{s}\right]$\,.
\end{enumerate}
Please note, if the PDFs are fitted in a basis of Chebyshev polynomials,
as proposed in \cite{Bonvini:2012sh,Bonvini:2014joa,Diehl:2021gvs},
and thereby $f_{i/n}$ and $f_{j/\nbar}$ can be then expressed in terms
of $\ln^k(\xi_n)/\xi_n^h$ and $\ln^k(\xi_\nbar)/\xi_\nbar^h$, respectively,
the second assumption is immediately fulfilled.

Equipped with the above ansatz, we are able to expand
Eq.~\eqref{eq:qcd:qT:kpm} in the vicinity of $\qT=0$.
To this end, we start with the expansion of the kinematic variable
$\mT $ in the hard coefficient function
$\mathcal{H}^{\rho,\sigma}_{[\kappa],\{\beta\}}$.
In the small \qT~domain where $\qT\ll m_H$, we can perform an
expansion of the
transverse mass \mT\ of the Higgs boson around its invariant
mass $m_H$,
\begin{align}\label{eq:asy:mom:mT}
  \mT
  \,=\,
    m_H\,\cdot\,
    \sum_{h=0}^{\infty}\,\frac{1}{h!}\,
    \frac{\Gamma[\tfrac{3}{2}]}{\Gamma[\frac{3}{2}-h]}\,
    \left(\frac{\qT^2}{m_H^2}\right)^h\,.
\end{align}
We can use this result to expand the hard coefficient
$\mathcal{H}^{\rho,\sigma}_{[\kappa],\{\beta\}}$ around $\qT=0$,
giving
\begin{equation}\label{eq:asy:mom:c}
  \begin{split}
    \mathcal{H}_{[\kappa],\{\beta\}}^{\rho,\sigma}&(\mT,Y_H,s)\\
    =&\,
      \sum_{h=0}^{\infty}\frac{(\mT-m_H)^h}{h!}\,
      \left\{
        \frac{\partial^h}{\partial \mT^h}\,
        \mathcal{H}_{[\kappa],\{\beta\}}^{\rho,\sigma}(\mT,Y_H,s)
        \bigg|_{\mT\to m_H}
      \right\}\, \\
    =&\,
      \sum_{h,l=0}^{\infty}\,\sum_{g=0}^h\,\frac{(-1)^{h-g}}{g!\,l!\,(h-g)!}\,
      \frac{\Gamma[\frac{g}{2}+1]}{\Gamma[\frac{g}{2}-l+1]}\,
      \left(\frac{\qT^2}{m^2_H}\right)^l\,m_H^h\,
      \mathcal{H}_{[\kappa],\{\beta\}}^{(h),\rho,\sigma}(m_H,Y_H,s)\,,
  \end{split}
\end{equation}
where we have introduced
$\mathcal{H}_{[\kappa],\{\beta\}}^{(h),\rho,\sigma}(m_H,Y_H,s)
\equiv\frac{\partial^h}{\partial \mT^h}\,\mathcal{H}_{[\kappa],\{\beta\}}^{\rho,\sigma}(\mT,Y_H,s)\bigg|_{\mT\to m_H}$.
An analogous expansion can also be applied to the derivatives of the
scaled PDFs ${F}^{(\alpha_n)}_{i/n,\beta_n}$ and
${F}^{(\alpha_\nbar)}_{j/\nbar,\beta_\nbar}$.
To this end, we re-express their arguments below in order to isolate
all \qT~dependences,
\begin{align}\label{eq:asy:mom:pdfs:arg}
  k_{\pm}+\mT e^{\mp Y_H}
  \to
    \left[k_{\pm}+m_He^{\mp Y_H}\right]
    +\left(\mT -m_H\right)\,e^{\mp Y_H}.
\end{align}
In proximity to $\qT=0$, the second term on the right handed side (r.h.s)
is always smaller in magnitude with respect to the first one.
Therefore, we can expand
${F}^{(\alpha_n)}_{i/n,\beta_n}(k_{+}+\mT e^{- Y_H})$ and
${F}^{(\alpha_\nbar)}_{j/\nbar,\beta_\nbar}(k_{-}+\mT e^{+ Y_H})$
around $(k_{+}+m_He^{- Y_H})$ and $(k_{-}+m_He^{+ Y_H})$,
respectively,
\begin{equation}\label{eq:asy:mom:pdfs}
  \begin{split}
    F^{(0)}_{i/n,\beta_n}&\left(k_{+}+\mT e^{- Y_H}\right)\\
    &=\,
      \sum_{h=0}^{\infty}\,\frac{(\mT-m_H)^h}{h!}\,(e^{-Y_H})^h\,
      F^{(h)}_{i/n,\beta_n}\left(k_{+}+m_He^{- Y_H}\right)\\
    &=\,
      \sum_{h,l=0}^{\infty}\sum_{g=0}^h\,
      \frac{(-1)^{h-g} }{g! \,l! \,(h-g)!}\,
      \frac{\Gamma[\frac{g}{2}+1]}{\Gamma[\frac{g}{2}-l+1]}\,
      \left(\frac{\qT^2}{m^2_H}\right)^l\,
      \left( m_H e^{-Y_H} \right)^h\,
      F^{(h)}_{i/n,\beta_n}\left(k_{+}+m_He^{- Y_H}\right)\,,\\
    F^{(0)}_{j/\nbar,\beta_\nbar}&\left(k_{-}+\mT e^{+Y_H}\right)\\
    &=\,
      \sum_{h=0}^{\infty}\,\frac{(\mT-m_H)^h}{h!}\,(e^{+Y_H})^h\,
      F^{(h)}_{j/\nbar,\beta_\nbar}\left(k_{-}+m_He^{+Y_H}\right)\\
    &=\,
      \sum_{h,l=0}^{\infty}\sum_{g=0}^h\,
      \frac{(-1)^{h-g} }{g! \,l! \,(h-g)!}\,
      \frac{\Gamma[\frac{g}{2}+1]}{\Gamma[\frac{g}{2}-l+1]}\,
      \left(\frac{\qT^2}{m^2_H}\right)^l\,
      \left( m_H e^{+Y_H} \right)^h\,
      F^{(h)}_{j/\nbar,\beta_\nbar}\left(k_{-}+m_He^{+Y_H}\right)\,.
  \end{split}
\end{equation}
Inserting the expressions of Eqs.~\eqref{eq:asy:mom:c}-\eqref{eq:asy:mom:pdfs}
into Eq.~\eqref{eq:qcd:qT:kpm}, the \qT~spectrum now takes the following form,
\begin{align}\label{eq:qcd:qT:kpm:exp1}
  &\frac{\done \sigma_H}{\done Y_H\, \done \qT^2}
  \,=\,
  \lambda_t^2\,
  \sum_{\omega}\,
  \left({\qT^2}\right)^{\omega}\,
  \sum_{[\kappa]}\,
  \sum_{\{\alpha,\beta\}}\,\sum_{\rho,\sigma }\,
  \widetilde{\mathcal{H}}^{(\omega ),\rho,\sigma}_{[\kappa],\{\alpha,\beta\}}
  \left(m_H,Y_H,s\right) \,
  \bigg\{
    \widetilde{\mathcal{I}}^{ \rho,\sigma}_{[\kappa],\{\alpha,\beta\}}
    \,+\,
    \Delta\widetilde{\mathcal{I}}^{ \rho,\sigma}_{[\kappa],\{\alpha,\beta\}}
  \bigg\}\;.
\end{align}
Therein,
$\widetilde{\mathcal{H}}^{(\omega),\rho,\sigma }_{[\kappa],\{\alpha,\beta\}}$
is introduced to absorb the factorials, gamma functions, as well as powers
of the hard scales $m_H$ and $\sqrt{s}$ emerging from
Eqs.~\eqref{eq:asy:mom:c}-\eqref{eq:asy:mom:pdfs},
\begin{equation}\label{eq:qcd:defHtilde}
  \begin{split} 
    \widetilde{\mathcal{H}}^{(\omega),\rho,\sigma }_{[\kappa],\{\alpha,\beta\}}
    \,=&\;  
          \left(\frac{1}{m^2_H}\right)^{\omega}\,
    \sum_{l_n=0}^{\omega}\,
    \sum_{l_\nbar=0}^{\omega-l_n}\,
     \sum_{\alpha_h=0}^{\infty}\,
      \prod_{x=\{h,n,\nbar\}} 
      \left\{
      \sum_{g_x=0}^{\alpha_x}\frac{(-1)^{\alpha_x-g_x} }{g_x! \,l_x! \,(\alpha_x-g_x)!}\,
      \frac{\Gamma[\frac{g_x}{2}+1]}{\Gamma[\frac{g_x}{2}-l_x+1]}\,
      \right\}\\
    &\times  \left( m_H e^{-Y_H} \right)^{\alpha_n}\,
            \left( m_H e^{+Y_H} \right)^{\alpha_\nbar}\,
             \left( m_H   \right)^{\alpha_h}\,
        {\mathcal{H}}^{ (\alpha_h),\rho,\sigma }_{[\kappa],\{\beta\}}\,,
  \end{split}
\end{equation}
where $l_h=\omega-l_n-l_\nbar\ge0$. 
The \qT~dependences have been factored out from
$\widetilde{\mathcal{H}}^{(\omega ),\rho,\sigma}_{[\kappa],\{\alpha,\beta\}}$
and the arguments of ${F}^{(\alpha_n)}_{i/n,\beta_n}$ and
${F}^{(\alpha_\nbar)}_{j/\nbar,\beta_\nbar}$.
The remaining ones participate through boundary conditions as
defined in Eq.~\eqref{eq:def:b.c.}.
To address them as well, we divide the phase space integral of
Eq.~\eqref{eq:qcd:qT:kpm:exp1} into two sectors,
$\widetilde{\mathcal{I}}^{ \rho,\sigma}_{[\kappa],\{\alpha,\beta\}}$ and
$\Delta\widetilde{\mathcal{I}}^{ \rho,\sigma}_{[\kappa],\{\alpha,\beta\}}$,
with
\begin{align}
  \label{eq:def:I:central}
  \widetilde{\mathcal{I}}^{ \rho,\sigma}_{[\kappa],\{\alpha,\beta\}}
  =&\,
    \int^{\tilde{k}^{\mathrm{max}}_+}_{\tilde{k}^{\mathrm{min}}_+}\,
    \frac{\done k_+}{k_+}  (k_-)^{\rho}  \, (k_+)^{\sigma} \,
    {F}^{(\alpha_n)}_{i/n,\beta_n}\left(k_++m_He^{-Y_H}\right)\,
    {F}^{(\alpha_\nbar)}_{j/\nbar,\beta_\nbar}\left(k_-+m_He^{+Y_H} \right)\,,\\
  \label{eq:def:I:boundary}
  \Delta\widetilde{\mathcal{I}}^{ \rho,\sigma}_{[\kappa],\{\alpha,\beta\}}
  =&\,
    \left(
      \int_{\tilde{k}^{\mathrm{max}}_+}^{{k}^{\mathrm{max}}_+}\!\!\!\!
      +\int^{\tilde{k}^{\mathrm{min}}_+}_{{k}^{\mathrm{min}}_+}
    \right)
    \frac{\done k_+}{k_+}  (k_-)^{\rho} \, (k_+)^{\sigma} \,
    {F}^{(\alpha_n)}_{i/n,\beta_n}\!\!\left(k_++m_He^{-Y_H}\right)
    {F}^{(\alpha_\nbar)}_{j/\nbar,\beta_\nbar}\!\!\left(k_-+m_He^{+Y_H} \right)\,,\!\!\!\!\!\!
\end{align}
by means of the following boundaries,
\begin{align}
  \tilde{k}_+^{\mathrm{max}}
  =
    \sqrt{s}-m_H e^{-Y_H}\,,\quad\quad\tilde{k}_+^{\mathrm{min}}
  =
    {\qT^2}/(\sqrt{s}-m_H e^{+Y_H}) \,.
\end{align}
Consequently,
$\Delta\widetilde{\mathcal{I}}^{ \rho,\sigma}_{[\kappa],\{\alpha,\beta\}}$
contains all $\mT$ dependences.
In analysing the scaling behaviour, we find that
within the central region $e^{\pm Y_H}\sim\mathcal{O}(1)$, the integration
variables $k_{\pm}$ exhibit unambiguous scaling in each integration region.
Therefore, as will be illustrated in~Sec.~\ref{sec:red:qT:boundary},
the power expansion of
$\Delta\widetilde{\mathcal{I}}^{ \rho,\sigma}_{[\kappa],\{\alpha,\beta\}}$
follows an analogous pattern to those in Eqs.~\eqref{eq:asy:mom:c}-\eqref{eq:asy:mom:pdfs}.
Nevertheless,
$\widetilde{\mathcal{I}}^{ \rho,\sigma}_{[\kappa],\{\alpha,\beta\}}$
behaves differently.
From its expression in Eq.~\eqref{eq:def:I:central}, a number of different
scales have been enclosed in the integration range of
$\widetilde{\mathcal{I}}^{ \rho,\sigma}_{[\kappa],\{\alpha,\beta\}}$,
starting from the ultrasoft scale
${k^{\mathrm{min}}_+}\sim\mathcal{O}(\qT^2/m_H)$,
through the soft one $k_{\pm}\sim\mathcal{O}(\qT)$, and ending up with
the hard scale ${k^{\mathrm{max}}_+}\sim\mathcal{O}(m_H)$.
They all correspond to a distinct way of expanding the integrands.
To this end, in Sec.~\ref{sec:red:qT:central}, we will make use of
a set of momentum cutoffs to unambiguously treat them.

\subsection{Power corrections on the boundary}\label{sec:red:qT:boundary} 
 
We have mentioned above that besides the integrand itself,
the integration boundaries are other sources for power corrections.
In the following, we will thus expand the integral
$\Delta\widetilde{\mathcal{I}}^{\rho,\sigma}_{[\kappa],\{\alpha,\beta\}}$
within the central region $e^{\pm Y_H}\sim \mathcal{O}(1)$.
As illustrated in Eq.~\eqref{eq:def:I:boundary},
$\Delta\widetilde{\mathcal{I}}^{\rho,\sigma}_{[\kappa],\{\alpha,\beta\}}$
consists of the integral of $k_+$ over two intervals,
${k}_+ \in[\tilde{k}_+^{\mathrm{max}}, {k}_+^{\mathrm{max}}]$
and ${k}_+ \in[  {k}_+^{\mathrm{min}}, \tilde{k}_+^{\mathrm{min}}]$.
In each of them, the momentum $k^{\mu}$ of the emitted particle is
subject to an unambiguous scaling,
\begin{align}
  \label{eq:def:kc:boundary}
 {k}_+ \in[\tilde{k}_+^{\mathrm{max}}, {k}_+^{\mathrm{max}}]\;:\quad k_+\sim  \mathcal{O}(m_H)\,,\quad k_-\sim  \mathcal{O}(\qT^2/m_H)\,, \quad \kTvec=-\qTvec\,,\\
  \label{eq:def:kcbar:boundary}
 {k}_+ \in[  {k}_+^{\mathrm{min}}, \tilde{k}_+^{\mathrm{min}}]\;:\quad k_+\sim  \mathcal{O}(\qT^2/m_H)\,,\quad k_-\sim  \mathcal{O}(m_H)\,, \quad \kTvec=-\qTvec\,,
\end{align}
where the transverse recoil $\kTvec$ is determined by the momentum conservation. Considering that the momenta $k_{\pm}$ in
Eqs.~\eqref{eq:def:kc:boundary}-\eqref{eq:def:kcbar:boundary} are nearly corresponding to the hardest emission along the beam direction that
are kinematically allowed
(in a logarithmic sense), henceforth, we dub them the $n$-ultra-collinear
and $\nbar$-ultra-collinear modes, respectively.
In the following, we will make use of those two scaling laws to expand
$\Delta\widetilde{\mathcal{I}}^{\rho,\sigma}_{[\kappa],\{\alpha,\beta\}}$
within the low \qT~regime.

We first consider the $n$-ultra-collinear contribution,
applying Eq.~\eqref{eq:def:kc:boundary} to Eq.~\eqref{eq:def:I:boundary}
yields
\begin{equation}
  \label{eq:asyexp:deltaI:uc}
  \begin{split}
    \Delta \widetilde{\mathcal{I}}^{\rho,\sigma}_{[\kappa],\{\alpha,\beta\}}\Bigg|_{uc}
    \,=\,&
      \int_{\tilde{k}^{\mathrm{max}}_+}^{ {k}^{\mathrm{max}}_+}\,
      \frac{\done k_+}{k_+}\,  (k_-)^{\rho}  \, (k_+)^{\sigma} \,
      {F}^{(\alpha_n)}_{i/n,\beta_n}\left(k_++m_He^{-Y_H}\right)\,
      {F}^{(\alpha_\nbar)}_{j/\nbar,\beta_\nbar}\left( k_-+m_He^{+Y_H} \right)\\
    =\,&
      (\qT^2)^{\rho}\,
      \int_{0}^{ (m_H-\mT)e^{-Y_H}}\,
      \done\hat{k}_+\,
      (\hat{k}_+ + \tilde{k}^{\mathrm{max}}_+)^{\sigma-\rho-1} \\
    &\hspace*{20mm}\times\,
      {F}^{(\alpha_n)}_{i/n,\beta_n}\left(\hat{k}_++\sqrt{s}\right)\,
      {F}^{(\alpha_\nbar)}_{j/\nbar,\beta_\nbar}
      \left(
        \frac{\qT^2}{\hat{k}_+
        +\tilde{k}^{\mathrm{max}}_+}+ m_He^{+Y_H}
      \right)\\
    =\,&
      \sum_{\tilde{\alpha}_n,\tilde{\alpha}_\nbar=0}^{\infty}\,
      \frac{(\qT^2)^{\rho+\tilde{\alpha}_\nbar}}{\tilde{\alpha}_n!\,\tilde{\alpha}_\nbar!} \,
      {F}^{(\alpha_n+\tilde{\alpha}_n)}_{i/n,\beta_n}\left(\sqrt{s}\right)\,
      {F}^{(\alpha_\nbar+\tilde{\alpha}_\nbar)}_{j/\nbar,\beta_\nbar}\left( m_He^{+Y_H} \right)\,
      \mathcal{B}^{\rho,\sigma}_{+,\{\alpha,\beta\}} \,,
  \end{split}
\end{equation}
where in the second step the integration variable is transformed such that
the power suppressed residual terms can be extracted, i.e.\
\begin{equation}\label{eq:trans:kp}
  k_+
  \,\to\,
  \underbrace{\quad\hat{k}_+\quad}_{\sim\mathcal{O}(\qT^2/m_H)}
  \,+\,
  \underbrace{\quad\tilde{k}_+^{\mathrm{max}}\quad}_{\sim\mathcal{O}(m_H)}\,.
\end{equation}
Based on this scaling behaviour, the functions
${F}^{(\alpha_n)}_{i/n,\beta_n}$ and
${F}^{(\alpha_\nbar)}_{j/\nbar,\beta_\nbar}$ can be expanded
around $\sqrt{s}$ and $m_He^{+Y_H}$, respectively, in the last
step of Eq.~\eqref{eq:asyexp:deltaI:uc}, thereby removed from
the $\hat{k}_+$-integral.
The remaining $\hat{k}_+$ dependences are collected in
the function $\mathcal{B}^{\rho,\sigma}_{+,\{\alpha,\beta\}}$,
defined through
\begin{equation}\label{eq:def:Gp}
  \begin{split}
    \mathcal{B}^{\rho,\sigma}_{+,\{\alpha,\beta\}}
    \equiv&\;
      \int_{0}^{ (m_H-\mT)e^{-Y_H}}\done\hat{k}_+\;
      (\hat{k}_+ + \tilde{k}^{\mathrm{max}}_+)^{\sigma-\rho-\tilde{\alpha}_\nbar-1}
      (\hat{k}_+)^{\tilde{\alpha}_n}\\
    =&\;
      \int_{0}^{ (m_H-\mT)e^{-Y_H}}\done\hat{k}_+\;
      (\hat{k}_+)^{\tilde{\alpha}_n}\,
      \left(\tilde{k}^{\mathrm{max}}_+\right)^{\sigma-\rho-\tilde{\alpha}_\nbar-1} \\
    &\hspace*{10mm}\times
      \Bigg\{
        \theta(\sigma-\rho-\tilde{\alpha}_\nbar-1)\,
        \sum_{\gamma=0}^{\sigma-\rho-\tilde{\alpha}_\nbar-1}\,
        \frac{1}{\gamma!}\,
        \left(\frac{\hat{k}_+}{\tilde{k}^{\mathrm{max}}_+}\right)^{\gamma}\,
        \frac{\Gamma[\sigma-\rho-\tilde{\alpha}_\nbar]}
            {\Gamma[\sigma-\rho-\tilde{\alpha}_\nbar-\gamma]}\\
    &\hspace*{17mm}{}+\,
        \bar{\theta}(\sigma-\rho-\tilde{\alpha}_\nbar-1)\,
        \sum_{\gamma=0}^{\infty}\,
        \frac{(-1)^{\gamma}}{\gamma!}\,
        \frac{\Gamma[\gamma-\sigma+\rho+\tilde{\alpha}_\nbar+1]}{\Gamma[-\sigma+\rho+\tilde{\alpha}_\nbar+1]}\,
        \left(\frac{\hat{k}_+}{\tilde{k}^{\mathrm{max}}_+}\right)^{\gamma}\,
      \Bigg\} \,.
  \end{split}
\end{equation}
Herein, we use the Heaviside theta function $\theta(x)$,
with $\theta(x)=1$ for $x\ge0$ and $0$ otherwise, and
its counter part $\bar{\theta}(x)\equiv1-\theta(x)$ to account
for the expansion of the integrand.
Even though within the central rapidity region
$e^{\pm Y_H}\sim\mathcal{O}(1)$ and thus
$\tilde{k}^{\mathrm{max}}_+\sim\mathcal{O}(m_He^{\pm Y_H}) $,
the integrand of Eq.~\eqref{eq:def:Gp} has already admitted a
formally asymptotic series in
$({\hat{k}_+}/{\tilde{k}^{\mathrm{max}}_+})$, the participation
of $\mT$ into the upper boundary condition can invoke
inhomogeneous behaviour after completing the phase space integral.
To this end, a further expansion is necessary with the
aid of Eq.~\eqref{eq:asy:mom:mT}.
It follows that
\begin{equation}
  \begin{split}
    \label{eq:asyexp:deltaI:ec:int}
    \int_{0}^{(m_H-\mT)e^{-Y_H}}\,\done\hat{k}_+\,
    \hat{k}_+^{\tilde{\alpha}_n+\gamma}\,
    =&\,
      \sum_{\gamma_2=0}^{\infty}\,
      \sum_{\gamma_1=0}^{1+\gamma+\tilde{\alpha}_n}\,
      \frac{(-1)^{\gamma_1}}{\gamma_1!\,\gamma_2!}\,
      \left(\frac{\qT^2}{m_H^2}\right)^{\gamma_2}\,(e^{-Y_H}m_H)^{1+\tilde{\alpha}_n+\gamma }\\
    &\hspace*{20mm}\times\,
      \frac{\Gamma[1+\tfrac{\gamma_1}{2}]}
           {\Gamma[1-\gamma_2+\tfrac{\gamma_1}{2}]}\,
      \frac{\Gamma[\tilde{\alpha}_n+\gamma+1]}
           {\Gamma[\tilde{\alpha}_n+\gamma-\gamma_1+2]}\,.
  \end{split}
\end{equation}
Combining Eqs.~\eqref{eq:def:Gp}-\eqref{eq:asyexp:deltaI:ec:int}
with Eq.~\eqref{eq:asyexp:deltaI:uc}, we then accomplish the
expansion of the $n$-ultra-collinear sector in the small parameter
$(\qT/m_H)$.
An analogous procedure can be similarly applied onto the
$\nbar$-ultra-collinear case, with the exception of the
conversion of the integration variable $k_+$ to $k_-$
by means of the on-shell condition $k_+k_-=\qT^2$
and the substitution of $e^{+Y_H}$ for $e^{-Y_H}$ in
Eq.~\eqref{eq:asyexp:deltaI:ec:int}, as appropriate.

From those results, we observe that the boundary corrections
here are always associated with PDFs at the opposite end
and therefore are a priori expected to play an negligible role
in comparison to
$\widetilde{\mathcal{I}}^{\rho,\sigma}_{[\kappa],\{\alpha,\beta\}}$
in Eq.\ \eqref{eq:def:I:central}.
While this is expected to hold for the first few terms in the
power expansion, it is possible that the higher power corrections
from $\widetilde{\mathcal{I}}^{\rho,\sigma}_{[\kappa],\{\alpha,\beta\}}$
become of a similar size as the boundary corrections from
$\Delta\widetilde{\mathcal{I}}^{\rho,\sigma}_{[\kappa],\{\alpha,\beta\}}$,
such that both parts of contributions are essential to reproduce
the desired asymptotic behaviour of the \qT~spectrum.
In Sec.~\ref{sec:asyexp:numeric}, we will deliver a quantitative
assessment thereof.

\subsection{Power corrections over the interior domain}
\label{sec:red:qT:central} 
   
\subsubsection{Asymptotic expansion in momentum space}
\label{sec:asyexp:qT:mom}

We can now perform  the power expansion on the integral
$\widetilde{\mathcal{I}}^{\rho,\sigma}_{[\kappa],\{\alpha,\beta\}}$
in the low $\qT$~area within the central rapidity region
$e^{\pm Y_H}\sim \mathcal{O}(1)$.
As exhibited in Eq.~\eqref{eq:def:I:central},
$\widetilde{\mathcal{I}}^{\rho,\sigma}_{[\kappa],\{\alpha,\beta\}}$
encompasses a variety of scales along the integration path,
including the ultra-soft scale $k_+\sim \mathcal{O}(\qT^2/m_H)$,
the soft one $k_+\sim \mathcal{O}(\qT)$, and
the hard one $k_+\sim\mathcal{O}(m_H)$.
Every one of those momentum modes in practice prompts a
distinct expansion of
$\widetilde{\mathcal{I}}^{\rho,\sigma}_{[\kappa],\{\alpha,\beta\} }$.
Hence, in the spirit of the expansion by regions method
\cite{Beneke:1997zp,Smirnov:1998vk,Smirnov:1999bza,
  Smirnov:2002pj,Jantzen:2011nz}
and also according to the pattern of the resulting power
series, we categorise the phase space as follows~\footnote{%
  It is interesting to note that the analogous strategy in
  refining the dynamic regions has also been utilised previously
  in the literature~\cite{Balitsky:2009xg,Balitsky:2013fea,
  Balitsky:2017flc,Balitsky:2019ayf,Balitsky:2022vnb,Balitsky:2023hmh}
  in probing the $k_{\mathrm{T}}$ factorisation in the small-$x$
  area.},
\begin{equation}\label{eq:def:col-n:soft:col-nbar:cutoff}
  \begin{split}
    n\text{-collinear~mode}:\;\;
    \;&
      \nu_n<k_+\le \tilde{k}^{\mathrm{max}}_+;\,\vphantom{\frac{\qT^2}{\nu_{\nbar}}}\\
    \text{transitional range}:\;\;
    \;&
      \frac{\qT^2}{\nu_{\nbar}}<k_+\le \nu_n;\,\\
    \nbar\text{-collinear~mode}:\;\;
    \;&
      \tilde{k}^{\mathrm{min}}_+\le k_+\le  \frac{\qT^2}{\nu_{\nbar}} \;  .
  \end{split}
\end{equation}
Here, a pair of auxiliary scales $\{\nu_{n},\nu_{\nbar}\}$ are
introduced to separate the $n$- and $\nbar$-collinear regions
from the moderate range in between.
Throughout this paper, $\{\nu_{n},\nu_{\nbar}\}$ are chosen to
be of similar magnitudes to the hard scales $m_H e^{ \pm Y_H}$
but always small enough that in the intermediate domain of
Eq.~\eqref{eq:def:col-n:soft:col-nbar:cutoff} the light-cone components
$k_+$ and $k_-$ are both well contained within the convergence
radii of ${F}^{(\alpha_n)}_{i/n,\beta_n}$ and
${F}^{(\alpha_\nbar)}_{j/\nbar,\beta_\nbar}$, as stipulated in
Sec.~\ref{sec:setups}, more specifically,
\begin{align}\label{eq:def:scaling:nun:nunbar}
  m_H e^{-Y_H}\gtrsim\nu_n\sim \mathcal{O}(m_H)\gg\qT\,,
  \quad\quad
  m_H e^{Y_H}\gtrsim\nu_{\nbar}\sim \mathcal{O}(m_H)\gg\qT\,.
\end{align}
Alternative choices of those two auxiliary boundaries can be used
to examine the asymptotic behaviour of
$\widetilde{\mathcal{I}}^{\rho,\sigma}_{[\kappa],\{\alpha,\beta\}}$.
For example, the scales proposed in~\cite{Balitsky:2023hmh}
may impact the expressions of the individual sectors
defined in Eq.~\eqref{eq:def:col-n:soft:col-nbar:cutoff}
but will leave the resulting power series unchanged upon
summation of all ingredients.
Having chosen $\{\nu_{n},\nu_{\nbar}\}$ according to
Eq.~\eqref{eq:def:scaling:nun:nunbar}, the momentum
$k^{\mu}$ obeys the unambiguous scaling law
$k_+\sim\mathcal{O}(m_H)\gg k_-\sim \mathcal{O}(\qT^2/m_H)$
in the $n$-collinear mode, from which we can expand
${F}^{(\alpha_\nbar)}_{j/\nbar,\beta_\nbar}$ around the intrinsic
scale $m_H e^{+Y_H}$,
\begin{align}\label{eq:def:asyexp:c:mom}
  \widetilde{\mathcal{I}}^{\rho,\sigma}_{[\kappa],\{\alpha,\beta\} }\Bigg|_{c}\,
  =\,
    \sum_{\omega=\rho}^{\infty}\,(\qT^2)^{\omega}\,
    \widetilde{\mathcal{I}}^{\rho,\sigma,(\omega)}_{[\kappa],\{\alpha,\beta\}}
    \Bigg|_{c} \,,
\end{align}
where 
\begin{align}\label{eq:def:asyexp:c:mom:at:omega}
  \widetilde{\mathcal{I}}^{\rho,\sigma,(\omega)}_{[\kappa],\{\alpha,\beta\}}
  \Bigg|_{c} \,
  \equiv\,
    \int^{\tilde{k}^{\mathrm{max}}_+}_{\nu_n}\,
    \frac{\done k_+}{k_+}\,
    \frac{(k_+)^{\sigma-\omega}}{(\omega-\rho)!} \,
    {F}^{(\alpha_n)}_{i/n,\beta_n}\left(k_++m_H e^{-Y_H}\right)\,
    {F}^{(\alpha_\nbar+\omega-\rho)}_{j/\nbar,\beta_\nbar}\left(m_H e^{+Y_H}\right)\,.
\end{align}
Here, all $\qT$ dependences have been extracted into the prefactor
of Eq.~\eqref{eq:def:asyexp:c:mom} to express the manifest power accuracy
of the contribution.
$\widetilde{\mathcal{I}}^{\rho,\sigma,(\omega)}_{[\kappa],\{\alpha,\beta\}}$
collects all hard scales and thus is always of $\mathcal{O}(1)$.
It merits noting that in the central rapidity region
$e^{\pm Y_H}\sim\mathcal{O}(1)$, the lower boundary $\nu_n$ in
Eq.~\eqref{eq:def:asyexp:c:mom:at:omega} is well separated from
the upper one and thus there is no need to utilise a further expansion
here, in contrast to Eq.~\eqref{eq:asyexp:deltaI:uc}.

The same method can also be applied to the $\nbar$-collinear mode,
after transforming the integration variable from $k_+$ to $k_-$
through the on-shell condition $k_+ k_-=\qT^2$.
It evaluates to,
\begin{align}\label{eq:def:asyexp:cbar:mom}
  \widetilde{\mathcal{I}}^{\rho,\sigma}_{[\kappa],\{\alpha,\beta\}}
  \Bigg|_{\bar{c}}\,
  =\,
    \sum_{\omega=\sigma}^{\infty}\,(\qT^2)^{\omega}\,
    \widetilde{\mathcal{I}}^{\rho,\sigma,(\omega)}_{[\kappa],\{\alpha,\beta\}}
    \Bigg|_{\bar{c}}\,,
\end{align}
with
\begin{align}\label{eq:def:asyexp:cbar:mom:at:omega}
  \widetilde{\mathcal{I}}^{\rho,\sigma,(\omega)}_{[\kappa],\{\alpha,\beta\}}
  \Bigg|_{\bar{c}}\,
  \equiv\,
    \int^{\frac{\qT^2}{\tilde{k}^{\mathrm{min}}_+}}_{\nu_{\nbar}}\,
    \frac{\done k_-}{k_-}\,
    \frac{(k_-)^{\rho-\omega}}{(\omega-\sigma)!} \,
    {F}^{(\alpha_n+\omega-\sigma)}_{i/n,\beta_n}\left(m_H e^{-Y_H}\right)\,
    {F}^{(\alpha_\nbar)}_{j/\nbar,\beta_\nbar}\left(k_-+m_H e^{+Y_H}\right)\,.
\end{align}

This leaves the transitional domain, as defined in
Eq.~\eqref{eq:def:col-n:soft:col-nbar:cutoff}, to be calculated.
In light of the scaling rules of Eq.~\eqref{eq:def:scaling:nun:nunbar},
we can expand both ${F}^{(\alpha_n)}_{i/n,\beta_n}$ and
${F}^{(\alpha_\nbar)}_{j/\nbar,\beta_\nbar}$ around the
scales $m_H e^{\pm Y_H}$, more specifically,
\begin{equation}\label{eq:def:asyexp:s:mom1}
  \begin{split}
    \widetilde{\mathcal{I}}^{\rho,\sigma}_{[\kappa],\{\alpha,\beta\}}\Bigg|_{t}\,
    =\,&
      \sum_{\lambda,\eta=0}^{\infty}\,
      \int^{\nu_n}_{\frac{\qT^2}{\nu_{\nbar}}}\,
      \frac{\done k_+}{k_+}\,
      \frac{(k_-)^{\rho+\lambda}\,(k_+)^{\sigma+\eta}}{\lambda!\,\eta!} \,
      {F}^{(\alpha_n+\eta)}_{i/n,\beta_n}\left(m_H e^{-Y_H}\right)\,
      {F}^{(\alpha_\nbar+\lambda)}_{j/\nbar,\beta_\nbar}\left(m_H e^{+Y_H}\right)
      \,\\
    =\,&
      \sum_{\lambda,\eta=0}^{\infty}\,
      \frac{{F}^{(\alpha_n+\eta)}_{i/n,\beta_n}\left(m_H e^{-Y_H}\right)\,
            {F}^{(\alpha_\nbar+\lambda)}_{j/\nbar,\beta_\nbar}\left(m_H e^{+Y_H}\right)}
           {\lambda!\,\eta!} \,\\
    &\hspace*{8mm}\times\,
      \Bigg\{
        \underbrace{
          \frac{(\qT^2)^{\rho+\lambda}\,\nu_n^{\sigma+\eta-\rho-\lambda}}
               {\sigma+\eta-\rho-\lambda}\,
          \bar{\delta}^{\sigma+\eta}_{\rho+\lambda}
        }_{n\text{-col}}
        +\underbrace{
           \frac{(\qT^2)^{\sigma+\eta }\,\nu_{\nbar}^{\rho+\lambda-\sigma-\eta}}
                {\rho+\lambda-\sigma-\eta}\,
           \bar{\delta}^{\sigma+\eta}_{\rho+\lambda}
         }_{\nbar\text{-col}}
        +\underbrace{
           (\qT^2)^{\sigma+\eta}\,
           \ln\left[\frac{\nu_n\nu_{\nbar}}{\qT^2}\right]\,
           {\delta}^{\sigma+\eta}_{\rho+\lambda}
         }_{n\text{-col}\,,~\nbar\text{-col}\,,~\text{s}}\,
      \Bigg\}\,,\hspace*{-10mm}
  \end{split}
\end{equation}
where we have already grouped the expression according to their powers in
$\nu_n$ and $\nu_{\nbar}$.
$\delta^{\alpha}_{\beta}$ is the Kronecker Delta,
$\delta^{\alpha}_{\beta}=1$ if $\alpha=\beta$ and zero otherwise,
$\bar{\delta}^{\alpha}_{\beta}\equiv1-\delta^{\alpha}_{\beta}$ is
its complement.
We note that even for fixed values of $\lambda$ and $\eta$,
the phase space integral results in terms with a variety of
power accuracies.
For instance, in the first term on r.h.s.\ the power precision is
determined by the exponent of $\qT$, which is identical to that
of $k_-$ in the integrand.
Hence, the scaling of the emitted momentum $k^{\mu}$ here is
determined by the $n$-collinear mode $k_-\sim\mathcal{O}(\qT^2/m_H)$
and $k_+\sim\mathcal{O}(m_H)$.
Similarly, applying this scaling analysis to the second term,
its power precision is given by the exponent of $k_+$ and
the $\nbar$-collinear mode with $k_-\sim\mathcal{O}(m_H)$ and
$k_+\sim\mathcal{O}(\qT^2/m_H)$.
The interpretation of the scaling of the integration variable of
the last term of Eq.~\eqref{eq:def:asyexp:s:mom1} is more flexible
due to the presence of the constraint $\rho+\lambda=\sigma+\eta$.
Hence, the emitted parton can be constituted of either
the $n(\nbar)$-collinear particle or the soft one
$k_-\sim k_+\sim\mathcal{O}(\qT)$.
This inhomogeneity stems from our choice of the momentum cutoffs
$\nu_n$ and $\nu_{\nbar}$ in Eq.~\eqref{eq:def:scaling:nun:nunbar},
from which multiple dynamic modes are allowed to participate in
the transitional area, such as the $n$-collinear one,
$\nbar$-collinear one, or the soft one.
In principle, reducing $\nu_n$ and $\nu_{\nbar}$ down to
$\mathcal{O}(\qT)$ will filter out the collinear modes in the
transitional domain rendering it homogeneous again, at the price
of additional soft scales in both collinear sectors through the
boundary condition, necessitating a second expansion in the
small parameters $\nu_n/m_H$ and $\nu_\nbar/m_H$ within there.
It remains to be noted, while alternative choices of $\nu_n$ and
$\nu_{\nbar}$ may modify the dynamics in each sector, their combined
power series is invariant under this choice.
Hence, in this work we choose the prescription of
Eq.~\eqref{eq:def:scaling:nun:nunbar}, i.e.\ homogenous collinear
sectors, for simplicity.
 
By collecting and refactoring terms of a common power in \qT\
we can recast Eq.~\eqref{eq:def:asyexp:s:mom1} as
\begin{align}\label{eq:def:asyexp:s:mom2}
  \widetilde{\mathcal{I}}^{\rho,\sigma}_{[\kappa],\{\alpha,\beta\} }\Bigg|_{t}\,
  =\,
    \left(
      \sum_{\omega=\rho}^{\infty}\,
      (\qT^2)^{\omega}\,
      \widetilde{\mathcal{I}}^{\rho,\sigma,(\omega)}_{[\kappa],\{\alpha,\beta\}}
      \Bigg|_{ cs}\,
    \right)
    +
    \left(
      \sum_{\omega=\sigma}^{\infty}\,
      (\qT^2)^{\omega}\,
      \widetilde{\mathcal{I}}^{\rho,\sigma,(\omega)}_{[\kappa],\{\alpha,\beta\}}
      \Bigg|_{ \bar{c}s}\,
    \right)\,,
\end{align}
where
\begin{equation}\label{eq:def:asyexp:cs:mom}
  \begin{split}
  &\widetilde{\mathcal{I}}^{\rho,\sigma,(\omega)}_{[\kappa],\{\alpha,\beta\}}
  \Bigg|_{cs}
  =
    \sum_{\eta=0}^{\infty}
    \frac{{F}^{(\eta+\alpha_n)}_{i/n,\beta_n}\left(m_H e^{-Y_H}\right)
          {F}^{(\alpha_\nbar+\omega-\rho)}_{j/\nbar,\beta_\nbar}\left(m_H e^{+Y_H}\right)\,}
         {(\omega-\rho)!\,\eta!}
    \Bigg\{
      \frac{\nu_n^{\sigma+\eta-\omega}}{\sigma+\eta-\omega}
      \bar{\delta}^{\sigma+\eta}_{\omega}
      +\ln\left[\frac{\nu_n}{\qT}\right]{\delta}^{\sigma+\eta}_{\omega}\,
    \Bigg\}\,, \\
  &\widetilde{\mathcal{I}}^{\rho,\sigma,(\omega)}_{[\kappa],\{\alpha,\beta\}}
  \Bigg|_{\bar{c}s}
  =
    \sum_{\lambda=0}^{\infty}
    \frac{{F}^{(\alpha_n+\omega-\sigma)}_{i/n,\beta_n}\left(m_H e^{-Y_H}\right)
          {F}^{(\alpha_\nbar+\lambda)}_{j/\nbar,\beta_\nbar}\left(m_H e^{+Y_H}\right)}
          {(\omega-\sigma)!\,\lambda!}
    \Bigg\{
      \frac{\nu_{\nbar}^{\rho+\lambda-\omega}}{\rho+\lambda-\omega}
      \bar{\delta}^{\rho+\lambda}_{\omega}
      +\ln\left[\frac{\nu_{\nbar}}{\qT}\right]{\delta}^{\rho+\lambda}_{\omega}
    \Bigg\}\,.
  \end{split}
\end{equation}
Here we have re-organized the logarithmic terms in
Eq.~\eqref{eq:def:asyexp:s:mom1} into the $n$-collinear-soft ($cs$)
and $\nbar$-collinear-soft ($\bar{c}s$) sectors based on their
scale dependences.
At variance from the results in Eq.~\eqref{eq:def:asyexp:c:mom:at:omega} and
Eq.~\eqref{eq:def:asyexp:cbar:mom:at:omega} in the $n$- and $\nbar$-collinear
sectors, where the coefficients at each power consist of only the hard scales
$k_\pm$, respectively, Eq.~\eqref{eq:def:asyexp:cs:mom} here introduces
an additional dependence on $\ln\left({\nu_{n(\nbar)}}/{\qT}\right)$.
In order to facilitate numerical calculations of these coefficients,
we rewrite the infinite series in Eq.~\eqref{eq:def:asyexp:cs:mom} in
terms of the integrals over $k_\pm$, in analogy to
Eqs.~\eqref{eq:def:asyexp:c:mom:at:omega} and
\eqref{eq:def:asyexp:cbar:mom:at:omega}.
If $\omega<\sigma$ in Eq.~\eqref{eq:def:asyexp:cs:mom}, this
transformation is immediate,
\begin{equation}\label{eq:def:asyexp:cs:mom:reg}
  \begin{split}
    &
    \bar{\theta}(\omega-\sigma)\,
    \sum_{ \eta =0}^{\infty}\,
    \frac{{F}^{(\eta+\alpha_n)}_{i/n,\beta_n}\left(m_H e^{-Y_H}\right)\,
          {F}^{(\alpha_\nbar+\omega-\rho)}_{j/\nbar,\beta_\nbar}\left(m_H e^{+Y_H}\right)\,}
         {(\omega-\rho)!\,\eta!} \,
    \frac{\nu_n^{\sigma+\eta-\omega}}{\sigma+\eta-\omega}\,\\
    &\qquad\qquad\qquad
    \Longrightarrow\,
    \frac{\bar{\theta}(\omega-\sigma)}{(\omega-\rho)!}\,
    \int^{\nu_n}_0\,\frac{\done k_+}{k_+}\,
    k_+^{\sigma-\omega}\,
    {F}^{(\alpha_n)}_{i/n,\beta_n}\left(k_++m_H e^{-Y_H}\right)\,
    {F}^{(\alpha_\nbar+\omega-\rho)}_{j/\nbar,\beta_\nbar}\left(m_H e^{+Y_H}\right) \,.
  \end{split}
\end{equation}
Here we have removed $\bar{\delta}^{\sigma+\eta}_{\omega}$
from the expressions since it always equals unity while
$\omega<\sigma$ and $\eta\ge0$.
The same reasoning can also be used to neglect the logarithmic
contributions in Eq.~\eqref{eq:def:asyexp:cs:mom}.
However, it is not straightforward to apply
Eq.~\eqref{eq:def:asyexp:cs:mom:reg} in the remaining
space with $\omega\ge\sigma$ as the integrand is singular
in the limit $k_+\to0$.
To this end, we make use of the higher-ranked star distribution,
\begin{align}\label{eq:def:sDist}
  \int_0^{\Lambda}\,\done x\,
  \left[\frac{1}{x^m}\right]^{\nu}_{*}\,f(x)
  \equiv
  \int_{\nu}^{\Lambda}\,\done x\,\frac{f(x)}{x^m}
  +
  \int_0^{\nu}\,\done x\,\frac{1}{x^m}\,
  \left[f(x)-\sum_{n=0}^{m-1}\frac{x^n}{n!}f^{(n)}(0)\right]\,,
\end{align}
to maintain the end point analyticity.
Therein, $f(x)$ is assumed to be always differentiable at $x=0$.
With $m=1$, Eq.~\eqref{eq:def:sDist} reduces to the customary
star distribution proposed in \cite{DeFazio:1999ptt,Bosch:2004th}.
Equipped with eq.~\eqref{eq:def:sDist}, we are now able to convert
the $n$-collinear-soft contribution in the case $\omega\ge\sigma$
and $\eta> \omega-\sigma$,
\begin{equation}\label{eq:def:asyexp:cs:mom:dist}
  \begin{split}
    &{\theta}( \omega-\sigma)\,
    \sum_{\eta =\omega-\sigma+1}^{\infty}\,
    \frac{{F}^{(\eta+\alpha_n)}_{i/n,\beta_n}\left(m_H e^{-Y_H}\right)\,
          {F}^{(\alpha_\nbar+\omega-\rho)}_{j/\nbar,\beta_\nbar}\left(m_H e^{+Y_H}\right)\,}{(\omega-\rho)!\,\eta!}\,
    \frac{\nu_n^{\sigma+\eta-\omega}}{\sigma+\eta-\omega}\,\\
    &
    \qquad\qquad\Longrightarrow\,
    \frac{{\theta}(\omega-\sigma)}{(\omega-\rho)!}\,
    \int^{\nu_n}_0\,\done k_+\,
    \left[\frac{1}{k_+^{\omega-\sigma+1}}\right]^{\nu_n}_{*}\,
    {F}^{(\alpha_n)}_{i/n,\beta_n}\left( k_++m_H e^{-Y_H}\right)\,
    {F}^{(\alpha_\nbar+\omega-\rho)}_{j/\nbar,\beta_\nbar}\left(m_H e^{+Y_H}\right)\,.
  \end{split}
\end{equation}
The remaining contributions in Eq.~\eqref{eq:def:asyexp:cs:mom}
include the logarithms $\ln\left({\nu_{n}}/{\qT}\right)$
as well as the first $(\omega-\sigma)$ terms of the Taylor series.
We will refrain from transforming those contributions here since,
at a given power precision, the number of remaining terms are always
finite.
 
Combining above results we can recast the $n$-collinear-soft
contribution of Eq.~\eqref{eq:def:asyexp:cs:mom} as follows
\begin{equation}\label{eq:def:asyexp:cs:mom:int}
  \begin{split}
    \widetilde{\mathcal{I}}^{\rho,\sigma,(\omega)}_{[\kappa],\{\alpha,\beta\}}\Bigg|_{cs}
    =
    &\phantom{{}+}
      \frac{\bar{\theta}( \omega-\sigma)}{(\omega-\rho)!}\,
      \int^{\nu_n}_0\,\frac{\done k_+}{k_+}\,k_+^{\sigma-\omega}\,
      {F}^{(\alpha_n)}_{i/n,\beta_n}\left(k_++m_H e^{-Y_H}\right)\,
      {F}^{(\alpha_\nbar+\omega-\rho)}_{j/\nbar,\beta_\nbar}\left(m_H e^{+Y_H}\right)  \,\\
    &{}+
      \frac{{\theta}(\omega-\sigma)}{(\omega-\rho)!}\,
      \int^{\nu_n}_0\,\done k_+\,
      \left[\frac{1}{k_+^{\omega-\sigma+1}}\right]^{\nu_n}_{*}\,
      {F}^{(\alpha_n)}_{i/n,\beta_n}\left(k_++m_H e^{-Y_H}\right)\,
      {F}^{(\alpha_\nbar+\omega-\rho)}_{j/\nbar,\beta_\nbar}\left(m_H e^{+Y_H}\right)\,\\
    &{}+
      \frac{{\theta}(\omega-\sigma-1)}{(\omega-\rho)!}\,
      \sum_{\eta=0}^{\omega-\sigma-1}\,
      \frac{{F}^{(\alpha_n+\eta)}_{i/n,\beta_n}\left(m_H e^{-Y_H}\right)\,
            {F}^{(\alpha_\nbar+\omega-\rho)}_{j/\nbar,\beta_\nbar}\left(m_H e^{+Y_H}\right)\,}
           {\eta!} \,
      \frac{\nu_n^{\sigma+\eta-\omega}}{\sigma+\eta-\omega}\,\\
    &{}+
      \frac{{\theta}(\omega-\sigma)}{(\omega-\rho)!}\,
      \ln\left[\frac{\nu_n}{\qT}\right]\,
      \frac{{F}^{(\alpha_n+\omega-\sigma)}_{i/n,\beta_n}\left(m_H e^{-Y_H}\right)\,
            {F}^{(\alpha_\nbar+\omega-\rho)}_{j/\nbar,\beta_\nbar}\left(m_H e^{+Y_H}\right)\,}{(\omega-\sigma)!}  \,.
  \end{split}
\end{equation}
By analogy, the contribution of the $\nbar$-collinear-soft sector
in Eq.~\eqref{eq:def:asyexp:cs:mom} can be transformed to finite
integrals following an analogous path, giving
\begin{equation}\label{eq:def:asyexp:cbars:mom:int}
  \begin{split}
    \widetilde{\mathcal{I}}^{\rho,\sigma,(\omega)}_{[\kappa],\{\alpha,\beta\}}\Bigg|_{\bar{c}s}
    =
    &\phantom{{}+}
      \frac{\bar{\theta}(\omega-\rho)}{(\omega-\sigma)!}\,
      \int^{\nu_{\nbar} }_0\,\frac{\done k_-}{k_-}\,k_-^{\rho-\omega}\,
      {F}^{(\alpha_n+\omega-\sigma)}_{i/n,\beta_n}\left(m_H e^{-Y_H}\right)\,
      {F}^{(\alpha_\nbar)}_{j/\nbar,\beta_\nbar}\left(k_-+m_H e^{+Y_H}\right)\,\\
    &{}+
      \frac{{\theta}(\omega-\rho)}{(\omega-\sigma)!}\,
      \int^{\nu_{\nbar}}_0\,\done k_-\,
      \left[\frac{1}{k_-^{\omega-\rho+1}}\right]^{\nu_{\nbar}}_{*}\,
      {F}^{(\alpha_n+\omega-\sigma)}_{i/n,\beta_n}\left(m_H e^{-Y_H}\right)\,
      {F}^{(\alpha_\nbar)}_{j/\nbar,\beta_\nbar}\left(k_-+m_H e^{+Y_H}\right)\,\\
    &{}+
      \frac{{\theta}(\omega-\rho-1)}{(\omega-\sigma)!}\,
      \sum_{\lambda=0}^{\omega-\rho-1}\,
      \frac{{F}^{(\alpha_n+\omega-\sigma)}_{i/n,\beta_n}\left(m_H e^{-Y_H}\right)\,
            {F}^{(\alpha_\nbar+\lambda)}_{j/\nbar,\beta_\nbar}\left(m_H e^{+Y_H}\right)\,}{\lambda!} \,
      \frac{\nu_{\nbar}^{\rho+\lambda-\omega}}{\rho+\lambda-\omega}\,\\
    &{}+
      \frac{{\theta}(\omega-\rho)}{(\omega-\sigma)!}\,
      \ln\left[\frac{\nu_{\nbar}}{\qT}\right]\,
      \frac{{F}^{(\alpha_n+\omega-\sigma)}_{i/n,\beta_n}\left(m_H e^{-Y_H}\right)\,
            {F}^{(\alpha_\nbar+\omega-\rho)}_{j/\nbar,\beta_\nbar}\left(m_H e^{+Y_H}\right)\,}{(\omega-\rho)!}\,.
  \end{split}
\end{equation}
Using the above result we can now combine the
$n$-collinear sector of Eq.~\eqref{eq:def:asyexp:c:mom},
the $\nbar$-collinear sector of Eq.~\eqref{eq:def:asyexp:cbar:mom},
and the transitional sector of Eqs.~\eqref{eq:def:asyexp:cs:mom:int}
and \eqref{eq:def:asyexp:cbars:mom:int},
arriving at the expansion of
$\widetilde{\mathcal{I}}^{\rho,\sigma}_{[\kappa],\{\alpha,\beta\}}$
of Eq.~\eqref{eq:def:I:central} in powers of $\qT^2$ in the small \qT~domain,
\begin{align}\label{eq:def:asyexp:central:power:omega}
  \widetilde{\mathcal{I}}^{\rho,\sigma}_{[\kappa],\{\alpha,\beta\} }
  =
    \sum_{\omega}(\qT^2)^{\omega}\,
    \widetilde{\mathcal{I}}^{\rho,\sigma,(\omega)}_{[\kappa],\{\alpha,\beta\}}
  \equiv
    \sum_{\omega}(\qT^2)^{\omega}
    \Bigg\{
      \widetilde{\mathcal{I}}^{\rho,\sigma,(\omega)}_{[\kappa],\{\alpha,\beta\}}\Bigg|_{c}
      +
      \widetilde{\mathcal{I}}^{\rho,\sigma,(\omega)}_{[\kappa],\{\alpha,\beta\}}\Bigg|_{\bar{c}}
      +
      \widetilde{\mathcal{I}}^{\rho,\sigma,(\omega)}_{[\kappa],\{\alpha,\beta\}}\Bigg|_{cs}
      +
      \widetilde{\mathcal{I}}^{\rho,\sigma,(\omega)}_{[\kappa],\{\alpha,\beta\}}\Bigg|_{\bar{c}s}
    \Bigg\}\,.
\end{align}
At this point it merits reminding that the pair of auxiliary scales,
$\nu_n$ and $\nu_{\nbar}$, that we have employed to define and separate
the dynamic regions in Eq.~\eqref{eq:def:col-n:soft:col-nbar:cutoff}
appear explicitly in the expressions of all three individual domains.
For consistency's sake, it is of the essence to examine whether the
combination of all domains also exhibits such a dependence, or whether
this dependence, in fact, vanishes after summing over all the ingredients.
To this end, we take the derivatives of the r.h.s.\ of
Eq.~\eqref{eq:def:asyexp:central:power:omega} with respect to
$\nu_n$, and it evaluates to
\begin{equation}\label{eq:def:asyexp:central:power:omega:nundiff}
  \begin{split}
    &
      \frac{\partial}{\partial \nu_n}
      \widetilde{\mathcal{I}}^{\rho,\sigma,(\omega)}_{[\kappa],\{\alpha,\beta\}}\Bigg|_{c}
    =
      -\,\frac{\partial }{\partial \nu_n}
      \widetilde{\mathcal{I}}^{\rho,\sigma,(\omega)}_{[\kappa],\{\alpha,\beta\}}\Bigg|_{cs}
    =
      -\frac{(\nu_n)^{\sigma-\omega-1}}{(\omega-\rho)}
      {F}^{(\alpha_n)}_{i/n,\beta_n}\left(\nu_n+m_H e^{-Y_H}\right)
      {F}^{(\alpha_\nbar+\omega-\rho)}_{j/\nbar,\beta_\nbar}\left(m_H e^{+Y_H}\right),\\
    &
      \frac{\partial}{\partial \nu_n}
      \widetilde{\mathcal{I}}^{\rho,\sigma,(\omega)}_{[\kappa],\{\alpha,\beta\}}\Bigg|_{\bar{c}}
    =
      \frac{\partial}{\partial \nu_n}
      \widetilde{\mathcal{I}}^{\rho,\sigma,(\omega)}_{[\kappa],\{\alpha,\beta\}}\Bigg|_{\bar{c}s}
    =0\,,
  \end{split}
\end{equation}
where the results of Eqs.~\eqref{eq:def:asyexp:c:mom:at:omega}
and \eqref{eq:def:asyexp:cs:mom:int} are utilised to derive
the r.h.s.\ of Eq.~\eqref{eq:def:asyexp:central:power:omega:nundiff}.
In addition, the terms arising from the derivative of the
star distribution, defined in Eq.~\eqref{eq:def:sDist}, have been
cancelled against the $\nu_n$ dependences arising from the
logarithmic term and the truncated Taylor polynomials in
Eq.~\eqref{eq:def:asyexp:cs:mom:int}.
A similar result can also be found for the $\nu_\nbar$ dependence
of Eq.~\eqref{eq:def:asyexp:central:power:omega}, giving
\begin{equation}\label{eq:def:asyexp:central:power:omega:nunbardiff}
  \begin{split}
    &
      \frac{\partial}{\partial \nu_\nbar}
      \widetilde{\mathcal{I}}^{\rho,\sigma,(\omega)}_{[\kappa],\{\alpha,\beta\}}\Bigg|_{c}
    =
      \frac{\partial}{\partial \nu_\nbar}
      \widetilde{\mathcal{I}}^{\rho,\sigma,(\omega)}_{[\kappa],\{\alpha,\beta\}}\Bigg|_{cs}
    =
      0\,,\\
    &
      \frac{\partial}{\partial \nu_\nbar}
      \widetilde{\mathcal{I}}^{\rho,\sigma,(\omega)}_{[\kappa],\{\alpha,\beta\} }\Bigg|_{\bar{c}}
    =
      -\frac{\partial}{\partial \nu_\nbar}
      \widetilde{\mathcal{I}}^{\rho,\sigma,(\omega)}_{[\kappa],\{\alpha,\beta\} }\Bigg|_{\bar{c}s}
    =-
      \frac{(\nu_\nbar)^{\rho-\omega-1}}{(\omega-\sigma)!}
      {F}^{(\alpha_n+\omega-\sigma)}_{i/n,\beta_n}\left(m_H e^{-Y_H}\right)
      {F}^{(\alpha_\nbar)}_{j/\nbar,\beta_\nbar}\left(\nu_\nbar+m_H e^{+Y_H}\right) .
  \end{split}
\end{equation}
Combining the results of Eqs.~\eqref{eq:def:asyexp:central:power:omega:nundiff}
and \eqref{eq:def:asyexp:central:power:omega:nunbardiff}, respectively, we can
conclude that the coefficient
$\widetilde{\mathcal{I}}^{\rho,\sigma,(\omega)}_{[\kappa],\{\alpha,\beta\}}$
in Eq.~\eqref{eq:def:asyexp:central:power:omega} is indeed independent
of the choice of the separators $\nu_n$ and $\nu_\nbar$ at each power.
This agrees with the expectation from a direct expansion of
$\widetilde{\mathcal{I}}^{\rho,\sigma }_{[\kappa],\{\alpha,\beta\}}$,
currently beyond our capabilities as the PDFs are not rigorously
known analytically, that it should be a function of $\qT$, $s$, and $m_H$
only and independent of $\nu_n$ or $\nu_\nbar$.

\subsubsection{Rapidity regularisation and zero-bin subtraction}\label{sec:asyexp:qT:extrapl}
 
In the previous section, with the help of the auxiliary scales
$\nu_n$ and $\nu_\nbar$, we have derived the power series for
$\widetilde{\mathcal{I}}^{\rho,\sigma}_{[\kappa],\{\alpha,\beta\}}$
in Eq.~\eqref{eq:def:asyexp:central:power:omega} in the vicinity
$\qT=0$ at central rapidities, $e^{\pm Y_H}\sim\mathcal{O}(1)$.
Even though this strategy has accomplished an asymptotic expansion of
$\widetilde{\mathcal{I}}^{\rho,\sigma}_{[\kappa],\{\alpha,\beta\}}$,
and in turn the \qT~distribution, at NLO, it is not straightforward
to generalise this method to higher perturbative orders.
The presence of subdivisions of phase space through the
auxiliary scales $\nu_n$ and $\nu_\nbar$, generally inducing additional
scales to phase space and loop integrals, complicates the
calculations substantially.
Further, the absence of additional
constraints on the integration paths is also one of prerequisites
to establish the factorisation in a SCET-based analysis.
To this end, we will discuss the rearrangement of the ingredients
defined in Eqs.~\eqref{eq:def:asyexp:c:mom:at:omega} and
\eqref{eq:def:asyexp:cbar:mom:at:omega} as well as
Eqs.~\eqref{eq:def:asyexp:cs:mom:int} and
\eqref{eq:def:asyexp:cbars:mom:int} such that the dependence
on the auxiliary scales to subdivide the phase space is removed.
In particular, recalling our findings regarding the
$\nu_{n(\nbar)}$-independence of
$\widetilde{\mathcal{I}}^{\rho,\sigma}_{[\kappa],\{\alpha,\beta\}}$,
but not its components in
Eqs.~\eqref{eq:def:asyexp:central:power:omega:nundiff} and
\eqref{eq:def:asyexp:central:power:omega:nunbardiff},
such a rearrangement is warranted.
Although in the discussion below, we will still restrict ourselves
to the NLO constituent
$\widetilde{\mathcal{I}}^{\rho,\sigma}_{[\kappa],\{\alpha,\beta\}}$
extracted from Eq.~\eqref{eq:msq} as an illustrative example,
it is expected that the conclusion here can provide the theoretical
baseline for an analysis in more general situations.

We start by examining the collinear sectors.
Here, eliminating the
dependences of the auxiliary scales $\nu_n$ and $\nu_\nbar$
amounts to reducing the lower boundaries in
Eq.~\eqref{eq:def:asyexp:c:mom:at:omega} and
Eq.~\eqref{eq:def:asyexp:cbar:mom:at:omega} from their intrinsic
domains defined in Eq.~\eqref{eq:def:scaling:nun:nunbar} to the
origin.
To be precise, we split
\begin{align}\label{eq:def:asyexp:c:cbar:diff}
  \widetilde{\mathcal{I}}^{\rho,\sigma,(\omega)}_{[\kappa],\{\alpha,\beta\}}\Bigg|_{c} \,
  =\,
    \widetilde{\mathcal{G}}^{\rho,\sigma,(\omega)}_{[\kappa],\{\alpha,\beta\}}\Bigg|_{c} \,
    -\,
    \widetilde{\mathcal{I}}^{\rho,\sigma,(\omega)}_{[\kappa],\{\alpha,\beta\}}\Bigg|_{c0} \,,
  \qquad\quad
  \widetilde{\mathcal{I}}^{\rho,\sigma,(\omega)}_{[\kappa],\{\alpha,\beta\}}\Bigg|_{\bar{c}}\,
  =\,
    \widetilde{\mathcal{G}}^{\rho,\sigma,(\omega)}_{[\kappa],\{\alpha,\beta\}}\Bigg|_{\bar{c}}\,
    -\,
    \widetilde{\mathcal{I}}^{\rho,\sigma,(\omega)}_{[\kappa],\{\alpha,\beta\}}\Bigg|_{\bar{c}0}\,,
\end{align}
where
$\widetilde{\mathcal{G}}^{\rho,\sigma,(\omega)}_{[\kappa],\{\alpha,\beta\}}\big|_{c}$ and
$\widetilde{\mathcal{G}}^{\rho,\sigma,(\omega)}_{[\kappa],\{\alpha,\beta\}}\big|_{\bar{c}}$
collect the collinear contributions, integrating
the light-cone momentum over the entire range.
Their expressions read
\begin{equation}\label{eq:def:asyexp:c:at:omega:nured}
  \begin{split}
    \widetilde{\mathcal{G}}^{\rho,\sigma,(\omega)}_{[\kappa],\{\alpha,\beta\} }\Bigg|_{c}
    &\equiv
      \lim_{\tau\to0}\int^{\tilde{k}^{\mathrm{max}}_+}_{0}
      \frac{\done k_+}{k_+}
      \mathcal{R}(k_-,k_+,\tau)
      \frac{(k_+)^{\sigma-\omega}}{(\omega-\rho)!}
      {F}^{(\alpha_n)}_{i/n,\beta_n}\left(k_++m_H e^{-Y_H}\right)
      {F}^{(\alpha_\nbar+\omega-\rho)}_{j/\nbar,\beta_\nbar}\left(m_H e^{+Y_H}\right), \\
    \widetilde{\mathcal{G}}^{\rho,\sigma,(\omega)}_{[\kappa],\{\alpha,\beta\} }\Bigg|_{\bar{c}}
    &\equiv
    \lim_{\tau\to0}\int^{\frac{\qT^2}{\tilde{k}^{\mathrm{min}}_+}}_{0}
    \frac{\done k_-}{k_-}
    \mathcal{R}(k_-,k_+,\tau)
    \frac{(k_-)^{\rho-\omega}}{(\omega-\sigma)!}
    {F}^{(\alpha_n+\omega-\sigma)}_{i/n,\beta_n}\left( m_H e^{-Y_H}\right)
    {F}^{(\alpha_\nbar)}_{j/\nbar,\beta_\nbar}\left(k_-+m_H e^{+Y_H}\right).
  \end{split}
\end{equation}
Therein, in order to regularise the rapidity divergence
in the limit $k_{\pm}\to0$, we multiply the original
integrand by $\mathcal{R}(k_-,k_+,\tau)$, which is a
function of the light-cone momenta $k_{\pm}$ and a
real parameter $\tau$, satisfying,
\begin{align}\label{rap:def:feat}
  \lim_{\tau\to0}\,
  \mathcal{R}(k_-,k_+,\tau)\,
  =\,
    \left\{
    \begin{matrix}
      1 & \text{$k_{\pm}\neq 0$}\,,\\
      0 & \text{$k_{+}=0$~or~$k_{-}=0$}\,.
    \end{matrix}
    \right.
\end{align}
For the regions $k_{\pm}\neq 0$, the integrands in
Eq.~\eqref{eq:def:asyexp:c:at:omega:nured} are uniformly
convergent, which permits reducing the rapidity regulator
$\mathcal{R}(k_-,k_+,\tau)$ in the limit $\tau\to0$ and
restoring the integrands to their original forms.
Nevertheless, in the asymptotic area $k_{+}=0$~or~$k_{-}=0$,
the role of the rapidity regulator becomes indispensable,
ensuring that the phase space integrals in
Eq.~\eqref{eq:def:asyexp:c:at:omega:nured} are always well
defined.
The discussion in this section will emphasise rapidity
regulators that are able to preserve their expressions
in both the $n$-collinear and $\nbar$-collinear sectors
for all powers, such as
the analytic regulator of \cite{Becher:2010tm,Becher:2011dz},
the exponential regulator of \cite{Li:2016axz,Li:2016ctv},
and the pure rapidity regulator of \cite{Ebert:2018gsn,Moult:2019vou}.
We will dub them the conservative rapidity regulator
or conservative regularisation scheme (CRa) hereafter.
Other than that, there are also alternative proposals,
including the $\Delta$-regulator \cite{Chiu:2009yx,
  Echevarria:2011epo,Echevarria:2015usa,Echevarria:2015byo,
  Echevarria:2016scs}
and the $\eta$-regulator \cite{Chiu:2011qc,Chiu:2012ir}.
At this point, either the rapidity regulators themselves
or the regulator-dressed propagators will participate in
the power expansion such that in practice the rapidity
divergences can be regularised in varying strategies in
different ingredients from power to power.
They will be called dissipative rapidity regulator (DRa)
in this work.
We will postpone their investigation for now but discuss
the possible prescriptions in App.~\ref{sec:dis:reg}.
 
In Eq.\ \eqref{eq:def:asyexp:c:cbar:diff}, we have also
introduced
$\widetilde{\mathcal{I}}^{\rho,\sigma,(\omega)}_{[\kappa],\{\alpha,\beta\}}\big|_{c0}$ and
$\widetilde{\mathcal{I}}^{\rho,\sigma,(\omega)}_{[\kappa],\{\alpha,\beta\} }\big|_{\bar{c}0}$ in order to restore the equality after.
Their expressions read,
\begin{equation}\label{eq:def:asyexp:c:at:omega:nured:diff}
  \begin{split}
    &\widetilde{\mathcal{I}}^{\rho,\sigma,(\omega)}_{[\kappa],\{\alpha,\beta\} }\Bigg|_{c0}
    =
      \lim_{\tau\to0}\int^{\nu_n}_{0}
      \frac{\done k_+}{k_+}
      \mathcal{R}(k_-,k_+,\tau)
      \frac{(k_+)^{\sigma-\omega}}{(\omega-\rho)!}
      {F}^{(\alpha_n)}_{i/n,\beta_n}\left(k_++m_H e^{-Y_H}\right)
      {F}^{(\alpha_\nbar+\omega-\rho)}_{j/\nbar,\beta_\nbar}\left(m_H e^{+Y_H}\right), \\
    &\widetilde{\mathcal{I}}^{\rho,\sigma,(\omega)}_{[\kappa],\{\alpha,\beta\} }\Bigg|_{\bar{c}0}
    =
      \lim_{\tau\to0}\int^{\nu_{\nbar}}_{0}
      \frac{\done k_-}{k_-}
      \mathcal{R}(k_-,k_+,\tau)
      \frac{(k_-)^{\rho-\omega}}{(\omega-\sigma)!}
      {F}^{(\alpha_n+\omega-\sigma)}_{i/n,\beta_n}\left( m_H e^{-Y_H}\right)
      {F}^{(\alpha_\nbar)}_{j/\nbar,\beta_\nbar}\left(k_-+m_H e^{+Y_H}\right).
  \end{split}
\end{equation}
Since these functions are always associated with the $k_{\pm}$
integrals from the origin point up to the cutoffs $\nu_{n(\nbar)}$,
we will make use of the subscripts $``c0"$ $(``\bar{c}0")$ to
represent them hereafter.
From Eq.~\eqref{eq:def:asyexp:c:at:omega:nured:diff}, noting that
the integration variables $k_{\pm}$ are both well situated within
the convergence radii of ${F}_{i/n,\beta_n}$ and
${F}_{j/\nbar,\beta_\nbar}$, according to Sec.~\ref{sec:setups},
we can now perform the power expansion in $k_{\pm}$.
We then have,
\begin{equation}\label{eq:def:asyexp:c0:at:omega:nured:diff:decomp}
  \begin{split}
    &\widetilde{\mathcal{I}}^{\rho,\sigma,(\omega)}_{[\kappa],\{\alpha,\beta\} }\Bigg|_{c0}\,
    =\,
      \frac{{F}^{(\alpha_\nbar+\omega-\rho)}_{j/\nbar,\beta_\nbar}\left(m_H e^{+Y_H}\right)}{(\omega-\rho)!}\,
      \left\{
        \widetilde{\mathcal{I}}^{\rho,\sigma,(\omega)}_{[\kappa],\{\alpha,\beta\} }\Bigg|_{c0r}\,
        +\,
        \theta\left(\omega-\sigma\right)\,
        \widetilde{\mathcal{I}}^{\rho,\sigma,(\omega)}_{[\kappa],\{\alpha,\beta\} }\Bigg|_{c0d}\,
      \right\}\,,\\
    &\widetilde{\mathcal{I}}^{\rho,\sigma,(\omega)}_{[\kappa],\{\alpha,\beta\} }\Bigg|_{\bar{c}0}\,
    =\,
      \frac{{F}^{(\alpha_n+\omega-\sigma)}_{i/n,\beta_n}\left( m_H e^{-Y_H}\right)}{(\omega-\sigma)!}\,
      \left\{
        \widetilde{\mathcal{I}}^{\rho,\sigma,(\omega)}_{[\kappa],\{\alpha,\beta\} }\Bigg|_{\bar{c}0r}\,
        +\,
        \theta\left(\omega-\rho\right)\,
        \widetilde{\mathcal{I}}^{\rho,\sigma,(\omega)}_{[\kappa],\{\alpha,\beta\} }\Bigg|_{\bar{c}0d}\,
      \right\}\,,
  \end{split}
\end{equation}
where
\begin{equation}\label{eq:def:asyexp:c0r:at:omega:nured:diff}
  \begin{split}
    \widetilde{\mathcal{I}}^{\rho,\sigma,(\omega)}_{[\kappa],\{\alpha,\beta\}}\Bigg|_{c0r}
    &=\,
      \sum^{\infty}_{\eta=\mathrm{max}\{0,(\omega-\sigma+1)\}}
      \frac{\nu^{\sigma-\omega+\eta}_n}{\sigma-\omega+\eta}
      \frac{{F}^{(\alpha_n+\eta)}_{i/n,\beta_n}\left(m_H e^{-Y_H}\right)}{\eta!}\,,
      \\
    \widetilde{\mathcal{I}}^{\rho,\sigma,(\omega)}_{[\kappa],\{\alpha,\beta\}}\Bigg|_{\bar{c}0r}
    &=\,
      \sum^{\infty}_{\lambda=\mathrm{max}\{0,(\omega-\rho+1)\}}
      \frac{\nu^{\rho-\omega+\lambda}_\nbar}{\rho-\omega+\lambda}
      \frac{{F}^{(\alpha_\nbar+\lambda)}_{j/\nbar,\beta_\nbar}\left(m_H e^{+Y_H}\right)}{\lambda!}\,,
      \\
  \end{split}
\end{equation}
are the regular components, in which the regulator could be safely
removed by the means of Eq.\ \eqref{rap:def:feat}, and
\begin{equation}\label{eq:def:asyexp:c0d:at:omega:nured:diff}
  \begin{split}
    \widetilde{\mathcal{I}}^{\rho,\sigma,(\omega)}_{[\kappa],\{\alpha,\beta\}}\Bigg|_{c0d}
    &=\,
      \sum_{\eta=0}^{\omega-\sigma}\,
      \frac{{F}^{(\alpha_n+\eta)}_{i/n,\beta_n}\left(m_H e^{-Y_H}\right)}{\eta!}\,
      \left\{
        \lim_{\tau\to0}\int^{\nu_n}_{0}
        \frac{\done k_+}{k_+}
        \mathcal{R}(k_-,k_+,\tau)
        (k_+)^{\sigma-\omega+\eta}
      \right\}\,,
      \\
    \widetilde{\mathcal{I}}^{\rho,\sigma,(\omega)}_{[\kappa],\{\alpha,\beta\}}\Bigg|_{\bar{c}0d}
    &=\,
      \sum_{\lambda=0}^{\omega-\rho}\,
      \frac{{F}^{(\alpha_\nbar+\lambda)}_{j/\nbar,\beta_\nbar}\left(m_H e^{+Y_H}\right)}{\lambda!}\,
      \left\{
        \lim_{\tau\to0}\int^{\nu_\nbar}_{0}
        \frac{\done k_-}{k_-}
        \mathcal{R}(k_-,k_+,\tau)
        (k_-)^{\rho-\omega+\lambda}
      \right\}\,,
  \end{split}
\end{equation}
contain the (regulated) rapidity divergences at $k_{\pm}\to0$.
Here, the regulator $\mathcal{R}(k_-,k_+,\tau)$ is essential
and the respective integrals are therefore dependent on the
choice of the regularisation scheme.
Substituting Eq.~\eqref{eq:def:asyexp:c0r:at:omega:nured:diff}
into Eq.~\eqref{eq:def:asyexp:c0:at:omega:nured:diff:decomp},
we observe that the results for
$\widetilde{\mathcal{I}}^{\rho,\sigma,(\omega)}_{[\kappa],\{\alpha,\beta\}}\big|_{c0}$ and
$\widetilde{\mathcal{I}}^{\rho,\sigma,(\omega)}_{[\kappa],\{\alpha,\beta\}}\big|_{c0}$
still explicitly depend on the auxiliary scales $\nu_{n(\nbar)}$.
However, according to Eq.~\eqref{eq:def:asyexp:central:power:omega:nundiff},
all those dependences will drop out 
upon combining them with the contributions from the transitional
region of Eq.~\eqref{eq:def:asyexp:cs:mom}, and it follows that
\begin{align}\label{eq:id:mom:rsp}
  \widetilde{\mathcal{I}}^{\rho,\sigma,(\omega)}_{[\kappa],\{\alpha,\beta\} }\Bigg|_{ cs}
  +
  \widetilde{\mathcal{I}}^{\rho,\sigma,(\omega)}_{[\kappa],\{\alpha,\beta\} }\Bigg|_{\bar{c}s}
  -
  \widetilde{\mathcal{I}}^{\rho,\sigma,(\omega)}_{[\kappa],\{\alpha,\beta\} }\Bigg|_{c0}
  -
  \widetilde{\mathcal{I}}^{\rho,\sigma,(\omega)}_{[\kappa],\{\alpha,\beta\} }\Bigg|_{\bar{c}0}
  =
  \widetilde{\mathcal{G}}^{\rho,\sigma,(\omega)}_{[\kappa],\{\alpha,\beta\} }\Bigg|_{comb},
\end{align}
where on the r.h.s.,
$\widetilde{\mathcal{G}}^{\rho,\sigma,(\omega)}_{[\kappa],\{\alpha,\beta\} }\big|_{comb}$, comprises a set of one-fold integrals of $k_+$ or $k_-$ over
the range $[0,+\infty]$.
Its precise form will be specified below.

In the literature, there are two preferences
for expressing this combined result.
The first one, used extensively when using the
method of expansion by regions \cite{Beneke:1997zp,
  Smirnov:1998vk,Smirnov:1999bza,Smirnov:2002pj,
  Jantzen:2011nz} and
SCET~\cite{Bauer:2001yt,Bauer:2001ct,Bauer:2000yr,
  Bauer:2000ew,Bauer:2002nz,Beneke:2002ph,Beneke:2002ni,
  Bauer:2002aj,Lange:2003pk,Beneke:2003pa},
comprises the soft interaction $k_{\pm}\sim\mathcal{O}(\qT)$
in full as well as the subtraction terms to remove the overlap
with the collinear regions.
Hereafter, we will refer to the results from this formulation
as the full soft \FSmeth prescription.
It yields that
\begin{align}\label{eq:def:s:ccs:fs:rap}
  \widetilde{\mathcal{G}}^{\rho,\sigma,(\omega)}_{[\kappa],\{\alpha,\beta\} }\Bigg|_{comb}\,
  =\,
    \widetilde{\mathcal{G}}^{\rho,\sigma,(\omega)}_{[\kappa],\{\alpha,\beta\} }\Bigg|^{\FSmeth}_{s}\,
    -\,
    \widetilde{\mathcal{G}}^{\rho,\sigma,(\omega)}_{[\kappa],\{\alpha,\beta\} }\Bigg|^{\FSmeth}_{c\bar{c}s}\,.
\end{align}
Here
$\widetilde{\mathcal{G}}^{\rho,\sigma,(\omega)}_{[\kappa],\{\alpha,\beta\}}\big|^{\FSmeth}_{s}$
stands for the soft contribution, derived by expanding the
integrand in Eq.~\eqref{eq:def:I:central} in accordance with
the soft scaling $k_{\pm}\sim\mathcal{O}(\qT)$.
At the $\omega$th-power, the result reads,
\begin{align}\label{eq:def:s:fs:rap}
  \widetilde{\mathcal{G}}^{\rho,\sigma,(\omega)}_{[\kappa],\{\alpha,\beta\} }\Bigg|^{\FSmeth}_{s}\,
  =&\,
    \theta(\omega-\rho-\sigma)\,
    \sum_{\eta=0}^{(\omega-\rho-\sigma)}\,
    {F}^{(\alpha_n+\omega-\rho-\sigma-\eta)}_{i/n,\beta_n}\left(m_He^{-Y_H}\right)\,
    {F}^{(\alpha_\nbar +\eta)}_{j/\nbar,\beta_\nbar}\left(m_He^{+Y_H}\right)\,
    \lim_{\tau\to0}\,
    \int^{\infty}_{0}\,
    \frac{\done k_+}{k_+}  \, \nnb\\
  &\times
    \mathcal{R}(k_-,k_+,\tau) \,
    \frac{(k_-)^{\rho+\eta-\omega}}{\eta !}  \,
    \frac{ (k_+)^{ -\rho -\eta }}{(\omega-\rho-\sigma-\eta)!}\,,
\end{align}
where we have extracted the factor of $(\qT^{2})^{\omega}$
from the integrals in Eq.~\eqref{eq:def:s:fs:rap}, in line
with the definition of the power series in
Eq.~\eqref{eq:def:asyexp:central:power:omega}.
Given the expression in Eq.~\eqref{eq:def:s:fs:rap}, we can
obtain an expression for
$\widetilde{\mathcal{G}}^{\rho,\sigma,(\omega)}_{[\kappa],\{\alpha,\beta\}}\big|^{\FSmeth}_{c\bar{c}s}$
by comparing Eq.~\eqref{eq:def:s:ccs:fs:rap} with
Eq.~\eqref{eq:id:mom:rsp}.
As this subtrahend at LP is intimately associated with the
soft limit of the collinear functions and thus the zeroth-bin
of the label momentum within the position-momentum space hybrid representation,
$\widetilde{\mathcal{G}}^{\rho,\sigma,(\omega)}_{[\kappa],\{\alpha,\beta\}}\big|^{\FSmeth}_{c\bar{c}s}$
is frequently dubbed the zero-bin contribution
\cite{Manohar:2006nz,Idilbi:2007yi,Idilbi:2007ff,Chiu:2009yx}
or the soft subtraction term \cite{Ji:2004xq,Ji:2004wu,
  Idilbi:2007yi,Idilbi:2007ff,Echevarria:2011epo}.
 
On the other hand, there are also alternative proposals in
the literature~\cite{Vladimirov:2021hdn,Vladimirov:2023aot,
  Goerke:2017ioi,Inglis-Whalen:2020rpi,Inglis-Whalen:2021bea}.
There, a decomposition of Eq.~\eqref{eq:id:mom:rsp} without
an explicit soft sector is chosen.
The virtue of such a scheme is that in absence of the soft
contribution the soft-collinear interaction vertices do not
need to be expanded using multiple scaling regimes, and the factorisation of the distinct dynamic modes is thus
considerably simplified.
In this work, we will dub this approach the one in the
non-soft \NSmeth prescription. It follows that,
\begin{align}
\label{eq:def:s:ccs:ns:rap}
 \widetilde{\mathcal{G}}^{\rho,\sigma,(\omega)}_{[\kappa],\{\alpha,\beta\} }\Bigg|_{comb}\,
 =\,
-\,
  \widetilde{\mathcal{G}}^{\rho,\sigma,(\omega)}_{[\kappa],\{\alpha,\beta\} }\Bigg|^{\NSmeth}_{c\bar{c}}\,.
\end{align}
Hence, the interior contribution
$\widetilde{\mathcal{I}}^{\rho,\sigma,(\omega)}_{[\kappa],\{\alpha,\beta\}}$
in Eq.~\eqref{eq:def:I:central} consists of only the collinear
sectors in Eq.~\eqref{eq:def:asyexp:c:at:omega:nured} as well as
$\widetilde{\mathcal{G}}^{\rho,\sigma,(\omega)}_{[\kappa],\{\alpha,\beta\}}\big|^{\NSmeth}_{c\bar{c}}$.
Since
$\widetilde{\mathcal{G}}^{\rho,\sigma,(\omega)}_{[\kappa],\{\alpha,\beta\}}\big|^{\NSmeth}_{c\bar{c}}$ here effectively removes the overlap
between the $n$-collinear and $\nbar$-collinear contributions,
it is also named the overlap subtraction term in
\cite{Goerke:2017ioi,Inglis-Whalen:2020rpi,Inglis-Whalen:2021bea}.
Notwithstanding, in view of the functional similarity of
$\widetilde{\mathcal{G}}^{\rho,\sigma,(\omega)}_{[\kappa],\{\alpha,\beta\}}\big|^{\NSmeth}_{c\bar{c}}$ and
$\widetilde{\mathcal{G}}^{\rho,\sigma,(\omega)}_{[\kappa],\{\alpha,\beta\}}\big|^{\FSmeth}_{c\bar{c}s}$,
we will not further distinguish the terminologies in this paper,
such as the overlapping subtraction, the zero-bin contribution, or
the soft remover.
Instead, they will be all regarded to be conceptually equivalent
but evaluated within different prescriptions.

Comparing Eq.~\eqref{eq:def:s:ccs:ns:rap} with
Eq.~\eqref{eq:def:s:ccs:fs:rap}, we note that the subtraction
terms in those two schemes possess the following relation,
\begin{align}\label{eq:def:ns:fs:trans}
  \widetilde{\mathcal{G}}^{\rho,\sigma,(\omega)}_{[\kappa],\{\alpha,\beta\}}\Bigg|^{\NSmeth}_{c\bar{c}}\,
  =\,
    \widetilde{\mathcal{G}}^{\rho,\sigma,(\omega)}_{[\kappa],\{\alpha,\beta\}}\Bigg|^{\FSmeth}_{c\bar{c}s}\,
    -\,
    \widetilde{\mathcal{G}}^{\rho,\sigma,(\omega)}_{[\kappa],\{\alpha,\beta\}}\Bigg|^{\FSmeth}_{s}\,.
\end{align}
In the following, we will be focused on the derivation of
$\widetilde{\mathcal{G}}^{\rho,\sigma,(\omega)}_{[\kappa],\{\alpha,\beta\}}\big|^{\NSmeth}_{c\bar{c}}$.
The expressions within the \FSmeth prescription can be obtained
by means of the relationship above.


We begin with the transformation of
$\widetilde{\mathcal{I}}^{\rho,\sigma,(\omega)}_{[\kappa],\{\alpha,\beta\}}\big|_{cs}$ and
$\widetilde{\mathcal{I}}^{\rho,\sigma,(\omega)}_{[\kappa],\{\alpha,\beta\}}\big|_{ \bar{c}s}$.
As illustrated in Eq.~\eqref{eq:def:asyexp:cs:mom},
$\widetilde{\mathcal{I}}^{\rho,\sigma,(\omega)}_{[\kappa],\{\alpha,\beta\}}\big|_{cs}$ comprises the logarithmic terms in the case of
${\omega}={(\sigma+\eta)}$ and powers of the auxiliary scales
$\nu_{n(\nbar)}$ otherwise.
The latter contribution can be further categorised based
on the relation between $\omega$ and $(\sigma+\eta)$,
more specifically,
\begin{equation}\label{eq:reexp:sc1}
  \begin{split}
    &\hspace*{-5mm}\sum_{\eta =0}^{\infty}
    \frac{{F}^{(\eta+\alpha_n)}_{i/n,\beta_n}\left( m_H e^{-Y_H}\right)
          {F}^{(\alpha_\nbar+\omega-\rho)}_{j/\nbar,\beta_\nbar}\left(m_H e^{+Y_H}\right)\,}
         {(\omega-\rho)!\,\eta!}
    \frac{\nu_n^{\sigma+\eta-\omega}}{\sigma+\eta-\omega}
    \bar{\delta}^{\sigma+\eta}_{\omega}\\
    =\,&
      \frac{{F}^{(\alpha_\nbar+\omega-\rho)}_{j/\nbar,\beta_\nbar}\left(m_H e^{+Y_H}\right)}{(\omega-\rho)!}
      \Bigg\{
        \sum_{\eta=\mathrm{max}\{0,(\omega-\sigma+1)\}}^{\infty}
        \frac{{F}^{(\alpha_n+\eta)}_{i/n,\beta_n}\left( m_H e^{-Y_H}\right)}
             {\eta!}
        \frac{\nu_n^{\sigma+\eta-\omega}}{\sigma+\eta-\omega}\\
    &\hspace*{30mm}{}
        -\theta(\omega-\sigma-1)\sum_{\eta=0}^{\omega-\sigma-1}
        \frac{{F}^{(\alpha_n+\eta)}_{i/n,\beta_n}\left(m_H e^{-Y_H}\right)}
             {\eta!\,}
        \lim_{\tau\to0}\int^{\infty}_{\nu_n}\frac{\done k_+}{k_+}
        \mathcal{R}(k_-,k_+,\tau)(k_+)^{\sigma-\omega+\eta}
      \Bigg\}\,.\hspace*{-20mm}
  \end{split}
\end{equation}
Therein, within the curly brackets, the first term collects
all the contributions with $\omega<(\sigma+\eta)$, whilst the
second term collects all contributions with $\omega>(\sigma+\eta)$
and has been transformed into its integral form.
Aiming at a representation for
$\widetilde{\mathcal{I}}^{\rho,\sigma,(\omega)}_{[\kappa],\{\alpha,\beta\}}\big|_{cs}$
that can manifestly cancel against the $\nu_n$ dependences in
Eq.~\eqref{eq:def:asyexp:c0r:at:omega:nured:diff} and
Eq.~\eqref{eq:def:asyexp:c0d:at:omega:nured:diff}, the rapidity
regulator $\mathcal{R}(k_-,k_+,\tau)$ is introduced here for
bookkeeping purposes.
Once the relationship $\omega<(\sigma+\eta)$ is satisfied,
$\mathcal{R}(k_-,k_+,\tau)$ always stays inactive in the
limit $\tau\to0$.

An analogous rearrangement is also applicable to
$\widetilde{\mathcal{I}}^{\rho,\sigma,(\omega)}_{[\kappa],\{\alpha,\beta\} }\big|_{ \bar{c}s}$
in Eq.~\eqref{eq:def:asyexp:cs:mom}.
It yields that
\begin{equation}\label{eq:reexp:scbar1}
  \begin{split}
    &\hspace*{-5mm}\sum_{\lambda =0}^{\infty}
    \frac{{F}^{(\alpha_n+\omega-\sigma)}_{i/n,\beta_n}\left( m_H e^{-Y_H}\right)
          {F}^{(\alpha_\nbar+\lambda)}_{j/\nbar,\beta_\nbar}\left(m_H e^{+Y_H}\right)}
         {(\omega-\sigma)!\,\lambda!}
    \frac{\nu_{\nbar}^{\rho+\lambda-\omega}}{\rho+\lambda-\omega}
    \bar{\delta}^{\rho+\lambda}_{\omega}\\
    =\,&
      \frac{{F}^{(\alpha_n+\omega-\sigma)}_{i/n,\beta_n}\left( m_H e^{-Y_H}\right)}{(\omega-\sigma)!}\,
      \Bigg\{
        \sum_{\lambda=\mathrm{max}\{0,(\omega-\rho+1)\}}^{\infty}
        \frac{{F}^{(\alpha_\nbar+\lambda)}_{j/\nbar,\beta_\nbar}\left(m_H e^{+Y_H}\right)}
             {\lambda!}
        \frac{\nu_{\nbar}^{\rho+\lambda-\omega}}{\rho+\lambda-\omega}\\
    &\hspace*{30mm}{}
        -\theta(\omega-\rho-1)\sum_{\lambda=0}^{\omega-\rho-1}\,
        \frac{{F}^{(\alpha_\nbar+\lambda)}_{j/\nbar,\beta_\nbar}\left(m_H e^{+Y_H}\right)}
            {\lambda!}
        \lim_{\tau\to0}\int^{\infty}_{\nu_\nbar}\frac{\done k_-}{k_-}
        \mathcal{R}(k_-,k_+,\tau)(k_-)^{\rho+\lambda-\omega}
      \Bigg\}\,.\hspace*{-20mm}
  \end{split}
\end{equation}
We can now recast the logarithmic contributions of
$\widetilde{\mathcal{I}}^{\rho,\sigma,(\omega)}_{[\kappa],\{\alpha,\beta\}}\big|_{cs}$ and
$\widetilde{\mathcal{I}}^{\rho,\sigma,(\omega)}_{[\kappa],\{\alpha,\beta\}}\big|_{\bar{c}s}$
in integral form,
as presented in Eqs.~\eqref{eq:def:asyexp:s:mom2}
and \eqref{eq:def:asyexp:cs:mom}.
It is worth noting at this point that the logarithmic terms
from both sectors always and only appear together once
$\omega\ge\rho$ and $\omega\ge\sigma$ simultaneously.
Therefore, in the following, we cope with them simultaneously, i.e., 
\begin{equation}\label{eq:reexp:sc:scbar:log}
  \begin{split}
  &\hspace*{-5mm}
  \frac{{F}^{(\alpha_n+\omega-\sigma)}_{i/n,\beta_n}\left(m_H e^{-Y_H}\right)
        {F}^{(\alpha_\nbar+\omega-\rho)}_{j/\nbar,\beta_\nbar}\left(m_H e^{+Y_H}\right)\,}{(\omega-\rho)!\,(\omega-\sigma)!}
  \Bigg\{
    \ln\left[\frac{\nu_n}{\qT}\right]\,
    +\,
    \ln\left[\frac{\nu_{\nbar}}{\qT}\right]
  \Bigg\}
  \\
  =\,&
    \frac{{F}^{(\alpha_n+\omega-\sigma)}_{i/n,\beta_n}\left(m_H e^{-Y_H}\right)
          {F}^{(\alpha_\nbar+\omega-\rho)}_{j/\nbar,\beta_\nbar}\left(m_H e^{+Y_H}\right)\,}{(\omega-\rho)!\,(\omega-\sigma)!}\\
  &\times\;
    \Bigg\{
      -\lim_{\tau\to0}\int^{\infty}_{\nu_n}\frac{\done k_+}{k_+}\,
       \mathcal{R}(k_-,k_+,\tau)
      +\lim_{\tau\to0}\int^{\infty}_{0}\frac{\done k_+}{k_+}\,
       \mathcal{R}(k_-,k_+,\tau)
      -\lim_{\tau\to0}\int^{\infty}_{\nu_\nbar}\frac{\done k_-}{k_-}\,
       \mathcal{R}(k_-,k_+,\tau)
    \Bigg\}\,,
  \end{split}
\end{equation}
where in the last step we reconvert the logarithms into their
integral form and employ $\mathcal{R}(k_-,k_+,\tau)$ to regulate
their possibly singular behaviour in the limit $k_{\pm}\to 0$.
 
Comparing the above expressions with
Eq.~\eqref{eq:def:asyexp:c0r:at:omega:nured:diff}, we observe
that terms containing powers of the auxiliary scale $\nu_{n(\nbar)}$
take the same form in Eqs.~\eqref{eq:reexp:sc1} and \eqref{eq:reexp:scbar1}
and Eq.~\eqref{eq:def:asyexp:c0r:at:omega:nured:diff}, and
hence cancel upon combination.
Similarly, the dependences of the integral on the auxiliary scales
through their boundaries in Eqs.~\eqref{eq:reexp:sc1}-\eqref{eq:reexp:sc:scbar:log} can be assimilated in their entirety
by combining them with the corresponding integrals in
Eq.~\eqref{eq:def:asyexp:c0d:at:omega:nured:diff}.
Therefore, we arrive at,
\begin{align}\label{eq:def:c0:cbar0:ns}
  \widetilde{\mathcal{G}}^{\rho,\sigma,(\omega)}_{[\kappa],\{\alpha,\beta\}}\Bigg|^{\NSmeth}_{c\bar{c}}\,
  =\,
    \widetilde{\mathcal{G}}^{\rho,\sigma,(\omega)}_{[\kappa],\{\alpha,\beta\}}\Bigg|^{\NSmeth}_{c0}\,
    +\,
    \widetilde{\mathcal{G}}^{\rho,\sigma,(\omega)}_{[\kappa],\{\alpha,\beta\}}\Bigg|^{\NSmeth}_{\bar{c}0}\,,
\end{align}
where
\begin{equation}\label{eq:def:c0:ns}
  \begin{split}
    \widetilde{\mathcal{G}}^{\rho,\sigma,(\omega)}_{[\kappa],\{\alpha,\beta\}}\Bigg|^{\NSmeth}_{c0}\,
    =\,&
      \theta\left(\omega-\sigma\right)\,
      \sum_{\eta=0}^{\omega-\sigma}\,
      \sum_{\lambda=0}^{\omega-\rho}\,
      \frac{{F}^{(\alpha_n+\eta)}_{i/n,\beta_n}\left(m_H e^{-Y_H}\right)\,
            {F}^{(\alpha_\nbar+\lambda)}_{j/\nbar,\beta_\nbar}\left(m_H e^{+Y_H}\right)}
           {\eta!\,\lambda!}\\
    &\hspace*{30mm}\times\,
      \lim_{\tau\to0}\int^{\infty}_{0}\frac{\done k_+}{k_+}
      \left(1-\frac{\delta_{\omega}^{\sigma+\eta}}{2}\right)\,
      \mathcal{R}(k_-,k_+,\tau)\,(k_+)^{\sigma-\omega+\eta}\,
      \delta^{\omega}_{\lambda+\rho}\,,\\
    \widetilde{\mathcal{G}}^{\rho,\sigma,(\omega)}_{[\kappa],\{\alpha,\beta\}}\Bigg|^{\NSmeth}_{\bar{c}0}\,
    =\,&
      \theta\left(\omega-\rho\right)\,
      \sum_{\eta=0}^{\omega-\sigma}\,
      \sum_{\lambda=0}^{\omega-\rho}\,
      \frac{{F}^{(\alpha_n+\eta)}_{i/n,\beta_n}\left( m_H e^{-Y_H}\right)
            {F}^{(\alpha_\nbar+\lambda)}_{j/\nbar,\beta_\nbar}\left(m_H e^{+Y_H}\right)}
           {\eta!\,\lambda!}\\
    &\hspace*{30mm}\times\,
      \lim_{\tau\to0}\int^{\infty}_{0}\frac{\done k_-}{k_-}
      \left(1-\frac{\delta_{\omega}^{\rho+\lambda}}{2}\right)\,
      \mathcal{R}(k_-,k_+,\tau)\,(k_-)^{\rho-\omega+\lambda}\,
      \delta^{\omega}_{\eta+\sigma}\,.
  \end{split}
\end{equation}
Here, we have divided
$\widetilde{\mathcal{G}}^{\rho,\sigma,(\omega)}_{[\kappa],\{\alpha,\beta\}}\big|^{\NSmeth}_{c\bar{c}}$
into two pieces,
$\widetilde{\mathcal{G}}^{\rho,\sigma,(\omega)}_{[\kappa],\{\alpha,\beta\}}\big|^{\NSmeth}_{c0}$ and
$\widetilde{\mathcal{G}}^{\rho,\sigma,(\omega)}_{[\kappa],\{\alpha,\beta\}}\big|^{\NSmeth}_{\bar{c}0}$,
inheriting the structure from
Eq.~\eqref{eq:def:asyexp:c0d:at:omega:nured:diff}.
At variance with
Eq.~\eqref{eq:def:asyexp:c0d:at:omega:nured:diff}, however,
additional factors $\delta_{\omega}^{\sigma+\eta}$ and
$\delta_{\omega}^{\rho+\lambda}$ are present in the integrands
now, induced by the unbounded integrals of
Eq.~\eqref{eq:reexp:sc:scbar:log}.
During the derivation, we have evenly distributed these terms
into the $n$-collinear and $\nbar$-collinear sectors symmetrically,
incurring a factor $\tfrac{1}{2}$ in Eq.~\eqref{eq:def:c0:ns}.
Alternative assignments are in principle possible, which may
impact the expressions in the individual sectors but will leave
the sum in Eq.~\eqref{eq:def:c0:cbar0:ns} invariant.

Equipped with the collinear functions of
Eq.~\eqref{eq:def:asyexp:c:at:omega:nured} and the subtrahends of
Eq.~\eqref{eq:def:c0:ns}, we are now able to re-express the
interior contribution
$\widetilde{\mathcal{I}}^{\rho,\sigma }_{[\kappa],\{\alpha,\beta\} }$
as follows,
\begin{align}
  \label{eq:def:asyexp:central:power:omega:rap}
  \widetilde{\mathcal{I}}^{\rho,\sigma }_{[\kappa],\{\alpha,\beta\}}
  =\,
    \sum^{\infty}_{\omega=\rho}
    \left(\qT^2\right)^{\omega}
    \left[\widetilde{\mathcal{G}}^{\rho,\sigma,(\omega)}_{[\kappa],\{\alpha,\beta\} }\Bigg|_{c}
    -
    \widetilde{\mathcal{G}}^{\rho,\sigma,(\omega)}_{[\kappa],\{\alpha,\beta\}}\Bigg|^{\NSmeth}_{c0}\right]
    +
    \sum^{\infty}_{\omega=\sigma}
    \left(\qT^2\right)^{\omega}
    \left[\widetilde{\mathcal{G}}^{\rho,\sigma,(\omega)}_{[\kappa],\{\alpha,\beta\} }\Bigg|_{\bar{c}}
    -
    \widetilde{\mathcal{G}}^{\rho,\sigma,(\omega)}_{[\kappa],\{\alpha,\beta\}}\Bigg|^{\NSmeth}_{\bar{c}0}\right].
\end{align}
Herein, we only show the expressions in the NS scheme for
brevity, which can be straightforwardly converted into the
FS ones through Eq.~\eqref{eq:def:ns:fs:trans}.
Referring back to Eq.~\eqref{eq:def:asyexp:central:power:omega},
we have now found a formulation in which the individual terms
on the r.h.s.\ of Eq.~\eqref{eq:def:asyexp:central:power:omega:rap}
are free of any dependence on the auxiliary scales $\nu_{n(\nbar)}$
along the integration path.
However, in return, they are now subject to the choice of
rapidity regularisation scheme implemented through
$\mathcal{R}(k_-,k_+,\tau)$.
As will be discussed in Secs.~\ref{sec:app:prap} and \ref{sec:app:exp},
the regulator $\mathcal{R}(k_-,k_+,\tau)$ also depends on
scales $\tilde{\nu}_n$ and $\tilde{\nu}_\nbar$ in practice,
which bears resemblance to those in
Eq.~\eqref{eq:def:col-n:soft:col-nbar:cutoff} and can
effectively concentrate the integrand in each sector onto
its intrinsic domain, akin to the conventional dimensional
regularisation~\cite{tHooft:1972tcz}.

Finally, we can now analyse the scaling of $k_{\pm}$ in each
sector on the r.h.s.\ of
Eq.~\eqref{eq:def:asyexp:central:power:omega:rap}.
Since the collinear functions here are derived by extrapolating
the lower boundaries of Eq.~\eqref{eq:def:asyexp:c:mom:at:omega}
and Eq.~\eqref{eq:def:asyexp:cbar:mom:at:omega}, the integration
variables in Eq.~\eqref{eq:def:asyexp:c:at:omega:nured} observe
the same scaling rule as those in momentum space.
More explicitly, we have $k_+\sim\mathcal{O}(1)$ and
$k_-\sim\mathcal{O}(\qT^2/m_H)$ for the $n$-collinear element
and
$k_+\sim\mathcal{O}(\qT^2/m_H)$ and
$k_-\sim\mathcal{O}(1)$ in the $\nbar$-collinear case.
\changed{
To determine the scaling in the subtraction terms of Eq.~\eqref{eq:def:c0:ns},
it merits noting that the integration variables $k_{\pm}$ therein
have been expanded in the arguments of both
${F}_{i/n,\beta_n}$ and ${F}_{j/\nbar,\beta_\nbar}$, and also that
the function ${F}_{j/\nbar,\beta_\nbar}$ (${F}_{i/n,\beta_n}$) in
the ``$c0$" (``$\bar{c}0$") sector observes the same scaling pattern
as that in the $n$-($\nbar$-)collinear contribution of
Eq.~\eqref{eq:def:asyexp:c:at:omega:nured}.
Therefore, we can interprete the variables $k_{\pm}$ in
Eq.~\eqref{eq:def:c0:ns} from
the dual scaling, expanding first the integrand with
the collinear scaling and fixing the $\omega$th-power correction,
and then applying a second power expansion in line with soft scaling
$k_{\pm}\sim\mathcal{O}(\qT)$ retaining all contributions below or
equal to the $(2\omega)$th-power.
}

\begin{figure}[h!]
  \centering
  \begin{subfigure}{0.49\textwidth}
    \centering
    \includegraphics[width=.6\linewidth, height=0.4\linewidth]{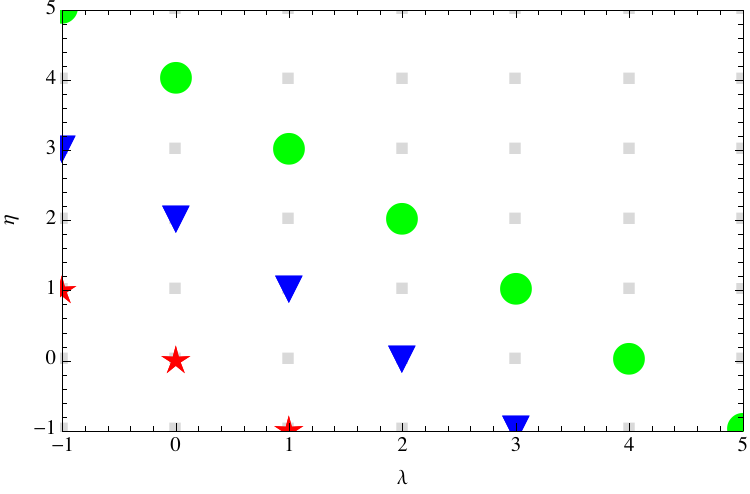}
     \caption{Subtrahends in the soft scaling}
    \label{fig:zbsub:soft}
  \end{subfigure}
  \begin{subfigure}{0.49\textwidth}
    \centering
    \includegraphics[width=.6\linewidth, height=0.4\linewidth]{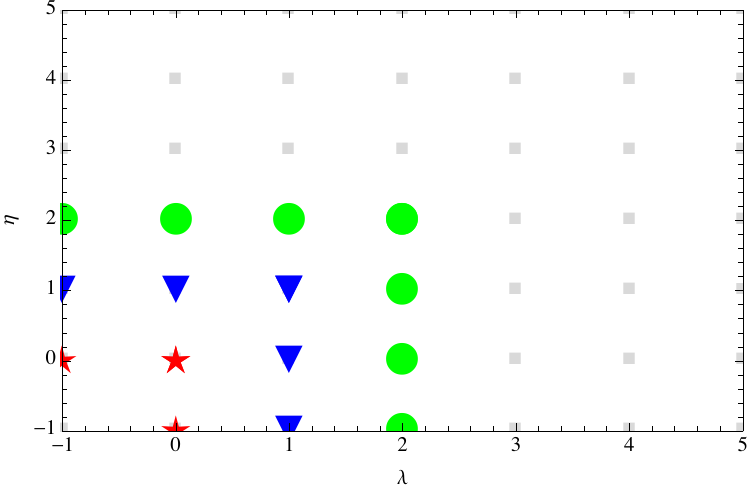}
     \caption{Subtrahends in the dual scaling}
    \label{fig:zbsub:duo}
      \end{subfigure}
  \caption{
    Organisation of zero-bin subtrahends in different scalings.
    Every point in the map stands for a index-pair
    $(\lambda,\eta)$ in Eq.~\eqref{eq:def:c0:ns} which characterises
    the subtraction procedure entailed by the expansion of
    $\widetilde{\mathcal{I}}^{\lambda,\eta}_{[\kappa],\{\alpha,\beta\}}$
    with integer $\lambda,\eta\ge -1$ in the small \qT~domain.
    As examples, the red stars highlight the subleading (\NLP) power
    corrections, the blue triangles represent the sub-subleading
    (\NNLP) ones, and the green dots indicate the sub-sub-subleading
    (\NNNLP) ones.
  }
  \label{fig:zbsub:scalings}
\end{figure}

It should be emphasised that at a lower power accuracy,
the dual scaling can be reduced to the soft-only one
especially in the case $\rho=\sigma=\chi$ with integer $\chi\ge -1$,
which however does not hold in general.
As an example, Fig.\ 1 illustrates the case $\chi=-1$, which
is the minimum of $\chi$ allowed by the squared amplitudes
in Eq.~\eqref{eq:msq} and thus concerns the leading power
approximation in the small \qT~domain.
We observe that, on the lowest level, both of schemes consist
of the origin $\lambda=\eta=-1$ and thus can be considered to
be equivalent here.
However, with the power accuracy growing, the soft scaling
forces the indices $\lambda$ and $\eta$ to align along the
diagonals, whereas the dual scaling organises the indices
of the zero-bin subtrahend along rectangular edges,
clearly separating the two.
Those distinct patterns can subsequently evaluate to different
zero-bin subtrahends for general choices of rapidity regulators,
which therefore defies our interpretation of the integration
variables of Eq.~\eqref{eq:def:c0:ns} in a purely soft scaling
prescription.

In order to systematically implement these scaling rules,
we introduce the power expansion operator
$\widehat{\mathbf{T}}_{i}^{(\omega)}$, with the subscript
$i$ running over the sectors $\{c,\bar{c},s\}$, acting on
the object following it and projecting out the contribution
at the $\omega$th-power in line with the scalings indicated
by $i$.
In this way, the components of
Eq.~\eqref{eq:def:asyexp:central:power:omega:rap} can be
recast into,
\begin{equation}\label{eq:def:asyexp:c:at:omega:nured:Texp}
  \begin{split}
    \left(\qT^2\right)^{\omega}\,
    \widetilde{\mathcal{G}}^{\rho,\sigma,(\omega)}_{[\kappa],\{\alpha,\beta\}}\Bigg|_{c} \,
    &=\,
      \lim_{\tau\to0}\,\int^{\tilde{k}^{\mathrm{max}}_+}_{0}\,
      \frac{\done k_+}{k_+}\,
      \mathcal{R}(k_-,k_+,\tau)\,
      \widehat{\mathbf{T}}_{c}^{(\omega)}\,
      I^{ \rho,\sigma}_{[\kappa],\{\alpha,\beta\}}\,,
    \\
    \left(\qT^2\right)^{\omega}\,
    \widetilde{\mathcal{G}}^{\rho,\sigma,(\omega)}_{[\kappa],\{\alpha,\beta\}}\Bigg|_{\bar{c}} \,
    &=\,
      \lim_{\tau\to0}\,\int^{\frac{\qT^2}{\tilde{k}^{\mathrm{min}}_+}}_{0}\,
      \frac{\done k_-}{k_-}\,
      \mathcal{R}(k_-,k_+,\tau)\,
      \widehat{\mathbf{T}}_{\bar{c}}^{(\omega)}\,
      I^{ \rho,\sigma}_{[\kappa],\{\alpha,\beta\}}\,,
  \end{split}
\end{equation}
and
\begin{equation}\label{eq:def:c0:ns:Texp}
  \begin{split}
    \left(\qT^2\right)^{\omega}\,
    \widetilde{\mathcal{G}}^{\rho,\sigma,(\omega)}_{[\kappa],\{\alpha,\beta\}}\Bigg|_{c0} \,
    & =\,
      \theta\left(\omega-\sigma \right)
      \lim_{\tau\to0}\int^{\infty}_{0}
      \frac{\done k_+}{k_+}\;
      \mathcal{R}(k_-,k_+,\tau)
      \sum_{\overline{\omega}=(\rho+\sigma) }^{2\omega}
      \left(1-\frac{\delta_{\overline{\omega}}^{2\omega}}{2}\right)
      \widehat{\mathbf{T}}_{s}^{(\overline{\omega})}
      \widehat{\mathbf{T}}_{c}^{(\omega)}
      I^{\rho,\sigma}_{[\kappa],\{\alpha,\beta\}}\,,\\
    \left(\qT^2\right)^{\omega}\,
    \widetilde{\mathcal{G}}^{\rho,\sigma,(\omega)}_{[\kappa],\{\alpha,\beta\}}\Bigg|_{\bar{c}0} \,
    & =\,
      \theta\left(\omega-\rho \right)
      \lim_{\tau\to0}\int^{\infty}_{0}
      \frac{\done k_-}{k_-}\;
      \mathcal{R}(k_-,k_+,\tau)
      \sum_{\overline{\omega}=(\rho+\sigma) }^{2\omega}
      \left(1-\frac{\delta_{\overline{\omega}}^{2\omega}}{2}\right)
      \widehat{\mathbf{T}}_{s}^{(\overline{\omega})}
      \widehat{\mathbf{T}}_{\bar{c}}^{(\omega)}
      I^{\rho,\sigma}_{[\kappa],\{\alpha,\beta\}}\,,
  \end{split}
\end{equation}
where $I^{\rho,\sigma}_{[\kappa],\{\alpha,\beta\}}$ is the
integrand of
$\widetilde{\mathcal{I}}^{ \rho,\sigma}_{[\kappa],\{\alpha,\beta\}}$
defined in Eq.~\eqref{eq:def:I:central},
\begin{align}\label{eq:def:subjt}
  I^{\rho,\sigma}_{[\kappa],\{\alpha,\beta\}}\equiv\,
  (k_-)^{\rho}\,(k_+)^{\sigma}\,
  {F}^{(\alpha_n)}_{i/n,\beta_n}\left(k_++m_He^{-Y_H}\right)\,
  {F}^{(\alpha_\nbar)}_{j/\nbar,\beta_\nbar}\left(k_-+m_He^{+Y_H}\right)\,.
\end{align}
We note that after the re-arrangements of
Eqs.~\eqref{eq:def:asyexp:c:at:omega:nured:diff}-\eqref{eq:id:mom:rsp} and
Eqs.~\eqref{eq:reexp:sc1}-\eqref{eq:reexp:sc:scbar:log}, the results in
Eqs.~\eqref{eq:def:asyexp:c:at:omega:nured:Texp}-\eqref{eq:def:c0:ns:Texp}
present a form that is distinct from that observed in 
Eqs.~\eqref{eq:def:asyexp:c:mom:at:omega}-\eqref{eq:def:asyexp:cbar:mom:at:omega}
as well as
Eqs.~\eqref{eq:def:asyexp:cs:mom:int}-\eqref{eq:def:asyexp:cbars:mom:int}
obtained via the momentum cutoffs $\nu_n$ and $\nu_\nbar$.
This is relevant in particular for the ``$sc$" and ``$s\bar{c}$"
sectors in
Eqs.~\eqref{eq:def:asyexp:cs:mom:int}-\eqref{eq:def:asyexp:cbars:mom:int},
despite hybrid scaling having been used in the transitional range.
As illustrated in Eq.~\eqref{eq:def:asyexp:s:mom1}, the resulting coefficient
$\widetilde{\mathcal{I}}^{\rho,\sigma,(\omega)}_{[\kappa],\{\alpha,\beta\}}\big|_{t}$
invokes only the integrands at the same power accuracy once the scaling
for the integration variable is given.
Contrarily, the situation in Eq.~\eqref{eq:def:c0:ns:Texp}
is quite different, where, after applying the soft scaling,
the results
$\widetilde{\mathcal{G}}^{\rho,\sigma,(\omega)}_{[\kappa],\{\alpha,\beta\}}\big|_{c0}$ and
$\widetilde{\mathcal{G}}^{\rho,\sigma,(\omega)}_{[\kappa],\{\alpha,\beta\}}\big|_{\bar{c}0}$
comprise not only corrections at the $\omega$th-power, but
also those on a lower level.

In order to interpret this structural difference, it merits recalling
that in deriving Eqs.~\eqref{eq:def:asyexp:c:mom:at:omega}
and \eqref{eq:def:asyexp:cbar:mom:at:omega} as well as
Eqs.~\eqref{eq:def:asyexp:cs:mom:int} and \eqref{eq:def:asyexp:cbars:mom:int}
we have introduced a set of dynamic regions, see
Eq.~\eqref{eq:def:col-n:soft:col-nbar:cutoff}, such that
within these restricted phase space domains the assigned
scalings are always effective before and after the phase
space integration.
However, in calculating the subtrahends in Eq.~\eqref{eq:def:c0:ns:Texp},
in particular after the combination with
Eqs.~\eqref{eq:reexp:sc1}-\eqref{eq:reexp:sc:scbar:log}, all
boundaries that separate the different scalings of
the integration variables cancel out.
As a result, the ``$c0$" and ``$ \bar{c}0$" expressions of
Eq.~\eqref{eq:def:c0:ns:Texp} mainly comprise contributions
from the rapidity extremities $k_{\pm}\to\pm0$, correlated
to the same positions of the collinear functions in
Eq.~\eqref{eq:def:asyexp:c:at:omega:nured:Texp},
rather than emphasising any integration segments in the physical
domain.
In light of this, the scaling laws implemented in
Eq.~\eqref{eq:def:c0:ns:Texp} should be conceived more of a
prescription guiding us to organise the zero-bin subtrahends
in order to subdue any unphysical asymptotic behaviour in
the collinear sectors and averting
the possibility of these sectors generating any non-trivial
remainders for the resulting power series.

It is worth noting that the integrands in
Eq.~\eqref{eq:def:c0:ns:Texp} appear to contradict the
homogeneity condition from the method of expansion by
regions~\cite{Beneke:1997zp,Smirnov:1998vk,Smirnov:1999bza,
  Smirnov:2002pj,Jantzen:2011nz}.
In this method, the asymptotic behaviour of the Feynman
or phase space integrals is associated with a set of regions
along the integration path.
From this one is able to expand the integrands before
completing the integration and at a given power precision,
the resulting power coefficients only concern the expanded
integrands of the same power accuracy, more specifically,
\begin{align}\label{eq:def:asyexp:homoexp}
 \widetilde{\mathcal{F}}^{(\omega)}\sim \sum_{i}\int\done\Phi\, \widehat{\mathbf{T}}_{i}^{(\omega)}\,f\,,
\end{align}
where $\widetilde{\mathcal{F}}^{(\omega)}$ denotes the
$\omega$th term in the power series, with $f$ standing for
a given integrand and $\Phi$ encoding a group of integration
variables.
In this paper, such a one-to-one correspondence between
the power accuracy of the integrands and that of
$\widetilde{\mathcal{F}}^{(\omega)}$ will be referred to as
the homogeneity of the asymptotic expansion hereafter.

On the contrary, a region analysis where
the dimensional regulator uniquely governs the divergences
incurred by the reduced integrands, such homogeneity is found to
generally hold in a variety of kinematical limits, such as the
off-shell large-momentum expansion and the large-mass limits
with either an internal or external heavy parton~\cite{Smirnov:2002pj}.
The reason comes in part from the fact that the overlapping
sectors for those asymptotic regimes, according to
\cite{Jantzen:2011nz}, admit the unbounded integrals over the
multiple-expanded integrands, which, using dimensional
regularisation, are all scaleless and thus vanish.
In principle, if the homogeneity is also desired for the
small~\qT\ expansion, one can embed a special group of
rapidity operators into Eq.~\eqref{eq:def:c0:ns:Texp},
such as the analytic \cite{Becher:2010tm,Becher:2011dz}
and pure rapidity \cite{Ebert:2018gsn,Moult:2019vou} cases,
from which, mimicking the dimensional regulator, the integrals
in the zero-bin subtrahends of Eq.~\eqref{eq:def:c0:ns:Texp}
become scaleless and are thus eliminated.
Subsequently, homogeneity is restored in the small~\qT\
expansion and the power coefficients are thereby recast as,
\begin{align}
 \label{eq:def:asyexp:central:power:omega:rap:hom}
 \widetilde{\mathcal{I}}^{\rho,\sigma,(\omega)}_{[\kappa],\{\alpha,\beta\}}\,
 =\,
  \widetilde{\mathcal{G}}^{\rho,\sigma,(\omega)}_{[\kappa],\{\alpha,\beta\}}\Bigg|^{\langle\mathrm{hom.rap.}\rangle}_{c}\,
  +\,
  \widetilde{\mathcal{G}}^{\rho,\sigma,(\omega)}_{[\kappa],\{\alpha,\beta\}}
  \Bigg|^{\langle\mathrm{hom.rap.}\rangle}_{\bar{c}}\,.
\end{align}
This work will dub this set of rapidity regulators the
homogeneous ones, as labeled in the superscripts above.
However, the above form is not generic.
For instance, once the exponential regulator
\cite{Li:2016axz,Li:2016ctv} is put in place, both the
``$c0$" and ``$\bar{c}0$" sectors in
Eq.~\eqref{eq:def:c0:ns:Texp} will make non-trivial and
indispensable contributions, from which all unphysical
singularities generated in the collinear sectors of
Eq.~\eqref{eq:def:asyexp:c:at:omega:nured:Texp} can be
eliminated at each power, according to
Eq.~\eqref{eq:def:asyexp:c:cbar:diff}.
In this sense, the decomposition in
Eqs.~\eqref{eq:def:asyexp:central:power:omega:rap:hom}
can only be appropriate for the asymptotic expansion
with a homogeneous regulator,  while the generalised
recipe, that will hold on a generic choice of rapidity
regulators, is presented in
Eq.~\eqref{eq:def:asyexp:central:power:omega:rap}.

\subsubsection{Comparison with existing results}
\label{sec:asyexp:comp}
      
We will now confront the zero-bin subtrahend derived in
Eqs.~\eqref{eq:def:asyexp:central:power:omega:rap}-\eqref{eq:def:c0:ns:Texp}
with those proposed in the literature.

At \LP, the zero-bin contribution at NLO involves only the
master integral
$\widetilde{\mathcal{I}}^{\rho,\sigma,(\omega)}_{[\kappa],\{\alpha,\beta\}}$
with $\rho=\sigma=\omega=-1$ in the process $pp\to H+X$,
and it has been demonstrated in~\cite{Manohar:2006nz,Idilbi:2007yi,
  Idilbi:2007ff,Chiu:2009yx,Echevarria:2011epo} that the
zero-bin subtrahend can be constructed by taking the soft
limit of the integrand.
It follows that,
\begin{align}\label{eq:def:c0:cbar0:ns:lp}
  \frac{1}{\qT^2}
  \widetilde{\mathcal{G}}^{-1,-1,(-1)}_{[\kappa],\{\alpha,\beta\} }\Bigg|^{\NSmeth}_{c\bar{c}}
  \xRightarrow[]{\text{~~soft~~}}
  \lim_{\tau\to0}\,\int^{\infty}_{0}\,
  \frac{\done k_-}{k_-}\,\mathcal{R}(k_-,k_+,\tau)\,
  \frac{
    {F}^{(\alpha_n)}_{i/n,\beta_n}\left(m_He^{-Y_H}\right)
    {F}^{(\alpha_\nbar)}_{j/\nbar,\beta_\nbar}\left(m_He^{+Y_H} \right)
  }
  {k_-\,k_+}\,.
\end{align}
This result can also be reproduced from
Eq.~\eqref{eq:def:c0:ns:Texp}.
At \LP, the repeated application of the expansion operators in
Eq.~\eqref{eq:def:c0:ns:Texp} does not produce any power of the
integration variables $k_{\pm}$ and therefore yields the same
powers of $k_\pm$ as that in Eq.~\eqref{eq:def:c0:cbar0:ns:lp}.
Moreover, owing to the relation $2\omega=\overline{\omega}=-2$,
the $\delta$-functions in Eq.~\eqref{eq:def:c0:ns:Texp} are
both contributing a factor of $(1/2)$ from each sector, which
add up to unity after combining the results.
This thus leads to the exact same expression of
Eq.~\eqref{eq:def:c0:cbar0:ns:lp}.

An alternative scheme to calculate the zero-bin subtraction
up to \NLP was proposed in \cite{Inglis-Whalen:2022vyn,
  Inglis-Whalen:2021bea}.
In this method, the asymptotic behaviour of the small
\qT~regime is assumed to be entirely governed by the
$n$- and $\nbar$-collinear momenta.
Subsequently, in order to remove the redundant overlapping
contributions, each collinear ingredient is expanded in
accordance with the scaling of the other one in the opposite
direction and then their sum is averaged, more specifically,
\begin{equation}\label{eq:def:c0:cbar0:ns:oppcol}
  \begin{split}
    \left(\qT^2\right)^{\omega}\,
    \widetilde{\mathcal{G}}^{\rho,\sigma,(\omega)}_{[\kappa],\{\alpha,\beta\}}\Bigg|^{\NSmeth}_{c\bar{c}}\,
    &\xRightarrow[]{\text{~~opp.col.~~}}\,
      \frac{\theta\left(\omega-\rho \right)\theta\left(\omega-\sigma\right)}{2}\,
      \lim_{\tau\to0}\,\int^{\infty}_{0}\,
      \frac{\done k_+}{k_+}\,\mathcal{R}(k_-,k_+,\tau)\,
    \\
    &\hspace*{20mm}\times\;
      \Bigg\{
        \left(
          \sum_{\overline{\omega}=\sigma}^{\omega}\,
          \widehat{\mathbf{T}}_{\bar{c}}^{(\overline{\omega})}
        \right)
        \widehat{\mathbf{T}}_{c}^{(\omega)}
        +
        \left(
          \sum_{\overline{\omega}=\rho}^{\omega}\,
          \widehat{\mathbf{T}}_{c}^{(\overline{\omega})}
        \right)
        \widehat{\mathbf{T}}_{\bar{c}}^{(\omega)}
    \\
    &\hspace*{28mm}{}
        +
        \widehat{\mathbf{T}}_{c}^{(\omega)}\,
        \left(
          \sum_{\overline{\omega}=\sigma}^{\omega-1}
          \widehat{\mathbf{T}}_{\bar{c}}^{(\overline{\omega})}
        \right)
        +
        \widehat{\mathbf{T}}_{\bar{c}}^{(\omega)}
        \left(
          \sum_{\overline{\omega}=\rho}^{\omega-1}\,
          \widehat{\mathbf{T}}_{c}^{(\overline{\omega})}
        \right)
      \Bigg\}\;
      I^{\rho,\sigma}_{[\kappa], \{\alpha,\beta\}}
      \,.
  \end{split}
\end{equation}
Using this result to calculate
$I^{-1,-1 }_{[\kappa],\{\alpha,\beta\}}$, we observe that
only the first two terms in the curly brackets can give
non-vanishing contributions at \LP as the range
of summation in the last two terms is more restricted.
In addition, due to the fact that the product of two
collinear expansion operators in the opposite collinear
directions is equivalent to one single soft expansion
operator in this case, Eq.~\eqref{eq:def:c0:cbar0:ns:oppcol}
evaluates to the identical expression to that in
Eq.~\eqref{eq:def:c0:cbar0:ns:lp}.


At \NLP, all terms of Eq.~\eqref{eq:def:c0:cbar0:ns:oppcol}
contribute to the zero-bin subtraction.
While in the first two terms in the curly brackets
the \NLP collinear sectors have been expanded in line with
the counting rule in the opposite direction, the last two terms recover
the power corrections that did not contribute to the \LP
result, but re-enter the zero-bin
subtraction procedure here.
Thus, retaining all terms up to \NLP, this yields,
\begin{equation}\label{eq:def:c0:cbar0:ns:nlp}
  \begin{split}
    \widetilde{\mathcal{G}}^{-1,-1,(0)}_{[\kappa],\{\alpha,\beta\}}\Bigg|^{\NSmeth}_{c\bar{c}}\,
    \xRightarrow[]{\text{~~opp.col.~~}}
    \,&
      \theta\left(\omega-\rho \right)\,
      \theta\left(\omega-\sigma \right)\,
      \lim_{\tau\to0}\,\int^{\infty}_{0}\,
      \frac{\done k_+}{k_+}\,\mathcal{R}(k_-,k_+,\tau)\,
      \bigg\{
        1\,+\,\frac{1}{k_-}\,+\frac{1}{k_+}
      \bigg\}\,
    \\
    &\hspace*{40mm}\times\;
      {F}^{(\alpha_n)}_{i/n,\beta_n}\left(m_He^{-Y_H}\right)\,
      {F}^{(\alpha_\nbar)}_{j/\nbar,\beta_\nbar}\left( m_He^{+Y_H} \right)\,.
  \end{split}
\end{equation}
This result agrees with the expectation from the dual scaling.
Please note, \LP-like terms have now entered the subtraction term at \NLP,
constituting the inhomogeneity we alluded to earlier.
More explicitly, substituting the integrand
$I^{-1,-1 }_{[\kappa],\{\alpha,\beta\}}$ into
Eq.~\eqref{eq:def:c0:ns:Texp}, we obtain,
\begin{equation}\label{eq:def:c0:ns:Texp:nlp}
  \begin{split}
    \widetilde{\mathcal{G}}^{-1,-1,(0)}_{[\kappa],\{\alpha,\beta\}}\Bigg|_{c0} \,
    =\,&
      \theta\left(\omega-\sigma \right)\,
      \lim_{\tau\to0}\,\int^{\infty}_{0}\,
      \frac{\done k_+}{k_+}\,\mathcal{R}(k_-,k_+,\tau)\,
      \bigg\{
        \frac{1}{2}\,+\,\frac{1}{k_+}
      \bigg\}\,
      {F}^{(\alpha_n)}_{i/n,\beta_n}\left(m_He^{-Y_H}\right)\,
      {F}^{(\alpha_\nbar)}_{j/\nbar,\beta_\nbar}\left( m_He^{+Y_H}\right),
    \\
    \widetilde{\mathcal{G}}^{-1,-1,(0)}_{[\kappa],\{\alpha,\beta\}}\Bigg|_{\bar{c}0} \,
    =\,&
      \theta\left(\omega-\rho \right)\,
      \lim_{\tau\to0}\,\int^{\infty}_{0}\,
      \frac{\done k_-}{k_-}\,\mathcal{R}(k_-,k_+,\tau)\,
      \bigg\{
        \frac{1}{2}\,+\,\frac{1}{k_-}
      \bigg\}\,
      {F}^{(\alpha_n)}_{i/n,\beta_n}\left(m_He^{-Y_H}\right)\,
      {F}^{(\alpha_\nbar)}_{j/\nbar,\beta_\nbar}\left( m_He^{+Y_H}\right).
  \end{split}
\end{equation}
Recalling that additional constraints on the exponents
$\rho$ and $\sigma$ have already been imposed in the
definition of the power series in
Eq.~\eqref{eq:def:asyexp:central:power:omega:rap}, combining
both contributions above precisely reproduces the results
of Eq.~\eqref{eq:def:c0:cbar0:ns:nlp}.

Furthermore, even though Eq.~\eqref{eq:def:c0:cbar0:ns:oppcol}
is obtained by summarising the asymptotic properties of the
\NLP ingredients in \cite{Inglis-Whalen:2022vyn,
  Inglis-Whalen:2021bea}, it is very interesting to note that
Eq.~\eqref{eq:def:c0:cbar0:ns:oppcol} is in fact still
useful in organising the overlap removal for N$^2$LP and beyond.
To see this, we note that the collinear expansion operators
$\widehat{\mathbf{T}}_{c} $ and $\widehat{\mathbf{T}}_{\bar{c}}$
commute.
Hence, we can drop the factor of $\tfrac{1}{2}$ and recast the
r.h.s.\ of Eq.~\eqref{eq:def:c0:cbar0:ns:oppcol} below,
\begin{equation}\label{eq:def:c0:cbar0:ns:oppcol2}
  \begin{split}
    \left(\qT^2\right)^{\omega}\,
    \widetilde{\mathcal{G}}^{\rho,\sigma,(\omega)}_{[\kappa],\{\alpha,\beta\}}\Bigg|^{\NSmeth}_{c\bar{c}}\,
    \xRightarrow[]{\text{~~opp.col.~~}}
    &\,
      \theta\left(\omega-\rho \right)\theta\left(\omega-\sigma \right)\,
      \lim_{\tau\to0}\,\int^{\infty}_{0}\,
      \frac{\done k_+}{k_+}\,\mathcal{R}(k_-,k_+,\tau)\,
    \\
    &\times
      \Bigg\{
        \widehat{\mathbf{T}}_{\bar{c}}^{(\omega)}
        \widehat{\mathbf{T}}_{c}^{(\omega)}
        +
        \widehat{\mathbf{T}}_{c}^{(\omega)}\,
        \left(
          \sum_{\overline{\omega}=\sigma}^{\omega-1}\,
          \widehat{\mathbf{T}}_{\bar{c}}^{(\overline{\omega})}\,
        \right)
        +
        \widehat{\mathbf{T}}_{\bar{c}}^{(\omega)}\,
        \left(
          \sum_{\overline{\omega}=\rho}^{\omega-1}\,
          \widehat{\mathbf{T}}_{c}^{(\overline{\omega})}\,
        \right)
      \Bigg\}
      I^{\rho,\sigma}_{[\kappa],\{\alpha,\beta\}} \,.\hspace*{-20mm}
  \end{split}
\end{equation}
Plugging the expressions of Eq.~\eqref{eq:def:subjt} into  Eq.~\eqref{eq:def:c0:cbar0:ns:oppcol2}, we then obtain,
\begin{equation}\label{eq:def:c0:cbar0:ns:oppcol3}
  \begin{split}
    \mathrm{r.h.s.~of~Eq.~\eqref{eq:def:c0:cbar0:ns:oppcol2}}
    =&\;
      \theta\left(\omega-\rho \right) \theta\left(\omega-\sigma \right)\,
      \lim_{\tau\to0}\,
      \int^{\infty}_{0}\,
      \frac{\done k_+}{k_+}\,
      \mathcal{R}(k_-,k_+,\tau)\\
    &\times
      \Bigg\{
        k_-^{\omega}\,
        k_+^{\omega}\,
        \frac{
          {F}^{(\alpha_n+\omega-\sigma)}_{i/n,\beta_n}\left(m_H e^{-Y_H}\right)
        }{(\omega-\sigma)!}
        \frac{
          {F}^{(\alpha_\nbar+\omega-\rho)}_{j/\nbar,\beta_\nbar}\left(m_H e^{+Y_H}\right)
        }{(\omega-\rho)!}
      \\
    &\hspace*{6mm}{}
        +
        k_-^{\omega}\,
        \sum_{\eta=0}^{\omega-\sigma-1 }\,
        k_+^{\sigma+\eta}\,
        \frac{
          {F}^{(\alpha_n+\eta)}_{i/n,\beta_n}\left(m_H e^{-Y_H}\right)\,
          {F}^{(\alpha_\nbar+\omega-\rho)}_{j/\nbar,\beta_\nbar}\left(m_H e^{+Y_H}\right)}
        {\eta!\,(\omega-\rho)!}\,
    \\
    &\hspace*{6mm}{}
        +
        k_+^{\omega}\,
        \sum_{\eta=0}^{\omega-\sigma-1 }\,
        k_-^{\rho +\lambda}\,
        \frac{
          {F}^{(\alpha_n+\omega-\sigma)}_{i/n,\beta_n}\left( m_H e^{-Y_H}\right)
          {F}^{(\alpha_\nbar+\lambda)}_{j/\nbar,\beta_\nbar}\left(m_H e^{+Y_H}\right)
        }{\lambda!(\omega-\sigma)!}\,
      \Bigg\}\,.
  \end{split}
\end{equation}
Extracting a common factor of $(\qT^2)^{\omega}$ from the curly
bracket above, we observe its
coincidence with the sum of Eq.~\eqref{eq:def:c0:ns} and thus the equivalence with Eq.~\eqref{eq:def:c0:ns:Texp}.

We will now compare our results with the results obtained through employing
the method of expansion by regions~\cite{Beneke:1997zp,Smirnov:1998vk,
  Smirnov:1999bza,Smirnov:2002pj,Jantzen:2011nz} and focus in particular
on the formalism proposed in~\cite{Jantzen:2011nz}.
Therein, the mathematical foundation behind the expansion by regions
has been discussed via examples for the one-loop integrals in different
kinematic limits.
Of them, the analysis of the Feynman integrals in the Sudakov
limit~\cite{Jantzen:2011nz} is intimately related to the \qT~spectra
within the asymptotic regime.
Therefore, after appropriate adaptations, the techniques proposed
in~\cite{Jantzen:2011nz} can also be exploited to compute the zero-bin
subtraction here.
In the following, we will elucidate this application.
 
The starting point of the formalism in~\cite{Jantzen:2011nz} is to work
out a group of dynamic regions that are able to encompass the entire phase
space and also separately accommodate the contributions from all involved
scales.
In our case, this goal can be accomplished by the dynamic modes presented
in Eq.~\eqref{eq:def:col-n:soft:col-nbar:cutoff}.
The next task is to extrapolate the bounded integrals contained in each
region so as to remove all auxiliary cutoff scales along the integration
paths.
This procedure can be illustrated by using the expansion operators
introduced in Eqs.~\eqref{eq:def:asyexp:c:at:omega:nured:Texp} and
\eqref{eq:def:c0:ns:Texp}, i.e.\
\begin{align}
  \widetilde{\mathcal{I}}^{\rho,\sigma }_{[\kappa],\{\alpha,\beta\} }
  =
    \sum_{\omega}
    \left[
      \int^{\tilde{k}^{\mathrm{max}}_+}_{\nu_n}
      \widehat{\mathbf{T}}_{c}^{(\omega)}\,
      +
      \int^{\nu_n}_{\frac{\qT^2}{\nu_\nbar}}\,
      \widehat{\mathbf{T}}_{s}^{(\omega)}\,
      +
      \int^{\frac{\qT^2}{\nu_\nbar}}_{ \tilde{k}^{\mathrm{min}}_+ }\,
      \widehat{\mathbf{T}}_{\bar{c}}^{(\omega)}
    \right]\,
    \frac{\done k_+}{k_+}\,
    I^{\rho,\sigma }_{[\kappa],\{\alpha,\beta\} }\,,
\end{align}
and then extrapolating the integral boundaries $\nu_n$ and $\nu_\nbar$,
\begin{equation}\label{eq:def:expbreg:c1}
  \begin{split}
    \sum_{\omega_c}\int^{\tilde{k}^{\mathrm{max}}_+}_{\nu_n}
    \widehat{\mathbf{T}}_{c}^{(\omega_c)}  &
    \rightarrow
      \lim_{\tau\to0}
      \left[
        \sum_{\omega_c}\int^{\tilde{k}^{\mathrm{max}}_+}_{0}
        \mathcal{R}\,
        \widehat{\mathbf{T}}_{c}^{(\omega_c)}
        -
        \sum_{\omega_c,\omega_s}\int^{\nu_n}_{\frac{\qT^2}{\nu_\nbar}}
        \mathcal{R}\,
        \widehat{\mathbf{T}}_{s}^{(\omega_s)}
        \widehat{\mathbf{T}}_{c}^{(\omega_c)}
        -
        \sum_{\omega_c,\omega_{\bar{c}}}\int^{\frac{\qT^2}{\nu_\nbar}}_{0}  \mathcal{R}\,
        \widehat{\mathbf{T}}_{\bar{c}}^{(\omega_{\bar{c}})}
        \widehat{\mathbf{T}}_{c}^{(\omega_c)}
      \right],
    \\
    \sum_{\omega_s}\int^{\nu_n}_{\frac{\qT^2}{\nu_\nbar}}
    \widehat{\mathbf{T}}_{s}^{(\omega_s)}  &
    \rightarrow
      \lim_{\tau\to0}
      \left[
        \sum_{\omega_s}\int^{\infty}_{0}
        \mathcal{R}\,
        \widehat{\mathbf{T}}_{s}^{(\omega_s)}
        -
        \sum_{\omega_c,\omega_s}\int^{\infty}_{\nu_n}
        \mathcal{R}\,
        \widehat{\mathbf{T}}_{c}^{(\omega_c)}
        \widehat{\mathbf{T}}_{s}^{(\omega_s)}
        -
        \sum_{\omega_s,\omega_{\bar{c}}}\int^{\frac{\qT^2}{\nu_\nbar}}_{0}
        \mathcal{R}\,
        \widehat{\mathbf{T}}_{\bar{c}}^{(\omega_{\bar{c}})}
        \widehat{\mathbf{T}}_{s}^{(\omega_s)}
      \right],
    \\
    \sum_{\omega_{\bar{c}}}\int^{\frac{\qT^2}{\nu_\nbar}}_{\tilde{k}^{\mathrm{min}}_+ }
    \widehat{\mathbf{T}}_{\bar{c}}^{(\omega_{\bar{c}})}  &
    \rightarrow
      \lim_{\tau\to0}
      \left[
        \sum_{\omega_{\bar{c}}}\int^{\infty}_{ \tilde{k}^{\mathrm{min}}_+ }
        \mathcal{R}\,
        \widehat{\mathbf{T}}_{\bar{c}}^{(\omega_{\bar{c}})}
        -
        \sum_{\omega_{\bar{c}},\omega_s}\int^{\nu_n}_{\frac{\qT^2}{\nu_\nbar}}
        \mathcal{R}\,
        \widehat{\mathbf{T}}_{s}^{(\omega_s)}
        \widehat{\mathbf{T}}_{\bar{c}}^{(\omega_{\bar{c}})}
        -
        \sum_{\omega_c,\omega_{\bar{c}}}\int^{\infty}_{\nu_n}
        \mathcal{R}\,
        \widehat{\mathbf{T}}_c^{(\omega_{c})}
        \widehat{\mathbf{T}}_{\bar{c}}^{(\omega_{\bar{c}})}
      \right].
  \end{split}
\end{equation}
Therein, in order to render the integrals still well-defined after
extrapolating the boundaries, the rapidity regulator $\mathcal{R}$
has been put in place.
Since the discussion here is focused on the conservative rapidity
regularisation prescription that preserves the expression of
$\mathcal{R}$ in all the involved sectors, in writing
Eq.~\eqref{eq:def:expbreg:c1}, we pull all regulators $\mathcal{R}$
in front of the expansion operators.
We note that the r.h.s.\ of Eq.~\eqref{eq:def:expbreg:c1} still
depends on the cutoff scales $\nu_n$ and $\nu_\nbar$.
This dependence can be eliminated through the following identity,
\begin{equation}\label{eq:def:expbreg:rel:c:s:cbar}
  \begin{split}
    \sum_{\omega_c,\omega_{\bar{c}},\omega_s}\int^{\infty}_{0}
    \mathcal{R}\,
    \widehat{\mathbf{T}}_{c}^{(\omega_c)}
    \widehat{\mathbf{T}}_{s}^{(\omega_s)}
    \widehat{\mathbf{T}}_{\bar{c}}^{(\omega_{\bar{c}})}
    \rightarrow&
      \sum_{\omega_c,\omega_s}\int^{\frac{\qT^2}{\nu_\nbar}}_{0}
      \mathcal{R}\,
      \widehat{\mathbf{T}}_{s}^{(\omega_s)}
      \widehat{\mathbf{T}}_{c}^{(\omega_c)}
      +
      \sum_{\omega_c,\omega_{\bar{c}}}\int_{\frac{\qT^2}{\nu_\nbar}}^{ \nu_n}
      \mathcal{R}\,
      \widehat{\mathbf{T}}_{\bar{c}}^{(\omega_{\bar{c}})}
      \widehat{\mathbf{T}}_{c}^{(\omega_c)}
    \\
    &{}+
      \sum_{\omega_{\bar{c}},\omega_s}\int^{\infty}_{\nu_n}
      \mathcal{R}\,
      \widehat{\mathbf{T}}_{s}^{(\omega_s)}
      \widehat{\mathbf{T}}_{\bar{c}}^{(\omega_{\bar{c}})}\,.
  \end{split}
\end{equation}
In its derivation, we have used the commutativity of the
power projection operators $\widehat{\mathbf{T}}_{c}$,
$\widehat{\mathbf{T}}_{s}$, and $\widehat{\mathbf{T}}_{\bar{c}}$.
Combining Eq.~\eqref{eq:def:expbreg:rel:c:s:cbar}  with
Eq.~\eqref{eq:def:expbreg:c1}, we have,
\begin{equation}\label{eq:def:asyexp:jantzen}
  \begin{split}
    \widetilde{\mathcal{I}}^{\rho,\sigma }_{[\kappa],\{\alpha,\beta\} }
    =&\;
      \lim_{\tau\to0}
      \bigg[
        \sum_{\omega_c}
        \int^{\tilde{k}^{\mathrm{max}}_+}_{0}
        \mathcal{R}\,
        \widehat{\mathbf{T}}_{c}^{(\omega_c)}\,
        +
        \sum_{\omega_s}
        \int^{\infty}_{0}
        \mathcal{R}\,
        \widehat{\mathbf{T}}_{s}^{(\omega_s)}
        +
        \sum_{\omega_{\bar{c}}}
        \int^{\infty}_{\tilde{k}^{\mathrm{min}}_+}
        \mathcal{R}\,
        \widehat{\mathbf{T}}_{\bar{c}}^{(\omega_{\bar{c}})}
    \\
    &\hspace*{10mm}{}
        -
        \sum_{\omega_c,\omega_s}
        \int^{\infty}_{0}
        \mathcal{R}\,
        \widehat{\mathbf{T}}_{c}^{(\omega_c)}\,
        \widehat{\mathbf{T}}_{s}^{(\omega_s)}
        -
        \sum_{\omega_c,\omega_{\bar{c}}}
        \int^{\infty}_{0}
        \mathcal{R}\,
        \widehat{\mathbf{T}}_{c}^{(\omega_c)}\,
        \widehat{\mathbf{T}}_{\bar{c}}^{(\omega_{\bar{c}})}
        -
        \sum_{\omega_s,\omega_{\bar{c}}}
        \int^{\infty}_{0}
        \mathcal{R}\,
        \widehat{\mathbf{T}}_{s}^{(\omega_s)}\,
        \widehat{\mathbf{T}}_{\bar{c}}^{(\omega_{\bar{c}})}
    \\
    &\hspace*{10mm}{}
        +
        \sum_{\omega_c,\omega_s,\omega_{\bar{c}}}
        \int^{\infty}_{0}
        \mathcal{R}\,
        \widehat{\mathbf{T}}_{c}^{(\omega_c)}\,
        \widehat{\mathbf{T}}_{s}^{(\omega_s)}\,
        \widehat{\mathbf{T}}_{\bar{c}}^{(\omega_{\bar{c}})}\,
      \bigg]\,
      \frac{\done k_+}{k_+}\,
      I^{\rho,\sigma }_{[\kappa],\{\alpha,\beta\} }\,.
  \end{split}
\end{equation}
Taking into account the facts that expanding in the collinear
scalings following the soft power-projection operator
$\widehat{\mathbf{T}}_{s} $ can only result in delta functions
as well as that two consecutive expansions in the collinear
scalings are equivalent to one single expansion in the soft
scaling upon having summed all power contributions, we see that
the doubly and triply expanded terms in Eq.~\eqref{eq:def:asyexp:jantzen}
are equal to an expansion in the soft scaling.
To this end, we obtain,
\begin{align}\label{eq:def:asyexp:jantzen:red}
  \widetilde{\mathcal{I}}^{\rho,\sigma }_{[\kappa],\{\alpha,\beta\}}
  =&\;
    \sum_{\omega}\,
    \lim_{\tau\to0}\,
    \bigg[
      \int^{\tilde{k}^{\mathrm{max}}_+}_{0}
      \mathcal{R}\,
      \widehat{\mathbf{T}}_{c}^{(\omega)}\,
      +
      \int^{\infty}_{\tilde{k}^{\mathrm{min}}_+}
      \mathcal{R}\,
      \widehat{\mathbf{T}}_{\bar{c}}^{(\omega)}
      -
      \int^{\infty}_{0}
      \mathcal{R}\,
      \widehat{\mathbf{T}}_{s}^{(\omega)}
    \bigg]\,
    \frac{\done k_+}{k_+}\,
    I^{\rho,\sigma }_{[\kappa],\{\alpha,\beta\} }\,.
\end{align}
Comparing with Eqs.~\eqref{eq:def:asyexp:c:at:omega:nured:Texp} and
\eqref{eq:def:c0:ns:Texp}, we observe that the first two terms in
Eq.~\eqref{eq:def:asyexp:jantzen:red} and the collinear functions in
Eq.~\eqref{eq:def:asyexp:c:at:omega:nured:Texp} are equivalent.
To explore the relationship of the third term in
Eq.~\eqref{eq:def:asyexp:jantzen:red} and the sum of
Eq.~\eqref{eq:def:c0:ns:Texp}, we note that after bringing in the
expression in Eq.~\eqref{eq:def:subjt} the soft sector in
Eq.~\eqref{eq:def:asyexp:jantzen:red} evaluates to an infinite sum
of the unbounded integrals, i.e.,
\begin{equation}
  \begin{split}
    \sum_{\eta,\lambda=0}^{\infty}\,
    \lim_{\tau\to0}\,
    \int_0^{\infty}\,
    \frac{\done k_+}{k_+}\,
    \mathcal{R}\left(k_-,k_+,\tau\right)\,
    \frac{k_+^{\sigma+\eta}\, k_-^{\rho+\lambda}}{\eta!\,\lambda!}\,
    {F}^{(\alpha_n+\eta)}_{i/n,\beta_n}\left(m_H e^{-Y_H}\right)\,
    {F}^{(\alpha_\nbar+\lambda)}_{j/\nbar,\beta_\nbar}\left(m_H e^{+Y_H}\right)\,,
  \end{split}
\end{equation}
which is exactly equal to the result from Eq.~\eqref{eq:def:c0:ns:Texp}
upon summing up all power contributions.
Therefore, we can conclude at least the correspondence of
Eq.~\eqref{eq:def:asyexp:jantzen:red} and
Eqs.~\eqref{eq:def:asyexp:c:at:omega:nured:Texp} and \eqref{eq:def:c0:ns:Texp}
on the cumulative level.
                
Notwithstanding, once a power-by-power comparison is being considered
between both results, the analysis becomes more subtle.
In~\cite{Jantzen:2011nz}, the analytic regulator is applied throughout
the calculation of the Sudakov form factors.
Implementing this scheme into Eq.~\eqref{eq:def:asyexp:jantzen}
(or Eq.~\eqref{eq:def:asyexp:jantzen:red}), the integrals that are
expanded in the soft scaling or comprise double or triple expansions
all vanish.
Then, Eq.~\eqref{eq:def:asyexp:jantzen} and
Eq.~\eqref{eq:def:asyexp:jantzen:red} both reduce into the same
decomposition we found in
Eq.~\eqref{eq:def:asyexp:central:power:omega:rap:hom} induced by
the homogeneous regulators.
However, if the inhomogeneous regularisation scheme is of one's
particular interest, the soft function and the doubly and triply
expanded constituents of Eq.~\eqref{eq:def:asyexp:jantzen} can
make non-trivial contributions at a given power accuracy, for
which an unambiguous counting rule is required in order to produce
the correct power coefficient
$\widetilde{\mathcal{I}}^{\rho,\sigma,(\omega) }_{[\kappa],\{\alpha,\beta\}}$.

\subsection{Discussion and extrapolation}
\label{sec:asyexp:recap}

In the previous subsections, we have carried out the power
expansion of the double-differential observable
${\done \sigma_H}/{\done Y_H \done \qT^2}$ in the low \qT~domain.
The result reads,
\begin{align}\label{eq:asyexp:tot}
  &\frac{\done \sigma_H}{\done Y_H \done \qT^2}
  \,=\,
    \lambda_t^2\,
    \sum_{\omega,\overline{\omega}}\,
    \left( {\qT^2} \right)^{\omega+\overline{\omega}}\,
    \sum_{[\kappa]}\,
    \sum_{ \{\alpha,\beta\},\rho,\sigma}\,
    \widetilde{\mathcal{H}}^{ (\omega),\rho,\sigma }_{[\kappa], \{\alpha,\beta\}  }\left(m_H,Y_H,s\right) \,
    \bigg\{
      \widetilde{\mathcal{I}}^{ \rho,\sigma,(\overline{\omega}) }_{[\kappa], \{\alpha,\beta\}  }
      \,+\,
      \Delta\widetilde{\mathcal{I}}^{ \rho,\sigma,(\overline{\omega}) }_{[\kappa], \{\alpha,\beta\}  }
    \bigg\}\,.
\end{align}
Therein,
$\widetilde{\mathcal{H}}^{(\omega),\rho,\sigma }_{[\kappa],\{\alpha,\beta\}}$
is a function of the hard scales $m_H$ and $s$.
Its expression has been presented in Eq.~\eqref{eq:qcd:defHtilde} on
a power-by-power basis.
$\Delta\widetilde{\mathcal{I}}^{\rho,\sigma,(\overline{\omega}) }_{[\kappa],\{\alpha,\beta\}}$
accounts for the power corrections induced by the boundary conditions
of the phase space integral.
The results in the $n$-ultra-collinear limit can be extracted from
Eqs.~\eqref{eq:asyexp:deltaI:uc}-\eqref{eq:asyexp:deltaI:ec:int},
while those in the $\nbar$-direction can be derived by exchanging the
light-cone coordinates in
Eqs.~\eqref{eq:def:Gp}-\eqref{eq:asyexp:deltaI:ec:int}, as appropriate.
Finally, Eq.~\eqref{eq:asyexp:tot} contains
$\widetilde{\mathcal{I}}^{\rho,\sigma,(\overline{\omega})}_{[\kappa],\{\alpha,\beta\}}$
which comprises the contribution from the bulk of the phase space.
In Sec.~\ref{sec:asyexp:qT:mom} and Sec.~\ref{sec:asyexp:qT:extrapl},
we made use of two strategies to evaluate the expansion coefficients
$\widetilde{\mathcal{I}}^{\rho,\sigma,(\overline{\omega})}_{[\kappa],\{\alpha,\beta\}}$.
In Sec.~\ref{sec:asyexp:qT:mom}, owing to the fact that our integration
region comprises multiple scales, we re-categorised the integration path
via two of auxiliary cutoff scales, $\nu_n$ and $\nu_\nbar$.
A dedicated method is applied for each subregion to perform the expansion
within the low \qT~domain.
The quoted power series following this approach is illustrated in
Eq.~\eqref{eq:def:asyexp:central:power:omega}.
Then, in order to pave the way for future investigations of the power
expansion at \NNLO and beyond, we recombined the individual terms of
Eq.~\eqref{eq:def:asyexp:central:power:omega}, eliminating any dependence
on the auxiliary scales.
The rearranged expansion coefficients are given in
Eq.~\eqref{eq:def:asyexp:central:power:omega:rap}.
Equipped therewith, we are able to calculate the power series of
${\done \sigma_H}/{\done Y_H\, \done \qT^2}$ at an arbitrary power
accuracy.

During our analysis, in order to facilitate the establishment of
the relevant scaling relations, we concentrated on the central
rapidity region where $e^{\pm Y_H}\sim\mathcal{O}(1)$.
As it turns out, the expression in Eq.~\eqref{eq:asyexp:tot}
can also be extrapolated to the domain
\begin{align}\label{eq:def:YH:etp}
  \mathrm{min}\{\sqrt{s}-m_He^{Y_H},\sqrt{s}-m_He^{-Y_H}\}>\mathrm{max}\{m_H e^{ Y_H},m_H e^{- Y_H}\}\,.
\end{align}
This extrapolation is straightforward to achieve in the kinematical
reduction in Sec.~\ref{sec:setups}, since the expansion parameter
therein is $(\qT^2/m_H^2)$.

Regarding $\Delta\widetilde{\mathcal{I}}^{\rho,\sigma}_{[\kappa],\{\alpha,\beta\}}$,
it is worth reminding the reader that the power expansion in
Sec.~\ref{sec:red:qT:boundary} primarily builds on the hierarchies,
\begin{align}\label{eq:def:b.c.:YH:etp}
  \sqrt{s}-m_He^{\pm Y_H}\gg \left| m_H-\mT\right|e^{\pm Y_H}\,,\quad   m_He^{\pm Y_H}\gg \frac{\qT^2}{\sqrt{s}-m_He^{\mp Y_H}}\,.
\end{align}
Both relations are maintained by Eq.~\eqref{eq:def:YH:etp} in
the low \qT~domain, and therefore the power expansion of
$\Delta\widetilde{\mathcal{I}}^{\rho,\sigma}_{[\kappa],\{\alpha,\beta\}}$
proceeds as before.
A similar situation can be found in the interior contribution
$\widetilde{\mathcal{I}}^{\rho,\sigma}_{[\kappa],\{\alpha,\beta\}}$,
in which the categorisation and the subsequent expansion base on
the following relationship,
\begin{align}\label{eq:def:soft:YH:etp}
  \sqrt{s}-m_He^{- Y_H}>   m_H e^{- Y_H} \gtrsim \nu_{n}  \gg  \frac{\qT^2}{\nu_\nbar} \gtrsim \frac{\qT^2}{m_He^{Y_H}}>\frac{\qT^2}{\sqrt{s}-m_He^{Y_H}}   \,.
\end{align}
These inequalities ensure the integration variable $k_{\pm}$ is always
comparable to $m_H e^{\mp Y_H}$ in magnitude in the collinear sector,
while that in the transitional domain is small enough to perform the
power expansion.
Eq.~\eqref{eq:def:soft:YH:etp} is manifestly satisfied by the domain of
Eq.~\eqref{eq:def:YH:etp} for the small~\qT~regime, from which there
exists a sufficiently wide window to introduce the auxiliary cutoff
scales $\nu_n$ and $\nu_\nbar$.
Hence, the subsequent power series of
$\widetilde{\mathcal{I}}^{\rho,\sigma}_{[\kappa],\{\alpha,\beta\}}$ can
be derived as in Sec.~\ref{sec:red:qT:central}.

Following the above consideration, we can implement the expansion of
Eq.~\eqref{eq:asyexp:tot} within the full range of Eq.~\eqref{eq:def:YH:etp}.
At the LHC with a colliding energy $\sqrt{s}=13\,\text{TeV}$, this
interval corresponds to the rapidity region $|Y_H|\lesssim4$ and, thus,
enables a reliable power expansion in the bulk of the accessible phase space.
Moving from $|Y_H|\approx0$ to $|Y_H|\lesssim4$, it is possible that novel hierarchies emerge from
$\widetilde{\mathcal{H}}^{(\omega),\rho,\sigma }_{[\kappa],\{\alpha,\beta\}}$
or any of the contributions in
$\widetilde{\mathcal{I}}^{ \rho,\sigma}_{[\kappa],\{\alpha,\beta\}}$.
For instance, $m_H e^{Y_H}\gg \qT\gg m_H e^{-Y_H}$ at $Y_H=3.5$ and
$\qT=30\,\text{GeV}$, which may alter the relative size of the coefficients
at each power.
It will not, however, impact the convergence of the~power~series of
Eq.~\eqref{eq:asyexp:tot} as a whole.

Eventually, it should be emphasised that our derivation of the power
series expansion of the process $pp\to H+X$ primarily replies on the
factorisation of the light-cone momenta in Eq.~\eqref{eq:qcd:def:H}.
An analogous situation can also be found in the NLO squared amplitudes
for the Drell-Yan process viewed in terms of the decomposed scalar
products in Eq.~\eqref{eq:def:sji}.
Therefore, our results for
$\widetilde{\mathcal{I}}^{\rho,\sigma}_{[\kappa],\{\alpha,\beta\}}$ and
$\Delta\widetilde{\mathcal{I}}^{\rho,\sigma}_{[\kappa],\{\alpha,\beta\}}$
obtained in this section are straightforwardly applicable onto the
Drell-Yan process.
After combining with its respective hard sector $\widetilde{\mathcal{H}}$,
it will in turn generate a power series via Eq.~\eqref{eq:asyexp:tot} in
the low \qT\ domain.
Beside those two process classes, producing a single colour singlet boson,
after appropriate generalisation of the kinematics, Eq.~\eqref{eq:asyexp:tot}
can in part also be used to describe the hadroproduction of the
multiple colourless bosons, $pp\to B_1+B_2\dots B_n+X(n\ge2)$, as well.
We will elaborate on this procedure in App.~\ref{sec:asyexp:Bs}.

\section{Implementation up to \texorpdfstring{\NNLP}{NNLP}}
\label{sec:asyexp:outputs}

In this section, we will
analyse the asymptotic behaviour of the \qT~spectrum on the
process $pp\to H+X$, and present the analytic expressions
for the power corrections up to \NNLP.
As illustrated in Eq.~\eqref{eq:asyexp:tot}, the power
expansion within the small \qT~domain comprises two sectors,
the interior contribution
$\widetilde{\mathcal{I}}^{ \rho,\sigma }_{[\kappa], \{\alpha,\beta\}}$
and the boundary corrections
$\Delta\widetilde{\mathcal{I}}^{ \rho,\sigma }_{[\kappa],\{\alpha,\beta\}}$.
In the following, we will make use of three different methods
to evaluate the interior sector.
In Sec.~\ref{sec:app:cfs}, the expressions in
Eqs.~\eqref{eq:def:asyexp:c:mom}-\eqref{eq:def:asyexp:cbar:mom:at:omega}
and \eqref{eq:def:asyexp:cs:mom:int}-\eqref{eq:def:asyexp:cbars:mom:int}
that are derived via the momentum cutoffs scales $\nu_{n(\nbar)}$ in
Eq.~\eqref{eq:def:col-n:soft:col-nbar:cutoff} will be implemented.
The results of Eq.~\eqref{eq:def:asyexp:c:at:omega:nured}
and Eq.~\eqref{eq:def:c0:ns} using a homogeneous regulator
will be detailed in Sec.~\ref{sec:app:prap} while an
inhomogeneous  prescription is used in Sec.~\ref{sec:app:exp}.
It will be demonstrated that all three methods result in
identical expressions for the interior contribution after
combining all relevant terms.
We expect the comparative study here can help the interpretation
of the rapidity-divergence regularisation and present a viable
prescription in the resummation beyond the leading power in the
future.

\subsection{Power expansion with momentum cutoffs}
\label{sec:app:cfs}
  
In line with the categorisation of phase space in
Eq.~\eqref{eq:def:col-n:soft:col-nbar:cutoff}, the
interior contributions are decomposed in terms of
the collinear and transitional sectors, from which
we can re-write Eq.~\eqref{eq:asyexp:tot} here,
\begin{align}\label{eq:def:asyexp:mcfs}
  \frac{\done \sigma_H}{\done Y_H \done \qT^2}\,
  =\,
    \frac{\done \sigma^{\langle \mathrm{m.c.}\rangle}_H}
         {\done Y_H \done \qT^2}
    \Bigg|_{c}\,
    +\,
    \frac{\done \sigma^{\langle \mathrm{m.c.}\rangle}_H}
         {\done Y_H \done \qT^2}
    \Bigg|_{cs}\,
    +\,
    \frac{\done \sigma^{\langle \mathrm{m.c.}\rangle}_H}
         {\done Y_H \done \qT^2}
    \Bigg|_{\bar{c}s}\,
    +\,
    \frac{\done \sigma^{\langle \mathrm{m.c.}\rangle}_H}
         {\done Y_H \done \qT^2}
    \Bigg|_{\bar{c}}\,
    +\,
    \frac{\done \sigma_H}{\done Y_H \done \qT^2}
    \Bigg|_{b.c.}\,.
\end{align}
Therein, the ingredients with the subscripts $``c"$ and
$``\bar{c}"$ account for the collinear pieces defined in
Eqs.~\eqref{eq:def:asyexp:c:mom:at:omega} and
\eqref{eq:def:asyexp:cbar:mom:at:omega}, respectively,
while those indicated by $``cs"$ and $``\bar{c}s"$ encode
the transitional elements from
Eqs.~\eqref{eq:def:asyexp:cs:mom:int}-\eqref{eq:def:asyexp:cbars:mom:int}.
In all these four terms, the superscript $``\mathrm{m.c.}"$
is introduced to manifest the fact that the momentum cutoffs
$\nu_n$ and $\nu_\nbar$ are utilised during the power expansion.
At last, we take account of the boundary corrections derived
in Sec.~\ref{sec:red:qT:boundary}, termed $``b.c."$.

In the last section, the power expansions on those sectors
are performed directly in the momentum space.
To render their expressions more compact, we here recast
them in terms of dimensionless parameters,
\begin{align}\label{eq:def:kin:xn:xnbar}
  x_n\,
  \equiv\,
    \frac{m_H\,e^{-Y_H}}{\sqrt{s}}\,\equiv 1-\bar{x}_n\,,
  \qquad\quad
  x_{\nbar}\,
  \equiv\,
    \frac{m_H\,e^{+Y_H}}{\sqrt{s}}\,\equiv 1-\bar{x}_{\nbar}\,,
\end{align}
and those related to the integration variables,       
\begin{align}\label{eq:def:kin:zn:znbar}
    k_+\,
    &\equiv\,
      m_H\,e^{-Y_H}\,\left(\frac{1-z_n}{z_n} \right)\,,
    \qquad\quad
    k_-\,
    \equiv\,
      m_H\,e^{+Y_H}\,\left(\frac{1-z_{\nbar}}{z_{\nbar}} \right)\,,
      \\
    \label{eq:def:kin:znt:znbart}
    \nu_n\,
    &\equiv\,
      m_H\,e^{-Y_H}\,\left(\frac{1-\tilde{z}_n}{\tilde{z}_n} \right)\,,
    \qquad\quad
    \nu_{\nbar}\,
    \equiv\,
      m_H\,e^{+Y_H}\,\left(\frac{1-\tilde{z}_{\nbar}}{\tilde{z}_{\nbar}} \right)
      \,.
\end{align}
Implementing this parameterisation is straightforward in
the boundary corrections of Eq.~\eqref{eq:asyexp:deltaI:uc},
the collinear sectors of
Eqs.~\eqref{eq:def:asyexp:c:mom:at:omega} and
\eqref{eq:def:asyexp:cbar:mom:at:omega},
and the regular terms in the transitional contributions of
Eqs.~\eqref{eq:def:asyexp:cs:mom:int}-\eqref{eq:def:asyexp:cbars:mom:int}.
For contributions involving the star-distribution, e.g.\
Eq.~\eqref{eq:def:sDist}, we first make use of the method
of integration by parts to reduce the higher ranked
contributions to lower ones, and then introduce the
dimensionless kinematic variables defined above to
convert the star-distribution to the customary
plus-distribution.
Up to \NNLP, the identities relevant to the reduction
of the star-distributions read,
\begin{equation}\label{eq:plus2:convertor}
  \begin{split}
    \int_0^{\Lambda}\,\done x\,\left[\frac{1}{x^2}\right]^{\nu}_{*}\,f(x)\,
    =\,&
      \int_0^{\Lambda}\,\done x\,\left[\frac{1}{x}\right]^{\nu}_{*}\,f'(x)
      \,-\,
      \frac{f(\Lambda)}{\Lambda}
      \,+\,
      f'(0)
      \,+\,
      \frac{f(0)}{\nu}
      \,,\\
    \int_0^{\Lambda}\,\done x\,\left[\frac{1}{x^3}\right]^{\nu}_{*}\,f(x)\,
    =\,&
      \frac{1}{2}\,
      \int_0^{\Lambda}\,\done x\,\left[\frac{1}{x}\right]^{\nu}_{*}\,f''(x)
      \,-\,
      \frac{f(\Lambda)}{2\,\Lambda^2}-\,\frac{f'(\Lambda)}{2\,\Lambda}
      \,+\,
      \frac{3}{4}\,f''(0)
      \,+\,
      \frac{f'(0)}{\nu}
      \,+\,
      \frac{f(0)}{2\nu^2}\,,
  \end{split}
\end{equation}
where $f(x)$ stands for a generic function sufficiently
differentiable at $x=0$.
Regarding the transformation of the star-distributions
to the plus distribution, the following relationships
are used during our calculation,
\begin{equation}\label{eq:trans:plusD:starD:n}
  \begin{split}
    \int^{\sqrt{s}\bar{x}_{n}}_0\done k_+\,
    \left[\frac{1}{k_+}\right]^{\nu_n}_{*}\,f(k_+)
    &\,=\,
      \int^1_{x_n}\done z_n\,\left[\frac{1}{1-z_n}\right]_+ \,f(z_n)
      \,+\,
      \int^1_{x_n} \frac{\done z_n}{z_n}\,f(z_n)
      \,+\,
      \ln\left[\frac{\sqrt{s}\,x_n}{\nu_n}\right]\,f(0)
      \,,\\
    \int^{\sqrt{s}\bar{x}_{\nbar}}_0\done k_-\,
    \left[\frac{1}{k_-}\right]^{\nu_{\nbar}}_{*}\,f(k_-)
    &\,=\,
      \int^1_{x_{\nbar}}\done z_{\nbar}\,
      \left[\frac{1}{1-z_{\nbar}}\right]_+ \,f(z_{\nbar})
      \,+\,
      \int^1_{x_{\nbar}} \frac{\done z_{\nbar}}{z_{\nbar}}\,f(z_{\nbar})
      \,+\,
      \ln\left[\frac{\sqrt{s}\,x_{\nbar}}{\nu_{\nbar}}\right]\,f(0)\,,
  \end{split}
\end{equation}
where the plus-distribution is defined as,
\begin{align}\label{eq:deef:plusD}
  \int^1_x\done z \,\left[\frac{1}{1-z}\right]_+\,f(z)
  \,=\,
    \int^1_x\done z \, \frac{f(z)-f(1)}{1-z}
    \,-\int^x_0 \done z\,\frac{f(1)}{1-z} \,.
\end{align}

\subsection*{Transitional domain}
We have now all tools in place to present the respective
expressions for each term on the r.h.s.\ of
Eq.~\eqref{eq:def:asyexp:mcfs}.
We begin with the transitional contributions that are
associated with $n$-collinear and soft scalings,
\begin{equation}\label{eq:def:asyexp:mcfs:cs}
  \begin{split}
    \frac{\done \sigma^{\langle \mathrm{m.c.}\rangle}_H}
         {\done Y_H \done \qT^2}\Bigg|_{cs}
    \,\equiv\,&
      \frac{\alpha_s^3C_t^2}{192\pi^2s v^2}\,
      \sum_{i,j=\{g,q,\bar{q}\}}\,
      \sum_{\omega=-1}^{\infty}\,
      \left(\frac{\qT^2}{m_H^2}\right)^{\omega}\,
      \int^{1}_{\tilde{z}_n}\,\done z_n\,
      \left[\mathbf{F}_{i/n}\left(\frac{x_n}{z_n}\right)\right]^{\mathbf{T}}
    \\
    &
      \,\times\,
      \left\{
        \mathbf{R}_{cs}^{(\omega),ij}(z_n)
        \,+\,
        \mathbf{P}_{cs}^{(\omega),ij}\,
        \left[\frac{1}{1-z_n}\right]_+\!\!
        \,+\,
        \mathbf{D}_{cs}^{(\omega),ij}\,
        \delta\left(1-z_n\right)
        \,+\,
        \mathbf{B}_{cs}^{(\omega),ij}(\tilde{z}_n)\,
        \delta\left(\tilde{z}_n-z_n\right)
      \right\}
    \\
    &
      \,\times\,
      \mathbf{F}_{j/\nbar}(x_{\nbar})\,.
  \end{split}
\end{equation}
Therein, the \qT~spectrum is expanded in the small parameter
$(\qT^2/m^2_H)$ on the r.h.s., and at each power of the
resulting series, the coefficient comprises the convolution
of the PDFs and the reduced squared amplitudes over the domain
$z_n\in[\tilde{z}_n,1]$.
To facilitate the discussion, we here express the power series
for each PDF in array form, more explicitly,
\begin{equation}\label{eq:PDF:N:ary}
  \begin{split}
    \left[ \mathbf{F}_{i/n} (\xi_n)\right]^{\mathbf{T}}
    \,\equiv\,
      \Big[
        f_{i/n}(\xi_n)\,,
        \xi_n\,f'_{i/n}(\xi_n)\,,
        \frac{(\xi_n)^2}{2!}\,f''_{i/n}(\xi_n)\,,
        \ldots\,,
        \frac{(\xi_n)^k}{k!}\,f^{(k)}_{i/n}(\xi_n)\,,
        \ldots
      \Big]\,,
    \\
    \mathbf{F}_{j/\nbar} (\xi_{\nbar})
    \,\equiv\,
      \Big[
        f_{j/\nbar}(\xi_\nbar)\,,
        \xi_\nbar\,f'_{j/\nbar}(\xi_\nbar)\,,
        \frac{(\xi_\nbar)^2}{2!}\,f''_{j/\nbar}(\xi_\nbar)\,,
        \ldots
        \frac{(\xi_\nbar)^k}{k!}\,f^{(k)}_{j/\nbar}(\xi_\nbar)\,,
        \ldots
    \Big]^{\mathbf{T}}\,,
  \end{split}
\end{equation}
where the superscript ${\mathbf{T}}$ denotes the transposition operation.
The shorthands $f^{(k)}_{i/n}$ and $f^{(k)}_{j/\nbar}$ are
employed to represent the $k$-th derivative of the PDFs,
\begin{align}
  f^{(k)}_{i/n}(\tilde{\xi}_n)\,
  \equiv\,
    \frac{\partial^k}{\partial \xi_n^k}\,
    f_{i/n }(\xi_n)\bigg|_{\xi_n=\tilde{\xi}_n}\,,\qquad
  f^{(k)}_{j/\nbar}(\tilde{\xi}_{\nbar})\,
  \equiv\,
    \frac{\partial^k}{\partial \xi_{\nbar}^k}\,
    f_{j/\nbar }(\xi_{\nbar})\bigg|_{\xi_{\nbar}=\tilde{\xi}_{\nbar}}\,.
\end{align}
Further, the reduced squared amplitudes $\mathbf{R}_{cs}^{(\omega),ij}$
are matrices in the rank of the PDF derivative encoded in $
\mathbf{F}_{i/n}$. 
They characterise the $\omega$th-power regular contributions
from the transitional domain initiated by the partons $i$ and $j$.
For the partonic channel $gg\to  Hg$, the results up to N$^2$LP read,
\begin{equation}\label{eq:Rcs:lp_nlp_nnlp:gg}
  \begin{split}
&\mathbf{R}_{cs}^{(-1),gg} =
 -z_n+1-\frac{1}{z_n}+\frac{1}{z_n^2}\,,
\\
& \mathbf{R}_{cs}^{(0),gg} =
 \left[
\begin{array}{ccc}  
 -\frac{3 z_n^2}{2}-4 z_n-\frac{1}{2}+\frac{1}{2 z_n^2}
  &
\begin{smallmatrix}
3 z_n^2\\
+\frac{5 z_n}{2}+\frac{3}{2}+\frac{1}{2 z_n}+\frac{1}{2 z_n^2} 
 \end{smallmatrix}
 \\[2.5ex]
 \frac{z_n^2}{2}+\frac{3 z_n}{2}+\frac{1}{2 z_n}
  &
   -z_n^2-z_n-1    
\end{array}
\right]\,,
\\
&\mathbf{R}_{cs}^{(1),gg}=
\left[ 
\begin{array}{ccc}
\begin{smallmatrix}
 -\frac{27 z_n^3}{2}+\frac{3 z_n^2}{8}-\frac{15 z_n}{4}-\frac{19}{8}\\-\frac{1}{8 z_n^2} 
 \end{smallmatrix}
 &
 \begin{smallmatrix}
  \frac{27 z_n^2}{4}+\frac{45z_n}{8}+\frac{29}{8}+\frac{5}{8 z_n}\\
 +\frac{1}{8 z_n^2}
  \end{smallmatrix}
   &
    \begin{smallmatrix}
    \frac{3 z_n}{4}-\frac{3}{4}+\frac{3}{4
   z_n}+\frac{1}{4 z_n^2} \\
   -6 z_n^3+3 z_n^2
     \end{smallmatrix}
     \\[4.ex]
      \begin{smallmatrix}
 \frac{27 z_n^3}{4}+\frac{13 z_n^2}{8}+\frac{25 z_n}{8}+\frac{7}{4}\\
 +\frac{1}{8 z_n} 
      \end{smallmatrix}
      & 
      -\frac{9 z_n^2}{4}-\frac{11 z_n}{4}-3+\frac{1}{4z_n} 
      &
      3 z_n^3
      \\[4.ex]
 -\frac{9 z_n^3}{4}-\frac{7 z_n^2}{4}-\frac{5 z_n}{2}-2 
 &
 0
 & -z_n^3-z_n^2-z_n-1 
 \\
\end{array}
\right]\,.
  \end{split}
\end{equation}
Here, all the entries that will vanish in the following are
omitted for brevity.
In the case of $i=q(\bar{q})$ and $j=g$, we have,
\begin{equation}\label{eq:Rcs:lp_nlp_nnlp:qg}
  \begin{split}
&\mathbf{R}_{cs}^{(-1),q(\bar{q})g} =
 \frac{2}{9}-\frac{4}{9 z_n}+\frac{4}{9 z_n^2} \,,
\\
&\mathbf{R}_{cs}^{( 0),q(\bar{q})g} =
\left[
\begin{array}{ccc}
 -\frac{2 z_n}{9}+\frac{1}{3}+\frac{2}{9 z_n^2}
  &
-\frac{1}{9}+\frac{2}{9 z_n}+\frac{2}{9 z_n^2} \\[2.ex] 
 \frac{z_n}{9}-\frac{2}{9}+\frac{2}{9 z_n}
  & 
  0   \\
\end{array}
\right] \,,\\
&\mathbf{R}_{cs}^{( 1),q(\bar{q})g} =
\left[
\begin{array}{ccc}
 \frac{z_n^2}{6}-\frac{z_n}{6}+\frac{1}{36}-\frac{1}{18 z_n^2}
  & 
  \frac{5 z_n}{9}+\frac{1}{4}+\frac{5}{18 z_n}+\frac{1}{18 z_n^2}
   &
   \frac{4 z_n}{9}+\frac{5}{18}+\frac{1}{3 z_n}+\frac{1}{9 z_n^2} \\[2.ex]
 -\frac{z_n^2}{9}+\frac{5 z_n}{36}-\frac{1}{18}+\frac{1}{18 z_n}
  &
   -\frac{5 z_n}{18}-\frac{2}{9}+\frac{1}{9 z_n}
    & 
    -\frac{2 z_n}{9}-\frac{2}{9}  \\[2.ex]
 \frac{z_n^2}{18}-\frac{z_n}{9}+\frac{1}{9}
  &
   0
    & 
    0 \\  
\end{array}
\right] \,.
  \end{split}
\end{equation}
Finally, those for the $q\bar{q}$ initial state read as follows,
\begin{equation}\label{eq:Rcs:qq}
  \begin{split}
&\mathbf{R}_{cs}^{(-1),q \bar{q} } =0\,,
\;\;\;\quad\quad\quad
\mathbf{R}_{cs}^{(0),q \bar{q} } = \frac{16}{27 z_n}-\frac{16}{27}\,,
\;\;\;\quad\quad\quad
\mathbf{R}_{cs}^{(1),q \bar{q} }=
\left[
\begin{array}{ccc}
 \frac{16 z_n}{27}-\frac{16}{27} 
 &
  \frac{8}{27}+\frac{8}{27 z_n}  \\[2.ex]
 \frac{8}{27}-\frac{8 z_n}{27} 
 & 
 0  
\end{array}
\right]
\,.
  \end{split}
\end{equation}
We next advance to the singular terms on the r.h.s.\ of Eq.~\eqref{eq:def:asyexp:mcfs}.
$\mathbf{P}^{(\omega)}_{cs}$, consisting of only the constant
coefficients, captures the contribution associated with the
plus-distribution in Eq.~\eqref{eq:deef:plusD}.
Up to \NNLP, they evaluate to,
\begin{equation}
  \begin{split}
   \label{eq:Pcs:gg_gq_qq}
\mathbf{P}_{cs}^{(-1),gg}&=1\,,\quad\quad\quad\quad\quad
\mathbf{P}_{cs}^{(0),gg} =
\left[
\begin{matrix}
 4 & -\frac{1}{2}\\[2.ex]
-\frac{1}{2} & 1\\
\end{matrix}
\right]
\,,\quad\quad\quad\quad\quad
\mathbf{P}_{cs}^{(1),gg} =
\left[
\begin{array}{ccc}
  3 & -\frac{13}{8} & \frac{9}{4} \\[2.ex]
  -\frac{13}{8} & \frac{13}{4} & 0   \\[2.ex]
  \frac{9}{4} & 0 & 1  
\end{array}
\right]\,,
\\
\mathbf{P}_{cs}^{(-1),q(\bar{q})g}&=0\,,\quad\quad\quad\quad\quad
\mathbf{P}_{cs}^{(0),q(\bar{q})g} =
\left[
\begin{matrix} 
 \frac{2}{9} &  \frac{2}{9} 
\end{matrix}
\right]
\,,\quad\quad\quad\quad\quad\quad
\mathbf{P}_{cs}^{(1),q(\bar{q})g} =
\left[
\begin{array}{ccc}
 0 & \frac{2}{9} & \frac{2}{9}\\[2.ex]
 \frac{1}{9} & \frac{1}{3} & \frac{2}{9}  
\end{array}
\right]\,,
\\
\mathbf{P}_{cs}^{(-1),q \bar{q} } &=0\,,\quad\quad\quad\quad\quad
\mathbf{P}_{cs}^{(0),q \bar{q} } =0\,,\quad\quad\quad\quad\quad\quad\quad\quad\quad\quad
\mathbf{P}_{cs}^{(1),q \bar{q} } =0\,.
  \end{split}
\end{equation}
Herein, due to the finiteness of the squared amplitude
of Eq.~\eqref{eq:msq} (qq) in the limit $k_{\pm}\to0$, the
$\mathbf{P}_{cs}^{(\omega),q\bar{q}}$ all vanish in the
first few power corrections.

The remaining singular terms of Eq.~\eqref{eq:msq} are
governed by the delta function and the coefficients
$\mathbf{D}_{cs}^{(\omega)}$.
Their expressions for \LP, \NLP, and \NNLP are,
\begin{equation}\label{eq:Dcs:lp_nlp_nnlp}
  \begin{split}
\mathbf{D}_{cs}^{(-1),gg}&=\,L_{\mathrm{cs}}\,\mathbf{P}_{cs}^{(-1),gg}\,,\quad \qquad \qquad \qquad \qquad \quad \;
\mathbf{D}_{cs}^{(-1),q(\bar{q})g}\,=\,\mathbf{D}_{cs}^{(-1),q \bar{q} }\,=\,\mathbf{D}_{cs}^{(0),q \bar{q} } \,=\,\mathbf{D}_{cs}^{(1),q \bar{q} } \,=\,0\,,\\
\mathbf{D}_{cs}^{(0),gg} &=\,
L_{\mathrm{cs}}\,
\mathbf{P}_{cs}^{(0),gg}\,
+\,
\left[
\begin{array}{ccc}
  3 &  -3  \\ 
  -1 &  +1 
\end{array}
\right]\,,\qquad\qquad\,\,
\mathbf{D}_{cs}^{(0),q(\bar{q})g} =\,L_{\mathrm{cs}}\,\mathbf{P}_{cs}^{(0),q(\bar{q})g}\,,\\
\mathbf{D}_{cs}^{(1),gg} &=\,
L_{\mathrm{cs}}\,\mathbf{P}_{cs}^{(1),gg}\,
+\,
\left[
\begin{array}{ccc}
  9 &  -9 & 0 \\[2.ex]
 -\frac{13}{2} &  +\frac{7}{2} & -\frac{3}{2} \\[2.ex]
  \frac{7}{2} & -\frac{1}{2} &  \frac{3}{2} 
\end{array}
\right]\,,\quad
\mathbf{D}_{cs}^{(1),q(\bar{q})g}=\,L_{\mathrm{cs}}\,\mathbf{P}_{cs}^{(1),q(\bar{q})g}\,+\,
\left[
\begin{array}{ccc}
 0 &  -\frac{4}{9} &  -\frac{4}{9} \\[2.ex]
 0 &  \frac{2}{9} &  \frac{2}{9} 
\end{array}
\right]\,,\hspace*{-20mm}
  \end{split}
\end{equation}
where the logarithm $L_{\mathrm{cs}}\equiv\ln[\sqrt{s}x_n/\qT]$
results from the logarithmic contributions in the transitional
sectors of Eqs.~\eqref{eq:def:asyexp:cs:mom:int} and \eqref{eq:trans:plusD:starD:n}.
We note that all $L_{\mathrm{cs}}$ dependences in
$\mathbf{D}_{cs}^{(\omega)}$ are correlated to the
coefficients $\mathbf{P}_{cs}^{(\omega)}$ at each power.
To interpret this phenomenon, it merits recalling that in
Eq.~\eqref{eq:def:asyexp:cs:mom:int}, the coefficient function
that is convoluted in the star-distribution reads
${F}^{(\alpha_n)}_{i/n,\beta_n}
 {F}^{(\alpha_\nbar+\omega-\rho)}_{j/\nbar,\beta_\nbar}$, which,
after reducing the rank with the help of
Eq.~\eqref{eq:plus2:convertor}, is recast into
${F}^{(\alpha_n+\omega-\sigma)}_{i/n,\beta_n}
 {F}^{(\alpha_\nbar+\omega-\rho)}_{j/\nbar,\beta_\nbar}$,
corresponding to the coefficient of the logarithmic term.
During the subsequent transformation in
Eq.~\eqref{eq:trans:plusD:starD:n}, this correspondence is
inherited by the coefficient associated with the plus distribution,
which, within the parametrisation of
Eq.~\eqref{eq:def:asyexp:mcfs:cs}, correlates
$\mathbf{P}_{cs}^{(\omega)}$ to the logarithmic terms in
$\mathbf{D}_{cs}^{(\omega)}$.
 
The last constituent in Eq.~\eqref{eq:def:asyexp:mcfs:cs} is
$\mathbf{B}_{cs}^{(\omega)}$.
It accommodates the boundary terms on the r.h.s.\ of
Eq.~\eqref{eq:plus2:convertor}, as a result of using an
integration by parts in its derivation.
The outputs for the first three orders read, 
\begin{equation}\label{eq:Bcs:lp_nlp_nnlp}
  \begin{split}
    &\mathbf{B}_{cs}^{(-1),gg}
    =\mathbf{B}_{cs}^{(-1),q(\bar{q})g}
    =\mathbf{B}_{cs}^{(-1),q \bar{q}}
    =\mathbf{B}_{cs}^{(0),q(\bar{q})g}
    =\mathbf{B}_{cs}^{(0),q \bar{q}}
    =\mathbf{B}_{cs}^{(1),q \bar{q}}
    =0\,,\\
& \mathbf{B}_{cs}^{(0),gg} =\left[
\begin{array}{ccc}
 -\frac{\tilde{z}_n^4}{\tilde{z}_n-1} & \frac{\tilde{z}_n^4}{\tilde{z}_n-1}  \\
\end{array}
\right]\,,\\
&\mathbf{B}_{cs}^{(1),gg}=
\left[
\begin{array}{ccc}
 \frac{\tilde{z}_n^4 [(11-6 \tilde{z}_n) \tilde{z}_n-8]}{2 (\tilde{z}_n-1)^2} & \frac{\tilde{z}_n^4 (5 \tilde{z}_n-4)}{2 (\tilde{z}_n-1)^2} & \frac{\tilde{z}_n^4
   [(8-3 \tilde{z}_n) \tilde{z}_n-6]}{2 (\tilde{z}_n-1)^2} \\[2.ex]
 \frac{\tilde{z}_n^5}{\tilde{z}_n-1} & 0 & \frac{\tilde{z}_n^5}{2 (\tilde{z}_n-1)} \\
\end{array}
\right]\,,
\quad
\mathbf{B}_{cs}^{(1),q(\bar{q})g}
=
\left[
\begin{array}{ccc}
 0 & \frac{2 \tilde{z}_n^3}{9 (\tilde{z}_n-1)} & \frac{2 \tilde{z}_n^3}{9 (\tilde{z}_n-1)} \\
\end{array}
\right]\,.\hspace*{-20mm}
  \end{split}
\end{equation}
Since this contribution primarily arises from the reduction
of the higher ranked star-distributions,
$\mathbf{B}_{cs}^{(\omega),q\bar{q}}$ vanishes for
$\omega\in[-1,0,1]$, similarly to the
$\mathbf{P}_{cs}^{(\omega),q \bar{q}}$ earlier.
The other part of the transitional region is subject to the
$\nbar$-collinear and soft scalings, which can be organised
in an analogous manner to Eq.~\eqref{eq:def:asyexp:mcfs:cs},
more specifically,
\begin{equation}\label{eq:def:asyexp:mcfs:cbars}
  \begin{split}
    \frac{\done \sigma^{\langle \mathrm{m.c.}\rangle}_H}{\done Y_H \done \qT^2}
    \Bigg|_{\bar{c}s}
    \,\equiv\,&
      \frac{\alpha_s^3C_t^2}{192\pi^2s v^2}\,
      \sum_{i,j=\{g,q,\bar{q}\}}\,
      \sum_{\omega=-1}^{\infty}\,
      \left(\frac{\qT^2}{m_H^2}\right)^{\omega}\,
      \int^{1}_{\tilde{z}_\nbar}\,\done z_\nbar\,
      \left[\mathbf{F}_{i/n}\left( {x_n} \right)\right]^{\mathbf{T}}
    \\
    &\,\times\,
      \left\{
        \mathbf{R}_{\bar{c}s}^{(\omega),ij}(z_\nbar)
        \,+\,
        \mathbf{P}_{\bar{c}s}^{(\omega),ij}\,
        \left[\frac{1}{1-z_\nbar}\right]_+\!\!
        \,+\,
        \mathbf{D}_{\bar{c}s}^{(\omega),ij}\,
        \delta\left(1-z_\nbar\right)
        \,+\,
        \mathbf{B}_{\bar{c}s}^{(\omega),ij}(\tilde{z}_\nbar)\,\delta\left(\tilde{z}_\nbar-z_\nbar\right)
      \right\}\hspace*{-5mm}
    \\
    &\,\times\,
      \mathbf{F}_{j/\nbar}\left(\frac{x_{\nbar}}{z_\nbar}\right)\,.
  \end{split}
\end{equation}
Herein, Eq.~\eqref{eq:def:asyexp:mcfs:cbars} comprises the
matrices $\mathbf{R}_{\bar{c}s}$, $\mathbf{P}_{\bar{c}s}$,
$\mathbf{D}_{\bar{c}s}$, and $\mathbf{B}_{\bar{c}s}$ to likewise
encapsulate the different facets of partonic contributions,
akin to those in Eq.~\eqref{eq:def:asyexp:mcfs:cs}.
Of them, the matrices for the partonic processes $gg\to Hg$
and $q\bar{q}\to Hg$ can be immediately derived from those
appearing in Eq.~\eqref{eq:def:asyexp:mcfs:cs}, as the squared
amplitudes for these channels in Eq.~\eqref{eq:msq} (gg and qq)
symmetric under the exchange of their initial states, more explicitly,
\begin{align}
\label{eq:Xcbars:nnlp}
\mathbf{X}_{\bar{c}s}^{(\omega),ij}=\left[\mathbf{X}_{cs}^{(\omega),ij}\right]^{\mathbf{T}}\Bigg|_{\substack {z_n\to z_\nbar\,,\; \tilde{z}_n\to \tilde{z}_\nbar\\[0.6ex] L_{\mathrm{cs}}\to L_{\mathrm{\bar{c}s}}}}\,,
\quad
\mathrm{where}
~\mathbf{X}\in\{\mathbf{R}, \mathbf{P},\mathbf{D},\mathbf{B}\}\;
\mathrm{and}\;
\{ij\}\in\{gg,q\bar{q}\}\,.
\end{align}
An analogous strategy can also be applied to evaluate the
matrices $\mathbf{P}_{\bar{c}s}^{(\omega),q(\bar{q})g}$ that
are induced by the initial state $q(\bar{q})g$.
As discussed before, $\mathbf{P}_{\bar{c}s}^{(\omega),q(\bar{q})g}$
is associated with the coefficient of the logarithmic term of
$\widetilde{\mathcal{I}}^{\rho,\sigma}_{[\kappa],\{\alpha,\beta\}}|_{\bar{c}s}$
in Eq.~\eqref{eq:def:asyexp:cbars:mom:int}, which, by comparison,
is same as that from
$\widetilde{\mathcal{I}}^{\rho,\sigma}_{[\kappa],\{\alpha,\beta\} }|_{cs}$
in Eq.~\eqref{eq:def:asyexp:cs:mom:int}.
To this end, we have,
\begin{align}
\label{eq:Pcbars:qg:nnlp} 
\mathbf{P}_{\bar{c}s }^{(\omega),q(\bar{q})g}=\mathbf{P}_{cs }^{(\omega),q(\bar{q})g}\,.
\end{align}
This correspondence also holds for the logarithmic contributions
in $\mathbf{D}_{\bar{c}s }^{(\omega),q(\bar{q})g}$,
\begin{equation}\label{eq:Dcbars:lp_nlp_nnlp}
  \begin{split}
\mathbf{D}_{\bar{c}s }^{(-1),q(\bar{q})g}&=\,0\,,\\
\mathbf{D}_{\bar{c}s }^{(0),q(\bar{q})g} &=\,
L_{\mathrm{\bar{c}s}}\,
\mathbf{P}_{cs }^{(0),q(\bar{q})g} \,
+
\left[
\begin{array}{ccc}
 -\frac{2}{9} &+\frac{2}{9} \\ 
\end{array}
\right]\,, \\
\mathbf{D}_{\bar{c}s }^{(1),q(\bar{q})g}&=\,
L_{\mathrm{\bar{c}s}}\,
\mathbf{P}_{cs }^{(1),q(\bar{q})g}\,
+\,
\left[
\begin{array}{ccc}
 0 &0 & \frac{2}{9} \\[2ex]
-\frac{2}{9} &\frac{2}{9} &\frac{1}{3} 
\end{array}
\right]\,,
  \end{split}
\end{equation}
where $L_{\mathrm{\bar{c}s}}\equiv\ln[\sqrt{s}x_\nbar/\qT ]$.
Comparing with Eq.~\eqref{eq:Dcs:lp_nlp_nnlp}, the constant
terms in Eq.~\eqref{eq:Dcbars:lp_nlp_nnlp} are very different,
mirroring the asymmetry of the matrix element of
$q(\bar{q})+g\to H+q(\bar{q})$ in Eq.~\eqref{eq:msq} (gq)
under the $n$-collinear and $\nbar$-collinear scalings.

Unsurprisingly, this asymmetric behaviour also emerges from
the regular coefficients,
\begin{equation}\label{eq:Rcbars:lp_nlp_nnlp}
  \begin{split}
&\mathbf{R}_{\bar{c}s }^{(-1),q(\bar{q})g} =0 \,,\\
&\mathbf{R}_{\bar{c}s}^{( 0),q(\bar{q})g} = 
\left[
\begin{array}{ccc}
 \frac{2}{9}+\frac{2}{9 {z}_\nbar} & -\frac{2}{9} \\
\end{array}
\right] \,,\\
&\mathbf{R}_{\bar{c}s}^{( 1),q(\bar{q})g} =
 \left[
\begin{array}{ccc}
 \frac{1}{9}-\frac{2 {z}_\nbar}{9} & \frac{2 {z}_\nbar}{9}-\frac{1}{9} & -\frac{2 {z}_\nbar}{9}-\frac{2}{9} \\[2ex]
 -\frac{2 {z}_\nbar}{9}+\frac{1}{3}+\frac{1}{9 {z}_\nbar} & \frac{2 {z}_\nbar}{9}-\frac{1}{3} & -\frac{2 {z}_\nbar}{9}-\frac{2}{9} 
\end{array}
\right]\,,
  \end{split}
\end{equation}
and the boundary corrections,
\begin{equation}\label{eq:Bcbars:lp_nlp_nnlp}
  \begin{split}
&\mathbf{B}_{\bar{c}s}^{(-1),q(\bar{q})g} =0\,,\quad
\mathbf{B}_{\bar{c}s}^{(0),q(\bar{q})g}=\frac{2 \tilde{z}_\nbar^2}{9 (\tilde{z}_\nbar-1)}\,,
\quad
\mathbf{B}_{\bar{c}s}^{(1),q(\bar{q})g}=
\left[
\begin{array}{ccc}
 -\frac{(\tilde{z}_\nbar-2) \tilde{z}_\nbar^2}{9 (\tilde{z}_\nbar-1)} & \frac{\tilde{z}_\nbar^3}{9 (\tilde{z}_\nbar-1)}   \\[2ex]
 -\frac{\tilde{z}_\nbar^2 [(\tilde{z}_\nbar-5) \tilde{z}_\nbar+5]}{9 (\tilde{z}_\nbar-1)^2} & \frac{\tilde{z}_\nbar^3}{9 (\tilde{z}_\nbar-1)}   
\end{array}
\right]\,.
  \end{split}
\end{equation}

\subsection*{Collinear and ultra-collinear regimes}
 
The collinear contributions governed by
Eq.~\eqref{eq:def:asyexp:c:mom:at:omega} and
Eq.~\eqref{eq:def:asyexp:cbar:mom:at:omega} are formally
aligned with the integrals from the transitional sectors
as exhibited in
Eqs.~\eqref{eq:def:asyexp:cs:mom:int}-\eqref{eq:def:asyexp:cbars:mom:int}.
This correspondence even holds after reducing their rank
using Eq.~\eqref{eq:plus2:convertor} and transforming them
with the help of Eq.~\eqref{eq:trans:plusD:starD:n} if the
plus distribution terms are treated to be regular functions
and the boundary terms emerging from the integration by parts
are all adapted to the collinear regimes as appropriate.
The results then read,
\begin{equation}\label{eq:def:asyexp:mcfs:c}
  \begin{split}
    \frac{\done \sigma^{\langle \mathrm{m.c.}\rangle}_H}{\done Y_H \done \qT^2}
    \Bigg|_c
    \,\equiv\,&
      \frac{\alpha_s^3C_t^2}{192\pi^2s v^2}\,
      \sum_{i,j=\{g,q,\bar{q}\}}\,
      \sum_{\omega=-1}^{\infty}\,
      \left(\frac{\qT^2}{m_H^2}\right)^{\omega}\,
      \int^{\tilde{z}_n}_{x_n}\,\done z_n\,
      \left[ \mathbf{F}_{i/n}\left( \frac{x_n}{z_n} \right)\right]^{\mathbf{T}}
    \\
    &\,\times\,
      \left\{
        \mathbf{R}_{cs}^{(\omega),ij}(z_n)
        \,+\,
        \frac{\mathbf{P}_{cs}^{(\omega),ij}}{1-z_n}
        \,+\,
        \mathbf{B}_{cs}^{(\omega),ij}(z_n)\,
        \Big[
          \delta\left(x_n-z_n\right)
          -
          \delta\left(\tilde{z}_n-z_n\right)
        \Big]
      \right\}
    \\
    &\,\times\,
      \mathbf{F}_{j/\nbar}\left(  x_\nbar \right)\,,
    \\
    \frac{\done \sigma^{\langle \mathrm{m.c.}\rangle}_H}{\done Y_H \done \qT^2}
    \Bigg|_{\bar{c}}
    \,\equiv\,&
      \frac{\alpha_s^3C_t^2}{192\pi^2s v^2}\,
      \sum_{i,j=\{g,q,\bar{q}\}}\,
      \sum_{\omega=-1}^{\infty}\,
      \left(\frac{\qT^2}{m_H^2}\right)^{\omega}\,
      \int^{\tilde{z}_\nbar}_{x_\nbar}\,\done z_\nbar\,
      \left[\mathbf{F}_{i/n}\left( {x_n} \right)\right]^{\mathbf{T}}
    \\
    &\,\times\,
      \left\{
        \mathbf{R}_{\bar{c}s}^{(\omega),ij}(z_\nbar)
        \,+\,
        \frac{\mathbf{P}_{\bar{c}s}^{(\omega),ij}}{1-z_\nbar}
        \,+\,
        \mathbf{B}_{\bar{c}s}^{(\omega),ij}(z_\nbar)\,
        \Big[
          \delta\left(x_\nbar-z_\nbar\right)
          -
          \delta\left(\tilde{z}_\nbar-z_\nbar\right)
        \Big]
      \right\}
    \\
    &\,\times\,
      \mathbf{F}_{j/\nbar}\left(\frac{x_{\nbar}}{z_\nbar}\right)\,.
  \end{split}
\end{equation}
Combining this result with the results from the moderate domain
in Eqs.~\eqref{eq:def:asyexp:mcfs:cs} and
\eqref{eq:def:asyexp:mcfs:cbars}, we observe that all the dependences
on the parameters $\tilde{z}_{n}$ and $\tilde{z}_{\nbar}$,
which are associated with auxiliary cutoff scales $\nu_n$ and
$\nu_{\nbar}$, drop out.
This cancellation is in agreement with the vanishing derivatives in
Eqs.~\eqref{eq:def:asyexp:central:power:omega:nundiff} and
\eqref{eq:def:asyexp:central:power:omega:nunbardiff} as well as
the expectation from a direct power expansion on the analytic
expressions of the \qT~spectrum (if they are known).

Finally, we move to the ultra-collinear contributions, which contains
the maximal longitudinal momentum of an emitted parton allowed by the
colliding energy.
As discussed in Sec.~\ref{sec:red:qT:boundary}, this sector stems from
the power expansion on the boundary conditions in
Eq.~\eqref{eq:def:b.c.}.
Therefore, the resulting expression will be proportional to the derivatives
of the respective PDFs at the end point.
Moreover, since, as defined in
Eqs.~\eqref{eq:def:kc:boundary} and \eqref{eq:def:kcbar:boundary}, the
phase space integrals of this region are all restricted in the boundary
strips of power-suppressed sizes, the ultra-collinear contribution will
not invoke any singular behaviour in the low~\qT~region but can be relevant
starting from \NLP as a non-logarithmic power correction.
The expression up to \NNLP can be organised as,
\begin{equation} \label{eq:def:asyexp:mcfs:uc:ucbar}
  \begin{split}
    \frac{\done \sigma_H}{\done Y_H \done \qT^2}\Bigg|_{b.c.}
    \,\equiv\,&
      \frac{\alpha_s^3C_t^2}{192\pi^2s v^2}\,
      \sum_{i,j=\{g,q,\bar{q}\}}\,
      \sum_{\omega=-1}^{\infty}\,
      \left(\frac{\qT^2}{m_H^2}\right)^{\omega}\,
      \Bigg\{\,
        \left[ \mathbf{F}_{i/n}\left(1\right)\right]^{\mathbf{T}}
        \,\cdot\,
        \mathbf{B}^{(\omega),ij}_{u {c}}(x_n)
        \,\cdot\,
        \mathbf{F}_{j/\nbar}\left(x_\nbar\right)
    \\
    &\hspace*{58mm}
        \,+\,
        \left[ \mathbf{F}_{i/n}\left(x_n\right)\right]^{\mathbf{T}}
        \,\cdot\,
        \mathbf{B}^{(\omega),ij}_{u\bar{c}}(x_\nbar)
        \,\cdot\, \mathbf{F}_{j/\nbar}\left(1\right)\,
      \Bigg\}\,.
  \end{split}
\end{equation}
Therein, the leading power coefficients all vanish,
\begin{align}
  \mathbf{B}^{(-1),ij}_{uc}=\mathbf{B}^{(-1),ij}_{u\bar{c}}=0\,.
\end{align}
The partonic matrices at the (sub-)subleading power
induced by the $\nbar$-collinear mode evaluate to,
\begin{equation}\label{eq:Bucbar:nlp_nnlp}
  \begin{split}
&\mathbf{B}^{(0),gg}_{u\bar{c}}(x_\nbar)= \frac{[({x}_\nbar-1) {x}_\nbar+1]^2}{2 ({x}_\nbar-1)}\,,\\
&\mathbf{B}^{(1),gg}_{u\bar{c}}(x_\nbar)= 
\left[
\begin{array}{ccc}
 \frac{-3 {x}_\nbar^6+11 {x}_\nbar^5-12 {x}_\nbar^4+17 {x}_\nbar^3-10 {x}_\nbar^2+{x}_\nbar-1}{8 ({x}_\nbar-1)^2} & \frac{{x}_\nbar
   [({x}_\nbar-1) {x}_\nbar+1]^2}{8 ({x}_\nbar-1)}  \\[2.0ex]
 -\frac{({x}_\nbar+1) [({x}_\nbar-1) {x}_\nbar+1]^2}{4 ({x}_\nbar-1)^2} & 0 \\
\end{array}
\right]\,,\\
&\mathbf{B}^{(0),q(\bar{q})g}_{u\bar{c}}(x_\nbar)=\mathbf{B}^{(0),q\bar{q}}_{u\bar{c}}(x_\nbar)=0\,,\\
&\mathbf{B}^{(1),q(\bar{q})g}_{u\bar{c}}(x_\nbar)=  -\frac{{x}_\nbar}{9 ({x}_\nbar-1)^2}\,,\\
&\mathbf{B}^{(1),q \bar{q} }_{u\bar{c}}(x_\nbar)=  \frac{8}{27} ({x}_\nbar-1) {x}_\nbar \,.
  \end{split}
\end{equation}
Those from the $n$-collinear mode can be derived from the
relation below,
\begin{align}
  \mathbf{B}^{(\omega),ij}_{uc}(x_n)
  =
    \left[\mathbf{B}^{(\omega),ij}_{u\bar{c}}(x_\nbar)\right]^{\mathbf{T}}
    \Bigg|_{x_\nbar\to x_n}\,,
  \quad
  \text{where}
  ~
  \omega\in\{-1,0,+1\}\;
  \text{and}\;
  \{ij\}\in\{gg,q\bar{q}\}\,,
\end{align}
together with,
\begin{equation}
  \begin{split}
&\mathbf{B}^{(0) q(\bar{q})g}_{uc} =\frac{1}{9} [-({x}_n-2) {x}_n-2]\,,\\
&\mathbf{B}^{(1) q(\bar{q})g}_{uc} =
\left[
\begin{array}{ccc}
 \frac{[{x}_n (2 {x}_n-5)+5] {x}_n^2+2}{36 ({x}_n-1)} & \frac{1}{18} \left({x}_n^2+\frac{2}{{x}_n-1}\right)  \\[2.ex]
 -\frac{1}{36} {x}_n [({x}_n-2) {x}_n+2] & 0  \\
\end{array}
\right]\,.
  \end{split}
\end{equation}
We have presented here the analytic expressions of the first
three terms in the power series of
${\done \sigma_H}/({\done Y_H \done \qT^2})$ in the vicinity
of $\qT=0$, using $pp \to H+X$ with the partonic channels
$gg\to Hg$, $q(\bar{q})g\to Hq(\bar{q})$, and $q\bar{q}\to Hg$
as an example.
In addition, the \qT~distribution also comprises the channels
$gq(\bar{q})\to H q(\bar{q})$ and $\bar{q}q\to H g$, for which
the partonic matrices from the corresponding sectors can be
derived from those in the $q(\bar{q})g\to Hq(\bar{q})$ and
$q\bar{q}\to Hg$ processes by exchanging the labels
$n\leftrightarrow \nbar$ as appropriate.
  
In the previous investigations on the process $pp\to H+X$,
the ingredients contributing to the leading singular behaviour
of the \qT~distributions are known up to \NNNLO~\cite{Catani:2007vq,
  Catani:2012qa,Catani:2022sgr,Catani:2011kr,Luo:2020epw,Luo:2019szz,
  Ebert:2020yqt,Gehrmann:2012ze,Gehrmann:2014yya,Echevarria:2016scs,
  Luo:2019hmp,Luo:2019bmw,Gutierrez-Reyes:2019rug,Chetyrkin:1997iv,
  Chetyrkin:1997un,Gehrmann:2014vha,Gehrmann:2010ue,Baikov:2009bg,
  Gehrmann:2005pd,Harlander:2000mg}.
Comparing the \NLO expressions in \cite{Li:2016ctv,Luo:2019bmw}
derived with the exponential rapidity regulator with the leading
terms of the sum of Eqs.~\eqref{eq:def:asyexp:mcfs:cs},
\eqref{eq:def:asyexp:mcfs:cbars}, and \eqref{eq:def:asyexp:mcfs:c},
we find perfect agreement between the two upon performing an inverse
Fourier transformation.
In addition, the \NLP power corrections have also been evaluated in
\cite{Ebert:2018gsn} via the $\eta$- and pure rapidity regularisation
schemes.
Here as well, we have verified that the sum of
Eqs.~\eqref{eq:def:asyexp:mcfs:cs}, \eqref{eq:def:asyexp:mcfs:cbars},
and \eqref{eq:def:asyexp:mcfs:c} can exactly reproduce the expressions
in \cite{Ebert:2018gsn},
\changed{
except for the boundary terms encoded in the $\mathbf{B}_{cs}$ matrix
in Eq.~\eqref{eq:def:asyexp:mcfs:cs}, $\mathbf{B}_{\bar{c}s}$ in
Eq.~\eqref{eq:def:asyexp:mcfs:cbars}, and
$\mathbf{B}_{uc,u\bar{c}}$ in Eq.~\eqref{eq:def:asyexp:mcfs:uc:ucbar}.
The reason for the difference in the boundary terms is that all PDFs
in \cite{Ebert:2018gsn} are assumed to vanish at end point, i.e.\
$f_{i/n}(1)= f_{j/\nbar}(1)=0$, thereby setting all boundary
corrections to zero by default.}
However, in this paper, all contributions from the boundary regions
are retained for completeness and generality, and we find
non-vanishing boundary terms at least in the PDF set \texttt{MSHT20nlo\_as118},
which is used in Sec.\ \ref{sec:asyexp:numeric}.
Further details of this comparison can be found in App.~\ref{sec:comp:nlp}.

\subsection{Power expansion with the pure  rapidity regulator}\label{sec:app:prap}
 
In this part, we re-derive the power series of
${\done \sigma_H}/({\done Y_H \done \qT^2})$ up to \NNLP with
the interior contribution evaluated from the rearranged ingredients,
as illustrated in Sec.~\ref{sec:asyexp:qT:extrapl}.
An appropriate regularisation scheme is required to be put in
place to tame the rapidity divergences emerging from the domains
$k_{\pm}\to0$.
In the following, the pure rapidity regularisation prescription
proposed in \cite{Ebert:2018gsn,Moult:2019vou} will be employed
to facilitate our calculation.
In this regard, the regulator $\mathcal{R}$ takes the form
\begin{align}\label{eq:def:p:rap}
  \mathcal{R}
  =\left|\frac{\tilde{\nu}_n}{k_+}\,\frac{k_-}{\tilde{\nu}_\nbar}\right|^{\tau}
  \,.
\end{align}
Here $\tilde{\nu}_n$ and $\tilde{\nu}_\nbar$ represent two
auxiliary scales, from which an effective cutoff can be imposed
upon the rapidity of the emitted partons~\cite{Ebert:2018gsn},
akin to dimensional regularisation.
In this paper, for simplicity, we take
$\tilde{\nu}_n=\tilde{\nu}_\nbar=m_H$ throughout the calculation.

Applying Eq.~\eqref{eq:def:p:rap} onto the $n$-collinear function
defined in Eq.~\eqref{eq:def:asyexp:c:at:omega:nured} concerns
only the integrand with $\sigma\le\omega$ in the limit $\tau\to0$,
more explicitly,
\begin{equation}\label{eq:def:asyexp:c:at:omega:nured:prap}
  \begin{split}
    \widetilde{\mathcal{G}}^{\rho,\sigma,(\omega)}_{[\kappa],\{\alpha,\beta\} }\Bigg|^{\langle \mathrm{p.ra.}\rangle}_{c}\!\!\!\!
    =\;&
      \frac{\bar{\theta}\left(\omega-\sigma\right)}{(\omega-\rho)!}
      \int^{\tilde{k}^{\mathrm{max}}_+}_{0}\!
      \frac{\done k_+}{k_+}\,(k_+)^{\sigma-\omega}
      {F}^{(\alpha_n)}_{i/n,\beta_n}\!\!\left(k_++m_H e^{-Y_H}\right)
      {F}^{(\alpha_\nbar+\omega-\rho)}_{j/\nbar,\beta_\nbar}\!\!\left(m_H e^{+Y_H}\right)
    \\
    &
      +
      \frac{\theta\left(\omega-\sigma\right)}{(\omega-\rho)!}
      \lim_{\tau\to0}
      \int^{\tilde{k}^{\mathrm{max}}_+}_{0}\!
      \frac{\done k_+}{k_+}\,(k_+)^{\sigma-\omega}
      \left|\frac{\qT^2}{k_+^2}\right|^{\tau}
      {F}^{(\alpha_n)}_{i/n,\beta_n}\!\!\left(k_++m_H e^{-Y_H}\right)
      {F}^{(\alpha_\nbar+\omega-\rho)}_{j/\nbar,\beta_\nbar}\!\!\left(m_H e^{+Y_H}\right)
      .\hspace*{-20mm}
  \end{split}
\end{equation}
To compute the second term on the r.h.s., we make use of the generalised
star-distribution of Eq.~\eqref{eq:def:sDist} to regularise all singular contributions, and then, within the pure rapidity regularisation scheme,
we complete the phase space integrals over those singular terms.
The result reads,
\begin{equation}\label{eq:def:asyexp:c:at:omega:nured:prap:2nd}
  \begin{split}
    \lefteqn{
      \theta\left(\omega-\sigma\right)
      \lim_{\tau\to0}
      \int^{\tilde{k}^{\mathrm{max}}_+}_{0}
      \frac{\done k_+}{k_+}\,(k_+)^{\sigma-\omega}
      \left|\frac{\qT^2}{k_+^2}\right|^{\tau}
      {F}^{(\alpha_n)}_{i/n,\beta_n}\left(k_++m_H e^{-Y_H}\right)
      {F}^{(\alpha_\nbar+\omega-\rho)}_{j/\nbar,\beta_\nbar}\left(m_H e^{+Y_H}\right)
    }\\
    &=\, \theta\left(\omega-\sigma\right)\int^{\tilde{k}^{\mathrm{max}}_+}_{0}
    {\done k_+}
      \left[\frac{1}{k_+^{\omega-\sigma+1}}\right]_*^{\nu_n}
      {F}^{(\alpha_n)}_{i/n,\beta_n}\left(k_++m_H e^{-Y_H}\right)
    {F}^{(\alpha_\nbar+\omega-\rho)}_{j/\nbar,\beta_\nbar}\left(m_H e^{+Y_H}\right)\\
    &\phantom{=}{}+\theta\left(\omega-\sigma-1\right)
    \sum_{\eta=0}^{\omega-\sigma-1}
    \frac{ {F}^{(\alpha_n+\eta)}_{i/n,\beta_n}\left( m_H e^{-Y_H}\right)
    {F}^{(\alpha_\nbar+\omega-\rho)}_{j/\nbar,\beta_\nbar}\left(m_H e^{+Y_H}\right)}{\eta!}
    \frac{\nu_n^{\sigma+\eta-\omega}}{\sigma+\eta-\omega}\\
    &\phantom{=}{}+\theta(  \omega-\sigma)\left\{ \ln\left[\frac{\nu_n}{\qT}\right]-\frac{1}{2\tau}\right\}\,
    \frac{  {F}^ {(\alpha_n+\omega-\sigma)}_{i/n,\beta_n}\left( m_H e^{-Y_H}\right)\,
        {F}^{ (\alpha_\nbar+\omega-\rho)}_{j/\nbar,\beta_\nbar}\left(m_H e^{+Y_H}\right)\,}{ (\omega-\sigma)!}\,,
  \end{split}
\end{equation}
where we have expanded in the small parameter $\tau$ and kept
only contributions up to $\mathcal{O}(\tau^0)$.
From Eqs.~\eqref{eq:def:asyexp:c:at:omega:nured:prap} and
\eqref{eq:def:asyexp:c:at:omega:nured:prap:2nd}, we note
that except for the pole term, the recombined collinear sector
here exactly reproduces the sum of
Eqs.~\eqref{eq:def:asyexp:c:mom:at:omega} and
\eqref{eq:def:asyexp:cs:mom:int} constructed via the explicit
cutoffs.
Likewise, the first term of Eq.~\eqref{eq:def:asyexp:c:at:omega:nured:prap},
proportional to $\bar{\theta}\left(\omega-\sigma\right)$, and the
term containing the star-distribution in
Eq.~\eqref{eq:def:asyexp:c:at:omega:nured:prap:2nd} are equivalent
to the sum of Eq.~\eqref{eq:def:asyexp:c:mom:at:omega} and the first
two lines of Eq.~\eqref{eq:def:asyexp:cs:mom:int}, as can be seen
with the help of Eq.~\eqref{eq:def:asyexp:c:at:omega:nured} and the
fact that the integrands in those sectors are all regular in the
limit $k_+\to0$.
In order to interpret the correspondence between the last two
lines of Eq.~\eqref{eq:def:asyexp:cs:mom:int} and
Eq.~\eqref{eq:def:asyexp:c:at:omega:nured:prap:2nd}, it is
beneficial to recall that the $\nu_n$-dependent terms in
Eq.~\eqref{eq:def:asyexp:cs:mom:int} in fact stem from the boundary
condition in the phase space integral of Eq.~\eqref{eq:def:asyexp:s:mom1},
which, through applying the star distribution in Eq.~\eqref{eq:def:sDist},
is entirely inherited by the singular terms in
Eq.~\eqref{eq:def:asyexp:c:at:omega:nured:prap:2nd} and then mostly
preserved in evaluating the $k_+$ integral within the pure rapidity
regularisation scheme.
Therefore, an analogous pattern is presented by the expressions
in Eqs.~\eqref{eq:def:asyexp:c:at:omega:nured:prap} and
\eqref{eq:def:asyexp:c:at:omega:nured:prap:2nd} and those in
Eq.~\eqref{eq:def:asyexp:c:mom:at:omega} and
Eq.~\eqref{eq:def:asyexp:cs:mom:int}.

A similar strategy can also be used to calculate the $\nbar$-collinear
term in Eq.~\eqref{eq:def:asyexp:c:at:omega:nured}, which evaluates to,
\begin{equation}\label{eq:def:asyexp:cbar:at:omega:nured:prap}
  \begin{split}
    \widetilde{\mathcal{G}}^{\rho,\sigma,(\omega)}_{[\kappa],\{\alpha,\beta\}}
    \Bigg|^{\langle \mathrm{p.ra.}\rangle}_{\bar{c}}
    =&\;
      \frac{\bar{\theta}( \omega-\rho)}{(\omega-\sigma)!}\,
      \int^{{\frac{\qT^2}{\tilde{k}^{\mathrm{min}}_+}} }_0\,
      \frac{\done k_-}{k_-}\,k_-^{\rho-\omega}\,
      {F}^{(\alpha_n+\omega-\sigma)}_{i/n,\beta_n}\left(m_H e^{-Y_H}\right)\,
      {F}^{(\alpha_\nbar)}_{j/\nbar,\beta_\nbar}\left(k_-+m_H e^{+Y_H}\right)
    \\
    &+
      \frac{{\theta}(  \omega-\rho)}{(\omega-\sigma)!}\,
      \int^{{\frac{\qT^2}{\tilde{k}^{\mathrm{min}}_+}}}_0\,
      \done k_-\,
      \left[\frac{1}{k_-^{\omega-\rho+1}}\right]^{\nu_{\nbar}}_{*}\,
      {F}^{(\alpha_n+\omega-\sigma)}_{i/n,\beta_n}\left(m_H e^{-Y_H}\right)\,
      {F}^{(\alpha_\nbar)}_{j/\nbar,\beta_\nbar}\left(k_-+m_H e^{+Y_H}\right)
      \hspace*{-20mm}
    \\
    &+
      \frac{{\theta}(\omega-\rho-1)}{(\omega-\sigma)!}\,\sum_{\lambda=0}^{\omega-\rho-1}\,
      \frac{
        {F}^{(\alpha_n+\omega-\sigma)}_{i/n,\beta_n}\left(m_H e^{-Y_H}\right)\,
        {F}^{(\alpha_\nbar+\lambda)}_{j/\nbar,\beta_\nbar}\left(m_H e^{+Y_H}\right)\,
      }{\lambda!} \,
      \frac{ \nu_{\nbar}^{\rho+\lambda-\omega}}{\rho+\lambda-\omega}
    \\
    &+
      \frac{ {\theta}(  \omega-\rho)}{(\omega-\sigma)!}\,
      \left\{\ln\left[\frac{\nu_{\nbar}}{\qT}\right]+\frac{1}{2\tau}\right\}\,
      \frac{
        {F}^{(\alpha_n+\omega-\sigma)}_{i/n,\beta_n}\left(m_H e^{-Y_H}\right)\,
        {F}^{(\alpha_\nbar+\omega-\rho)}_{j/\nbar,\beta_\nbar}\left(m_H e^{+Y_H}\right)\,
      }{(\omega-\rho)!}
      \,.
  \end{split}
\end{equation}
In comparison with Eq.~\eqref{eq:def:asyexp:c:at:omega:nured:prap:2nd},
we note that the sign in front of the pole term is reversed. 
This indicates that the rapidity regulator in appraising the
$\nbar$-collinear sector is in practice activated in a distinct
extremal region than in the $n$-collinear case.
Combining the $n$-collinear function in
Eq.~\eqref{eq:def:asyexp:c:at:omega:nured:prap:2nd} with the
$\nbar$-collinear piece of
Eq.~\eqref{eq:def:asyexp:cbar:at:omega:nured:prap}, it is immediate
to find that all poles cancel, fully reproducing our previous results
summing Eqs.~\eqref{eq:def:asyexp:c:mom:at:omega} and
\eqref{eq:def:asyexp:cbar:mom:at:omega} as well as  Eqs.~\eqref{eq:def:asyexp:cs:mom:int} and
\eqref{eq:def:asyexp:cbars:mom:int} derived using the momentum cutoffs.

In fact, this agreement suggests that within the pure-rapidity
regularisation scheme, the soft contribution of
Eq.~\eqref{eq:def:s:fs:rap} and zero-bin subtrahends of
Eq.~\eqref{eq:def:c0:ns} could be redundant.
To confirm this, it is worth noting that the phase space integrals
in these two functions are all unbounded by definition, such that
applying the pure rapidity regulator here only results in vanishing
quantities in the limit $\tau\to0$.
It thus leads to,
\begin{align}\label{eq:def:asyexp:s:c0:cbar0:prap}
  \widetilde{\mathcal{G}}^{\rho,\sigma,(\omega)}_{[\kappa],\{\alpha,\beta\} }\Bigg|^{\FSmeth,\langle\mathrm{p.ra.} \rangle}_{s}= \widetilde{\mathcal{G}}^{\rho,\sigma,(\omega)}_{[\kappa],\{\alpha,\beta\} }\Bigg|^{\langle \mathrm{NS}\rangle,\langle\mathrm{p.ra.}\rangle}_{c0} =\widetilde{\mathcal{G}}^{\rho,\sigma,(\omega)}_{[\kappa],\{\alpha,\beta\} }\Bigg|^{\langle \mathrm{NS}\rangle,\langle\mathrm{p.ra.} \rangle}_{\bar{c}0}=0\,.
\end{align}
In absence of the soft function and the zero-bin subtraction
within the pure-rapidity regularisation scheme, we can then
skip indulging in the inhomogeneous behaviour induced by the
doubly and triply projected integrands of
Eq.~\eqref{eq:def:asyexp:jantzen}.
Similarly, we can also freely switch from Jantzen's
formalism~\cite{Jantzen:2011nz} in
Eq.~\eqref{eq:def:asyexp:jantzen:red} to our proposal in
Eq.~\eqref{eq:def:asyexp:central:power:omega:rap:hom} to perform
the power expansion.
This forms in part the reason why $\mathcal{R}$ in
Eq.~\eqref{eq:def:p:rap} is categorised to be one of the
homogeneous regulators here.
 
Applying the expressions of
Eqs.~\eqref{eq:def:asyexp:c:at:omega:nured:prap}-%
\eqref{eq:def:asyexp:s:c0:cbar0:prap} to
Eq.~\eqref{eq:def:asyexp:central:power:omega:rap} or
Eq.~\eqref{eq:def:asyexp:central:power:omega:rap:hom},
we arrive at the expression of the interior contribution
and, in turn, the power expansion of
${\done \sigma_H}/(\done Y_H \done \qT^2)$.
To make their expressions more compact, we recast them
once again in terms of the dimensionless parameters defined in
Eq.~\eqref{eq:def:kin:xn:xnbar} by means of the identities in
Eq.~\eqref{eq:plus2:convertor}.
Up to \NNLP, the results can be organised below,
\begin{equation}\label{eq:def:asyexp:prap}
  \begin{split}
  \frac{\done \sigma_H}{\done Y_H \done \qT^2}
  \,=&\,
  \frac{\done \sigma_H}{\done Y_H \done \qT^2}\Bigg|_{b.c.}
  \,+\,
  \frac{\done \sigma^{\langle \mathrm{p.ra.}\rangle}_H}{\done Y_H \done \qT^2}
  \Bigg|_{c}
  \,+\,
  \frac{\done \sigma^{\langle \mathrm{p.ra.}\rangle}_H}{\done Y_H \done \qT^2}
  \Bigg|_{\bar{c}}  \,,
\end{split}
\end{equation}
where the first term on the r.h.s accounts for the boundary correction as illustrated in Eq.~\eqref{eq:def:asyexp:mcfs:uc:ucbar}. The second and third terms govern the $n$-collinear contributions in Eqs.~(\ref{eq:def:asyexp:c:at:omega:nured:prap}-\ref{eq:def:asyexp:c:at:omega:nured:prap:2nd}) and the $\nbar$-collinear case in Eq.~\eqref{eq:def:asyexp:cbar:at:omega:nured:prap}, respectively. 
Their expressions read, 
\begin{equation}
  \begin{split}
    \frac{\done \sigma^{\langle \mathrm{p.ra.}\rangle}_H}{\done Y_H \done \qT^2}\Bigg|_c
    \,\equiv\,&
      \frac{\alpha_s^3C_t^2}{192\pi^2s v^2}\,
      \sum_{i,j=\{g,q,\bar{q}\}}\,
      \sum_{\omega=-1}^{\infty}\,
      \left(\frac{\qT^2}{m_H^2}\right)^{\omega}\,
      \int^{1}_{x_n}\,\done z_n\,
      \left[\mathbf{F}_{i/n}\left(\frac{x_n}{z_n}\right)\right]^{\mathbf{T}}
    \\
    &\,\times\,
      \Bigg\{
        \mathbf{R}_{cs}^{(\omega),ij}(z_n)
        \,+\,
        \mathbf{P}_{cs}^{(\omega),ij}\,\left[\frac{1}{1-z_n}\right]_+
        +\,
        \left(\mathbf{S}^{(\omega),ij}_c+\mathbf{D}_{cs}^{(\omega),ij}\right)\,\delta\left(1-z_n\right)
    \\
    &\hspace*{8mm}
        \,+\,
        \mathbf{B}_{cs}^{(\omega),ij}(z_n)\,
        \delta\left(x_n-z_n\right)
      \Bigg\}\;
      \mathbf{F}_{j/\nbar}(x_{\nbar})\,,
  \end{split}
\end{equation}
and
\begin{equation}
  \begin{split}
    \frac{\done \sigma^{\langle \mathrm{p.ra.}\rangle}_H}{\done Y_H \done \qT^2}
    \Bigg|_{\bar{c}}
    \,\equiv\,&
      \frac{\alpha_s^3C_t^2}{192\pi^2s v^2}\,
      \sum_{i,j=\{g,q,\bar{q}\}}\,
      \sum_{\omega=-1}^{\infty}\,
      \left(\frac{\qT^2}{m_H^2}\right)^{\omega}\,
      \int^{1}_{x_\nbar}\,\done z_\nbar\,
      \left[\mathbf{F}_{i/n}\left(x_n\right)\right]^{\mathbf{T}}
    \\
    &\,\times\,
      \Bigg\{
        \mathbf{R}_{ \bar{c}s}^{(\omega),ij}(z_\nbar)
        \,+\,
        \mathbf{P}_{\bar{c}s}^{(\omega),ij}\,\left[\frac{1}{1-z_\nbar}\right]_+
        +\,
        \left(\mathbf{S}^{(\omega),ij}_{\bar{c}}
        \,+\,
        \mathbf{D}_{\bar{c}s}^{(\omega),ij}\right)\,
        \delta\left(1-z_\nbar\right)
    \\
    &\hspace*{8mm}
        \,+\,
        \mathbf{B}_{\bar{c}s}^{(\omega),ij}(z_\nbar)\,
        \delta\left(x_\nbar-z_\nbar\right)
      \Bigg\}\;
      \mathbf{F}_{j/\nbar}\left(\frac{x_{\nbar}}{z_\nbar}\right)\,,
  \end{split}
\end{equation}
where the matrices $\mathbf{R}$, $\mathbf{P}$, $\mathbf{D}$,
and $\mathbf{B}$ have been previously displayed in
Sec.~\ref{sec:app:cfs}.
The novel ingredients here are $\mathbf{S}_c$ and
$\mathbf{S}_{\bar{c}}$, characterising the pole terms in
Eqs.~\eqref{eq:def:asyexp:c:at:omega:nured:prap:2nd} and
\eqref{eq:def:asyexp:cbar:at:omega:nured:prap}.
The coefficients of these singularities in the
$n$-($\nbar$-)collinear sector are of opposite (same)
sign, but otherwise identical, to those associated with
the logarithmic contributions.
\changed{The} latter are, as discussed in Sec.~\ref{sec:app:cfs},
associated with the partonic matrices in front of the
plus distributions.
Hence, we can now express $\mathbf{S}_{c}$ and
$\mathbf{S}_{\bar{c}}$ in terms of $\mathbf{P}_{cs}$ in
Eqs.~\eqref{eq:Pcs:gg_gq_qq}, more specifically,
\begin{align}
  \label{eq:Sc:nnll}
  \mathbf{S}^{(\omega),ij}_{c}\,
  =\,
  -\,\mathbf{S}^{(\omega),ij}_{\bar{c}}\,
  =\,
  -\,\frac{1}{2\tau}\,
  \mathbf{P}^{(\omega),ij}_{cs}\,,
  \quad
  \mathrm{where}~
  \omega\in\{-1,0,1\}~
  \mathrm{and}~
  \{ij\}=\{gg,q(\bar{q})g,q \bar{q}\}\,.
\end{align}
  
%
%

\subsection{Power expansion with the exponential rapidity regulator}
\label{sec:app:exp}

In order to explore the pattern of the power expansion of
${\done \sigma_H}/({\done Y_H \done \qT^2})$ in a more generic
way, in the following we re-appraise the interior region with
the aid  of the exponential regulator~\cite{Li:2016axz,Li:2016ctv}.
As will be illustrated below, the zero-bin subtrahends at this
moment are non-trivial and play an important role in the
cancellation of the pole terms, at variance to those derived in
Sec.~\ref{sec:app:prap}.
We will use the \NSmeth prescription here and
make use of Eq.~\eqref{eq:def:c0:ns} to establish them.
  
The exponential regulator takes the form,
\begin{align}\label{eq:def:exp}
  \mathcal{R}=\exp\left(-\tau\,b_0\,k_0\right)\,,
\end{align}
where $b_0=2e^{-\gamma_{\mathrm{E}}}$.
At \LP, where non-Abelian exponentiation
\cite{Li:2016ctv,Luo:2019hmp} holds, this type of regularisation
scheme has been extensively utilised to calculate the fixed-order
ingredients related to \qT~resummation.
For instance, the leading power beam and soft functions are now
available up to \NNNLO~\cite{Li:2016ctv,Li:2016axz,Luo:2019hmp,
  Luo:2019bmw,Luo:2020epw,Luo:2019szz}, while the anomalous dimensions
are available at \NNNNLO accuracy~\cite{Duhr:2022yyp,Moult:2022xzt}.
In the following, we will implement this method to the
\NLO~\qT~spectrum up to \NNLP, for the first time.
Our deliberations may also be useful for future analyses of
power suppressed contribution at \NNLO and beyond, within an
inhomogeneous regularisation scheme.
 
We start our discussion by considering the collinear sectors
of Eq.~\eqref{eq:def:asyexp:c:at:omega:nured}.
Applying the exponential regularisation method here follows
an analogous pattern to our steps leading up to
Eqs.~\eqref{eq:def:asyexp:c:at:omega:nured:prap} and
\eqref{eq:def:asyexp:c:at:omega:nured:prap:2nd}.
First, we use the star-distribution of Eq.~\eqref{eq:def:sDist}
to separate the regular and singular terms in the limit
$k_{\pm}\to0$.
The regular terms maximally preserve the form of
Eq.~\eqref{eq:def:asyexp:c:at:omega:nured}, while the
singular ones contain rapidity divergences under the
phase space integral necessitating the exponential regulator.
Here we use the functions $\Upsilon_{n}^{(h)}$ and
$\Upsilon_{\nbar}^{(h)}$ to collect those singular contributions
in the $n$- and $\nbar$-collinear sectors, respectively.
Within the exponential regularisation scheme, their expressions read,
\begin{align}\label{eq:def:ups:exp}
  \Upsilon_{n}^{(h)}\,
  \equiv\,
  \lim_{\tau\to0}\,\int_0^{\nu_n} \done k_+\,\frac{e^{-\tau b_0k_0}}{k_+^{h+1}}\,,
  \qquad\qquad\qquad
  \Upsilon_{\nbar}^{(h)}\,
  \equiv\,
  \lim_{\tau\to0}\,\int_0^{\nu_\nbar} \done k_-\,\frac{e^{-\tau b_0k_0}}{k_-^{h+1}}\,
  =\,\Upsilon_{n}^{(h)}\Bigg|_{\nu_n\to\nu_\nbar}\,.
\end{align}
As for the first few ranks, $\Upsilon_{n}^{(h)}$ evaluates to, 
\begin{equation}\label{eq:def:ups:exp:1}
  \begin{split}
    \Upsilon_{n}^{(0)}\,&=\,\ln\left[\frac{\nu_n}{\qT}\right]\,-\,L_{\tau}\,,\quad\quad
    \Upsilon_{n}^{(1)}\,=\,\frac{1}{\qT^2\,\tilde{\tau}}\,-\,\frac{1}{\nu_n}\,,\quad\quad
    \Upsilon_{n}^{(2)}\,=\,\frac{1}{\qT^4\,\tilde{\tau}^2}\,-\,\frac{1}{2\,\nu^2_n}\,-\,\frac{1}{\qT^2}\,,\\
    \Upsilon_{n}^{(3)}\,&=\,\frac{2}{\qT^6\,\tilde{\tau}^3}\,-\,\frac{1}{3\,\nu^3_n}\,-\,\frac{1}{\qT^4\,\tilde{\tau}}\,,\quad\quad\dots\dots\,,
  \end{split}
\end{equation}
where $L_{\tau}=\ln(\qT\tau)$ and
$\tilde{\tau}=\tau e^{-\gamma_{\mathrm{E}}}$.
The expression of $\Upsilon_{\nbar}^{(h)}$ can be derived
from $\Upsilon_{n}^{(h)}$ by the replacement $n\leftrightarrow \nbar$.
Thus equipped, we are now poised to establish the collinear
contribution of Eq.~\eqref{eq:def:asyexp:c:at:omega:nured}
within the exponential regularisation scheme, more explicitly
\begin{align}\label{eq:def:asyexp:c:at:omega:nured:exp}
  \widetilde{\mathcal{G}}^{\rho,\sigma,(\omega)}_{[\kappa],\{\alpha,\beta\} }
  \Bigg|^{\langle \mathrm{exp}\rangle}_{c}
  =\,&
    \frac{\bar{\theta}\left(\omega-\sigma\right)}{(\omega-\rho)!}
    \int^{\tilde{k}^{\mathrm{max}}_+}_{0}
    \frac{\done k_+}{k_+}\,(k_+)^{\sigma-\omega}
    {F}^{(\alpha_n)}_{i/n,\beta_n}\left(k_++m_H e^{-Y_H}\right)
    {F}^{(\alpha_\nbar+\omega-\rho)}_{j/\nbar,\beta_\nbar}\left(m_H e^{+Y_H}\right)\nnb\\
  &
    +
    \frac{\theta\left(\omega-\sigma\right)}{(\omega-\rho)!}
    \int^{\tilde{k}^{\mathrm{max}}_+}_{0}{\done k_+}
    \left[\frac{1}{k_+^{\omega-\sigma+1}}\right]_*^{\nu_n}
    {F}^{(\alpha_n)}_{i/n,\beta_n}\left(k_++m_H e^{-Y_H}\right)
    {F}^{(\alpha_\nbar+\omega-\rho)}_{j/\nbar,\beta_\nbar}\left(m_H e^{+Y_H}\right)\nnb\\
  &
    +
    \frac{\theta\left(\omega-\sigma-1\right)}{(\omega-\rho)!}
    \sum_{\eta=0}^{\omega-\sigma-1}
    \frac{
      {F}^{(\alpha_n+\eta)}_{i/n,\beta_n}\left( m_H e^{-Y_H}\right)
      {F}^{(\alpha_\nbar+\omega-\rho)}_{j/\nbar,\beta_\nbar}\left(m_H e^{+Y_H}\right)
    }{\eta!}\,
    \Upsilon_{n}^{(\omega-\sigma-\eta)}
  \nnb\\
  &
    +
    \frac{\theta(  \omega-\sigma)}{(\omega-\rho)!} \,
    \frac{
      {F}^{(\alpha_n+\omega-\sigma)}_{i/n,\beta_n}\left( m_H e^{-Y_H}\right)\,
      {F}^{(\alpha_\nbar+\omega-\rho)}_{j/\nbar,\beta_\nbar}\left(m_H e^{+Y_H}\right)
    }{(\omega-\sigma)!}\,
     \Upsilon_{n}^{(0)}\,,\\
\label{eq:def:asyexp:cbar:at:omega:nured:exp} 
  \widetilde{\mathcal{G}}^{\rho,\sigma,(\omega)}_{[\kappa],\{\alpha,\beta\} }
  \Bigg|^{\langle \mathrm{exp}\rangle}_{\bar{c}}
  =&\,
    \frac{\bar{\theta}( \omega-\rho)}{(\omega-\sigma)!}\,
    \int^{{\frac{\qT^2}{\tilde{k}^{\mathrm{min}}_+}} }_0\,
    \frac{\done k_-}{k_-}\,k_-^{\rho-\omega}\,
    {F}^{(\alpha_n+\omega-\sigma)}_{i/n,\beta_n}\left(  m_H e^{-Y_H}\right)\,
    {F}^{(\alpha_\nbar)}_{j/\nbar,\beta_\nbar}\left(k_-+m_H e^{+Y_H}\right)  \,\nnb\\
  &
    +
    \frac{{\theta}(  \omega-\rho)}{(\omega-\sigma)!}\,
    \int^{{\frac{\qT^2}{\tilde{k}^{\mathrm{min}}_+}}}_{0}{\done k_-}
    \left[\frac{1}{k_-^{\omega-\rho+1}}\right]^{\nu_{\nbar}}_{*}\,
    {F}^{(\alpha_n+\omega-\sigma)}_{i/n,\beta_n}\left(m_H e^{-Y_H}\right)\,
    {F}^{(\alpha_\nbar)}_{j/\nbar,\beta_\nbar}\left(k_-+m_H e^{+Y_H}\right)\,\nnb\\
  &
    +
    \frac{{\theta}(\omega-\rho-1)}{(\omega-\sigma)!}\,
    \sum_{\lambda=0}^{\omega-\rho-1}\,
    \frac{
      {F}^{(\alpha_n+\omega-\sigma)}_{i/n,\beta_n}\left( m_H e^{-Y_H}\right)\,
      {F}^{(\alpha_\nbar+\lambda)}_{j/\nbar,\beta_\nbar}\left(m_H e^{+Y_H}\right)
    }{\lambda!} \,
    \Upsilon_{\nbar}^{(\omega-\rho-\lambda)}\nonumber\\
  &
    +
    \frac{{\theta}(\omega-\rho)}{(\omega-\sigma)!}\,
    \frac{
      {F}^{(\alpha_n+\omega-\sigma)}_{i/n,\beta_n}\left( m_H e^{-Y_H}\right)\,
      {F}^{(\alpha_\nbar+\omega-\rho)}_{j/\nbar,\beta_\nbar}\left(m_H e^{+Y_H}\right)
      }{(\omega-\rho)!} \,
      \Upsilon_{\nbar}^{(0)}\,.
\end{align}
By comparison with the expressions in
Eqs.~\eqref{eq:def:asyexp:c:at:omega:nured:prap}-%
\eqref{eq:def:asyexp:cbar:at:omega:nured:prap} evaluated
in the pure-rapidity regularisation scheme,
the above expressions possess the same regular contributions
in the first two lines, as the regulator has no relevance in the
absence of rapidity divergences.
In their third lines, aside from identical $\nu_n$ and
$\nu_\nbar$ dependences residing in $\Upsilon_{n(\nbar)}$ compared to those in
Eqs.~\eqref{eq:def:asyexp:c:at:omega:nured:prap:2nd}-%
\eqref{eq:def:asyexp:cbar:at:omega:nured:prap}, the power series
in terms of $\qT$ and $\tilde{\tau}$ is appearing.
This behaviour stems from the definition of the exponential
regulator in Eq.~\eqref{eq:def:exp}, comprising the temporal
component $k^0= k_++(\qT^2/k_+)$ as a whole.
In deriving the $n$-collinear function, for instance, its second
part $(\qT^2/k_+)$ suppresses the rapidity divergences, thereby
yielding the leading singularity in the limit $\tilde{\tau}\to0$,
whilst the first part $k_+$ gives rise to corrections of
$\mathcal{O}(\tilde{\tau})$ and in turn generates subleading
singular terms and the constant contributions, as exhibited in
Eqs.~\eqref{eq:def:ups:exp:1}.
Ultimately, confronting the fourth line of
Eqs.~\eqref{eq:def:asyexp:c:at:omega:nured:prap:2nd}-%
\eqref{eq:def:asyexp:cbar:at:omega:nured:prap} to the results in
Eqs.~\eqref{eq:def:asyexp:c:at:omega:nured:exp}-%
\eqref{eq:def:asyexp:cbar:at:omega:nured:exp}, we observe that
addition logarithms $L_{\tau}=\ln(\qT\tau)$ emerge in the
exponential regularisation scheme.
In the limit $\tau\to0$, those logarithms in practice play an
equivalent role to the pole term in
Eqs.~\eqref{eq:def:asyexp:c:at:omega:nured:prap:2nd}-%
\eqref{eq:def:asyexp:cbar:at:omega:nured:prap}.
 
Combining the collinear functions we have derived using the
exponential regulator above, a cancellation of all pole terms is
not found, at variance with our earlier findings for
Eqs.~\eqref{eq:def:asyexp:c:at:omega:nured:prap}-%
\eqref{eq:def:asyexp:cbar:at:omega:nured:prap}.
This therefore calls for the zero-bin subtraction procedure derived
in Eq.~\eqref{eq:def:c0:ns} to remove all overlapping contributions.
Implementing the exponential regulator in Eq.~\eqref{eq:def:c0:ns}
prompts the $k_{\pm}$-integrals over the interval $k_{\pm}\in[0,\infty]$,
which are encoded by the function $\Upsilon_{zb}^{(h)}$ in this paper,
more specifically,
\begin{align} \label{eq:def:ups:exp:zb}
  \Upsilon_{zb}^{(h)}\,
  \equiv\,
  \lim_{\tau\to0}\, \int^{\infty}_{0}\, \done k_{\pm} \,
  \frac{e^{-\tau b_0k_0}}{(k_{\pm})^{h+1}}\,.
\end{align}
The results for the first few ranks read, 
\begin{align}  \label{eq:def:ups:exp:zb:n2lp}
  \Upsilon_{zb}^{(0)}\,=\,-\,2\,L_{\tau}\,,\quad\;
  \Upsilon_{zb}^{(1)}\,=\,\frac{1}{\qT^2\,\tilde{\tau}}\,,\quad\;
  \Upsilon_{zb}^{(2)}\,=\,\frac{1}{\qT^4\,\tilde{\tau}^2}
    \,-\,\frac{1}{\qT^2}\,,
  \quad\;
  \Upsilon_{zb}^{(3)}\,&=\,\frac{2}{\qT^6\,\tilde{\tau}^3}
    \,-\, \frac{1}{\qT^4\,\tilde{\tau}}\,,  \quad\;\dots\dots \,.
\end{align}
With the help of Eq.~\eqref{eq:def:ups:exp:zb}, the zero-bin subtrahends
in the exponential regularisation can be written as,
\begin{equation}\label{eq:def:c0:ns:exp}
  \begin{split}
    \widetilde{\mathcal{G}}^{\rho,\sigma,(\omega)}_{[\kappa],\{\alpha,\beta\}}
    \Bigg|^{\NSmeth,  \langle \mathrm{exp} \rangle}_{c0}\,
    \!\!\!=\,&
      \theta\left(\omega-\sigma\right)
      \sum_{\eta=0}^{\omega-\sigma }
      \frac{
        {F}^{(\alpha_n+\eta)}_{i/n,\beta_n}\left(m_H e^{-Y_H}\right)\,
        {F}^{(\alpha_\nbar+\omega-\rho)}_{j/\nbar,\beta_\nbar}\left(m_H e^{+Y_H}\right)
      }{\eta!\, (\omega-\rho)!}
      \left(1-\frac{\delta_{\omega}^{\sigma+\eta}}{2}\right)
      \Upsilon_{zb}^{(\omega-\sigma-\eta)},\hspace*{-20mm}
    \\
    \widetilde{\mathcal{G}}^{\rho,\sigma,(\omega)}_{[\kappa],\{\alpha,\beta\} }
    \Bigg|^{\NSmeth,  \langle \mathrm{exp} \rangle}_{\bar{c}0}\,
    \!\!\!=\,&
      \theta\left(\omega-\rho\right)
      \sum_{\lambda=0}^{\omega-\rho}
      \frac{
        {F}^{(\alpha_n+\omega-\sigma)}_{i/n,\beta_n}\left( m_H e^{-Y_H}\right)
        {F}^{(\alpha_\nbar+\lambda)}_{j/\nbar,\beta_\nbar}\left(m_H e^{+Y_H}\right)
      }{\lambda!\,(\omega-\sigma)!}
      \left(1-\frac{\delta_{\omega}^{\rho+\lambda}}{2}\right)
      \Upsilon_{zb}^{(\omega-\rho-\lambda)}.\hspace*{-20mm}
  \end{split}
\end{equation}
At leading power $\omega=-1$,  where the integrand of
Eq.~\eqref{eq:def:subjt} with $\rho=\sigma=-1$ is the relevant
contributor, the sum of the contribution above amounts to the
leading soft function in Eq.~\eqref{eq:def:s:fs:rap}, agreeing
with the assertion made in \cite{Li:2016axz}.
Nevertheless, it is worth emphasising that this coincidence
will not hold in general.
At \NLP, for instance, the zero-bin subtrahends here comprise
the terms
\begin{equation}\label{eq:sub:nlp:eg}
  \begin{split}
    \widetilde{\mathcal{G}}^{-1,-1,(0)}_{[\kappa],\{\alpha,\beta\} }
    \Bigg|^{\NSmeth,  \langle \mathrm{exp} \rangle}_{c0}\,
    +\,
      \widetilde{\mathcal{G}}^{-1,-1,(0)}_{[\kappa],\{\alpha,\beta\} }
      \Bigg|^{\NSmeth,  \langle \mathrm{exp} \rangle}_{\bar{c}0}\,
    =\,&
      {F}^{(\alpha_n+1)}_{i/n }
      {F}^{(\alpha_\nbar+1)}_{j/\nbar }
      \Upsilon_{zb}^{(0)}\,\\
    &
      +\,
      {F}^{(\alpha_n )}_{i/n }
      {F}^{(\alpha_\nbar+1)}_{j/\nbar }
      \Upsilon_{zb}^{(1)}\,
      +\,
      {F}^{(\alpha_n+1)}_{i/n }
      {F}^{(\alpha_\nbar )}_{j/\nbar }
      \Upsilon_{zb}^{(1)}\,.
  \end{split}
\end{equation}
Recalling to the definition in Eq.~\eqref{eq:def:s:fs:rap}, the soft
sector at the same power accuracy reads,
\begin{equation}\label{eq:s:nlp:eg}
  \begin{split}
 \widetilde{\mathcal{G}}^{-1,-1,(0)}_{[\kappa],\{\alpha,\beta\} }\Bigg|^{\FSmeth}_{s}\,
 =\,
{F}^{(\alpha_n+1)}_{i/n }  {F}^{(\alpha_\nbar+1)}_{j/\nbar } \Upsilon_{zb}^{(0)}\,
 +\,
  \frac{{F}^{(\alpha_n )}_{i/n }  {F}^{(\alpha_\nbar+2)}_{j/\nbar }}{2} \Upsilon_{zb}^{(2)}\,
  +\,
   \frac{{F}^{(\alpha_n+2)}_{i/n }  {F}^{(\alpha_\nbar )}_{j/\nbar } }{2}\Upsilon_{zb}^{(2)}\,.
  \end{split}
\end{equation}
Comparing those two expressions, with the exception of the first term
on the r.h.s.\ of Eq.~\eqref{eq:s:nlp:eg}, the other constituents in
Eq.~\eqref{eq:s:nlp:eg} both differ from those in Eq.~\eqref{eq:sub:nlp:eg}.
This discrepancy highlights the structural differences between the soft
function and zero-bin subtrahends beyond leading power.
Indeed, both of them contain unbounded integrals over integrands expanded
in line with soft scaling.
However, the soft function consists of all ingredients at a given power,
whereas the zero-bin subtrahends concern only those associated with the
collinear functions from the lowest power accuracy up to the current power
under consideration.

Combining the results in Eq.~\eqref{eq:def:c0:ns:exp} with their
corresponding collinear ingredients in
Eqs.~\eqref{eq:def:asyexp:c:at:omega:nured:exp} and
\eqref{eq:def:asyexp:cbar:at:omega:nured:exp}, we observe that all
the $\tau$-dependences will drop out as they should, from which we
can separately reproduce the $n$- and $\nbar$-collinear contributions
in Eqs.~\eqref{eq:def:asyexp:c:mom:at:omega}-%
\eqref{eq:def:asyexp:cbar:mom:at:omega} and
Eqs.~\eqref{eq:def:asyexp:cs:mom:int}-%
\eqref{eq:def:asyexp:cbars:mom:int} derived via the momentum cutoffs,
more explicitly,
\begin{equation}
  \begin{split}
    \widetilde{\mathcal{G}}^{\rho,\sigma,(\omega)}_{[\kappa],\{\alpha,\beta\} }
    \Bigg|^{\langle \mathrm{exp}\rangle}_{c} \,
    -\,
    \widetilde{\mathcal{G}}^{\rho,\sigma,(\omega)}_{[\kappa],\{\alpha,\beta\} }
    \Bigg|^{\NSmeth,  \langle \mathrm{exp} \rangle}_{c0}\,
    =\,
    \widetilde{\mathcal{I}}^{\rho,\sigma,(\omega)}_{[\kappa],\{\alpha,\beta\} }
    \Bigg|_{c} \,
    +\,
    \widetilde{\mathcal{I}}^{\rho,\sigma,(\omega)}_{[\kappa],\{\alpha,\beta\} }
    \Bigg|_{cs} \,,
    \\
    \widetilde{\mathcal{G}}^{\rho,\sigma,(\omega)}_{[\kappa],\{\alpha,\beta\} }
    \Bigg|^{\langle \mathrm{exp}\rangle}_{\bar{c}} \,
    -\,
    \widetilde{\mathcal{G}}^{\rho,\sigma,(\omega)}_{[\kappa],\{\alpha,\beta\} }
    \Bigg|^{\NSmeth,  \langle \mathrm{exp} \rangle}_{\bar{c}0}\,
    =\,
    \widetilde{\mathcal{I}}^{\rho,\sigma,(\omega)}_{[\kappa],\{\alpha,\beta\} }
    \Bigg|_{\bar{c}} \,
    +\,
    \widetilde{\mathcal{I}}^{\rho,\sigma,(\omega)}_{[\kappa],\{\alpha,\beta\} }
    \Bigg|_{\bar{c}s} \,.
  \end{split}
\end{equation}
Now we are ready to use the zero-bin subtracted collinear contributions
to evaluate the internal region in
Eq.~\eqref{eq:def:asyexp:central:power:omega:rap} and in turn the
\qT~spectrum.
Akin to the previous subsections, to make our results more compact, we
rewrite the momentum space expressions in
Eqs.~\eqref{eq:def:asyexp:c:at:omega:nured:exp}-%
\eqref{eq:def:asyexp:cbar:at:omega:nured:exp} and
Eq.~\eqref{eq:def:c0:ns:exp} in terms of the dimensionless parameters
defined in Eq.~\eqref{eq:def:kin:xn:xnbar}.
Up to \NNLP, this kind of transformation can be always achieved
through Eqs.~\eqref{eq:plus2:convertor} and \eqref{eq:trans:plusD:starD:n}.
Collectively, we express the result below,
\begin{equation}\label{eq:asyexp:exp:sum}
  \begin{split}
    \frac{\done \sigma_H}{\done Y_H \done \qT^2}
    \,=&\,
    \frac{\done \sigma_H}{\done Y_H \done \qT^2}\Bigg|_{b.c.}
    \,+\,
    \frac{\done \sigma^{\langle \mathrm{exp}\rangle}_H}{\done Y_H \done \qT^2}\Bigg|_{c}
    \,-\,
    \frac{\done \sigma^{\langle \mathrm{exp}\rangle}_H}{\done Y_H \done \qT^2}\Bigg|_{c0}
    \,+\,
    \frac{\done \sigma^{\langle \mathrm{exp}\rangle}_H}{\done Y_H \done \qT^2}\Bigg|_{\bar{c}}
    \,-\,
    \frac{\done \sigma^{\langle \mathrm{exp}\rangle}_H}{\done Y_H \done \qT^2}\Bigg|_{\bar{c}0}
    \,,
  \end{split}
\end{equation}
where the term with the subscript $``b.c."$ indicates the
ultra-collinear contributions as presented in
Eq.~\eqref{eq:def:asyexp:mcfs:uc:ucbar}.
The following four terms in Eq.~\eqref{eq:asyexp:exp:sum},
marked by the superscripts $``\langle \mathrm{exp}\rangle"$, stand
for the collinear sectors and their corresponding zero-bin subtrahends
evaluated within the exponential regularisation scheme.
Of them, the constituents dressed with the subscripts $``c"$ and
$``\bar{c}"$ are induced by Eq.~\eqref{eq:def:asyexp:c:at:omega:nured:exp}
and Eq.~\eqref{eq:def:asyexp:cbar:at:omega:nured:exp}, respectively.
Their expressions read,
\begin{align}
  \label{eq:asyexp:exp:sum:c}
  \frac{\done \sigma^{\langle \mathrm{exp}\rangle}_H}{\done Y_H \done \qT^2}\Bigg|_{c}
  \,\equiv\,&
    \frac{\alpha_s^3C_t^2}{192\pi^2s v^2}
  \,
  \sum_{i,j=\{g,q,\bar{q}\}}
  \,
  \sum_{\omega=-1}^{\infty}
  \,
    \left(\frac{\qT^2}{m_H^2}\right)^{\omega}
  \,
  \int^{1}_{x_n}\,\done z_n
  \,
  \left[ \mathbf{F}_{i/n}\left(\frac{x_n}{z_n}\right)\right]^{\mathbf{T}}
  \nonumber
  \\
  &\,\cdot\,
  \left\{
  \mathbf{R}_{ cs}^{(\omega),ij}(z_n)\,+\, \mathbf{P}_{ cs }^{(\omega),ij} \,\left[\frac{1}{1-z_n}\right]_++\,\left(\mathbf{T}^{(\omega),ij}_c+ \mathbf{D}_{ cs }^{(\omega),ij}\right)\,\delta\left(1-z_n\right) +\, \mathbf{B}_{ c }^{(\omega),ij}(z_n)\,\delta\left(x_n-z_n\right)
  \right\}
  \nonumber\\
  &\,\cdot\, \mathbf{F}_{j/\nbar}(x_{\nbar})\,,\\
  \label{eq:asyexp:exp:sum:cbar}
  \frac{\done \sigma^{\langle \mathrm{exp}\rangle}_H}{\done Y_H \done \qT^2}\Bigg|_{\bar{c}}
  \,\equiv\,&
    \frac{\alpha_s^3C_t^2}{192\pi^2s v^2}
  \,
  \sum_{i,j=\{g,q,\bar{q}\}}
  \,
  \sum_{\omega=-1}^{\infty}
  \,
    \left(\frac{\qT^2}{m_H^2}\right)^{\omega}
  \,
  \int^{1}_{x_\nbar}\,\done z_\nbar
  \,
  \left[ \mathbf{F}_{i/n}\left( x_n \right)\right]^{\mathbf{T}}
  \nonumber
  \\
  &\,\cdot\,
  \left\{
  \mathbf{R}_{ \bar{c}s}^{(\omega),ij}(z_\nbar)\,+\, \mathbf{P}_{ \bar{c}s }^{(\omega),ij} \,\left[\frac{1}{1-z_\nbar}\right]_++\,\left(\mathbf{T}^{(\omega),ij}_{\bar{c}}+ \mathbf{D}_{ \bar{c}s }^{(\omega),ij}\right)\,\delta\left(1-z_\nbar\right) +\, \mathbf{B}_{ \bar{c}s}^{(\omega),ij}(z_\nbar)\,\delta\left(x_\nbar-z_\nbar\right)
  \right\}
  \nonumber\\
  &\,\cdot\, \mathbf{F}_{j/\nbar}\left(\frac{x_{\nbar}}{z_\nbar}\right)\,.
\end{align}
Here the partonic matrices $\mathbf{R}$, $\mathbf{P}$, $\mathbf{D}$,
and $\mathbf{B}$ are same as those in Eq.~\eqref{eq:def:asyexp:mcfs:cs}
and Eq.~\eqref{eq:def:asyexp:mcfs:cbars}.
The new ingredients are $\mathbf{T}_{c}$ and $\mathbf{T}_{\bar{c}}$,
accommodating all the $\tau$-dependences from $\Upsilon_n$ and
$\Upsilon_\nbar$ in Eq.~\eqref{eq:def:ups:exp} and \eqref{eq:def:ups:exp:1}.
Up to \NNLP, the results for $\mathbf{T}_{c}$ read,
\begin{equation}
  \begin{split}
&\mathbf{T}^{(-1),q(\bar{q})g}_{c}\,=\,\mathbf{T}^{(-1),q \bar{q} }_{c}\,=\,\mathbf{T}^{(0),q \bar{q} }_{c}\,=\,\mathbf{T}^{(1),q \bar{q} }_{c}\,=\,0\,,\\
&\mathbf{T}^{(-1),gg}_{c}\,=\,-\,L_{\tau}\,\mathbf{P}^{(-1),gg}_{c}\,,\qquad\qquad
\mathbf{T}^{(0),q(\bar{q})g}_{c}\,=\,-\,L_{\tau}\,\mathbf{P}^{(0),q(\bar{q})g}_{c}\,,\\
&
\mathbf{T}^{(0),gg}_{c}\,=\,-\,L_{\tau}\,\mathbf{P}^{(0),gg}_{c}\,+\,
\left[
\begin{array}{cc}
-\frac{r_n}{\qT \tilde{\tau}} &  \frac{r_n}{\qT \tilde{\tau}} 
\end{array}
\right]\,,\\
&\mathbf{T}^{(1),gg}_{c} \,=\,-\,L_{\tau}\,\mathbf{P}^{(1),gg}_{c}\,+\,
\left[
\begin{array}{ccc}
 \frac{3 r_n^2}{\qT^2 \tilde{\tau}^2}-3 r_n^2-\frac{7 r_n}{2 \qT \tilde{\tau}}  & -\frac{r_n^2}{\qT^2
   \tilde{\tau}^2}+r_n^2+\frac{7 r_n}{2 \qT \tilde{\tau}}  & \frac{r_n^2}{\qT^2 \tilde{\tau}^2 }-r_n^2  \\[2ex]
 +\frac{5 r_n}{2 \qT \tilde{\tau}} & -\frac{r_n}{2 \qT \tilde{\tau}} &
   \frac{r_n}{\qT \tilde{\tau}}  
 \end{array}
\right]\,,\\
&
\mathbf{T}^{(1),q(\bar{q})g}_{c}\,=\,-\,L_{\tau}\,\mathbf{P}^{(1),q(\bar{q})g}_{c}\,+\,
\left[
\begin{array}{ccc}
 0 & \frac{2 r_n}{9 \qT \tilde{\tau}}  & \frac{2 r_n}{9 \qT \tilde{\tau}}  \\
\end{array}
\right]\,,
  \end{split}
\end{equation}
where $r_n=x_n\,\sqrt{s}/\qT$ and ${r_\nbar}= x_\nbar\,\sqrt{s}/\qT$.
As for the partonic processes $gg\to Hg$ and $q\bar{q}\to Hg$, the
expressions for $\mathbf{T}_{\bar{c}}$ can be derived by $\mathbf{T}_{c}$
via the following relationship,
\begin{align}
  \mathbf{T}^{(\omega),gg}_{\bar{c}}=\left[\mathbf{T}^{(\omega),gg}_{c}\right]^{\mathbf{T}}\Bigg|_{r_n\to r_\nbar}\,,\quad\quad
  \mathbf{T}^{(\omega),q\bar{q} }_{\bar{c}}=\left[\mathbf{T}^{(\omega),q\bar{q} }_{c}\right]^{\mathbf{T}}\Bigg|_{r_n\to r_\nbar}\,.
\end{align}
For those induced by the channel $q(\bar{q})g\to Hq(\bar{q})$, they
evaluate to
\begin{equation}
  \begin{split}
&
\mathbf{T}^{(-1),q(\bar{q})g}_{\bar{c}}\,=\,0\,,\qquad\qquad
\mathbf{T}^{(0),q(\bar{q})g}_{\bar{c}}\,=\,-\,L_{\tau}\,\mathbf{P}^{(0),q(\bar{q})g}_{\bar{c}}\,+\,\frac{2 r_\nbar}{9 \qT \tilde{\tau}}\,,
\\
&
\mathbf{T}^{(1),q(\bar{q})g}_{\bar{c}}\,=\,-\,L_{\tau}\,\mathbf{P}^{(1),q(\bar{q})g}_{\bar{c}}\,+\,
\left[
\begin{array}{ccc}
 \frac{r_\nbar}{9 \qT \tilde{\tau}} & \frac{r_\nbar}{9 \qT \tilde{\tau}}    \\[2.ex]
 \frac{2 r_\nbar^2}{9 \qT^2 \tilde{\tau}^2}-\frac{2 r_\nbar^2}{9}+\frac{r_\nbar}{3 \qT \tilde{\tau}} &
   \frac{2 r_\nbar}{9 \qT \tilde{\tau}}   \\ 
\end{array}
\right]\,.
  \end{split}
\end{equation}
On the other hand, Eq.~\eqref{eq:asyexp:exp:sum} also includes the
components $``c0"$ and $``\bar{c}0"$, accounting for the zero-bin
subtraction terms in Eq.~\eqref{eq:def:c0:ns:exp}.
Their results are,
\begin{equation}\label{eq:asyexp:exp:sum:c0}
  \begin{split}
\frac{\done \sigma^{\langle \mathrm{exp}\rangle}_H}{\done Y_H \done \qT^2}\Bigg|_{c0}
 \,\equiv\,&
  \frac{\alpha_s^3C_t^2}{192\pi^2s v^2}
 \,
 \sum_{i,j=\{g,q,\bar{q}\}}
 \,
 \sum_{\omega=-1}^{\infty}
 \,
  \left(\frac{\qT^2}{m_H^2}\right)^{\omega}
 \,
\left[ \mathbf{F}_{i/n}\left( {x_n} \right)\right]^{\mathbf{T}}
\,\cdot\,
\mathbf{T}^{(\omega),ij}_c
\,\cdot\,
 \mathbf{F}_{j/\nbar}(x_{\nbar})\,,\\
\frac{\done \sigma^{\langle \mathrm{exp}\rangle}_H}{\done Y_H \done \qT^2}\Bigg|_{\bar{c}0}
 \,\equiv\,&
  \frac{\alpha_s^3C_t^2}{192\pi^2s v^2}
 \,
 \sum_{i,j=\{g,q,\bar{q}\}}
 \,
 \sum_{\omega=-1}^{\infty}
 \,
  \left(\frac{\qT^2}{m_H^2}\right)^{\omega}
 \,
\left[ \mathbf{F}_{i/n}\left( {x_n} \right)\right]^{\mathbf{T}}
\,\cdot\,
\mathbf{T}^{(\omega),ij}_{\bar{c}}
\,\cdot\,
 \mathbf{F}_{j/\nbar}(x_{\nbar})\,.
  \end{split}
\end{equation}
Since $\Upsilon_{zb}$ in Eqs.~\eqref{eq:def:ups:exp:zb} and
\eqref{eq:def:ups:exp:zb:n2lp} exactly reproduces all the
$\tau$-dependent contributions from $\Upsilon_n$ and
$\Upsilon_\nbar$ in Eq.~\eqref{eq:def:ups:exp} and
\eqref{eq:def:ups:exp:1}, respectively,
the partonic matrices $\mathbf{T}_{c}$ and $\mathbf{T}_{\bar{c}}$
governing the zero-bin subtrahends here are identical to those
in the collinear sectors in Eqs.~\eqref{eq:asyexp:exp:sum:c} and
\eqref{eq:asyexp:exp:sum:cbar}.

\section{Numerical results}
\label{sec:asyexp:numeric}

Having introduced all building blocks in the previous sections,
we will now present numerical results for the \qT~spectrum
for the process $pp\to H+X$ at LHC by the means of the power
expansion derived in Sec.~\ref{sec:asyexp:outputs}.
Since we have shown that this power series is equivalent whether
it is computed using momentum cutoffs to regulate rapidity
divergences, as in Eq.~\eqref{eq:def:asyexp:mcfs}, the
pure-rapidity regulator in Eq.~\eqref{eq:def:asyexp:prap}, or
the exponential rapidity regulator in Eq.~\eqref{eq:asyexp:exp:sum},
any one can be employed.
We would like to stress again, though, that although individual
constituents  are dependent on the rapidity regularisation
scheme, the sum of all sectors is identical in all three choices
with no scheme-dependent power-suppressed terms remaining.
Hence, it is scheme-independent sum of sectors that we implement.
 
During our calculation, we take the mass $m_H$ of the Higgs
boson and the vacuum expectation value $v$ from PDG
\cite{ParticleDataGroup:2022pth}.
As to Wilson coefficient $C_t$, although it is known up to
four-loop order \cite{Inami:1982xt,Djouadi:1991tk,Chetyrkin:1997iv,
  Chetyrkin:1997un,Chetyrkin:2005ia,Schroder:2005hy,Baikov:2016tgj},
we only consider its LO contribution here in accordance with the
perturbative accuracy of the amplitudes presented in Eqs.~\eqref{eq:msq}.
In this paper, we use the \texttt{MSHT20nlo\_as118} \cite{Bailey:2020ooq}
PDF set, with the associated value and evolution of $\alpha_s$,
interfaced through \LHAPDF \cite{Buckley:2014ana,Bothmann:2022thx}.
We set the renormalisation scale to $\mu_R=m_H$, entering our
calculation only as the scale we evaluate the strong coupling at.
In implementing  \texttt{MSHT20nlo\_as118} for
Eq.~\eqref{eq:def:asyexp:mcfs} particular attention should be paid
to the fact that the PDFs and their higher order derivatives are
both involved in the power correction expansion, as illustrated in
Eq.~\eqref{eq:PDF:N:ary}.
\changed{
As \LHAPDF does not offer direct access to the PDFs' derivatives,
in this work, we make use of the method of \cite{Bonvini:2014joa,Diehl:2021gvs}
and fit \texttt{MSHT20nlo\_as118} in terms of Chebyshev polynomials
at the factorisation scale $\mu_F=m_H$ first, from which the derivatives
of the PDFs can be evaluated analytically.
The virtue of this strategy is that the robustness of the PDFs'
derivatives is maintained in regards to the fitted PDFs throughout
our numerical calculation, which, as a matter of fact, is one of
prerequisites for our validation procedure.
To examine our fitted PDFs, we compare the $q_{\mathrm{T}}$
distributions generated by \Sherpa~\cite{Gleisberg:2003xi,
Gleisberg:2008ta,Sherpa:2019gpd} and \Rivet~\cite{Buckley:2010ar,
  Bierlich:2019rhm} with \LHAPDF  against those from the fitted ones,
finding only per-mill level discrepancies between them.
The above ansatz, however, proves challenging once dynamical
scales are considered and, potentially, different solutions
must be sought.}

Equipped with those inputs, we are capable of calculating the power
series of the \qT~distribution via Eq.~\eqref{eq:def:asyexp:mcfs}.
In the following, we will emphasis three types of results, i.e.,
\begin{equation}\label{eq:def:acc:lp_nlp_nnlp}
  \begin{split}
  \frac{\done \sigma^{\langle \mathrm{asy }\rangle}_H}{\done Y_H\, \done \qT^2}\Bigg|_{\text{\LP}}\,=&
  \,
  \sum_m\,   \frac{\Delta_{\text{\LP}}^{(m)}}{\qT^2}\, (L_H)^m\,,\\
  \frac{\done \sigma^{\langle \mathrm{asy }\rangle}_H}{\done Y_H\, \done \qT^2}\Bigg|_{\text{\NLP}}\,=&
  \,
  \sum_m\,   \frac{\Delta_{\text{\LP}}^{(m)}}{\qT^2}\, (L_H)^m \,
  +\,
  \sum_m\,   \Delta_{\text{\NLP}}^{(m)}\,  (L_H)^m  \,,\\
  \frac{\done \sigma^{\langle \mathrm{asy }\rangle}_H}{\done Y_H\, \done \qT^2}\Bigg|_{\text{\NNLP}}\,=&
  \,
  \sum_m\,   \frac{\Delta_{\text{\LP}}^{(m)}}{\qT^2}\, (L_H)^m \,
  +\,
  \sum_m\,   \Delta_{\text{\NLP}}^{(m)}\,  (L_H)^m \,
    +
  \,
   \sum_m\, \qT^2\,\Delta_{\text{\NNLP} }^{(m)}\,   (L_H)^m   \,,
  \end{split}
\end{equation}
where the superscript $``\mathrm{asy}"$ signifies the truncated
asymptotic series in the low~\qT~domain.
$\Delta_{\mathrm{N}^k\text{\LP}} $ stands for the partonic
contribution at the $k$-th power convoluted with PDFs and their
derivatives, which can be extracted from the power coefficients
in Eq.~\eqref{eq:def:asyexp:mcfs}, Eq.~\eqref{eq:def:asyexp:prap},
or Eq.~\eqref{eq:asyexp:exp:sum}.
In Eq.~\eqref{eq:def:acc:lp_nlp_nnlp}, the \qT~spectrum at \LP encodes the
most singular behavior in the low \qT~domain, consisting of the
first term of the power series in Eq.~\eqref{eq:def:acc}.
The \NLP result in Eq.~\eqref{eq:def:acc:lp_nlp_nnlp} includes also the
first power correction term, comprising the first two constituents
of Eq.~\eqref{eq:def:acc} in total.
Our most precise result is calculated at N$^2$LP, given in
the last line of Eq.~\eqref{eq:def:acc:lp_nlp_nnlp}.
It includes the contributions from $\Delta_{\text{\LP}}$
up to $\Delta_{\text{\NNLP}}$ in Eq.~\eqref{eq:def:acc}.
Moreover, in order to deliver a quantitative assessment of the
quality of these approximate results, we also showcase the exact
fixed-order \qT~distribution below as a benchmark.
It is derived based on Eq.~\eqref{eq:qcd:qT} and dubbed
``F.O." hereafter.

\begin{figure}[h!]
  \centering
  \begin{subfigure}{0.32\textwidth}
    \centering
    \includegraphics[height=.8\linewidth]{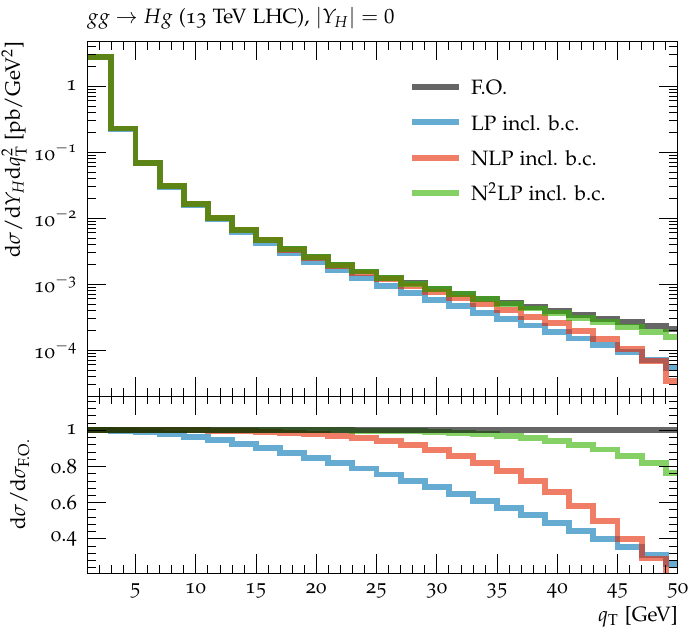}
    \caption{  }
    \label{fig:results:val:qTratios:asyexp:ggHg:YH0}
  \end{subfigure}
  \begin{subfigure}{0.32\textwidth}
    \centering
    \includegraphics[height=.8\linewidth]{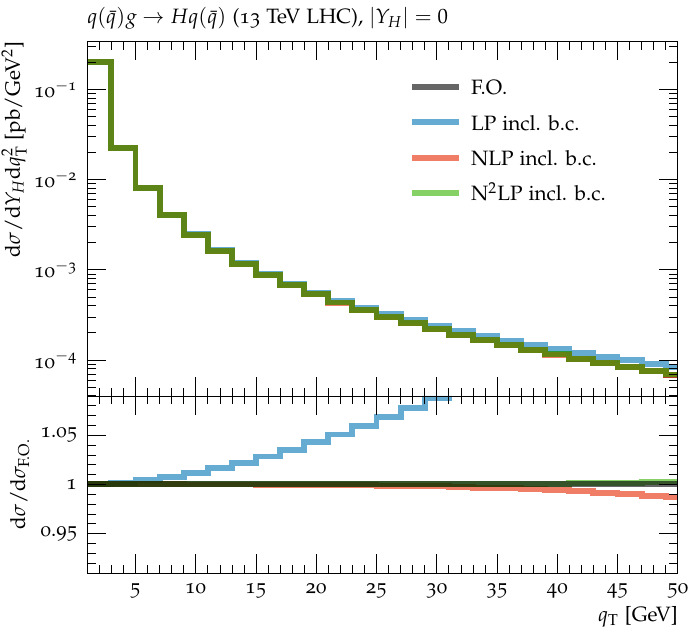}
    \caption{   }
    \label{fig:results:val:qTratios:asyexp:qgHq:YH0}
  \end{subfigure} 
  \begin{subfigure}{0.32\textwidth}
    \centering
    \includegraphics[height=.8\linewidth]{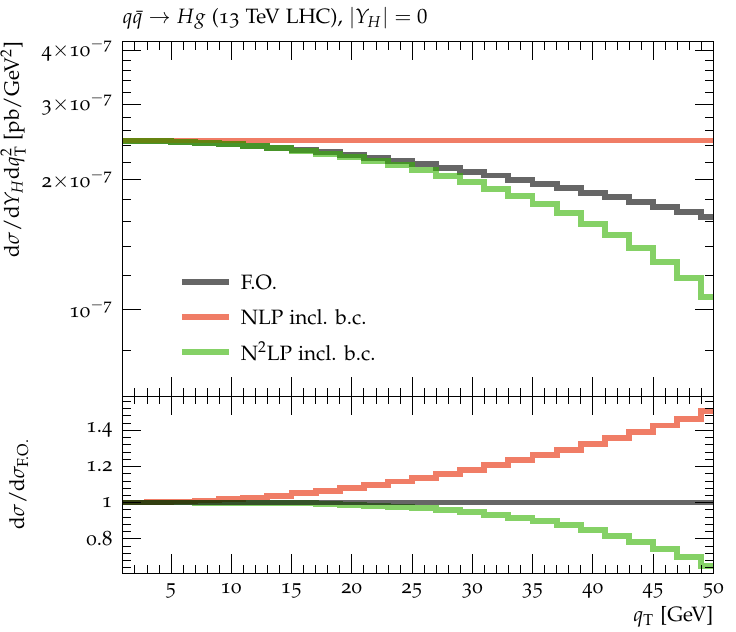}
    \caption{   }
    \label{fig:results:val:qTratios:asyexp:qqHg:YH0}
  \end{subfigure}
  \caption{
    Comparisons of the \qT\ spectra between the exact fixed-order
    calculation (black) and approximations constructed
    using the power expansion up to \LP (blue), \NLP (red), and
    \NNLP (green) for different partonic processes at $|Y_H|=0$.
    While the upper plot contains the absolute differential
    cross sections, the lower plot contains the ratios of the
    approximations w.r.t.\ the exact fixed-order result.
    Throughout we include the boundary corrections $\mathbf{B}_{cs}$,
    $\mathbf{B}_{\bar{c}s}$, and $\mathbf{B}_{uc(\bar{c})}$.
  }
  \label{fig:results:val:qTratios:asyexp:YH0}
\end{figure}

We start the discussion of our results by considering the \qT\ spectrum
at central rapidities, setting $Y_H=0$ for definitiveness.
Fig.~\ref{fig:results:val:qTratios:asyexp:YH0} therefore details a
comparison of our power-expanded approximate result of Secs.\
\ref{sec:asyexp:mom} and \ref{sec:asyexp:outputs} including terms
up to \LP (blue), \NLP (red), and \NNLP (green) to its exact fixed-order
counter-part (black) for all three partonic channels, $gg\to Hg$ (left),
$q(\bar{q})g\to Hq(\bar{q})$ (center), and $q\bar{q}\to Hg$ (right).
Generally, the main plot focusses on the absolute values of the \qT\
distribution, while the ratio plot shows the the ratio of each of the
approximate results to the exact one, allowing to judge the quality of
the individual approximation.
We will retain this pattern of visualisation throughout this section.

Examining at our results in more detail, we begin by discussing the
$gg\to Hg$ subprocess, shown in
Fig.~\ref{fig:results:val:qTratios:asyexp:ggHg:YH0}.
This process dominates the inclusive cross section and any improvements
in the quality of the power-expanded approximation will be exceedingly
beneficial to the description of the inclusive \qT\ spectrum.
Due to the presence of singular terms, see Eq.~\eqref{eq:def:acc}, the
\qT\ spectrum  is divergent as $\qT\to 0$, hence we limit our
deliberations to $\qT>1\,\text{GeV}$.
In this limit, the \LP contribution is, of course, dominant and the
approximated calculations, independent  of the number of higher-power
terms included, reproduce the exact result.
Nevertheless, departing from the low~\qT~domain towards higher values
of our observable, the power corrections manifest themselves progressively.
More specifically, the \LP-only result deteriorates in quality the quickest,
failing to describe the exact spectrum by about $4\%$ at $\qT=10\,\text{GeV}$
and increasing to about $30\%$ near $\qT=30\,\text{GeV}$.
Including the \NLP terms alleviates this deviation somewhat, decreasing it
to about $1\%$   at $\qT=10\,\text{GeV}$ and about $10\%$ at
$\qT=30\,\text{GeV}$.
The best approximation provided in this work, including terms up to \NNLP,
reproduces the \qT\ spectrum on level of 1\% up to values in excess of 30\,GeV,
leaving higher-power corrections to contribute more than percent level only
for $\qT\gtrsim40\,\text{GeV}$.

Fig.~\ref{fig:results:val:qTratios:asyexp:qgHq:YH0} now shows the
contributions from the partonic channel characterised by $q(\bar{q})g$
initial states.
This channel contributes on the 10\% level to the inclusive spectrum.
Our results here in general demonstrate an analogous behaviour to
those of $gg\to Hg$ in the vicinity of $\qT=0$~GeV, including the
presence of a divergence for $\qT\to 0$ as well as the mutual agreement
between the fixed-order and approximate curves in that limit.
However, advancing to the intermediate \qT~range, distinct features
are found between the two in regard to the relative magnitudes of
subleading and sub-subleading power corrections.
While in Fig.~\ref{fig:results:val:qTratios:asyexp:ggHg:YH0}, these
two terms observe convergent but still comparable magnitudes,
respectively accounting for$~20\%$ and$~10\%$ contributions of the
full theory in the vicinity of $\qT=30$~GeV, the \NLP terms in
Fig.~\ref{fig:results:val:qTratios:asyexp:qgHq:YH0} play a dominant
role and relegate the terms at \NNLP and beyond to an almost
insignificant role.

Moving onto the last partonic process contributing at our accuracy,
$q\bar{q}\to Hg$, depicted in
Fig.~\ref{fig:results:val:qTratios:asyexp:qqHg:YH0}, a very different
scenario presents itself.
For lack of the LP contributions and the associated singularity at
$\qT\to 0$, the fixed-order and approximate results here both approach
constants in the low~\qT~region.
In particular, the \NLP spectrum here takes the leading role and
remains constant throughout the \qT~interval of our interest.
The \NNLP corrections, meanwhile, exhibit the linear dependences on
the variable \qT\ as the logarithmic contributions in
Eqs.~\eqref{eq:Pcs:gg_gq_qq} and \eqref{eq:Dcs:lp_nlp_nnlp} vanish.
Comparing with the fixed-order spectrum, the \NLP and \NNLP results
both yield the correct regular behaviour in the small \qT~range but
give oscillatorily converging power corrections for moderate \qT.
For instance, the \NLP approximation overshoots the full theory by
about $20\%$ near $\qT=30\,\text{GeV}$, whilst incorporating \NNLP
corrections underestimates the exact one by a few percent here.

\begin{figure}[h!]
  \centering
  \begin{subfigure}{0.32\textwidth}
    \centering
    \includegraphics[width=.9\linewidth]{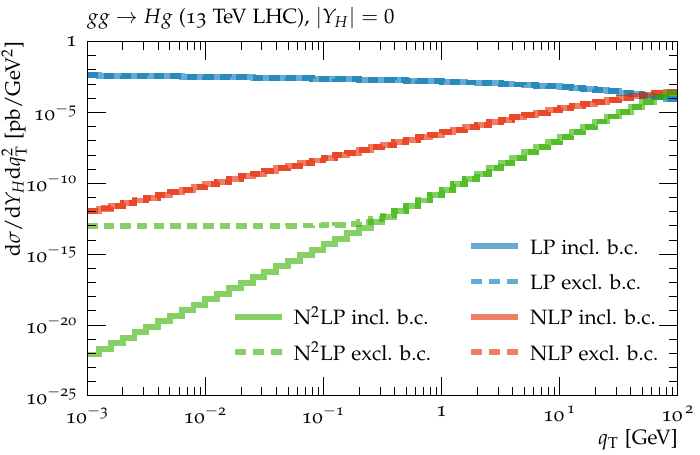}
    \caption{  }
    \label{fig:results:val:qTdist:asyexp:ggHg:YH0}
  \end{subfigure}
  \begin{subfigure}{0.32\textwidth}
    \centering
    \includegraphics[width=.9\linewidth]{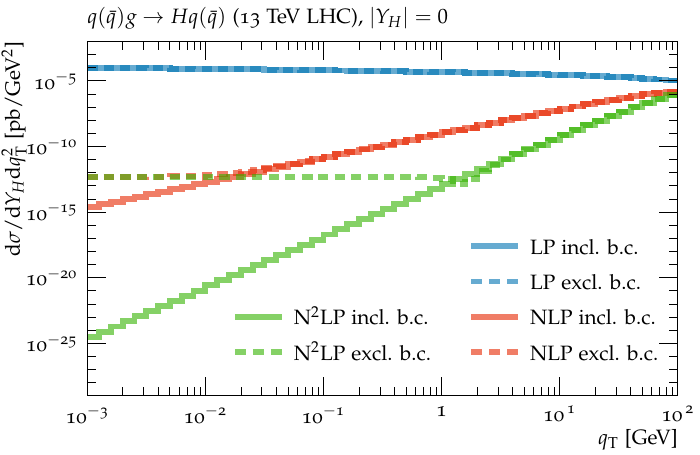}
    \caption{   }
    \label{fig:results:val:qTdist:asyexp:qgHq:YH0}
  \end{subfigure} 
  \begin{subfigure}{0.32\textwidth}
    \centering
    \includegraphics[width=.9\linewidth]{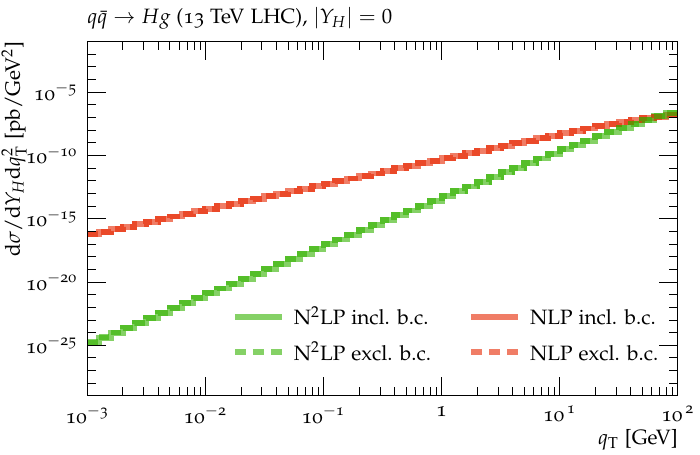}
    \caption{   }
    \label{fig:results:val:qTdist:asyexp:qqHg:YH0}
  \end{subfigure}
  \caption{
    Difference of the exact spectra and the approximations
    for the three different partonic processes at $|Y_H|=0$.
    The solid (dashed) lines present the results including (excluding)
    the boundary corrections $\mathbf{B}_{cs}$ in
    Eq.~\eqref{eq:def:asyexp:mcfs:cs}, $\mathbf{B}_{\bar{c}s}$ in
    Eq.~\eqref{eq:def:asyexp:mcfs:cbars}, and $\mathbf{B}_{uc(\bar{c})}$ in
    Eq.~\eqref{eq:def:asyexp:mcfs:uc:ucbar}.
    The blue, red, and green lines detail the \LP, \NLP, and \NNLP approximations, respectively.
  }
  \label{fig:results:val:qTdist:asyexp:YH0}
\end{figure}

In pursuit of further examining the higher-power correction terms
derived in Sec.~\ref{sec:asyexp:outputs}, we look into the differences
between the fixed order results and the approximate ones.
We note at this point that although we have already seen that their
respective ratios approach unity in the limit $\qT\to 0$, finite
differences in this limit play an important role when the approximants
are used to subtract or replace the exact spectrum in higher-order
calculations where the divergences cancel against the virtual corrections
in the $\qT=0$ bin.
Similarly, while the ratio tests above emphasise the relative sizes between
the full theory and the sum of all corrections to each power considered,
their differences allow to examine the individual term of the power series.
To this end,
according to the power series in Eq.~\eqref{eq:def:acc} as well as the
definitions in Eq.~\eqref{eq:def:acc:lp_nlp_nnlp}, the differences between the two
are expected to take the following asymptotic behaviour,
\begin{equation}\label{eq:def:acc:diff}
  \begin{split}
    \frac{\done \sigma^{\langle \mathrm{F.O. }\rangle}_H}{\done Y_H \done \qT^2}\,-\,
    \frac{\done \sigma^{\langle \mathrm{asy }\rangle}_H}{\done Y_H \done \qT^2}\Bigg|_{\text{\LP}}\,=&
    \,
    \sum_m\,   \Delta_{\text{\NLP}}^{(m)}\,  (L_H)^m \,+\dots\,,\\
      \frac{\done \sigma^{\langle \mathrm{F.O. }\rangle}_H}{\done Y_H \done \qT^2}\,-\,
    \frac{\done \sigma^{\langle \mathrm{asy }\rangle}_H}{\done Y_H \done \qT^2}\Bigg|_{\text{\NLP}}\,=&
    \,
    \sum_m\, \qT^2\,\Delta_{\text{\NNLP} }^{(m)}\,   (L_H)^m \,
  +\,\dots \,,\\
      \frac{\done \sigma^{\langle \mathrm{F.O. }\rangle}_H}{\done Y_H \done \qT^2}\,-\,
    \frac{\done \sigma^{\langle \mathrm{asy }\rangle}_H}{\done Y_H \done \qT^2}\Bigg|_{\text{\NNLP}}\,=&
    \,
      \sum_m\, \qT^4\,\Delta_{\mathrm{N}^3\text{\LP} }^{(m)}\,   (L_H)^m \,
  +\,\dots \,.
  \end{split}
\end{equation}
The numerical results for the magnitudes of these missing, or yet uncalculated,
higher-power corrections
are presented in Fig.~\ref{fig:results:val:qTdist:asyexp:YH0} for all
three partonic processes at $Y_H=0$.
Therein, the solid lines denote our results including both the
interior contribution from Sec.~\ref{sec:red:qT:central}
and the boundary correction discussed in Sec.~\ref{sec:red:qT:boundary}.
Due to the presence of the logarithmic contributions in Eqs.~\eqref{eq:Dcs:lp_nlp_nnlp} and \eqref{eq:Dcbars:lp_nlp_nnlp},
the differences between the exact \qT~spectra and its \LP approximations
(blue solid line)
experience mild enhancements in magnitude in the low~\qT~regime,
agreeing with the expectation in Eq.~\eqref{eq:def:acc:diff}.\footnote{%
  In fact, the difference between the \LP expression and the exact result
  forms an integrable divergence as $\qT\to 0$.
  This integrable divergence is removed once \NLP corrections are introduced.
}
After incorporating the \NLP corrections, as exhibited in the red solid
line, the difference between the full theory and our \NLP-correct
approximation starts to decrease as $\qT\to 0$ for all three partonic
channels.
This echoes again, our \NLP expectation of Eq.~\eqref{eq:def:acc:diff}
that, starting at this order, the difference vanishes in this limit.
This also demonstrates that at least up to \NLP, the power coefficients
of the~\qT~distribution can be reproduced as appropriate by the
expressions in Sec.~\ref{sec:asyexp:outputs}.
Similarly, including the power corrections up to \NNLP results, detailed
by the green line, results in the difference vanishing twice as fast.

On the other hand, Fig.~\ref{fig:results:val:qTdist:asyexp:YH0} also
depicts our results for which we have removed the boundary contributions of
Eq.~\eqref{eq:def:asyexp:mcfs:uc:ucbar}, induced by expanding the integral
boundaries of Eq.~\eqref{eq:def:b.c.} as well as those in
Eq.~\eqref{eq:def:asyexp:mcfs:cs} and Eq.~\eqref{eq:def:asyexp:mcfs:cbars}
which arise from integrating the preceding expressions by parts,
in dashed lines.
Since those boundary corrections will start at \NLP accuracy, our \LP
result (blue) remains unchanged.
At \NLP, these boundary impacts play only a small role in both the $gg$
and $q\bar{q}$ channels, but manifest themselves in the
$q(\bar{q})g$-initiated process for $\qT\lesssim\mathcal{O}(10^{-2})$.
The resulting expression then ceases to reproduce the entire \NLP coefficient
of Eq.~\eqref{eq:def:acc}, halting the $\propto\qT^2$ behaviour we
would expect the difference with the exact result to exhibit.
At \NNLP, the sensitivity of the result to the presence of these boundary
contributions is even larger, being noticeable  already for \qT\ of
$\mathcal{O}(10^{-1})$ in the $gg$-initiated and $\mathcal{O}(1)$ for
the $q(\bar{q})g$-initiated partonic channels.
In both cases, the respective power accuracy is lost in that domain,
and only \LP accuracy is maintained, exemplifying the necessity of
the boundary corrections.

It is worth noting that despite their indispensable effects in
reproducing the power coefficients at a given power, at least up
to \NNLP, those boundary terms can only generate constant corrections
of comparably small magnitude.
They may therefore not play a role of phenomenological interest when
the larger \qT\ region, where power corrections are larger by construction,
is being investigated, and our results may form the basis of a subleading
power resummation.
However, it should be stressed that, from a theoretical point of view,
the emergence of the boundary corrections highlights one of the main
differences between the asymptotic expansion of loop integrals
and phase space integrals in the small \qT~regime.
While the loop integrals generally involve unbounded integrals and,
thus, only incur relevant power expansions of the integrand within a
variety of scalings, the expansion of phase space integrals in the small,
but finite, \qT\ regime additionally induces phase space boundaries through
Eq.~\eqref{eq:def:b.c.}, from which the appearance of hierarchies $\qT\ll m_H$
(and the application of the integration by parts identities) will introduce
a novel type of power corrections associated with the PDFs at the end point,
as shown in Eqs.~\eqref{eq:asyexp:deltaI:uc} and \eqref{eq:trans:plusD:starD:n}.

\begin{figure}[t!]
  \centering
  \begin{subfigure}{0.32\textwidth}
    \centering
    \includegraphics[width=.9\linewidth]{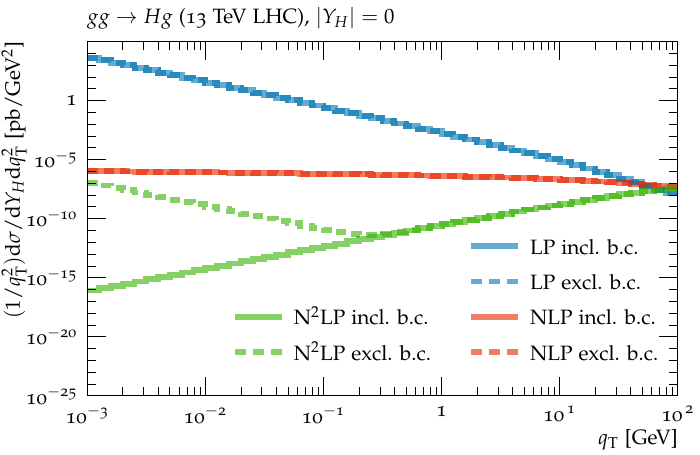}
    \caption{  }
    \label{fig:results:val:qTwt:asyexp:ggHg:YH0}
  \end{subfigure}
  \begin{subfigure}{0.32\textwidth}
    \centering
    \includegraphics[width=.9\linewidth]{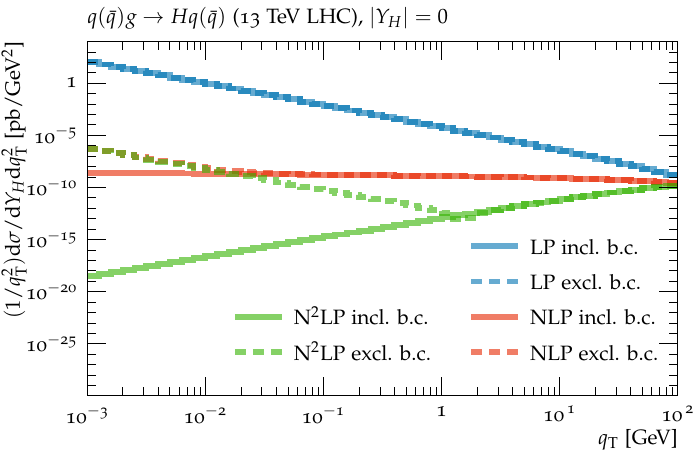}
    \caption{   }
    \label{fig:results:val:qTwt:asyexp:qgHq:YH0}
  \end{subfigure} 
  \begin{subfigure}{0.32\textwidth}
    \centering
    \includegraphics[width=.9\linewidth]{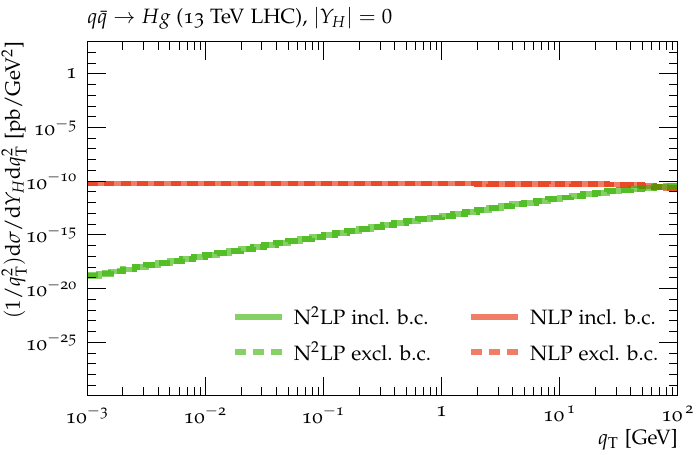}
    \caption{   }
    \label{fig:results:val:qTwt:asyexp:qqHg:YH0}
  \end{subfigure}
  \caption{
    Difference of the weighted exact spectra and the approximations
    for the three different partonic processes at $|Y_H|=0$.
    The solid (dashed) lines present the results including (excluding)
    the boundary corrections $\mathbf{B}_{cs}$ in
    Eq.~\eqref{eq:def:asyexp:mcfs:cs}, $\mathbf{B}_{\bar{c}s}$ in
    Eq.~\eqref{eq:def:asyexp:mcfs:cbars}, and $\mathbf{B}_{uc(\bar{c})}$ in
    Eq.~\eqref{eq:def:asyexp:mcfs:uc:ucbar}.
    The blue, green, and red lines detail the \LP, \NLP, \NNLP approximations,
    respectively.
  }
  \label{fig:results:val:qTwt:asyexp:YH0}
\end{figure}

To scrutinise the power correction at \NNLP, we investigate the
differences between the weighted \qT~distribution in the full theory
and the power-expanded approximations below.
From Eqs.~\eqref{eq:def:acc} as well as \eqref{eq:def:acc:lp_nlp_nnlp},
we expect these differences to exhibit the following asymptotic properties,
\begin{equation}\label{eq:def:acc:lp_nlp_nnlp:diff:wt}
  \begin{split}
    \frac{1}{\qT^2}\,\frac{\done \sigma^{\langle \mathrm{F.O. }\rangle}_H}{\done Y_H \done \qT^2}\,-\,
      \frac{1}{\qT^2}\,\frac{\done \sigma^{\langle \mathrm{asy }\rangle}_H}{\done Y_H \done \qT^2}\Bigg|_{\text{\LP}}\,=&
    \,
    \sum_m\,     \frac{1}{\qT^2}\,\Delta_{\text{\NLP}}^{(m)}\,  (L_H)^m \,+\dots\,,\\
        \frac{1}{\qT^2}\,\frac{\done \sigma^{\langle \mathrm{F.O. }\rangle}_H}{\done Y_H \done \qT^2}\,-\,
    \frac{1}{\qT^2}\, \frac{\done \sigma^{\langle \mathrm{asy }\rangle}_H}{\done Y_H \done \qT^2}\Bigg|_{\text{\NLP}}\,=&
    \,
    \sum_m\,\Delta_{\text{\NNLP} }^{(m)}\,   (L_H)^m \,
  +\,\dots \,,\\
      \frac{1}{\qT^2}\, \frac{\done \sigma^{\langle \mathrm{F.O. }\rangle}_H}{\done Y_H \done \qT^2}\,-\,
  \frac{1}{\qT^2}\,  \frac{\done \sigma^{\langle \mathrm{asy }\rangle}_H}{\done Y_H \done \qT^2}\Bigg|_{\text{\NNLP}}\,=&
    \,
      \sum_m\, \qT^2\,\Delta_{\mathrm{N}^3\text{\LP} }^{(m)}\,   (L_H)^m \,
  +\,\dots \,.
  \end{split}
\end{equation}
Comparing with Eqs.~\eqref{eq:def:acc:diff}, the remainders on the
r.h.s.\ of Eq.~\eqref{eq:def:acc:lp_nlp_nnlp:diff:wt} are generally
one power lower.
Their numeric results are displayed in
Fig.~\ref{fig:results:val:qTwt:asyexp:YH0}.
Our attention is immediately drawn to the observation that the
difference between the exact and approximate weighted spectrum diverges
at \LP as $\qT\to 0$ whereas it approaches a constant at \NLP.
Only once the \NNLP corrections are taken into account, including the
boundary terms, does the difference of the weighted spectra vanish
in the $\qT\to 0$ limit, in line with the \NNLP expectation of
Eq.~\eqref{eq:def:acc:lp_nlp_nnlp:diff:wt} justifying (at least the
first three terms of) our power series derived in Sec.~\ref{sec:asyexp:outputs}.
 
\begin{figure}[h!]
  \centering
  \begin{subfigure}{0.32\textwidth}
    \centering
    \includegraphics[height=.8\linewidth]{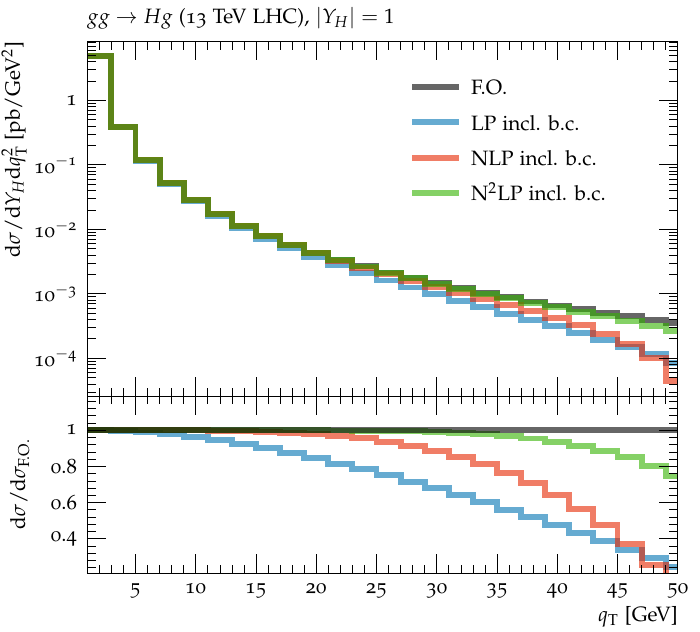}
    \caption{  }
    \label{fig:results:val:qTratios:asyexp:ggHg:YH1}
  \end{subfigure}
  \begin{subfigure}{0.32\textwidth}
    \centering
    \includegraphics[height=.8\linewidth]{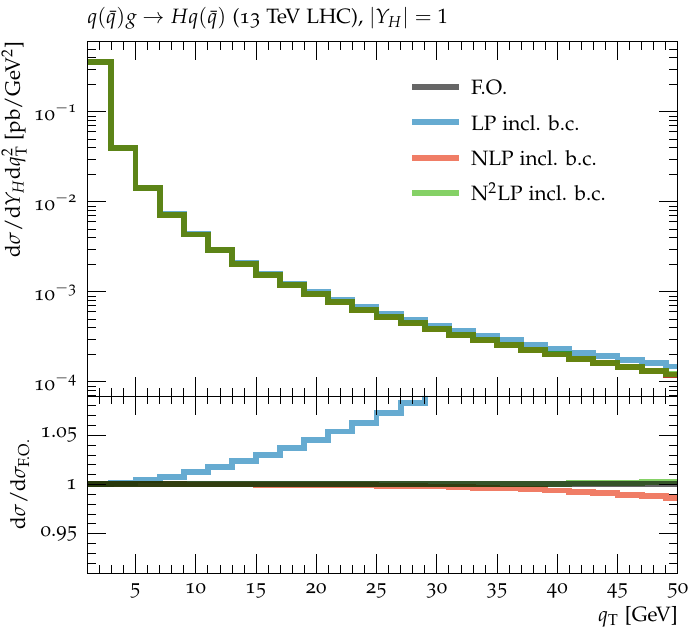}
    \caption{   }
    \label{fig:results:val:qTratios:asyexp:qgHq:YH1}
  \end{subfigure} 
  \begin{subfigure}{0.32\textwidth}
    \centering
    \includegraphics[height=.8\linewidth]{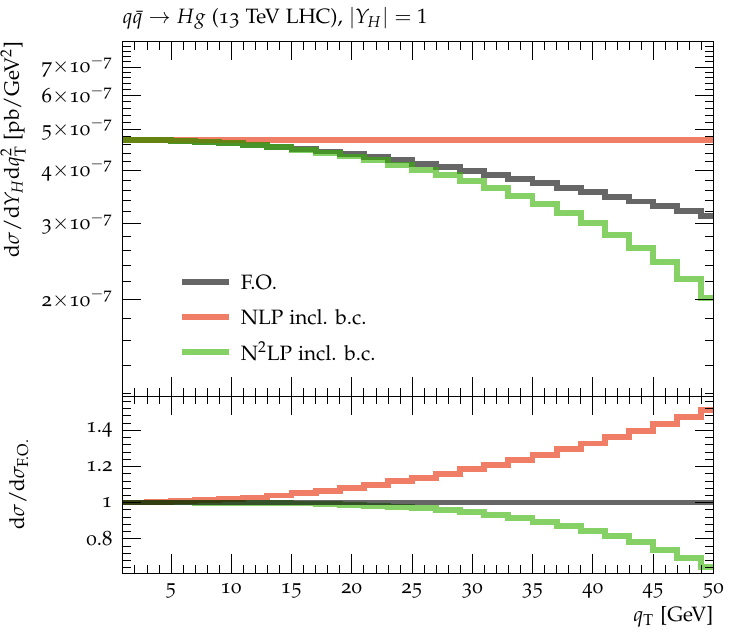}
    \caption{   }
    \label{fig:results:val:qTratios:asyexp:qqHg:YH1}
  \end{subfigure}\\[2mm]
  \centering
  \begin{subfigure}{0.32\textwidth}
    \centering
    \includegraphics[height=.8\linewidth]{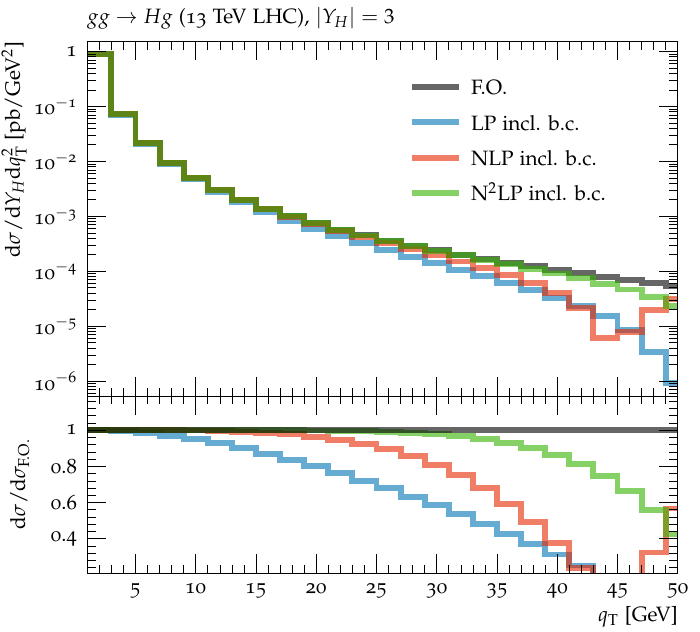}
    \caption{  }
    \label{fig:results:val:qTratios:asyexp:ggHg:YH3}
  \end{subfigure}
  \begin{subfigure}{0.32\textwidth}
    \centering
    \includegraphics[height=.8\linewidth]{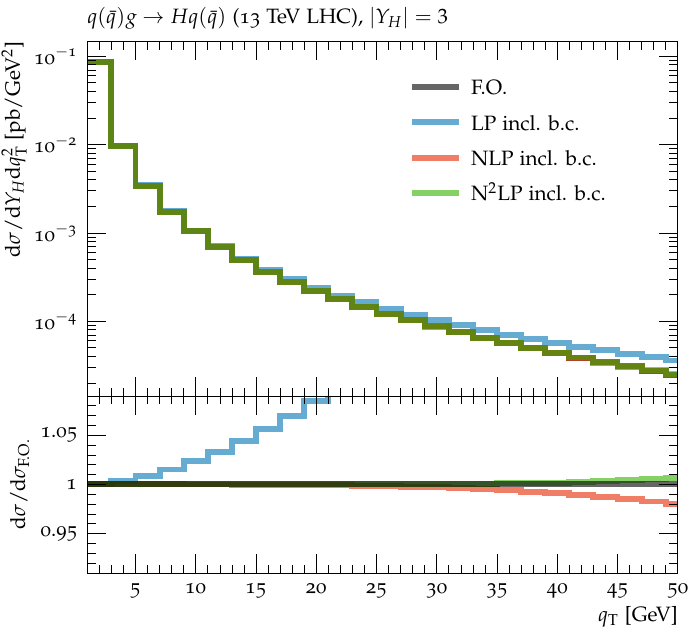}
    \caption{   }
    \label{fig:results:val:qTratios:asyexp:qgHq:YH3}
  \end{subfigure} 
  \begin{subfigure}{0.32\textwidth}
    \centering
    \includegraphics[height=.8\linewidth]{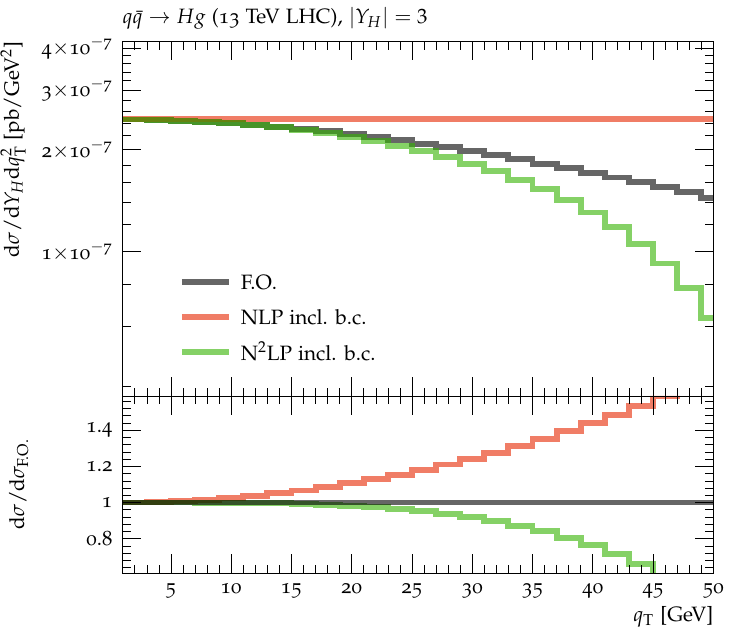}
    \caption{   }
    \label{fig:results:val:qTratios:asyexp:qqHg:YH3}
  \end{subfigure}
  \caption{
    Comparisons of the \qT\ spectra between the exact
    fixed-order calculation (black) and approximations
    constructed using the power expansion up to \LP (blue),
    \NLP (red), and \NNLP (green) for different partonic
    processes at $|Y_H|=1$ (top) and $|Y_H|=3$ (bottom).
    Throughout we include the boundary corrections
    $\mathbf{B}_{cs}$, $\mathbf{B}_{\bar{c}s}$, and
    $\mathbf{B}_{uc(\bar{c})}$.
  }
  \label{fig:results:val:qTratios:asyexp:YH3}
\end{figure} 
  
In the previous figures we have focussed on the \qT~spectra
at central rapidities, $|Y_H|=0$, where $e^{\pm Y_H}\sim\order(1)$.
We now increase the rapidity of the Higgs boson to affect the
scaling of the relevant light cone momenta $k_\pm$, associated
momentum fractions of Eq.\ \eqref{eq:def:kin:xn:xnbar}, and 
phase space boundaries of Eq.\ \eqref{eq:def:b.c.}.
To this end, Fig.~\ref{fig:results:val:qTratios:asyexp:YH3} displays
the \qT\ spectra at $|Y_H|=1$ and $|Y_H|=3$.
We observe that the \qT~distributions at $|Y_H|=1$ generally
observe similar patterns to those at $|Y_H|=0$ from
Fig.~\ref{fig:results:val:qTratios:asyexp:YH0} for all three
partonic channels, as $e^{\pm Y_H}$ is still more or less of $\order(1)$.
At $|Y_H|=3$, where $e^{|Y_H|}$ is now of $\order(10)$, even though
ability of the \LP approximation to reproduce the exact result is
further reduced in the moderate domain, the higher power corrections
are still convergent as before.
More specifically, nearing $\qT=30\,\text{GeV}$, the \LP contribution
to the process $gg\to Hg$ accounts for less than $60\%$ of the full
theory, which is raised to nearly $80\%$ after including the \NLP
corrections and about $97\%$ with the \NNLP terms.
A marked difference, even at \NNLP, w.r.t.\ the central rapidity region is only
observed at very high \qT\ around 50\,GeV.
We would like to note that the non-monotonous behaviour of the \NLP
result in Fig.~\ref{fig:results:val:qTratios:asyexp:ggHg:YH3} around
$\qT\sim45\,\text{GeV}$ is caused by a changing sign of the \NLP
approximants.

\begin{figure}[t!]
  \centering
  \begin{subfigure}{0.32\textwidth}
    \centering
    \includegraphics[width=.9\linewidth]{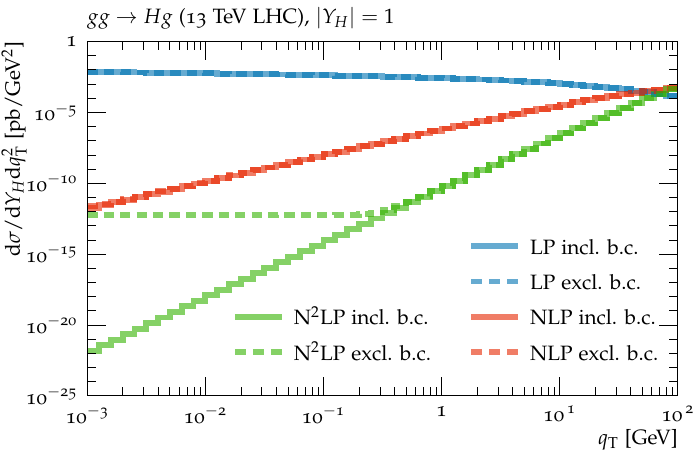}
    \caption{  }
    \label{fig:results:val:qTdist:asyexp:ggHg:YH1}
  \end{subfigure}
  \begin{subfigure}{0.32\textwidth}
    \centering
    \includegraphics[width=.9\linewidth]{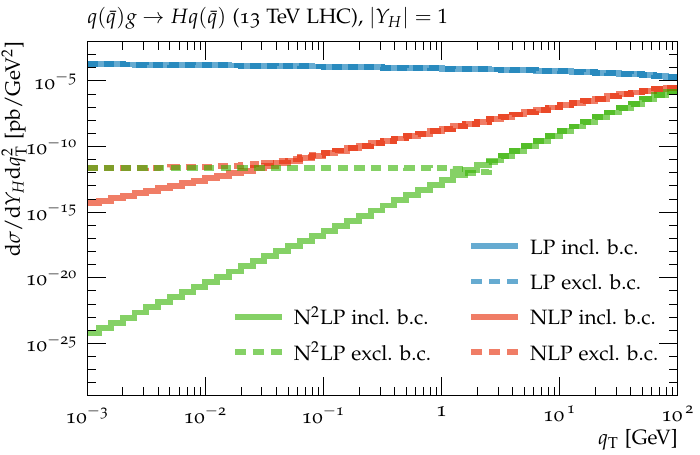}
    \caption{   }
    \label{fig:results:val:qTdist:asyexp:qgHq:YH1}
  \end{subfigure} 
  \begin{subfigure}{0.32\textwidth}
    \centering
    \includegraphics[width=.9\linewidth]{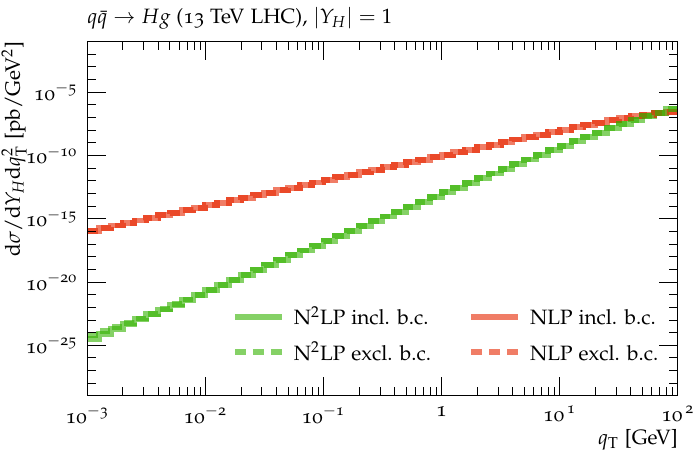}
    \caption{   }
    \label{fig:results:val:qTdist:asyexp:qqHg:YH1}
  \end{subfigure}
  \\[2mm]
  \centering
  \begin{subfigure}{0.32\textwidth}
    \centering
    \includegraphics[width=.9\linewidth]{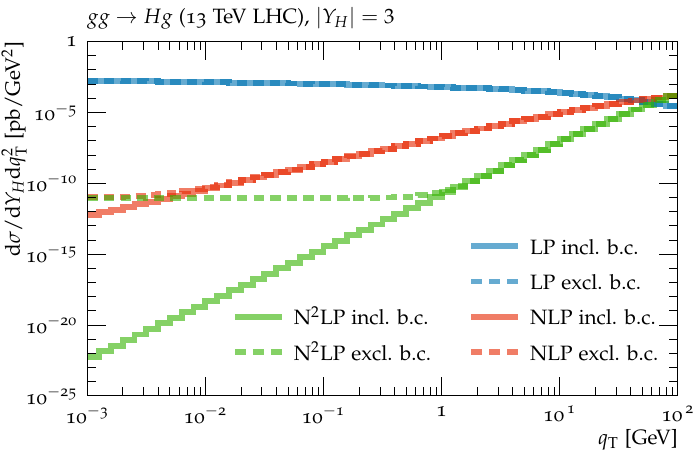}
    \caption{  }
    \label{fig:results:val:qTdist:asyexp:ggHg:YH3}
  \end{subfigure}
  \begin{subfigure}{0.32\textwidth}
    \centering
    \includegraphics[width=.9\linewidth]{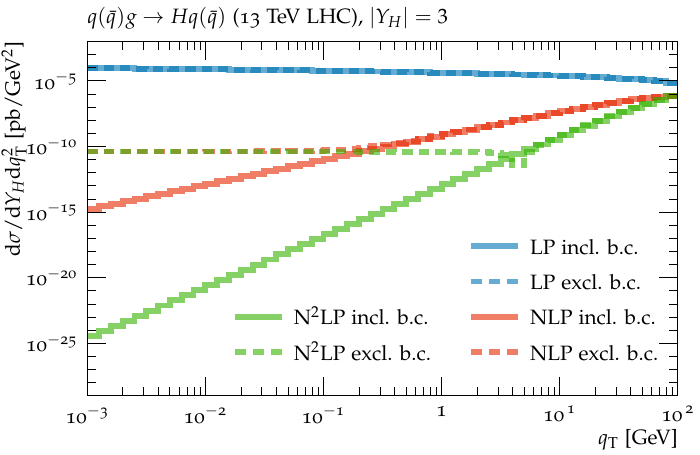}
    \caption{   }
    \label{fig:results:val:qTdist:asyexp:qgHq:YH3}
  \end{subfigure} 
  \begin{subfigure}{0.32\textwidth}
    \centering
    \includegraphics[width=.9\linewidth]{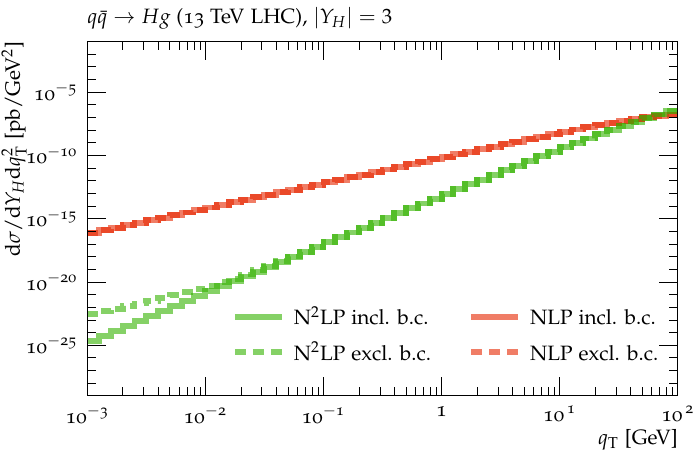}
    \caption{   }
    \label{fig:results:val:qTdist:asyexp:qqHg:YH3}
  \end{subfigure}
  \\[2mm]
  \centering
  \begin{subfigure}{0.32\textwidth}
    \centering
    \includegraphics[width=.9\linewidth]{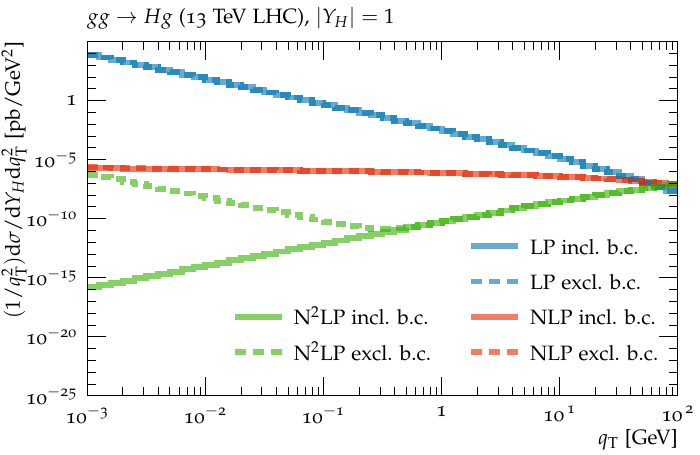}
    \caption{  }
    \label{fig:results:val:qTwt:asyexp:ggHg:YH1}
  \end{subfigure}
  \begin{subfigure}{0.32\textwidth}
    \centering
    \includegraphics[width=.9\linewidth]{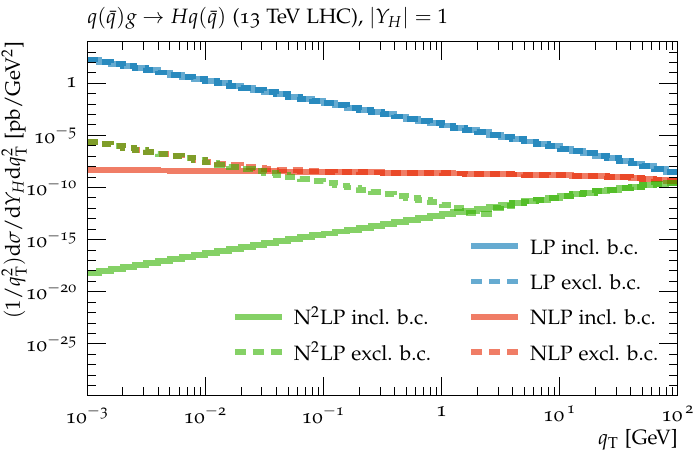}
    \caption{   }
    \label{fig:results:val:qTwt:asyexp:qgHq:YH1}
  \end{subfigure}
  \begin{subfigure}{0.32\textwidth}
    \centering
    \includegraphics[width=.9\linewidth]{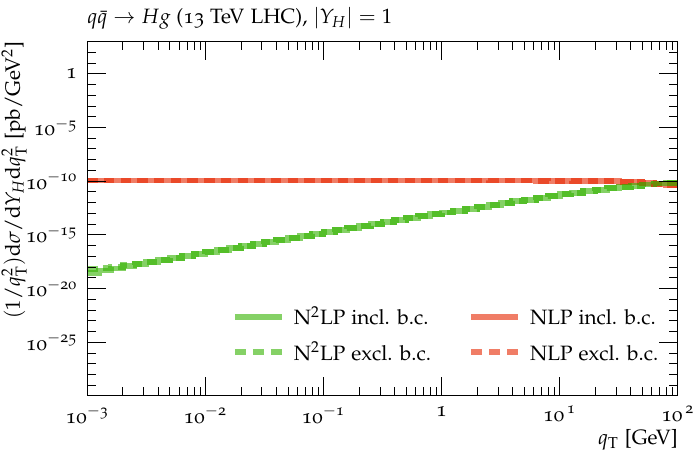}
    \caption{   }
    \label{fig:results:val:qTwt:asyexp:qqHg:YH1}
  \end{subfigure}
  \\[2mm]
  \centering
  \begin{subfigure}{0.32\textwidth}
    \centering
    \includegraphics[width=.9\linewidth]{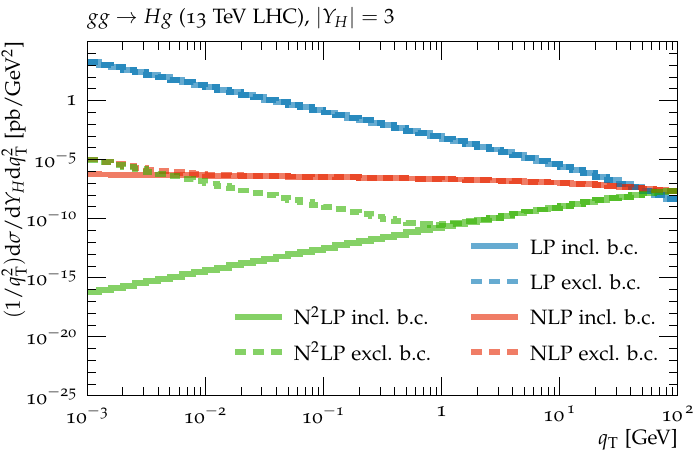}
    \caption{  }
    \label{fig:results:val:qTwt:asyexp:ggHg:YH3}
  \end{subfigure}
  \begin{subfigure}{0.32\textwidth}
    \centering
    \includegraphics[width=.9\linewidth]{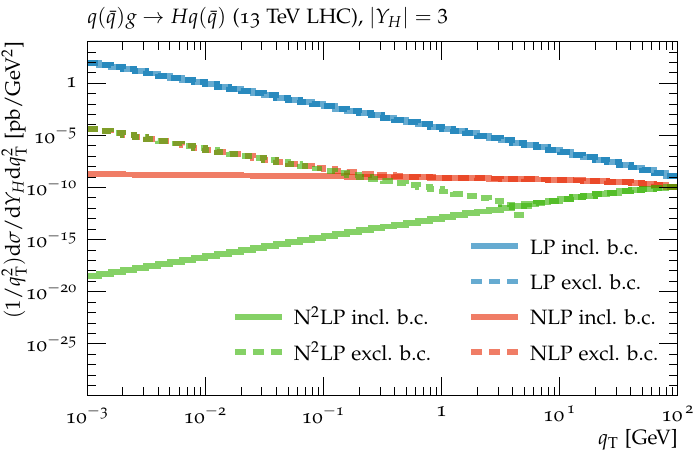}
    \caption{   }
    \label{fig:results:val:qTwt:asyexp:qgHq:YH3}
  \end{subfigure} 
  \begin{subfigure}{0.32\textwidth}
    \centering
    \includegraphics[width=.9\linewidth]{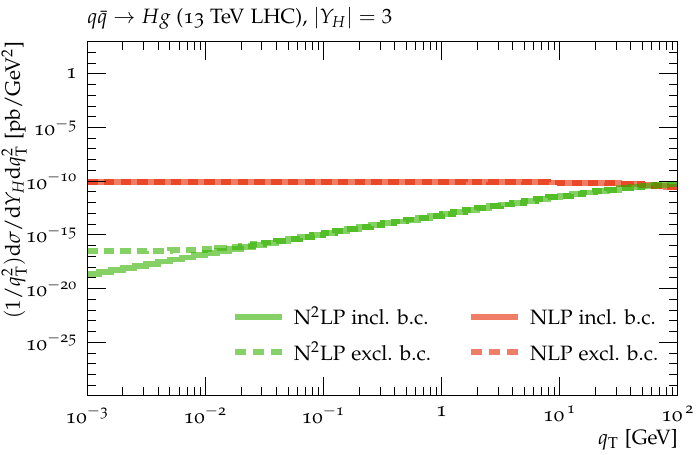}
    \caption{   }
    \label{fig:results:val:qTwt:asyexp:qqHg:YH3}
  \end{subfigure}
  \\
    \caption{
      Differences of the \qT\ ($1^\text{st}\,\&\,2^\text{nd}$ row)
      and weighted \qT\ ($3^\text{rd}\,\&\,4^\text{th}$ row)
      spectra between the exact fixed-order calculation and the
      approximations for the three different partonic processes
      at $|Y_H|=1$ ($1^\text{st}\,\&\,3^\text{rd}$ row) and
      $|Y_H|=3$ ($2^\text{nd}\,\&\,4^\text{th}$ row).
      The solid (dashed) lines present the results including (excluding)
      the boundary corrections $\mathbf{B}_{cs}$ in
      Eq.~\eqref{eq:def:asyexp:mcfs:cs}, $\mathbf{B}_{\bar{c}s}$ in
      Eq.~\eqref{eq:def:asyexp:mcfs:cbars}, and $\mathbf{B}_{uc(\bar{c})}$ in
      Eq.~\eqref{eq:def:asyexp:mcfs:uc:ucbar}.
      The blue, green, and red lines detail the \LP, \NLP, and
      \NNLP approximations, respectively.
    }
  \label{fig:results:val:qTwt:asyexp:YH1:YH3}
\end{figure}

Finally, to assess the coefficients at each power, we investigate
the difference between the exact (weighted) \qT~spectra and their
approximations on different levels of the power expansion in
Fig.~\ref{fig:results:val:qTwt:asyexp:YH1:YH3}.
We find that the power series appraised in Sec.~\ref{sec:asyexp:outputs}
is capable of reproducing the asymptotic behaviour of the full theory
for both $|Y_H|=1$ and $|Y_H|=3$.
Moreover, an interesting phenomenon can be observed here: the ranges
that are sensitive to the boundary corrections grow with increasing
rapidity $|Y_H|$ for all three partonic processes.
For instance, the \NNLP accurate computation, represented by the
green curves, shows discrepancies between the solid and dashed lines,
which are driven by the boundary corrections, arise below
$\qT\approx 0.4\,\text{GeV}$ in the $|Y_H|=1$ case in the $gg$ channel
but already below $\qT\approx 1\,\text{GeV}$ in the $|Y_H|=3$ case.
To interpret this behaviour, it merits recalling that in the central
rapidity regime where $e^{\pm Y_H}\sim\order(1)$, the momentum fractions
$x_n$ and $x_\nbar$ in Eq.~\eqref{eq:def:kin:xn:xnbar} both approach
zero.
Hence, while the interior contribution experiences no suppression from
the PDFs, the boundary correction in Eq.~\eqref{eq:asyexp:deltaI:uc} does,
suppressing it in the majority of the phase space.
However, nearing the rapidity extremes $Y_H\to\pm 4.5$, one of $x_n$ and $x_\nbar$ approaches unity.
Now, the interior contribution will be suppressed by one of the PDFs,
analogous to that in Eq.~\eqref{eq:asyexp:deltaI:uc}, thereby losing
its predominant role, relatively enlarging the boundary terms.

\clearpage
\section{Conclusions}
\label{sec:conclusions}

In this paper, we constructed a systematic and mathematically
well-defined framework to construct a small-\qT\ expansion at
\NLO up to arbitrary power accuracy.
This framework is applicable to all possible conservative regulators of
the emerging rapidity divergences, and we have shown that the
results are identical for three radically different choices.
In our power expansion we have refactored all \qT~dependences
from the transition amplitudes
and PDFs, except for those residing in the integration path and
phase space boundaries.
To achieve this, we have divided the phase space into two sectors,
the interior of the integration domain and the segments in the
vicinity of the integration boundaries.

The contributions originating from the integration boundaries,
which we have evaluated in this paper for the first time, are
always associated with the PDF of the beam at the opposite side
and, ultimately, give rise to power suppressed constant
corrections at \NLO.
We have summarised their analytic expressions in
Eqs.~\eqref{eq:def:Gp} and \eqref{eq:asyexp:deltaI:ec:int}.
The analysis of the contributions from the interior domain,
on the other hand, is more involved due to the
number of scales that are enclosed in the phase space integral.
To derive the asymptotic series of the interior contributions,
we have introduced a set of auxiliary cutoff scales to disentangle
the scale hierarchies and in turn perform the appropriate power expansion.
The recombination of all artificially separated contributions at
each power removes all dependence on the auxiliary scales, such that
the final result is independent of them.
In any case, the contributions from the interior domain are
comprised of three sectors, the $n$- and $\nbar$-collinear domains,
and the zero-bin subtrahends.
At a given $\omega$-th power, for example, the two collinear contributions
can be derived by expanding the integrands according to the respective
collinear scaling of the integration variable and projecting onto
the $\omega$-th power contributions.
Conversely, the zero-bin subtrahends are organised by dual-scalings,
in which a second expansion of the $\omega$-th power collinear functions
needs to be performed according to the soft scaling, giving contributions
from the lowest power precision up to the $\omega$-th power.
This algorithm is presented in
Eqs.~\eqref{eq:def:asyexp:c:at:omega:nured:Texp} and
\eqref{eq:def:c0:ns:Texp} in terms of the expansion operators.

To demonstrate the applicability of our algorithm, two additional
fundamentally different rapidity regulators, the pure-rapidity and
exponential regulators, in addition to the above cutoff scale regulator,
have been embedded into
Eqs.~\eqref{eq:def:asyexp:c:at:omega:nured:Texp}-\eqref{eq:def:c0:ns:Texp}
in order to calculate the power corrections from the interior
domain up to \NNLP accuracy, which we present here for the first time.
We found that even though substantial differences emerge from the
individual sectors in those two regularisation schemes, after summing
all contributions both results produce the same power series as we
have obtained via the momentum cutoffs.
Together with the correction terms induced by the kinematical variables
and the integral boundaries, we have derived analytic expressions
for the expanded \qT~distribution for Higgs hadroproduction up to \NNLP,
following Eq.~\eqref{eq:asyexp:tot}.

To validate our results, we confronted the resulting power
series approximation to the full QCD calculation at the same order.
We have shown that our expanded \qT~spectra can satisfactorily replicate
the desired qualities of the full theory for all contributing partonic
channels at three different rapidity slices, $|Y_H|=0,1,3$.
In particular, when including \NNLP corrections, the approximate
predictions can capture the exact ones within (sub)percent level accuracy
up to $\qT=30\,\text{GeV}$.
In addition, to explore the relationship between the sizes of the
different ingredients at a given power, we produced results without
accounting for contributions stemming from the integral boundaries,
a situation that may arise if PDFs and their derivatives are considered
to vanish at the opposite in these regions.
We found that the boundary contributions are generally small at
\qT\ larger than a few GeV, but can play a decisive
role as $\qT\to 0$.
Here, their omission effectively degrades the \NLP and \NNLP results
to \LP accuracy.
To give an example, the \NNLP corrections in the partonic process
$q(\bar{q})g\to H q(\bar{q})$ are driven by the boundary corrections
already at $\qT\lesssim 10\,\text{GeV}$ at $Y_H=3$.

At last, it is worth emphasising that although Higgs hadroproduction
is used as an example here to illustrate the capabilities and practicability
of our results in Eq.~\eqref{eq:asyexp:tot}, the algorithm developed in
this paper is expected to be directly applicable to the Drell-Yan family
of processes and other colour-singlet processes containing similar
denominators in the transition amplitudes.
Furthermore, during our investigation, except for the indispensable ansatz made on the analytic properties of the PDFs, the power expansions are all carried out following a mathematically well-defined manner.
Therefore, we expect the conclusions from this work can not only serve
as a robust recipe to derive the power series of the \qT\ distribution
at NLO, but also offer the theoretical baseline for exploring the
asymptotic properties of contributions at higher perturbative order.



\subsection*{Acknowledgements}

%
MS is funded by the Royal Society through a
University Research Fellowship (URF\textbackslash{}R1\textbackslash{}180549
and URF\textbackslash{}R\textbackslash{}231031) and a
Royal Society Enhancement Award
(RGF\textbackslash{}EA\textbackslash{}181033, CEC19\textbackslash{}100349,
and RF\textbackslash{}ERE\textbackslash{}210397).

\appendix
\section{Application and adaptation of the dissipative regulators}\label{sec:dis:reg}

Here we will demonstrate that, after suitable adaptation, our zero-bin subtrahends in Eqs.~\eqref{eq:def:c0:ns:Texp} can also be utilised in the power expansion governed by the dissipative regulators. In the followings, we will illustrate this by implementing the $\Delta$- and $\eta$-regulators.

\subsection{The~\texorpdfstring{$\Delta$}{Delta}-regulator}

The~$\Delta$-regulator was proposed in \cite{Chiu:2009yx} to regulate the rapidity divergence in the massive Sudakov factor, which was afterwards generalised in \cite{Echevarria:2011epo,Echevarria:2015usa,Echevarria:2015byo,Echevarria:2016scs} for fulfilling the restrictions from the non-abelian exponentiation theorem in higher-perturbative order calculation.
The method behind this regularisation scheme is to shift the mass of the mediating particles in the transition amplitudes by an infinitesimal amount, such that when approaching the rapidity extrema, i.e.\ $y_k\to\pm\infty$, the involved propagators can never go on-shell and thus are always insulated from the rapidity divergences.

The~$\Delta$-regulator is dedicated to the rapidity divergences at LP, as extra residual momenta can be generated during the expansion of the squared amplitudes  beyond LP~\cite{Inglis-Whalen:2021bea},  which are singular in the limit $y_k\to\pm\infty$ and stay unprotected by the $\Delta$-regulator. To this end, in the following discussion, we will focus on the LP contribution only.

At LP, the zero-bin subtraction for the small~\qT~expansion primarily concerns the process $gg\to Hg$ up to NLO, which, as presented in Eq.~\eqref{eq:msq}, corresponds to the integrand $ I^{ -1,-1 }_{[\kappa], \{\alpha,\beta\}  }$ based on the definition in Eq.~\eqref{eq:def:subjt}.  Applying the~$\Delta$-regulator onto Eq.~\eqref{eq:def:subjt} with $\rho=\sigma=-1$ amounts to re-weight the squared amplitudes by a factor of,
\begin{align} 
\label{eq:def:delta:rap}
      \mathcal{R} (k_-,k_+,\tau)\,=\,\frac{\qT^2}{(k_++\tau)\,(k_-+\tau)}\,,     
 \end{align}
where $\tau$ denotes an infinitesimal parameter, $\tau>0$.  It is seen that $\mathcal{R} $ in Eq.~\eqref{eq:def:delta:rap} indeed satisfies the criteria of Eq.~\eqref{rap:def:feat}, which keeps void for the rapidity-safe integrands but becomes activated for the singular transition amplitudes.

Substituting Eq.~\eqref{eq:def:delta:rap} into Eqs.~\eqref{eq:def:asyexp:c:at:omega:nured:Texp} and \eqref{eq:def:c0:ns:Texp} and then truncating out the corrections beyond LP, it yields,
 \begin{align}
\label{eq:def:asyexp:c:at:omega:nured:delta}
\widetilde{\mathcal{G}}^{-1,-1,(-1)}_{[\kappa],\{\alpha,\beta\} }\Bigg|^{\langle \Delta\rangle}_{c} 
=\,
&
\int^{\tilde{k}^{\mathrm{max}}_+}_{0}
{\done k_+}
  \left[\frac{1}{k_+}\right]_*^{\nu_n}
  {F}^{(\beta_2)}_{i/N,\beta_1}\left(k_++m_H e^{-Y_H}\right)
 {F}^{(\alpha_2)}_{j/\bar{N},\alpha_1}\left(m_H e^{+Y_H}\right)\nnb\\
&+  \,   
  {F}^ {(\beta_2 )}_{i/N,\beta_1}\left( m_H e^{-Y_H}\right)\,
     {F}^{ (\alpha_2 )}_{j/\bar{N},\alpha_1}\left(m_H e^{+Y_H}\right)\,
    \ln\left[\frac{\nu_n}{\tau}\right]\,,\\
\label{eq:def:asyexp:cbar:at:omega:nured:delta} 
 \widetilde{\mathcal{G}}^{-1,-1,(-1)}_{[\kappa],\{\alpha,\beta\} }\Bigg|^{\langle \Delta \rangle}_{\bar{c}} 
=\,
&
\int^{{\frac{\qT^2}{\tilde{k}^{\mathrm{min}}_+}}}_0\,
 \done k_-\,\left[\frac{1}{k_- }\right]^{\nu_{\bar{n}}}_{*}
 \,{F}^{ (\beta_2 )}_{i/N,\beta_1}\left(  m_H e^{-Y_H}\right)\,
     {F}^{(\alpha_2)}_{j/\bar{N},\alpha_1}\left(k_-+m_H e^{+Y_H}\right)\,\nonumber\\
&+ 
   {F}^{ (\beta_2 )}_{i/N,\beta_1}\left( m_H e^{-Y_H}\right)\,
     {F}^{ (\alpha_2 )}_{j/\bar{N},\alpha_1}\left(m_H e^{+Y_H}\right)\,
    \ln\left[\frac{\nu_\nbar}{\tau}\right]\,,   
    \end{align}
    and
         \begin{align}
  \label{eq:def:c0:cbar0:ns:delta}
   \widetilde{\mathcal{G}}^{-1,-1,(-1)}_{[\kappa],\{\alpha,\beta\} }\Bigg|^{\langle \mathrm{NS} \rangle,  \langle \Delta \rangle}_{c0}\, 
   =\,&
   \widetilde{\mathcal{G}}^{-1,-1,(-1)}_{[\kappa],\{\alpha,\beta\} }\Bigg|^{\langle \mathrm{NS} \rangle,  \langle \Delta \rangle}_{\bar{c}0}\,
  =\,  
 {F}^{(\beta_2 )}_{i/N,\beta_1}\left(m_H e^{-Y_H}\right)\,{F}^{(\alpha_2 )}_{j/\bar{N},\alpha_1}\left(m_H e^{+Y_H}\right)\,
 \ln\left[\frac{\qT}{\tau}\right]\,.
\end{align}

It is immediate to find that the difference between Eqs.~\eqref{eq:def:asyexp:c:at:omega:nured:delta}-\eqref{eq:def:asyexp:cbar:at:omega:nured:delta} and  Eq.~\eqref{eq:def:c0:cbar0:ns:delta} exactly replicates the leading power outputs in  Eqs.~\eqref{eq:def:asyexp:c:mom:at:omega}-\eqref{eq:def:asyexp:cbar:mom:at:omega} and Eqs.~\eqref{eq:def:asyexp:cs:mom:int}-\eqref{eq:def:asyexp:cbars:mom:int} derived via momentum cutoffs, those in Eqs.~\eqref{eq:def:asyexp:c:at:omega:nured:prap}-\eqref{eq:def:asyexp:cbar:at:omega:nured:prap} by means of the pure-rapidity regulator,  or those in Eqs.~\eqref{eq:def:asyexp:c:at:omega:nured:exp}-\eqref{eq:def:c0:ns:exp} from the exponential one, after setting $\omega=\rho=\sigma=-1$ therein.

\subsection{The~\texorpdfstring{$\eta$}{Eta}-regulator}

The $\eta$-regulator was devised by~\cite{Chiu:2011qc,Chiu:2012ir} to facilitate the establishment of the rapidity renormalisation group equations in the SCET$_{\mathrm{II}}$-based analysis of the jet broadening and \qT~distributions.
Akin to the analytic regularisation prescription \cite{Becher:2010tm,Becher:2011dz}, the $\eta$-regulator  concerns the momenta of the emitted partons raised by the power of $\tau$ as well.
However, in place of the light-cone component $k_+$ or $k_-$ entailed in \cite{Becher:2010tm,Becher:2011dz}, the $\eta$-regulator puts the magnitude of the longitudinal one into the base, namely, 
\begin{align} 
\label{eq:def:eta:rap}
      \mathcal{R}(k_-,k_+,\tau)\,=\,\left|k_--k_+\right| ^{-\tau}.
 \end{align}
Implementing the $\eta$-regularisation scheme is rather subtle.  In its original definition~\cite{Chiu:2011qc,Chiu:2012ir} and the recent review of \cite{Ebert:2018gsn}, the calculation on the collinear sector calls for the expansions of both the rapidity regulator in Eq.~\eqref{eq:def:eta:rap} and the integrand of Eq.~\eqref{eq:def:subjt}, which, based on the notation in Eq.~\eqref{eq:def:asyexp:c:at:omega:nured:Texp}, requires that $\mathcal{R}$ always appears to the right hand side of $ \widehat{\mathbf{T}}_{c}^{(\omega)}$ and $ \widehat{\mathbf{T}}_{\bar{c}}^{(\omega)}$. In doing this, a set of evanescent power corrections can be generated from Eq.~\eqref{eq:def:eta:rap}, which are proportional to $\tau$ (or its higher powers) and will make non-trivial contributions in combination with the pole terms.
Those evanescent influences have not been taken into account during the derivation of Eqs.~\eqref{eq:def:asyexp:c:at:omega:nured:Texp}-\eqref{eq:def:c0:ns:Texp}, and therefore, it is not straightforward to apply Eq.~\eqref{eq:def:eta:rap} here until appropriate adaptations are put in place.

In the following, we will introduce one of strategies to adapt the $\eta$-regulator to our formalism in Eqs.~\eqref{eq:def:asyexp:c:at:omega:nured:Texp}-\eqref{eq:def:c0:ns:Texp}.
Considering that the main obstacle hindering this application comes from  evanescent contributions, it could be beneficial to maintain the regulator in  Eq.~\eqref{eq:def:eta:rap} always to the left of the  operators $ \widehat{\mathbf{T}}_{c}^{(\omega)}$ and $ \widehat{\mathbf{T}}_{\bar{c}}^{(\omega)}$ in practical calculations.
In this way, Eq.~\eqref{eq:def:eta:rap}  abides by the criteria of Eq.~\eqref{rap:def:feat} and can still  serve as a qualified rapidity-divergence regularisation scheme at any power accuracy.
Furthermore, since the $\eta$-regulator at this moment does not participate into the expansion governed by $ \widehat{\mathbf{T}}_{c}^{(\omega)}$ and $ \widehat{\mathbf{T}}_{\bar{c}}^{(\omega)}$, one can save efforts in coping with the evanescent contribution. 

Now applying the $\eta$-regulator onto Eqs.~\eqref{eq:def:asyexp:c:at:omega:nured:Texp}-\eqref{eq:def:c0:ns:Texp} becomes immediate, which evaluates to,
\begin{align}
\label{eq:def:asyexp:c:at:omega:nured:etarap}
\widetilde{\mathcal{G}}^{\rho,\sigma,(\omega)}_{[\kappa],\{\alpha,\beta\} }\Bigg|^{\langle \eta\rangle}_{c} 
=\,
&
  \frac{\bar{\theta}\left(\omega-\sigma\right)}{(\omega-\rho)!} 
 \int^{\tilde{k}^{\mathrm{max}}_+}_{0}
 \frac{\done k_+}{k_+}
   (k_+)^{\sigma-\omega}
  {F}^{(\beta_2)}_{i/N,\beta_1}\left(k_++m_H e^{-Y_H}\right)
 {F}^{(\alpha_2+\omega-\rho)}_{j/\bar{N},\alpha_1}\left(m_H e^{+Y_H}\right)\nnb\\
 &\,+ \frac{\theta\left(\omega-\sigma\right)}{(\omega-\rho)!} \int^{\tilde{k}^{\mathrm{max}}_+}_{0}
{\done k_+}
  \left[\frac{1}{k_+^{\omega-\sigma+1}}\right]_*^{\nu_n}
  {F}^{(\beta_2)}_{i/N,\beta_1}\left(k_++m_H e^{-Y_H}\right)
 {F}^{(\alpha_2+\omega-\rho)}_{j/\bar{N},\alpha_1}\left(m_H e^{+Y_H}\right)\nnb\\
& +\frac{\theta\left(\omega-\sigma-1\right)}{(\omega-\rho)!} 
 \sum_{\eta=0}^{\omega-\sigma-1}
 \frac{ {F}^{(\beta_2+\eta)}_{i/N,\beta_1}\left( m_H e^{-Y_H}\right)
 {F}^{(\alpha_2+\omega-\rho)}_{j/\bar{N},\alpha_1}\left(m_H e^{+Y_H}\right)}{\eta!}\,
\frac{\nu_n^{\sigma+\eta-\omega}}{\sigma+\eta-\omega}
\nnb\\
&+   \frac{\theta(  \omega-\sigma)}{(\omega-\rho)!} \,   
 \frac{  {F}^ {(\beta_2+\omega-\sigma)}_{i/N,\beta_1}\left( m_H e^{-Y_H}\right)\,
     {F}^{ (\alpha_2+\omega-\rho)}_{j/\bar{N},\alpha_1}\left(m_H e^{+Y_H}\right)\,}{ (\omega-\sigma)!}\,
   \left\{\ln\left[\frac{\nu_{n}}{\qT^2} \right]+\frac{1}{\tau} \right\}\,,\\
\label{eq:def:asyexp:cbar:at:omega:nured:etarap} 
 \widetilde{\mathcal{G}}^{\rho,\sigma,(\omega)}_{[\kappa],\{\alpha,\beta\} }\Bigg|^{\langle \eta\rangle}_{\bar{c}} 
=&
\frac{\bar{\theta}( \omega-\rho)}{(\omega-\sigma)!}\,
\int^{{\frac{\qT^2}{\tilde{k}^{\mathrm{min}}_+}} }_0\,\frac{\done k_-}{k_-}\,
k_-^{\rho-\omega}\,
  {F}^{ (\beta_2+\omega-\sigma)}_{i/N,\beta_1}\left(  m_H e^{-Y_H}\right)\,
     {F}^{(\alpha_2)}_{j/\bar{N},\alpha_1}\left(k_-+m_H e^{+Y_H}\right)  \,\nonumber\\
&+  
\frac{{\theta}(  \omega-\rho)}{(\omega-\sigma)!}\,
\int^{{\frac{\qT^2}{\tilde{k}^{\mathrm{min}}_+}}}_0\, \done k_-\,\left[\frac{1}{k_-^{\omega-\rho+1}}\right]^{\nu_{\bar{n}}}_{*}\,{F}^{ (\beta_2+\omega-\sigma)}_{i/N,\beta_1}\left(  m_H e^{-Y_H}\right)\,
     {F}^{(\alpha_2)}_{j/\bar{N},\alpha_1}\left(k_-+m_H e^{+Y_H}\right)\,\nonumber\\
&+
\frac{{\theta}(  \omega-\rho-1)}{(\omega-\sigma)!}\,\sum_{\lambda=0}^{\omega-\rho-1}\,
   \frac{  {F}^{ (\beta_2+\omega-\sigma)}_{i/N,\beta_1}\left( m_H e^{-Y_H}\right)\,
     {F}^{  (\alpha_2+\lambda)}_{j/\bar{N},\alpha_1}\left(m_H e^{+Y_H}\right)\,}{ \lambda!} \,
 \frac{\nu_\nbar^{\rho+\lambda-\omega}}{\rho+\lambda-\omega}\nonumber\\
&+ 
\frac{ {\theta}(  \omega-\rho)}{(\omega-\sigma)!}\,
 \frac{  {F}^{ (\beta_2+\omega-\sigma)}_{i/N,\beta_1}\left( m_H e^{-Y_H}\right)\,
     {F}^{ (\alpha_2+\omega-\rho)}_{j/\bar{N},\alpha_1}\left(m_H e^{+Y_H}\right)\,}{ (\omega-\rho)!} \,
     \left\{\ln\left[\frac{\nu_{\nbar}}{\qT^2} \right]+\frac{1}{\tau} \right\}\,,
 \end{align}
together with the subtraction terms, 
  \begin{align} 
  \label{eq:def:c0:ns:etarap} 
   \widetilde{\mathcal{G}}^{\rho,\sigma,(\omega)}_{[\kappa],\{\alpha,\beta\} }\Bigg|^{\langle \mathrm{NS} \rangle,  \langle \eta \rangle}_{c0}\, 
  =\,&
  \theta\left(\omega-\sigma \right)
\frac{{F}^{(\beta_2+\omega-\sigma)}_{i/N,\beta_1}\left(m_H e^{-Y_H}\right)\,{F}^{(\alpha_2+\omega-\rho)}_{j/\bar{N},\alpha_1}\left(m_H e^{+Y_H}\right)}{(\omega-\sigma)!\, (\omega-\rho)!}
 \left\{\ln\left[\frac{1}{\qT } \right]+\frac{1}{\tau} \right\}
 ,\\
   \label{eq:def:cbar0:ns:etarap}
  \widetilde{\mathcal{G}}^{\rho,\sigma,(\omega)}_{[\kappa],\{\alpha,\beta\} }\Bigg|^{\langle \mathrm{NS} \rangle,  \langle \eta \rangle}_{\bar{c}0}\,
    =\,&
\theta\left(\omega-\rho\right) 
\frac{{F}^{(\beta_2+\omega-\sigma)}_{i/N,\beta_1}\left(m_H e^{-Y_H}\right)\,{F}^{(\alpha_2+\omega-\rho)}_{j/\bar{N},\alpha_1}\left(m_H e^{+Y_H}\right)}{(\omega-\sigma)!\, (\omega-\rho)!}
 \left\{\ln\left[\frac{1}{\qT } \right]+\frac{1}{\tau} \right\}\,.
\end{align}

Subtracting Eqs.~\eqref{eq:def:c0:ns:etarap}-\eqref{eq:def:cbar0:ns:etarap} from Eqs.~\eqref{eq:def:asyexp:c:at:omega:nured:etarap}-\eqref{eq:def:asyexp:cbar:at:omega:nured:etarap}, it reproduces the results in Eqs.~\eqref{eq:def:asyexp:c:mom:at:omega}-\eqref{eq:def:asyexp:cbar:mom:at:omega} and Eqs.~\eqref{eq:def:asyexp:cs:mom:int}-\eqref{eq:def:asyexp:cbars:mom:int} obtained from momentum cutoffs, and also those in Sec.~\ref{sec:asyexp:outputs} by means of pure-rapidity and exponential regularisation schemes.
 
\section{Comparison of the NLP results with the previous literature}\label{sec:comp:nlp}

Subleading power corrections of the process $pp\to H+X$ have been investigated in \cite{Ebert:2018gsn} as well by means of the $\eta$- and pure rapidity regulators.  Here we take the partonic process $q\bar{q}\to H g$ as an illustrative example to compare the expressions in \cite{Ebert:2018gsn} and those in Eq.~\eqref{eq:def:asyexp:mcfs:cs}, \eqref{eq:def:asyexp:mcfs:cbars}, and \eqref{eq:def:asyexp:mcfs:c}. Over the course, we will particularly emphasise the interior contributions encoded by the partonic matrices $\mathbf{P}$, $\mathbf{D}$, and $\mathbf{R}$, as the boundary corrections in $\mathbf{B}$ are related to the PDFs at the opposite end, which are all assumed to be vanishing in \cite{Ebert:2018gsn}.

In \cite{Ebert:2018gsn}, the analytic results for the channel $q\bar{q}\to H g$ entail the expanded amplitudes of Eq.~\eqref{eq:msq} sandwiched by the (derivatives of) PDFs. Recasting their expressions into matrix form according to the bases in Eq.~\eqref{eq:PDF:N:ary} and  also synchronising the pre-factors in line with Eq.~\eqref{eq:def:asyexp:mcfs:cs}, we obtain
  \begin{align}
    \label{eq:sing:qq:Ebert18}
     \widehat{\mathbf{P}}_{cs}^{(0),q \bar{q} } =&     \widehat{\mathbf{P}}_{\bar{c}s}^{(0),q \bar{q} } =  \widehat{\mathbf{D}}_{cs}^{(0),q \bar{q} }= \widehat{\mathbf{D}}_{\bar{c}s}^{(0),q \bar{q} }=0\,,\\
  \label{eq:Rcs:qq:Ebert18}
  \widehat{\mathbf{R}}_{cs}^{(0),q \bar{q} } =&
  \left[
\begin{array}{cc}
 -\frac{32}{27}+\frac{16}{27  z_n}+\frac{16}{27  z_n^2}  \\[2.ex]
 \frac{16}{27}-\frac{32}{27  z_n}+\frac{16}{27  z_n^2}   \\
\end{array}
\right]\,,\\
  \label{eq:Rcbars:qq:Ebert18}
  \widehat{\mathbf{R}}_{\bar{c}s}^{(0),q \bar{q} } = &
 \left[
\begin{array}{cc}
 -\frac{32}{27}+\frac{16}{27 z_\nbar}+\frac{16}{27 z_\nbar^2} & \frac{16}{27}-\frac{32}{27 z_\nbar}+\frac{16}{27 z_\nbar^2} \\
\end{array}
\right]\,,
  \end{align}
  where the hatted bold matrices encode the results of \cite{Ebert:2018gsn}.
It is immediate to see that the matrices in Eq.~\eqref{eq:sing:qq:Ebert18}~associated with the plus distribution and delta function are all zero, echoing the outputs from this work in Eqs.~\eqref{eq:Dcs:lp_nlp_nnlp}, \eqref{eq:Pcs:gg_gq_qq}, and \eqref{eq:Xcbars:nnlp}, whereas,  due the emergence of the extra off-diagonal entries in Eqs.~\eqref{eq:Rcs:qq:Ebert18}-\eqref{eq:Rcbars:qq:Ebert18},   the regular contributions from \cite{Ebert:2018gsn} appear to observe distinct patterns from ours in Eq.~\eqref{eq:Rcs:qq} and Eq.~\eqref{eq:Xcbars:nnlp}.

As a matter of fact, this kind of difference can be resolved by applying the method of integration by parts as appropriate. From Eq.~\eqref{eq:PDF:N:ary}, the off-diagonal terms in Eqs.~\eqref{eq:Rcs:qq:Ebert18}-\eqref{eq:Rcbars:qq:Ebert18} always invoke the derivatives of one of the PDFs, which,  via the integration by parts,  can be in turn transformed into the  PDFs multiplied by the derivatives of the original integrands, up to a few of boundary terms, more specifically,
\begin{align}
\int_{x_n}^{1}\, \done z_n\, 
 \frac{x_n}{z_n}f'_{i/N}\left( \frac{x_n}{z_n}\right)\,
 f_{j/\bar{N}}\left( x_\nbar \right)\,
g(z_n)\,
=&\,  
\int_{x_n}^{1}\,  \done z_n\, 
f_{i/N}\left( \frac{x_n}{z_n}\right)\,
 f_{j/\bar{N}}\left( x_\nbar \right)\,
\frac{\partial\,g(z_n)}{\partial\,\ln(z_n)}\,\nonumber\\
&\,-\,f_{i/N}\left(x_n\right)\, f_{j/\bar{N}}\left( {x_\nbar}\right)\,g_n(1)\,
+\,b.c.\,,\\
\int_{x_\nbar}^{1}\,  \done z_\nbar\, 
 f_{i/N}\left(   x_n \right)\,
 \frac{x_\nbar}{z_\nbar}f'_{j/N}\left( \frac{x_\nbar}{z_\nbar}\right)\,
g_n(z_\nbar)\,
=&\,  
\int_{x_\nbar}^{1}\,  \done z_\nbar\, 
f_{i/N}\left( x_n  \right)\,
 f_{j/\bar{N}}\left( \frac{x_\nbar}{z_\nbar}\right)\,
\frac{\partial\,g_\nbar(z_\nbar)}{\partial\,\ln(z_\nbar)}\,\nonumber\\
&\,-\,f_{i/N}\left(x_n\right)\, f_{j/\bar{N}}\left( {x_\nbar}\right)\,g_\nbar(1)\,
+\,b.c.\,,
  \end{align}
  where $g_n(z_n)$ and $g_\nbar(z_\nbar)$ represent the two regular functions with finite derivatives across the domains $z_n\in[x_n,1]$ and $z_\nbar\in[x_\nbar,1]$, respectively.

Applying those two relations onto Eqs.~\eqref{eq:Rcs:qq:Ebert18}-\eqref{eq:Rcbars:qq:Ebert18}, the off-diagonal entries therein are both eliminated, replicating our results in Eq.~\eqref{eq:Rcs:qq} and Eq.~\eqref{eq:Xcbars:nnlp}.
Analogous strategy can also be utilised in comparing the expressions of the processes $gg\to Hg$ and $q(\bar{q})g \to Hq(\bar{q})$.

 \section{Small \texorpdfstring{\qT}{qT} expansion for multi-boson hadroproduction}\label{sec:asyexp:Bs}
 
 Here we will demonstrate that, after appropriate generalisations, the method presented in this paper in Sec.~\ref{sec:asyexp:mom} can also be used in the power expansion on the hadroproduction of multiple colour-singlet bosons, i.e.\ $pp\to \{B_n\}+X$.  Here a set of electroweak (or Higgs) bosons are collected within the set $\{B_n\}\equiv\{B_1,B_2,B_3,\dots B_n\}$ $(n\ge2)$.
 
\begin{figure}[t!]
    \centering
    \includegraphics[width=.6\linewidth, height=0.15\linewidth]{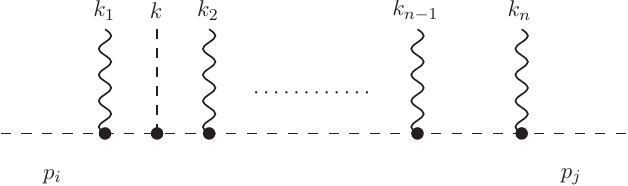}
     \caption{
   Representative diagram of the topology that the final colour-singlet bosons are all directly connected to the initial partons.
    Here the dashed lines stand for the colourful particles, while wave lines indicate the colour-singlet bosons.
   }
  \label{fig:Bn:IE}
\end{figure}

\begin{figure}[t!]
    \centering
    \includegraphics[width=.6\linewidth, height=0.15\linewidth]{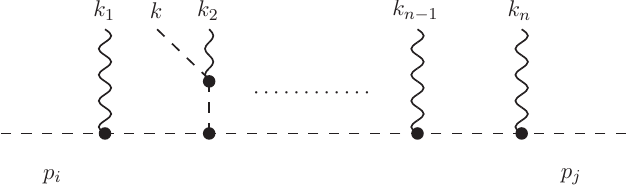}
     \caption{
      Representative diagram of the topology that the final colour-singlet bosons are in part connected to the finial QCD parton. Here the dashed lines stand for the colourful particles, while wave lines indicate the colour-singlet bosons.
  }
  \label{fig:Bn:FE}
\end{figure}

 We start with the analysis of the fixed-order results. According the QCD factorization, the differential \qT\ distributions  at NLO can be expressed as,
 \begin{align}\label{eq:qT:Bi}
  \frac{\done \sigma_B}{\done\Phi_n\, \done^2\qTvec}
  \;= 
  &
    \frac{1}{2^{4n}\,\pi^{3n-1}\, s^2}\sum_{i,j}\, 
    \int^{k_+^{\mathrm{max}}}_{k_+^{\mathrm{min}}}\,
    \frac{\done k_+}{k_+}
    \frac{f_{i/n}(\xi_n)}{\xi_n}
    \frac{f _{j/\nbar}(\xi_{\nbar})}{\xi_{\nbar}}\;
    \overline{\sum_{\mathrm{col},\mathrm{pol}} }
    \big|\mathcal{M}(i+j\to \{B_n\}+k)\big|^2\,.
\end{align}
Now $\qT$ represents the transverse momentum of the whole colourless system $\{B_n\}$. The differential $\done\Phi_n$ collects the differentials for the rapidity $y_i$ and transverse momentum $\vec{k}^i_{\bot}$ of the $i$th boson,
 \begin{align}\label{eq:dPhi:Bi}
     \done\Phi_n=\left(\prod_{i=1}^n\done y_i\right)\left(\prod_{i=1}^{n-1}\done^2\vec{k}^i_{\bot}\right)\,.
\end{align}
Here only the transverse momenta of the first $n-1$ partons have been spelled out, as the last one is subjected to the momentum conservation condition, 
 \begin{align} 
\vec{k}^n_{\bot}=\qTvec-\sum_{i=1}^{n-1}\vec{k}^i_{\bot}\,.
\end{align}
On the r.h.s of Eq.~\eqref{eq:qT:Bi}, the integrand concerns the  momentum fractions $\xi_n$ and $\xi_\nbar$, which can be determined by the momentum conservation condition on the longitudinal direction,  
 \begin{align}\label{eq:def:ppBn:XinXinbar}
 \xi_n\,=\,\frac{1}{\sqrt{s}}\,\left(k_+\,+\,\sum_{i=1}^n\mT^ie^{-y_i}\right)\,,\qquad \xi_\nbar\,=\,\frac{1}{\sqrt{s}}\,\left(k_-\,+\,\sum_{i=1}^n\mT^ie^{+y_i}\right)\,.
 \end{align}
Here  $k_{\pm}$ stands for  the light-cone components of the emitted colourful particle as defined in Eq.~\eqref{eq:def:lcone}. $\mT^i$ reveals the transverse mass for the $i$th boson.
According to the conditions $\xi_n\le1$ and $\xi_\nbar\le1$, we can then evaluate the boundaries for the $k_+$-integral in Eq.~\eqref{eq:qT:Bi},
  \begin{align}
 k_+^{\mathrm{max}}\,=\,{\sqrt{s}}\,-\,\sum_{i=1}^n\mT^ie^{-y_i} \,,\qquad  k_-^{\mathrm{max}}\,=\,  \frac{\qT^2}{k_+^{\mathrm{min}}}\,=\, {{\sqrt{s}}\,- \,\sum_{i=1}^n\mT^ie^{+y_i} }\,.
 \end{align}
 The calculation on Eq.~\eqref{eq:qT:Bi} also entails the squared amplitudes for the ensuing partonic process $i+j\to \{B_n\}+k$. 
 On the tree level, the Feynman diagrams presiding over this transition can be categorised into two groups: (A) topologies in which the final colour-singlet bosons \changed{are not directly connected to final colourful particle}; (B) topologies in which  at least one of the final colour-singlet bosons is emitted from the finial QCD parton.  The representative circumstances on those two topologies have been illustrated in Fig.~\ref{fig:Bn:IE} and Fig.~\ref{fig:Bn:FE}.
 The topology (A) is, for instance,  expected to govern the partonic channel $q\bar{q}\to \{B_n\}+k$ with $\{B_n\}$ comprising only the electroweak vector bosons.
In the other circumstances, the contributions from topology (B) can be relevant and therefore have to be addressed as appropriate. However, it merits noting that in topology (B), the emission off the QCD parton will leave all the propagators in the ensuing Feynman diagrams away from the threshold by $\mathcal{O}(m^i_{B})$, where $m^i_{B}$ indicates the mass of the $i$th final colourless boson, such that the leading singular behaviour of the process $pp\to\{B_n\}+X$ is entirely captured by the configuration (A).

 Without any loss of generality, the matrix elements induced by those two topologies can be recast in the following forms, 
   \begin{align}
   \label{eq:def:ppBn:M2:TA}
   &  \overline{\sum_{\mathrm{col},\mathrm{pol}} }
    \big|\mathcal{M}(i+j\to \{B_n\}+k)\big|^2\Bigg|_{\mathbf{(A)}}\, \nonumber\\
    =&\,  \sum_{F\in\mathbf{T}_{\mathbf{A}}}\,\lambda_{ijk}(F)\,\left[\prod_{ n=1}^{\mathbf{card}[I_{F}]}  \left(p_i-\sum_{l\in I_{F}^{(n)}} k_l\right)^{-2} \right]\,\left[\prod_{m=1}^{\mathbf{card}[J_{F}]}\   \left(p_j-\sum_{l\in J_F^{(m)}}k_l\right)^{-2}\right]\, 
  \sum_{\rho,\sigma}\,k^{\rho}_{-}\,k^{\sigma}_{+}\, \left( \widetilde{\mathcal{K}}_{\bot}{\cdot}\widetilde{\mathcal{E}}^{\rho\sigma}_{F} \right)\,,
  \\
     \label{eq:def:ppBn:M2:TB}
  &   \overline{\sum_{\mathrm{col},\mathrm{pol}} }
    \big|\mathcal{M}(i+j\to \{B_n\}+k)\big|^2\Bigg|_{\mathbf{(B)}}\, \nonumber\\
    =&\,  \sum_{F\in\mathbf{T}_{\mathbf{B}}}\,\lambda_{ijk}(F)\,\left[\prod_{ n=1}^{\mathbf{card}[I_{F}]}   \left(p_i-\sum_{l\in I_{F}^{(n)} }k_l\right)^{-2} \right]\,\left[\prod_{m=1}^{\mathbf{card}[J_{F}]}\,   \left(p_j-\sum_{l\in  J_F^{(m)} }k_l\right)^{-2}\right]\,\left[\prod_{h=1}^{\mathbf{card}[K_{F}]}    \left( k+\sum_{l\in K_F^{(h)} }k_l\right)^{-2} \right]\,\nonumber\\
&    \times\, 
  \sum_{\rho,\sigma}\,k^{\rho}_{-}\,k^{\sigma}_{+}\,  \left( \widetilde{\mathcal{K}}_{\bot}{\cdot}\widetilde{\mathcal{E}}^{\rho\sigma}_{F} \right) \,.
 \end{align}
 Here $F$ denotes the Feynman diagrams driving the process $i+j\to \{B_n\}+k$, which, at LO, belongs to either topology  $\mathbf{T}_{\mathbf{A}}$ or $\mathbf{T}_{\mathbf{B}}$. $\lambda_{ijk}(F)$ collects the coupling constants, colour factors, and average factors for the initial states and the final identical particles.  $I_F$ and $J_F$ encode two sets of colourless particles emitted from the initial partons $i$ and $j$, respectively. For instance, if the squared amplitudes are induced by the configuration in Fig.~\ref{fig:Bn:IE} and its complex conjugate, we have $I_F=\{I_F^{(1)}, I_F^{(2)}\}$ with $I_F^{(1,2)}=B_1$ and $J_F=\{J_F^{(1)}, J_F^{(2)}, J_F^{(3)}, J_F^{(4)},\dots \}$ with $J_F^{(1,2)}=B_n$, $J_F^{(3,4)}=\{B_{n-1},B_n\}$, etc. The number of the elements in each set is evaluated by the cardinality operator $\mathbf{card}$. With the help of $I_F$ and $J_F$, one can specify the expressions of the denominators in a given Feynman diagram as presented in the square brackets in Eqs.~(\ref{eq:def:ppBn:M2:TA}-\ref{eq:def:ppBn:M2:TB}).
  
  To express the numerators of the squared amplitudes, we introduce the arrays $\widetilde{\mathcal{K}}_{\bot}$ and $\widetilde{\mathcal{E}}^{\rho\sigma}_{F}$,   
   \begin{align}
   \label{eq:def:ppBn:KT}
   \widetilde{\mathcal{K}}_{\bot}=&\left[1,(\qTvec)_{i_1},(\qTvec)_{i_2}(\qTvec)_{j_2},(\qTvec)_{i_3}(\qTvec)_{j_3}(\qTvec)_{k_3},\dots\right]\,,\\
      \label{eq:def:ppBn:ET}
   \left(\widetilde{\mathcal{E}}^{\rho\sigma}_{F}\right)^{\mathbf{T}}=&\left[E^{\rho\sigma,(0)}_{F},\sum_l\, (\vec{k}_{\bot}^l)_{i_1}\,E^{\rho\sigma,(1_{\rm{S}})}_{F,l}+\sum_l\,(\tilde{k}_{\bot}^l)_{i_1}\,E^{\rho\sigma,(1_{\rm{A}})}_{F,l}, \epsilon^{i_2j_2}_{\bot}E^{\rho\sigma,(2_{\rm{AA}})}_{F}+\,\sum_{l_1,l_2}\, (\vec{k}_{\bot}^{l_1})_{i_2}\,(\vec{k}_{\bot}^{l_2})_{j_2}\,E^{\rho\sigma,(2_{\rm{SS}})}_{F,l_1,l_2}\,\right. \nonumber\\
   &
   \left. +\,\sum_{l_1,l_2}\, (\tilde{k}_{\bot}^{l_1})_{i_2}\,(\vec{k}_{\bot}^{l_2})_{j_2}\,E^{\rho\sigma,(2_{\rm{AS}})}_{F,l_1,l_2}\,+\,\sum_{l_1,l_2}\, (\vec{k}_{\bot}^{l_1})_{i_2}\,(\tilde{k}_{\bot}^{l_2})_{j_2}\,E^{\rho\sigma,(2_{\rm{SA}})}_{F,l_1,l_2}\,,\dots\right]\,.
    \end{align}
    Here $\widetilde{\mathcal{K}}_{\bot}$ consists of the tensors constructed by the momentum \qTvec\ in different ranks, while $\widetilde{\mathcal{E}}^{\rho\sigma}_{F}$ collects those formed by $k^{\mu}_{i}$ with $i\in[1,n]$. To accommodate the asymmetric behaviour induced by the Dirac traces that contain $\gamma_5$, we introduce $\epsilon^{\rho\sigma}_{\bot}=\epsilon^{\mu\nu\rho\sigma}n_{\mu}\nbar_{\nu}$ and $(\tilde{k}_{\bot})^{\rho}=\epsilon^{\rho\sigma}_{\bot}\vec{k}_{\bot,\sigma}$ in Eq.~\eqref{eq:def:ppBn:ET}, where $\epsilon^{\mu\nu\rho\sigma}$ denotes the customary anti-symmetric tensor. The coefficients $E^{\rho\sigma,(m)}_{F,\{l\}}$ in charge of tensor structures as specified in ``$(m)$" encode the hard scales $\mT^ie^{\pm y_i}$, $\left(\vec{k}^i_{\bot}{\cdot}\vec{k}^j_{\bot}\right)$, and $\left(\tilde{k}^i_{\bot}{\cdot}\vec{k}^j_{\bot}\right)$.
    
In the following, we will show that the \qT\ spectra induced by the topology (A)  can be all expanded following the method developed in Sec.~\ref{sec:asyexp:mom}. First of all, it should be noted that in Eq.~\eqref{eq:def:ppBn:M2:TA}, all the dependences on the integration variable $k_{\pm}$ have been taken out from the numerator, while the denominators, after evaluating the scalar products, present only the linear dependences on $k_{\pm}$, more explicitly,     
  \begin{align}
  \label{eq:def:ppBn:sp:kp}
   \left(p_i-\sum_{l\in  X}k_l\right)^{2}\equiv& \left(p_i-k_X\right)^{2}= -\mT^Xe^{y_X}\left(k_++\sum_{l\notin  X}\mT^l e^{-y_l}+\frac{|\vec{k}^X_{\bot}|^2}{\mT^Xe^{y_X}}\right)\,,\\
     \label{eq:def:ppBn:sp:km}
      \left(p_j-\sum_{l\in  X}k_l\right)^{2}\equiv& \left(p_j-k_X\right)^{2}= -\mT^Xe^{-y_X}\left(k_-+\sum_{l\notin  X}\mT^l e^{y_l}+\frac{|\vec{k}^X_{\bot}|^2}{\mT^Xe^{-y_X}}\right)\,.
 \end{align}
Therein, $k_X$ represents the total momentum  of the particles belong to the set $X$, from which we are able to calculate the transverse mass $\mT^X$ and rapidity $y_X$ for this colourless system.

The factorised form in Eqs.~(\ref{eq:def:ppBn:sp:kp}-\ref{eq:def:ppBn:sp:km}) then allows us to recategorise the dominators in Eq.~\eqref{eq:def:ppBn:M2:TA} according to their $k_{\pm}$-dependences, more explicitly,
\begin{equation} 
\label{eq:ppBn:qcd:qT:kpm}
  \begin{split}
      \frac{\done \sigma_B}{\done\Phi_n\, \done^2\qTvec}\Bigg|_{\mathbf{(A)}}
    \,=&\;
     \sum_{F\in\mathbf{T}_{\mathbf{A}}}\,    
      \sum_{\rho,\sigma } \,
 \left( \widetilde{\mathcal{K}}_{\bot}{\cdot}\widetilde{\mathcal{H}}^{\rho\sigma}_{F} \right) \,
  \int^{k_+^{\mathrm{max}}}_{k_+^{\mathrm{min}}}
     \frac{\done k_+}{k_+}\;
      (k_-)^{\rho} \, (k_+)^{\sigma} \,
      \widehat{F}^{\{0\}}_{i/n,\{\beta_n\}}\left(k_+,\{Q_n\},\{I_F\}\right)\,
    \\
    &\times\;     
           \widehat {F}^{\{0\}}_{j/\nbar,\{\beta_\nbar\}}\left(k_-,\{Q_\nbar\},\{J_F\}\right)\,,
  \end{split}
\end{equation}
where the novel array $\widetilde{\mathcal{H}}^{\rho\sigma}_{F}$ emerges, which is defined analogously to Eq.~\eqref{eq:def:ppBn:ET} but can absorb extra hard scales as a result of the denominator decomposition in Eqs.~(\ref{eq:def:ppBn:sp:kp}-\ref{eq:def:ppBn:sp:km}).  Within the  integral, all the $k_{\pm}$ reliances from the PDFs and denominators of the squared amplitudes have been collected in the generalised functions $\widehat{F}^{\{\alpha_n\}}_{i/n,\{\beta_n\}}$ and $ \widehat {F}^{\{\alpha_\nbar\}}_{j/\nbar,\{\beta_\nbar\}}$,
\begin{align} \label{eq:def:ppBn:FioN}
 & \widehat{F}^{\{\alpha_n\}}_{i/n,\{\beta_n\}}\left(k_+,\{Q_n\},\{I_F\}\right)\nonumber\\
  \equiv \,&
\frac{\partial^{\alpha^{(-1)}_n}}{\partial (k_+)^{\alpha^{(-1)}_n}}
\left\{   \prod_{l=0}^{\mathbf{card}[I_F]} \frac{\partial^{\alpha^{(l)}_n}}{\partial (Q^{(l)}_n)^{\alpha^{(l)}_n}}\right\}
    \left[
      \frac{ f_{i/n}\left(\frac{k_++Q^{(0)}_n}{\sqrt{s}}\right)}{k_++Q^{(0)}_n}   \right]\prod_{h=1}^{\mathbf{card}[I_F]}\left(k_++Q^{(h)}_n\right)^{-\beta^{(h)}_n}\,,\\
  & \widehat {F}^{\{\alpha_\nbar\}}_{j/\nbar,\{\beta_\nbar\}}\left(k_-,\{Q_\nbar\},\{J_F\}\right)\nonumber\\
  \label{eq:def:ppBn:FioNbar}
  \equiv \,&
\frac{\partial^{\alpha^{(-1)}_\nbar}}{\partial (k_-)^{\alpha^{(-1)}_\nbar}}
\left\{   \prod_{l=0}^{\mathbf{card}[J_F]} \frac{\partial^{\alpha^{(l)}_\nbar}}{\partial (Q^{(l)}_\nbar)^{\alpha^{(l)}_\nbar}}\right\}
    \left[
      \frac{ f_{j/\nbar}\left(\frac{k_-+Q^{(0)}_\nbar}{\sqrt{s}}\right)}{k_-+Q^{(0)}_\nbar}   \right]\prod_{h=1}^{\mathbf{card}[J_F]}\left(k_-+Q^{(h)}_\nbar\right)^{-\beta^{(h)}_\nbar}\,,
\end{align}
where $\{Q_n,Q_\nbar\}$ signify two sets of hard scales. $Q_n^{(0)}$ and $Q_\nbar^{(0)}$ can be extracted from the fractions in Eq.~\eqref{eq:def:ppBn:XinXinbar}, while $Q_n^{(l)}$ and $Q_\nbar^{(l)}$ with $l\ge1$ are derived by matching onto Eqs.~(\ref{eq:def:ppBn:sp:kp}-\ref{eq:def:ppBn:sp:km}). In comparison with the functions ${F}^{(\alpha_n)}_{i/n,\beta_n}$ and ${F}^{(\alpha_\nbar)}_{j/\nbar,\beta_\nbar}$ in Eq.~\eqref{eq:def:FioN:FjoNbar}, Eqs.~(\ref{eq:def:ppBn:FioN}-\ref{eq:def:ppBn:FioNbar}) here observe structural similarity but call for more indices to accommodate  the input scales, such as $\alpha^{(h)}_{n/\nbar}$ presiding over the orders of the derivatives of the kinematics variables and $\beta^{(h)}_{n/\nbar}$ controlling the exponents of the various denominators  $(k_{\pm}+Q^{(h)}_{n/\nbar})$. 
Equipped with those results, in Eq.~\eqref{eq:ppBn:qcd:qT:kpm}, we manage to transform the \qT\ distribution into its $k_{+}/k_-$-factorised formulation as in Eq.~\eqref{eq:qcd:qT:kpm}, for which the power expansion can be carried out by repeating the procedures in Sec.~\ref{sec:setups}, Sec.~\ref{sec:red:qT:boundary}, and Sec.~\ref{sec:red:qT:central}.

However, if the contribution related to $\mathbf{T}_{\mathbf{B}}$ is of one's concern, the small \qT\ expansion becomes more involved.  To see this, it is worth recalling that extra type of propagators appear from the third square bracket of Eq.~\eqref{eq:def:ppBn:M2:TB}, which  prompts the quadratic dependences on $k_{\pm}$ upon evaluating the scalar products, 
\begin{align}\label{eq:def:ppBn:sp:kX}
\left( k+\sum_{l\in  X}k_l\right)^{2}\equiv& \left(k+k_X\right)^{2}\,=\,k_+\mT^Xe^{y_X}+k_-\mT^Xe^{-y_X}+m_X^2-2\qTvec\cdot\vec{k}_{\bot}^X\,. 
\end{align}
At this moment,  it is not straightforward to derive the factorised form for the squared amplitudes of Eq.~\eqref{eq:def:ppBn:M2:TB} as in Eq.~\eqref{eq:ppBn:qcd:qT:kpm}, which therefore defies an immediate implementation of the frameworks in Sec.~\ref{sec:asyexp:mom}.
Promisingly, the power series of the $\mathbf{T}_{\mathbf{B}}$-induced contributions can also be derived via a set of dynamic regions from the phase space. 
However,  more considerations and efforts should be paid to work out a group of auxiliary cutoffs that are capable of unambiguously and consistently separating all the relevant scales embedded by Eq.~\eqref{eq:def:ppBn:sp:kX}.  Hence, we postpone this part of discussion to the further investigations.

\bibliographystyle{amsunsrt_mod}
\bibliography{refs}
\end{document}